\documentclass[%
 reprint,
superscriptaddress,
 amsmath,amssymb,
 aps,
pra,
]{revtex4-2}

\usepackage{graphicx}
\usepackage{dcolumn}
\usepackage{bm}
\usepackage[T1]{fontenc}
\usepackage{savesym}
\usepackage{amsmath}
\savesymbol{iint}
\usepackage{txfonts}
\restoresymbol{TXF}{iint}
\usepackage{amssymb}
\usepackage{verbatim}
\usepackage{mathrsfs}
\usepackage{amsfonts}
\usepackage{epsfig}
\usepackage{color}
\usepackage{soul}
\usepackage{hyperref}
\usepackage{subfigure}
\usepackage{longtable}
\usepackage{bigstrut}
\usepackage{makecell}
\usepackage{colortbl}
\usepackage{units}
\usepackage{natbib}
\usepackage{ctable}
\usepackage[version=3]{mhchem}
\DeclareRobustCommand{\VAN}[3]{#2}
\let\VANthebibliography\thebibliography
\def\thebibliography{\DeclareRobustCommand{\VAN}[3]{##3}\VANthebibliography}

\definecolor{orange}{rgb}{0.9,0.45,0}

\def\part_n{\partial_\perp}

\long\def\symbolfootnote[#1]#2{\begingroup%
\def\thefootnote{\fnsymbol{footnote}}\footnote[#1]{#2}\endgroup}
\begin{document}

\preprint{APS/123-QED}

\title[Accretion Efficiency Evolution of Central Supermassive Black Holes in Quasars]{Accretion Efficiency Evolution of Central Supermassive Black Holes in Quasars}

\author{Arta Khosravi}
\email{artakh10@gmail.com}
 \affiliation{PDAT Laboratory, Department of Physics, K. N. Toosi University of Technology, Tehran, P.O. Box 15875-4416, Iran.}
\author{Alireza Karamzadeh}
 \email{alirezakaramzadehkntu@gmail.com}
  \affiliation{PDAT Laboratory, Department of Physics, K. N. Toosi University of Technology, Tehran, P.O. Box 15875-4416, Iran.}
\affiliation{Department of Physics, Shahid Beheshti University, Tehran, P.O. Box 19839-69411, Iran.}
\author{Seyed Sajad Tabasi}
 \email{seyed.tabasi@monash.edu}
\affiliation{School of Physics and Astronomy, Monash University, Clayton, VIC 3800, Australia.}
\author{Javad T. Firouzjaee}
 \email{firouzjaee@kntu.ac.ir}
  \affiliation{PDAT Laboratory, Department of Physics, K. N. Toosi University of Technology, Tehran, P.O. Box 15875-4416, Iran.}
\affiliation{Department of Physics, K.N. Toosi University of Technology, Tehran, P.O. Box 15875-4416, Iran.}
\affiliation{School of Physics, Institute for Research in Fundamental Sciences (IPM), P.O. Box 19395-5531, Tehran, Iran}

\begin{abstract}
The ongoing debate regarding the most accurate accretion model for supermassive black holes at the center of quasars has remained a contentious issue in astrophysics. One significant challenge is the variation in calculated accretion efficiency, with values exceeding the standard range of $0.038 < \epsilon < 0.42$. 
This discrepancy is especially pronounced in high redshift supermassive black holes, necessitating the development of a comprehensive model that can address the accretion efficiency for supermassive black holes in both the low and high redshift ranges.
The selection effect was removed from model construction by creating a flux- and volume-limited sample, as the range of values for estimating the accretion efficiency factor varied through different redshifts.
In this study, we have focused on low redshift ($z < 0.5$) Palomar-Green quasars (79 quasars) and high redshift ($z \geq 3$) quasars with standard disks from the flux- and volume-limited QUOTAS+QuasarNET dataset (75 quasars) to establish a model for accretion efficiency.
By considering the QUOTAS+QuasarNET+DL11 dataset, a peak can be seen around $z \sim 2.708$, and it seems to be related to the peak of the star formation rate ($1 < z_{SFR} < 3$).
Consequently, the observed maximum and minimum values of accretion efficiency in standard disks, through the considered bond (3$\sigma$), display a significantly wider range than previously noted and differentiate over time. In redshifts higher than 2.708, the accretion efficiency shows patterns of increase as redshift decreases, while in redshifts lower than 2.708, accretion efficiency is seen to decrease with reducing redshift. 
This result can potentially lead to a more accurate correlation between the star formation rate in quasars and their relationship with the mass of the central supermassive black holes with a more comprehensive disk model in future studies.

\end{abstract}

\maketitle

\section{Introduction}
Within quasars, gas is drawn towards a central Supermassive Black Hole (SMBH) and generates electromagnetic radiation. The gas emits intense light as a result of gravitational and frictional interactions. This characteristic renders quasars among the most radiant entities in the cosmos, shining nearly a thousand times more brightly than the entire Milky Way \cite{lynden1969galactic,rees1984black,wang2007early,treister2010major,melia2019cosmological}.

SMBHs are massive objects with masses ranging from below $10^6$ to above $10^{10}$ solar mass. They are usually located at the center of quasars, which can be observed through various methods. One way is to measure the amount of energy and the duration of variability in a specific region of space, specifically at the center of quasars \cite{bromley2004high,kozlowski2009quantifying,cherepashchuk2016observing}. 
Another way is to observe objects that orbit around an invisible mass at high speeds \cite{milgrom1983modification,ghez2005stellar,li2023tracing}.

As previous discussions on SMBHs and their hosts, it is necessary to provide information about their mass relations. In this context, clarification is needed on certain parameters to gain a better understanding of SMBHs and the different components of their host galaxies. The host galaxy contains various elements, including a bulge, a disk, and other objects. It is evident that the total mass of the system, which includes the galaxy and everything within it, is equal to the sum of the bulge mass and the disk mass.
It can be shown that there are relations between the mentioned masses, which some works, such as Ref. \cite{davis2018black,davis2019black}, have derived accordingly.

\raggedbottom
High redshift quasars enable exploration of the early universe, less than 1 Gyr after the Big Bang. The early evolution of SMBHs can only be studied with the help of the early-forming quasars. 
The mass of supermassive black holes in quasars can be determined by observing the reverberation between the fluctuations in wide emission lines and the continuum \cite{osmer1982quasars,fan1999high,mclure2002measuring}. 
The X-ray emission follows a power-law spectral pattern resulting from inverse Compton scattering of photons from the accretion disk of relativistic electrons in the hot corona and a soft X-ray excess. The optical-to-ultraviolet continuum emission can be explained by a standard thin accretion disk extending down to the Innermost Stable Circular Orbit (ISCO) \cite{shakura1973black,ding2022accretion}.

From optical to near-infrared spectroscopy, measurements of quasars imply that SMBHs were already in place when the universe was merely 700 million years old \cite{kaspi2000reverberation,fan2003survey,shen2019gemini}. 
Numerous theoretical ideas, including the super-Eddington accretion process \cite{kawaguchi2004growth} and the use of primordial density seeds \cite{kawasaki2012primordial}, have been put forth to explain the existence of SMBHs.

Many works, such as Ref. \cite{pohlen2006structure,percival2007measuring}, use the Sloan Digital Sky Survey (SDSS) to utilize the data about quasars and SMBHs from the sky. SDSS has produced a colorful 3D map of the universe, providing data on the mass, luminosity, redshift, and other astrophysical parameters of various objects. This instrument makes high-resolution pictures with five different colors of a quarter of the sky using a specially built 2.5-meter telescope with a CCD camera located in New Mexico. After that, people measure the shape, brightness, and color of objects using advanced image processing software and then list the properties of the observed stars and galaxies \cite{lyke2020sloan}.
SDSS includes around 0.5 million quasars that have been found up to a redshift of approximately 7; consisting of a uniform sample of quasars having optical spectra. A distinct scientific objective drove every round of the SDSS quasar surveys. The SDSS DR7 quasar catalog contains 105,783 quasars confirmed through spectroscopic analysis. These quasars were identified during the SDSS-I/II survey, which focused on studying the Quasar Luminosity Function (QLF) and clustering features \cite{abazajian2009seventh}.
Each time new data is released, the quantity of identified quasars has significantly increased. For instance, SDSS DR14 has significantly expanded the number of quasars by a factor of 5 compared to the previous release, SDSS DR7. Additionally, it has become 1.5 magnitudes fainter, allowing for the investigation of quasar properties throughout a considerably more comprehensive range of luminosities \cite{richards2006sloan,abazajian2009seventh,abolfathi2018fourteenth}.
The SDSS-IV quasar catalog from Data Release 16 (DR16Q) of the Extended Baryon Oscillation Spectroscopic Survey (eBOSS) is one of the largest quasar catalogs. DR16Q catalog contains two sub-catalogs: a superset that includes all SDSS-IV/eBOSS objects containing 1,440,615 observations and a quasar-only including 750,414 quasars, of which 225,082 are new and appear for the first time. The quasar-only catalog is 99.8\% complete with 0.3\%-1.3\% contamination \cite{lyke2020sloan}. 

The other dataset that many other works, including Ref. \cite{beutler2017clustering,zhao2019clustering,alam2021completed,xu2023evidence}, have used is the Baryon Oscillation Spectroscopic Survey (BOSS) dataset.
BOSS aims to explore a large volume required for precise measurements of the Baryon Acoustic Oscillations (BAO) scale at the percent level. Achieving this level of precision provides important constraints on the source of cosmic acceleration \cite{eisenstein2005detection,schlegel2009baryon,aubourg2015cosmological}.
BOSS also has measured the redshifts of 1.5 million luminous galaxies and 160,000 high redshift quasars \cite{wang2017measurement}. 

In this paper, we have outlined the basic equations governing the physics of accretion and the factors influencing them in Sec. \ref{phys_acc}. We have delved into the importance of accretion disks and their types in Sec. \ref{acc_disks}. The primary datasets on which this work is focused have been introduced in Sec. \ref{QUOTAS}. 
The process of providing flux- and volume-limited samples has been explained in detail in Sec. \ref{fluxlim}.
Subsequently, we have analyzed the relationship between the accretion efficiency of Black Holes (BHs) and its correlation with their mass using both high and low redshift flux- and volume-limited datasets.
Ultimately, Sec. \ref{result} focuses on examining the finalized plots concerning accretion efficiency regarding BH mass and redshift, next to explaining the downsizing issue and its implications for the data.

\section{The Physics Behind Accretion}\label{phys_acc}

Bondi-Hoyle-Lyttleton (BHL) accretion describes the supersonic movement of a point mass through a gas cloud \cite{bondi1944mechanism}. Such a process has been extensively studied by Ref. \cite{bondi1952spherically} in the context of spherically symmetric accretion onto a point mass. Beyond the Bondi radius, the flow is subsonic, the density almost constant, and the gas there reaches supersonic speed and eventually settles into a freefall state. As a result of gravity, a portion of the gas may be drawn towards the material concentrated behind the point mass, a process known as BHL accretion \cite{edgar2004review}. 

The BHL model has been used in various studies, such as Ref. \cite{gregoris2023black} specifically focusing on the evolution of the accreted fluid \cite{xu2019bondi}, and Ref. \cite{tejeda2019relativistic} concentrating on the mass-radius relationship of the photon sphere beyond the Schwarzschild, unaffected by the accretion process.
Furthermore, the BHL model can also be used to study the accretion of Interstellar Medium (ISM) gas onto astrophysical objects, with the relativistic counterpart being the ability of a BH to accrete fluid \cite{cruz2023bondi}.

In addition, the work of Ref. \cite{font1999non} has been on the relativistic example of an ideal gas accreting onto a BH \cite{li2023accretion}. Their study has focused on two-dimensional cases, specifically axisymmetric accretion into a Schwarzschild BH \cite{lora2013axisymmetric}, and the thin-disc approximation of accretion onto a Kerr BH, due to computational limitations \cite{lora2015relativistic}. In the Newtonian model, supersonic flow resulted in the formation of a shock cone, which sometimes clung to the BH horizon. However, the subsonic flow has led to a smoother flow. 
Ref. \cite{font1999non} has also found that at far distances from the horizon, the accretion pattern of a Kerr BH was similar to that of a non-spinning BH. However, as one has approached the horizon, the shock cone, if present, has encircled the BH in a spin-dependent manner, with this effect diminishing further away from the horizon \cite{blakely2015relativistic}.

However, it has been noticeable in many works, such as Ref. \cite{ruffert1996non,fukue1999hoyle,fukue2001bondi,scicluna2014old,comerford2019bondi}, that how due to the angular momentum of infalling accreted materials and mass accretion, the eventual formation of accretion disks is only anticipated.
Hence, as a result of the expectation of the accretion disk forming through angular momentum acquisition, the continued calculations will consider the accretion disk model and the parameters in relation.
Thus, to calculate the accretion efficiency of quasars, one needs to be familiar enough with a few astrophysical parameters.
Therefore, in this work, it is critical to grasp the methods for computing values like bolometric luminosity, Eddington luminosity, Eddington ratio, and accretion efficiency. The correlation between accretion efficiency and astrophysical factors, such as SMBH mass and redshift, is essential. Hence, employing various methodologies to calculate accretion efficiency becomes essential, enabling the examination of how factors like bolometric luminosity, central SMBH mass, or even redshift individually impact the results.

Bolometric luminosity is the total power output across all electromagnetic radiation wavelengths of an astronomical body. 
The way to measure the bolometric luminosity can be obtained by \cite{torres2010use}
\begin{equation}
    L_{bol} = L_{\odot} \Big(\frac{4.74-2.5 log\big(\frac{L}{L_{\odot}}\big)}{m_{\odot}}\Big)^4 ,
\end{equation}
where $m_{\odot}$ is the solar magnitude and $L_{\odot}$ is the solar luminosity.

The Eddington luminosity, also referred to as the Eddington limit, is the maximum luminosity that an object can reach when there is a balance between the force of radiation acting outward and the gravitational force acting inward, called the hydrostatic equilibrium. 
Since most massive stars have luminosities far below the Eddington luminosity, their winds are driven mainly by the less intense line absorption \cite{wielgus2015stable}. 
The Eddington luminosity can be measured from \cite{rybicki1991radiative}
\begin{equation}\label{ledd}
    L_{edd} \simeq 1.28 \times 10^{38} (\frac{M_{BH}}{M_{\odot}}) ,
\end{equation}
where $M_\odot$ $\simeq 2.0 \times 10^{30}$ Kg.

Two critical factors define the accretion mechanism, which is a crucial component. The relationship between the bolometric luminosity and the BH growth rate may be established by looking at the radiative accretion efficiency, represented by $\epsilon$ \cite{zhang2017mean}. 
It should be mentioned that the Active Galactic Nuclei (AGN) bolometric luminosity is connected to the Eddington luminosity by the Eddington ratio \cite{eddington1988internal}
\begin{equation}\label{lambda_edd}
    \lambda_{edd} = \frac{ L_{bol}}{ L_{edd}}.
\end{equation}
In addition to all of these introduced parameters, the mass of the target BH is also very important for this work.

It is important to specify whether bolometric luminosity plays a crucial role in determining the accretion efficiency, as demonstrated by \cite{bardeen1972rotating,novikov1973astrophysics}
\begin{equation}
    \epsilon = \frac{L_{bol} c^2}{\dot{M}},
\end{equation}
where $\dot M$ is the accretion rate.

The Eddington accretion rate, $\dot M_ { edd } $, is determined by comparing the disk luminosity with the Eddington limit and calculating a mass flow rate \cite{abramowicz1988slim,heinzeller2007eddington}
\begin{equation}\label{Medd}
\dot M_{edd} = \frac{L_{edd}}{\epsilon c^2},
\end{equation}
where $c$ is the speed of light.

According to Ref. \cite{piotrovich2023estimate}, accretion efficiency is calculated for the Narrow Line Seyfert 1 (NLS1) galaxies data \cite{robson1996active,netzer2015revisiting}. 
It has been mentioned that the main model has been used based on Ref. \cite{du2014supermassive}, where the defined accretion efficiency is
\begin{equation}\label{epsilon}
\epsilon(a) = 0.105 \bigg(\frac{L_{bol}}{10^{46} erg/s}\bigg) \bigg(\frac{L_{5100}}{10^{45} erg/s}\bigg)^{-1.5} M_8 \mu^{1.5},
\end{equation}
where $L_{5100}$ is the luminosity of a BH in the wavelength of 5100 $\AA$, $M_8 = M_{BH}/(10^8 M_\odot)$, where $M_{BH}$ is the BH mass, and $\mu = \cos{i} = 0.7$, where the angle between the Line of Sight (LoS) and the axis of the accretion disk is denoted by $i$. Assuming an average angle is the standard practice due to the lack of information on the angles of most objects and the lack of evidence for a preferred direction in the orientation of galaxies \cite{czerny2011constraints}. 
Moreover, the radiative efficiency and the observable physical properties of AGNs are linked in many models, such as Ref. \cite{davis2011radiative,raimundo2012can,trakhtenbrot2014most,lawther2017catalogue}. 

In Ref. \cite{raimundo2012can}, a trend of $\epsilon \propto M_{BH}^{0.5}$ using a simulated sample corresponds to the distribution of the Palomar-Green (PG) quasars in the Ref. \cite{davis2011radiative} dataset (DL11) along an area in the accretion efficiency versus the SMBH mass. The DL11 dataset is set in the low redshift quasars and uses the PG quasar sample of Ref. \cite{schmidt1983quasar}. For the DL11 sample, the distribution of the optical luminosity and SMBH mass with respect to redshift is investigated

\begin{equation}\label{epsilon_opt}
\epsilon = 0.063 \Bigg(\frac{L_{bol}}{10^{46} erg/s}\bigg)^{0.99} \bigg(\frac{L_{opt}}{10^{45} erg/s}\bigg)^{-1.5} M_8^{0.89},
\end{equation}
where $L_{opt}$ is the optical luminosity in erg $s^{-1}$.

It is worth mentioning that the accretion efficiency can approximately be calculated with BH mass based on the data of PG quasars \cite{davis2011radiative}
\begin{equation}
    \epsilon = 1.08 \times 10^{-21} M_{BH}^{0.52}.
\end{equation}

All in all, different types of accretion disks directly impact the calculations of accretion efficiency. Therefore, it is necessary to consider the selected accretion disk models. Thus, in the next Section, the various types of accretion disks and their differences will be discussed in detail.

\section{Types of Accretion Disks}\label{acc_disks}

BHs are believed to be the sources of ultraviolet and soft X-ray emissions detected from AGNs due to the accretion disks surrounding them \cite{plotkin2016x}. The "big blue bumps" in the ultraviolet spectrum provide observational evidence for this \cite{shields1978thermal,shang2005quasars}. However, this model has faced challenges when confronted with data from multiple wavelengths \cite{liu2008tests}.

Empirical and theoretical analyses have shed light on the current state of this issue. As Ref. \cite{inoue2010lyman} has explained, the absence of observable Lyman discontinuity is a significant obstacle for theoretical models, which has been examined by several works including Ref. \cite{davidson1976some,shukla2016effects,sellwood2022spirals}. 

The two primary factors influencing the accretion process are the angular momentum of the material falling into the BH and the accretion rate. High angular momentum flow results in an accretion disk, and the accretion rate is expressed as a dimensionless value, which is the ratio of the accretion rate to the Eddington accretion rate \cite{chakrabarti1996accretion,bu2019real}.

The accretion efficiency of a thin disk is determined by the BH spin, and some energy is trapped in the disk, reducing the efficiency of the accretion flow \cite{bu2023black}. The value of the angular momentum at the outer disk, its connection to the local Keplerian angular momentum, as well as the Eddington ratio of the flow, is used to classify commonly employed accretion disk models \cite{he2022thin}.

Within the framework of three accretion disk models, the relationship of accretion efficiency is explored: Radiation Inefficient Accretion Flow (RIAF), slim disk, and standard disk \cite{czerny2011constraints}. 

\subsection{RIAF Accretion Disks}
At lower specific accretion rates ($ L_ { int } /L_ { Edd } \lesssim 10^ { -2 } $, where $ L_ { int } $ represents intrinsic accretion luminosity; or $ \dot m \lesssim 10^ { -2 } $), some AGNs are revealed without a Broad Line Region (BLR), while others lack broad lines and exhibit narrow line features \cite{trump2011accretion}. In the inner radius of the accretion disk, a growing RIAF is responsible for the disappearance of the broad emission lines. 
The presence of an RIAF simplifies the production of a radio outflow, causing AGNs with 
this accretion model
(narrow-line and line-less) to exhibit radio-to-optical/UV emission ratios approximately ten times larger than those with higher Eddington ratios (broad line) \cite {kollmeier2006black,trump2009observational,gaskell2013line}. 

In AGNs with low Eddington ratios, the IR torus signature often weakens or disappears, while additional mid-IR synchrotron emission associated with the RIAF may be present \cite {neri2011narrow}.
Ref. \cite{begelman1984theory,narayan1995explaining,yuan2004nature,narayan2008advection} all have anticipated the presence of RIAFs in AGNs with a $L_{int}/L_{Edd} \lesssim 0.01$. 
 
RIAFs are a kind of rotating accretion flow characterized by minimal radiative losses 
\cite{ichimaru1977bimodal,rees1982ion,narayan1994advection}. The heat energy is instead stored as gravitational potential energy released by turbulent strains in the accretion flow. The accreting gas is thus, very hot, with characteristic thermal energy similar to its gravitational potential energy; this implies $T \sim G M_{BH} m_p/3k_BR \sim 10^{12} K$ at the vicinity of the BH, where $T$ represents the temperature of the gas, $m_p$ represents the proton mass, $k_B$ is the Boltzmann constant, and $R$ corresponds to the distance from the BH \cite{quataert2003radiatively}.

\subsection{Slim Accretion Disks}
Slim disk model characterizes high Eddington ratio sources as having an optically dense, geometrically not very thin, quasi-Keplerian accretion flow onto a BH, as put out by Ref. \cite{katz1977x,begelman1978black,abramowicz1988slim}. Contrary to the traditional standard accretion disk model, in such a flow, a significant fraction of the energy wasted in the disk interior is conveyed radially with the flow rather than being re-emitted at the same radius \cite{lipunova2018standard}. 
Observations confirm the existence of optically thick disks in these sources. However, much discussion remains over the stability of these models and the required adjustments. 

In Ref. \cite{abramowicz1988slim}, the slim disk relies on the so-called viscosity assumption, whereas a completely self-consistent model should predict the viscous torque \cite{liu2022accretion}. Despite significant progress in this regard, the 3-dimensional numerical magneto-hydrodynamical models still lack key elements for realistically depicting the flow onto the BH. Semi-analytical models are now a highly effective tool for comparing the models to the observational data. The link between the limit cycle behavior expected by the slim disk model and the occurrence of heartbeat states in specific astronomical sources is a central concern \cite{yuan2014hot}.

Finally, as the accretion rate increases beyond the Eddington accretion rate in equation (\ref{Medd}), the assumptions of the standard model break down ($\dot m \gtrsim 1$) \cite{heinzeller2007eddington}. Since the BHs horizon allows energy to escape through inflow, local dissipation no longer controls the local emission. The accretion efficiency continues to lag behind that of a standard disk as the accretion rate rises \cite{bian2003accretion}. 
These models may account for the observed properties of Super-Eddington quasars, NLS1 galaxies \cite{zhou2017vizier}, and some phases of gamma-ray bursts. This is possible with some modest adjustments to the physics to allow for neutrino cooling and nuclear processes \cite{czerny2019slim}.

\subsection{Standard Accretion Disks}
The standard accretion disk serves as the fundamental framework for a radiatively efficient, geometrically thin, and optically thick disk that emits multi-color black-body radiation with effective temperatures between $10^5K$ and $10^7K$, as proposed by Ref. \cite{shakura1973black}. 
Besides Ref. \cite{shakura1973black,novikov1973astrophysics,lynden1974evolution} all contributed to the standard accretion disk model (for an accretion rate of $10^{-2}$ $\lesssim$ $\dot m$ $\lesssim 1$). 

In the standard representation, the accretion disk mainly emits thermal radiation within the optical-UV wavelength range for AGNs \cite{liu2022accretion}. AGNs with the mentioned accretion rate are foreseen to have accretion disk thermal characteristic time-frames spanning from a few months to a few years \cite{burke2021characteristic}. The central radiations from narrow accretion disks have been the subject of several works, including Ref. \cite{hanawa1989x,ebisawa1991application,li2005multitemperature,zimmerman2005multitemperature,pereyra2006characteristic}. 

While several investigations use standard disks as a basis \cite[e.g.][]{calvet1999evolution,caroline2007theory,lorenzin2009comparative,armijo2012accretion}, only a limited number of studies are dedicated to evaluating the validity of conventional accretion disk models \cite[e.g.][]{beloborodov2001accretion,khesali2013local}.

In BH X-ray Binaries (BHXRBs) and AGNs, it is generally agreed that gas accretion onto BHs is the primary source of radiation power \cite{smith2021timing}. Viscosity causes the caught gas to spiral inward toward the center of the BH, where it loses part of its angular momentum and converts part of its gravitational potential energy into heat. Depending on the specifics of the radiation mechanisms involved, a spectrum can be generated from either some or all of the viscous heat \cite{liu2022accretion}.
Four fundamental answers describe the accretion flows, which are the standard accretion disk \cite{he2022thin}, the optically thin two-temperature disk \cite{shapiro1976two}, the slim disk \cite{katz1977x,begelman1984theory,abramowicz1988slim}, and the advection dominated accretion flow \cite{ichimaru1977bimodal,rees1982ion}. A combination of these approaches is often used to solve problems with BHs.
In this work, only standard accretion disks have been utilized for final results. Similar to works done by Ref. \cite{bian2003accretion,kokubo2018constraints,yoneyama2023x}, our model is based on the standard accretion disk model as well.
Since the standard disk model can form a comprehensive theory of the accretion flow, it can be easily applied to observations \cite{montesinos2012accretion}.
For example, it completely agrees with the emission properties of dwarf-novae, X-ray binaries \cite{cannizzo1993accretion}, and the UV-soft X-ray emission of AGNs \cite{sun1986new}.
In the next Section, we introduce the main dataset used in this work.

\section{Dataset}\label{QUOTAS}

QUOTAS dataset is a revolutionary dataset allowing data-driven analysis of the SMBH population.
The importance of the QUOTAS dataset is emphasized when comparing correlations between observed SDSS quasars and their hosts to models like Ref. \cite{schneider2010sloan}; it is demonstrated that with the use of Machine Learning (ML), a template can be provided to have a more acceptable match of expanded simulated samples of quasars with the observational survey volumes \cite{natarajan2023quotas}. Another reason for the significance of the QUOTAS dataset is that with aggregating and co-locating the high redshift ($z\geq$ 3) quasars population with simulated data spanning at the same cosmic epochs, there is a possibility to examine and study SMBHs. 

The data from SDSS, which is fully standardized and homogenized, sets the current standard for the rest of the observational data collected for the QUOTAS dataset. As mentioned before, the data collected from the SDSS have spectroscopically confirmed quasars, mainly gathered from SDSS DR7 \cite{schneider2010sloan,mcconnell2013revisiting}.
Additionally, To create the QUOTAS dataset, the NASA-IPAC Extra-galactic Database (NED; Ref. \cite{cook2023completeness}) has been used to gather information. However, some data values are missing from the NED repository due to inaccurate photometric redshifts. 
In this study, the aim is to investigate the accretion efficiency of the central SMBHs, as introduced by the QUOTAS dataset, using the data provided by Ref. \cite{piotrovich2023probing}.

The observed sources in the QUOTAS dataset can be divided into two primary classes: Type I and Type II AGNs \cite{natarajan2023quotas}. 
They are classified based on the width of emission lines in their spectra, which tends to correlate with the viewing angle to the central BH. Type I AGNs have broad emission lines (1000–20,000 km/s) produced by clouds orbiting closely in Keplerian orbits within the BLR and the accretion disk \cite{Schmidt1963wkp}. Type II AGNs have narrow spectral emission lines (300–1000 km/s) originating from more distant, quiescent clouds in the Narrow Line Region (NLR). When viewed face-on, the BLR can be detected, but when viewed edge-on, the obscuring dusty torus blocks the direct view into the accretion disk, allowing only a view of the toroidal region \cite{Antonucci1985aa}. Type I AGNs are more luminous and commonly detected at high redshifts, while Type II AGNs, though less luminous, are expected to exist but may be heavily obscured and undetected.

The QUOTAS database includes various data from multiple surveys, encompassing a broad redshift range of $3.01 < z < 7.07$.
The data include 23,301 quasars from the BOSS DR9 survey ($2.2 < z < 3.5$) \cite{ross2013sdss}, 1,785 quasars from the SDSS DR7 survey ($3.7 < z < 4.7$) \cite{schneider2010sloan}, 103 quasars from the SDSS DR7 survey ($4.7 < z < 5.5$) \cite{mcgreer2013z}, and multiple works of various surveys all containing under 100 quasars, including NDWFS \cite{glikman2011faint}, DLS \cite{glikman2011faint}, SDSS+WISE \cite{yang2016survey}, SDSS DR7 Stripe 82 \cite{mcgreer2013z}, SDSS Main \cite{jiang2016final}, SDSS Main Overlap \cite{jiang2016final}, SDSS Main Stripe 82 \cite{jiang2016final}, CFHQS Deep \cite{willott2010canada}, CFHQS Very Wide \cite{willott2010canada}, Subaru High-z Quasar \cite{kashikawa2014subaru}, CANDELS GOODS-S \cite{giallongo2015faint}, UKIDSS \cite{mortlock2011luminous}, UKIDSS \cite{venemans2015identification}, and ALLWISE+UKIDSS+DECaLS \cite{banados2018800} ($3.6 < z < 7.4$) \cite{natarajan2023quotas}.
Additionally, the emission lines considered in the QUOTAS database include Ly$\alpha$, CIV, MgII, and H$\beta$ \cite{jiang2007gemini,kurk2007black,willott2010eddington,shen2011catalog,de2011evidence,trakhtenbrot2011black,de2014black,zuo2015black,shao2017gas,kozlowski2017virial,mazzucchelli2017physical,eilers2018first,shen2019gemini,shen2019sloan,matsuoka2019discovery,
farina2022x,matsuoka2022subaru}.

The theoretical models that attempt to depict the evolution of the populations of the BH assembly history over time rely on two functions that have been determined by observations: the BH Mass Function (BHMF) and the QLF. The BHMF represents the evolution of mass as a statistical measure of the distributed BH mass via redshifts. Statistically measuring the distribution of quasar luminosities across redshift, the BHMF is comparable to the QLF because it represents the accretion history, referring to the QUOTAS dataset \cite{natarajan2023quotas}.

An extensively utilized technique for determining the SMBH mass for many QUOTAS sources claims that the BLR is virialized and that the motion of the producing clouds corresponds to the gravitational potential of the central black hole.
SMBH masses and errors determined with BLR are included in QUOTAS. Deep Learning (DL) techniques have been utilized for quasars, such as the "changing look" quasars with no available uniform sampled variability data \cite{tachibana2020deep}.

A different approach for determining the SMBH mass is available, which does not rely on the premise that the BLR is in a state of virialization. Instead, this method utilizes the luminosities of various spectral wavelengths, such as X-ray, ultraviolet, infrared, and optical, to estimate the size of the BLR. To achieve this objective, continuum luminosities ($L_{cont}$) are frequently favored over line luminosities since they tend to exhibit a stronger correlation with the size of the BLR \cite{kaspi2005relationship,bentz2009radius}. Reverberation mapping has uncovered a strong relation between the size of the BLR and the continuum luminosity. Subsequently, this empirical scaling equation calculates the SMBH mass in QUOTAS.

On the other hand, the QuasarNET dataset is presented as a deep Convolutional Neural Network (CNN) that can accurately classify astrophysical spectra and estimate their redshift. These two activities are framed as a feature detection problem, where the presence or absence of spectral characteristics determines the class, and their wavelength identifies the redshift \cite{busca2018quasarnet}. The QuasarNET dataset establishes a sample purity and coverage of $99.51 \pm 0.03\%$ when run through the BOSS data \cite{alam2015eleventh} for quasar identification by emission lines, exceeding the requirements of many analyses. 

The data sample is based on a large and publicly available database of spectra with human-expert classifications and redshift determinations, composed of 627,751 \cite{paris2017sloan}, spectra of 546,856 objects from the Data Release 12 of the BOSS that was targeted as quasar candidates according to their color properties \cite{alam2015eleventh}.
The final sample consists of 491,797 spectra (449,013 unique objects), of which 192,925 spectra (176,618 unique objects) are annotated as STAR, 14,966 spectra (14,203 unique objects) as GALAXY, 274,967 spectra (249,762 unique objects) as QSO spectra. In comparison, the remaining 8,939 spectra (8,430 unique objects) do not have a firm identification. Among the QSO spectra, 26,211 spectra (23,992 unique objects) were tagged as exhibiting Broad Absorption Line (BAL) features \cite{weymann1981absorption,weymann1991comparisons}.

A separate line finder in the output layer identifies each emission line. These line-finders are specific to the emission line they detect. To implement this feature identification, the latest strategies for object detection in photos have been used as inspiration for a classification and regression problem. 
Seven line-finders covering wavelengths have been used from Ly$\alpha$, CIV, CIII, MgII, H$\alpha$, H$\beta$, and BAL CIV in a range of [121.6 nm--656.3 nm].
These lines are chosen for quasars with redshifts between 0 and 5.45 so that at least two of them are visible in the optical spectrograph of BOSS \cite{redmon2017yolo9000}.

With the QuasarNET dataset, the line-confusion problem that causes catastrophic redshift failures has been reduced to below 0.2\%. 
To clarify, the problem of line-confusion happens when spectral lines from varying redshifts may coincide in the same observed frequency, leading to potential confusion \cite{cheng2020phase}. It can occur in large-scale spectroscopic surveys, which may cause a shift and broadening of the BAO peak, potentially leading to errors in determining the position of the BAO peak \cite{massara2021line}. 

The BOSS database is used to train the QuasarNET dataset. This database has about 500,000 quasar-target spectra visually reviewed and annotated by human specialists \cite{paris2017sloan}. 
For the classification of spectra, including BAL features, the QuasarNET dataset will also be expanded, reaching an accuracy of $98.0 \pm 0.4\%$ for recognizing BAL and $97.0 \pm 0.2\%$ for rejecting non-BAL quasars. Spectra with a BAL from ongoing and prospective astrophysical surveys like eBOSS \cite{dawson2016sdss}, DESI \cite{aghamousa2016desia,aghamousa2016desib}, and 4MOST \cite{de20124most,alonso2016reconstructing} might be easily classified using the QuasarNET dataset. Since it was trained on low signal-to-noise and medium-resolution data, redshifts may also be reliably calculated using neural networks \cite{busca2018quasarnet}. 

As said before, with the use of ML (including random forest techniques), a template may be given for a better fit between the observational survey volumes and the generated samples of quasars \cite{breiman2001random}. Therefore, this is how QuasarNET network architecture works: approximately 4,400 flux pixels, evenly spaced in log-wavelength between 360 nm and 1 $\mu$m, make up the raw spectra. Using the same upper and lower bounds, downsample them to 443 evenly spaced pixels in long wavelengths. It is well-known that learning is slowed when neural networks are fed data with significant changes, as with the distribution of quasar brightness.
To alleviate this issue, spectra have been renormalized by removing the weighted mean from each flux and dividing the resulting value by the weighted root-mean-square for each spectrum. Spectra are downsampled and normalized before being fed into QuasarNET, where they are reprocessed by four convolutional layers of 100 filters of size 10 pixels and strides of 2 with Rectified Linear Unit (ReLU) activations, and then layer 5, a fully connected layer of 100 sigmoid-activation units, where they are encoded into a 100-dimensional vector. By adding batch normalization \cite{ioffe2015batch}, after each convolutional layer, improved training performance was seen. In this stage, the input spectra have been renormalized in the same manner as the layer outputs \cite{busca2018quasarnet}.

It is worth mentioning that the QuasarNET dataset is randomly divided into ten 80/20 training/validation sub-samples. For each iteration, the QuasarNET dataset is taught using the training sample, and its behavior is analyzed using the validation sample. Spectra that match targets also present in the training sample have been removed from the validation sample to reduce the likelihood of any correlations between the two sets of data.
Interestingly, the QuasarNET dataset predicts confidence and location for each emission line in each spectrum in the validation sample. 
If one or more emission lines are identified in a spectrum, it is very likely to be that of a quasar.
It takes roughly 12 minutes to train over a single epoch on the typical 24-CPU machine that has been utilized and about 20 hours to train over the whole set of 100 epochs \cite{busca2018quasarnet}.

To achieve a comprehensive model for quasars that aligns with both high and low redshift observations, data from datasets that include both is required. Combining the QUOTAS+QuasarNET dataset with a set of PG quasars enables us to attain the desired outcome.
In contrast to the QUOTAS+QuasarNET dataset, which uses higher redshift data, DL11 has also been utilized, with low redshift data values less than 1 ($z \sim 0.025-0.5$).
In Ref. \cite{davis2011radiative}, the absolute accretion rate may be inferred from standard thin accretion disk model spectral fits to the optical luminosity density in individual AGNs, given the SMBH mass. The ratio of the bolometric luminosity to the accretion rate is used to determine the accretion efficiency. 
This technique is used to find $ \epsilon $ in 80 PG quasars \cite {schmidt1983quasar} with known bolometric luminosity \cite {neugebauer1987continuum}. The accretion rate is defined by a standard accretion disk model fitted to the optical luminosity density, and SMBH mass is either given by the bulge stellar velocity dispersion or the BLR. 
DL11 includes 80 PG quasars and has a wide range of bands being covered, including optical to far-UV \cite{scott2004composite} and X-rays \cite{brandt2000nature}.
Although the unobservable EUV emission has significant uncertainty, this selection has obtained a robust estimate of the bolometric luminosity. Aside from the accretion rate estimates, these luminosities allow the accretion efficiency of each source in the population to be determined.
Additionally, for the DL11 dataset, the optical luminosity was sampled at 4861 $\AA$, and for the QUOTAS+QuasarNET dataset, the optical luminosity was sampled at 1350, 1700, 3000, and 5100 $\AA$.
It is noteworthy that the QUOTAS+QuasarNET dataset is utilized in this work because few datasets have this much SMBH data in high redshift, and QUOTAS+QuasarNET is one of the best options for that. The DL11 dataset is used because central SMBHs of PG quasars are known for their low redshifts; thus, they are useful for this work.

A validation dataset is required to estimate the accuracy and precision of this work's obtained results.
A total of 103 data have been gathered from numerous observatories regarding this matter:

I) The GEMINI observatory, or the Gemini Near-Infrared Spectrograph (GNIRS), is located in Maunakea, Hawaii. Observations have indicated 
Type 1 radio-quiet AGNs, or better said, flat spectrum radio quasars with prominent emission lines \cite{urry1995unified}, with a total of 9 $\gamma$-ray Fermi-detected blazars, demonstrated in the work of Ref. \cite{paliya2020blazars}. 
Broad line detections, MgII and H$\beta$, in the GNIRS, have been operated and have covered the full J-, H-, and K- bands \cite{burke2024gemini}.

II) In the work of Ref. \cite{dietrich2004implications}, 15 luminous high redshift quasars were recorded, where the observatories are the Very Large Telescope (VLT) based in Chile (SofI), CIV MGII
the New Technology Telescope (NTT) based in Chile (SofI), 
the Calar Alto (CA) Observatory based in Spain, 
the Cerro Tololo Inter-American Observatory (CTIO) based in Chile, 
and the W. M. Keck (KECK) Observatory based in Hawaii. Previous observations have been submitted in the works of Ref. \cite{dietrich1999spectroscopic,dietrich2000elemental,dietrich2002high,dietrich2003elemental}. Analyzing the high redshift quasar spectra was executed by utilizing a multicomponent fit approach, and broad line detections were found in the CIV, MgII, and H$\beta$ emission lines, with spectra recorded in the J-, H-, and K-band.

III) The New Technology Telescope (NTT) based in Chile (SofI) has displayed the near-infrared spectra data of 10 luminous, intermediate-redshift quasars. Having probed the emission line region of H$\beta$-[OIII] and covering the J- and H-bands, a multicomponent fit approach was used in order to identify the required components for gaining measurements of emission line properties \cite{dietrich2009black}.

IV) The selected data from the XXL survey presented in the work of \cite{duras2020universal}, has showcased a subsample of type 1 AGNs derived from the XXL-N survey in their work \cite{pierre2016xxl}.
From the 8445 point-like X-ray sources identified by Ref. \cite{liu2016x}, AGNs with 2–10 keV photon counts exceeding 50 were chosen. High bolometric luminosities were selected ($L_{bol} >$ $10^{46.5}$ erg $s^{-1}$), leading to 37 XXL sources with H$\beta$, MgII, and CIV broad lines. Subsequently, 6 RL sources were eliminated, resulting in the final XXL sample of 31 type 1 AGN, which after removing the NaN data, 29 final data are left.

V) The data of SDSS has been utilized from the DR7 \cite{zuo2015black} and the DR14 catalogs \cite{diana2022evolution}. 
For the DR7 data, the sample includes 32 luminous quasars with near-infrared observations of the H$\beta$ and MgII emission lines, with the spectroscopy in the J, H, and K bands, by employing the Palomar Hale 200-inch telescope and the Large Binocular Telescope. Additionally, targets were selected from the SDSS DR7 quasar catalog in Ref. \cite{schneider2010sloan,shen2011catalog}.
For the DR14 data, the study focuses on a sample from the DR14 catalog \cite{blanton2017sloan}, of 380 luminosity-selected flat spectrum radio quasars (FSRQs) gathered from the Cosmic Lens All Sky Survey (CLASS) \cite{browne2003cosmic,myers2003cosmic}. This work has gathered 19 of their high-redshift ($z > 4$) blazers while isolating and analyzing the CIV emission line, with a focus on the R-band.

\section{Creating Flux- and volume-limited samples}\label{fluxlim}

To work on the QUOTAS+QuasarNET+DL11 dataset, model a parameter through redshift, and eliminate biases, one needs to apply flux- and volume-limit correcting methods.
The correction of datasets, especially those including higher redshift objects, is crucial to ensuring the accuracy and reliability of the analyses and results. Implementing these methods allows relevant information to be accurately captured while avoiding overestimation or underestimation due to different biases.
In the following, the biases and problems that arise when working with higher redshift data are explained.

I) The selection of quasars based on their brightness or location can lead to a bias towards more luminous objects. As a result, the selection process is influenced by a preference for brighter objects \cite{green1986palomar,hewett1995large}.
II) Redshift bias occurs when quasars are either over-represented or under-represented in specific redshifts, which can compromise analyses due to their varied luminosity. Their low luminosity and faintness cause under-representation, and high luminosity or brightness cause over-representation \cite{york2000sloan}.
III) The distribution of varying luminosities is a sensitive matter; therefore, the luminosity function needs to be properly estimated. Estimating the luminosity function causes a bias due to its varied luminosity distribution through cosmic time \cite{schmidt1983quasar,boyle20002df}.
IV) It is possible that rare objects and quasars might not be detected in deep surveys or when using narrow selection criteria. However, these rare objects can provide vital information about the physical processes or beneficial details. Thus, failure to detect rare objects can cause a significant bias \cite{fan2003survey}.
V) Dust and gas can obscure quasars, leading to a bias toward brighter objects. This bias can ruin identifying quasars with different levels of extinction. This bias is also known as galactic extinction bias \cite{blakeslee2002early}.
VI) Quasars may be observed in more than one survey, leading to a bias because of the difference in selection criteria or coverage of surveys, thus leading to the survey bias \cite{york2000sloan}.
VII) The efficiency and completeness of the quasar sample near high redshift quasars are significantly reduced since their colors are similar to regular stars \cite{fan1999simulation,skrutskie2006two}. In other words, an issue of incompleteness is observed at the brighter end of galaxy magnitudes, which is attributed to the saturation of bright galaxies and blending by saturated stars \cite{tago2010groups}.

Flux-limiting the samples of quasars can significantly improve the results and eliminate most of the mentioned biases. By using the flux-limiting method, the selection bias can be resolved because it selects quasars based on their observed luminosity without any bias concerning brighter objects. Therefore, the bias that results from the preferred selection can be eliminated \cite{york2000sloan,croom20042df}.
Furthermore, to address the issue of redshift distribution bias, it is important to use a flux-limited sample. 
Such a sample helps to make the data more representative across different redshift ranges by including more distant quasars that may be under-represented in the sample due to their faintness and lower luminosity. This helps to reduce any bias in the redshift distribution by including quasars through varied redshifts \cite{paris2017sloan}.
Moreover, the bias towards estimating the luminosity function can be mitigated by flux-limiting a sample of quasars with varying luminosity distributed uniformly and balanced over cosmic time due to the uniform and balanced distribution of quasars in the flux-limited sample \cite{ilbert2004bias,cole2011maximum}.
Ultimately, The flux-limited sample can clear bias for detecting rare objects by including rare objects, such as high-redshift quasars, and identifying them accurately \cite{banados2018800}.

Therefore, limiting the utilized dataset with flux constraints is necessary for accurate results.
Firstly, to be able to use the flux-limiting method, one needs to extract the sample of the galaxy, extract the required columns from the catalogs, eliminate the galaxy duplicates, and erase galaxies with magnitude errors.
The QUOTAS+QuasarNET dataset contains 23,301 quasars from the BOSS SDSS-III dataset. 

To apply the flux-limiting method on a dataset, one should get familiar with the Friends of Friends (FoF) method.
The group discovery is based on the FoF method, which was proposed by Ref. \cite{turner1976groups}. The FoF algorithm has been the most widely used method for detecting clusters and groups in galactic redshift data. 
This approach uses a certain neighborhood radius, the Linking Length (LL), to connect objects into systems. It is assumed that all objects within the LL radius are part of the same system for any given object. The selected LL significantly impacts the richness and quantity of the identified groupings. Typically, LL is not fixed but rather permitted to change as a function of distance and other variables. It has been noted that the objectives of the particular investigation dictate the LL to be used. Finding a large number of groups with uniform distance-dependent general group attributes is the goal here \cite{cunningham2020hydrogen}. 

The best estimate for the LL is \cite{tago2010groups}

\begin{equation}
   \frac{LL}{LL_0} = 1 + a \arctan{(\frac{z}{z_*})} ,
   \label{LL}
\end{equation}
where $a = 1.00$ and $z_* = 0.050$ are free parameters, and $LL_0$ is the value of LL at the initial redshift of $z = 0$. Flux-limited samples can be made with constant or varied LL, and the size of the groups depends on the LL. The constant value for $LL_0$ is considered to be 250 km$s^{-1}$ ($\sim 0.25$ $h^{-1}$Mpc).

Only the brightest items may be seen while looking at distant objects. The parameters of a set of objects are found by calculating their absolute magnitude from their mean distance. If an object or a group member is too faint to be seen, this will delete them automatically. Applying the flux-limited approach and fixing the data are both made possible by this procedure. 
The downside of this procedure is that the remaining absolute magnitudes of the objects can exceed the range of magnitudes covered by the sample. In such a situation, the sample has to have those objects deleted and the groupings recreated. After this is completed, the revised group data must be used to recalculate the absolute magnitudes of the objects. 
It is important to note that there is no assurance for the end of this cycle \cite{tempel2014flux}.

The separation of galaxies along the LoS or the Plane of the Sky (PoS) is used to group them in the FoF percolation technique. All galaxies that are connected in pairs based on their distances are grouped. The two connecting lengths that identify FoF are the LoS and the PoS. The average distance between field galaxies is used to normalize these lengths \cite{duarte2014well}.

As mentioned, objects that do not meet flux-limited sample requirements are removed from groups after their absolute magnitudes and distances are calculated.
Afterward, the FoF technique is applied to the identified sub-sample of QSOs, keeping the LL constant. The process of group finding is finished with this step \cite{etherington2015measuring}.
Thus, to create the flux-limited sample, the FoF procedure needs to be carried out with a constant LL, and the LL scaling law with sample shifting and the minimal spanning tree method, which considers the luminosity-density relation in groups needs to be determined; and afterward, a modified FoF with a changing LL in accordance with the scaling law to find the final flux-limited sample of groups is utilized.

One advantage of this type of sample is that it utilizes as much observational data as possible. However, many research projects require creating volume-limited samples, which means that a large amount of data must be rejected. A flux-limited sample naturally has negative selection effects, even though attempts have been made to minimize them.
An extra issue comes up when it comes to Quasi-stellar objects (QSOs) \cite{ghirlanda2012impact}. QSOs must have a certain absolute magnitude to be included in the volume-limited sampling. The latter is reliant on the distance of a QSO. Initially, without group formation, the redshift of a QSO must be used to calculate its absolute magnitude. Once a group has been formed, the mean radial velocity (redshift) of the group determines the distance of the QSOs from one another \cite{best2024volume}.

Moreover, after creating the flux-limited sample, some biases remain unsolved, including the bias regarding galactic extinction, the survey bias, and the bias of incompleteness. Therefore, an additional method is pursued called volume-limiting.
After volume-limiting and considering varied extinction levels, the bias regarding galactic extinction can be solved. This reduces the effect of quasars less affected by extinction \cite{skrutskie2006two}.
Volume-limited samples can also solve the survey bias caused by selecting quasars based on distance, and they can be tailored to specific surveys \cite{york2000sloan,ross2012sdss}.
Additionally, the volume-limited sample addresses the incompleteness bias by including a greater number of quasars in a selected cosmic volume of space \cite{croom20042df}.
Therefore, it is necessary to use the volume-limiting method to solve biases.

For applying the volume-limiting method, it is important to note that by using the QUOTAS+QuasarNET dataset and considering that the minimum absolute magnitude of the BOSS dataset is approximately $M_i$ $\approx$ -24.5 at the redshift of $z = 2.2$, the mentioned information can be used to estimate the detectable absolute magnitude for quasars located at further distances \cite{ross2013sdss}.

To calculate the mean absolute magnitude for each co-moving distance, first, the minimum absolute magnitude at each redshift needs to be found. A plot can then be illustrated by fitting the minimums of absolute magnitudes, and any quasar with an absolute magnitude below the minimum is removed. To convert the apparent magnitude, "$m$", to the absolute magnitude, "$M$", at redshift $z = 0$, the mean redshift, "$z_{cl}$", and mean co-moving distance, "$d_{com}$", are utilized. For this purpose, the following equation can be used \cite{tempel2014flux}
\begin{equation} \label{dist_mod}
M_\lambda = m_\lambda - 25 - 5 log_{10}(d_L) - K_\lambda,
\end{equation}
where $d_L = d_{com}(1 + z_{cl})$ is the luminosity distance in units $h^{-1}$Mpc, $K_\lambda$ is the correction factor, and the index $\lambda$ refers to each of the $ugriz$ filters. For distance calculations in this work, a flat universe with zero curvature is considered.

Hence, to generate the volume-limited sub-samples, firstly, the limits of the volume-limited sub-samples in redshift, apparent magnitude, and absolute magnitude need to be determined. The volume-limited sub-samples of groups of galaxies using FoF with a constant LL within a sub-sample are to be searched for, while one scales LL between sub-samples using the luminosity-density scaling. Ultimately, catalogs of groups and galaxies for the flux-limited sample as well as for every volume-limited sub-sample are the output \cite{tago2010groups}.

Overall, the flux- and volume-limiting procedure has been applied for a sub-sample group with a limit of 2.5 absolute magnitude, resulting in a single sub-sample. 
The 2.5 absolute magnitude constraint is not made too rigid because the aim is to eliminate as little data as possible.
In this work, after applying both flux- and volume-limiting methods, 12,359 data from the QUOTAS+QuasarNET dataset remain for calculation.

Moreover, the effect of flux- and volume-limiting on low redshift data is little to none. Therefore, after removing the accretion efficiencies higher than ten from the DL11 dataset, 79 quasars remain.

\section{$\epsilon$ -- $M_{BH}$ relationship} \label{comparison}

Previous works have demonstrated how luminosity, redshift, and mass of quasars can significantly impact the relationship between the SMBH accretion efficiency and mass. 
Therefore, the QUOTAS+QuasarNET dataset is employed to calculate the accretion efficiency of each object using Eq. (\ref{epsilon_opt}) and try to plot the accretion efficiency parameter in terms of SMBHs mass.
The flexibility to find a more precise and general result is provided by the fact that the SMBHs mass in the QUOTAS+QuasarNET dataset spans from $10^8$ to $10^{11} M_\odot$.
This work aims to expand the results of Ref. \cite{raimundo2012can} by modeling the accretion efficiency of SMBHs in low and high redshifts altogether.

A related figure can be plotted to investigate the relationship between the SMBH accretion efficiency and mass. By drawing lines that represent the maximum and minimum luminosity, the boundaries for the distribution of SMBH accretion efficiency can be defined.
Moreover, it is possible to plot the Eddington ratio, which includes maximums and minimums, as well as to surround the data with the four lines of $L_{max}$, $L_{min}$, $\lambda_{max}$, and $\lambda_{min}$.

In Sec. \ref{QUOTAS}, the corrections required in the considered dataset have been discussed. The first step in the data cleaning process has been to remove the unrealistic values for data points, including zero mass or negative numbers. Then, the accretion efficiency for each wavelength contained in the dataset is calculated, which includes $1350$, $1700$, and $3000 \AA$. To determine the accretion efficiencies for the QUOTAS+QuasarNET dataset, Eq. (\ref{epsilon_opt}) is used.
As previously stated, the DL11 dataset is included in order to create a general and comprehensive model for low and high redshift values.

The smallest value from the three calculated accretion efficiencies needs to be selected to identify the maximum Eddington ratio leading to the minimum accretion efficiency. This important step allows the data to be filtered and the most relevant information to be selected.
The required data for satisfying all parameters of Eq. (\ref{epsilon_opt}) by flux and volume-limiting the QUOTAS+QuasarNET dataset are presented in Table \ref{tab:qn-tab-fin}.

\begin{figure}[h!]
\includegraphics[width=0.5\textwidth]{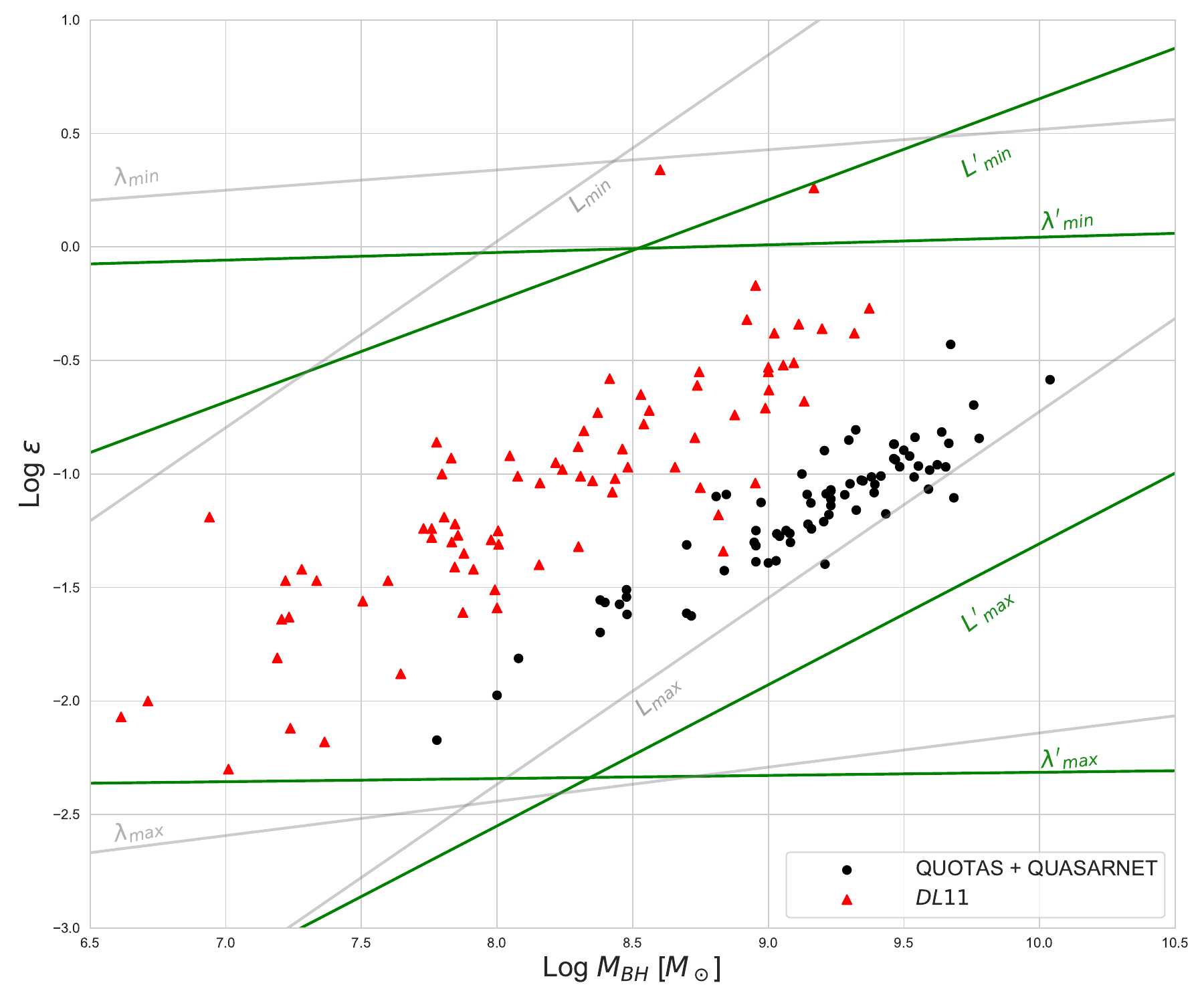}
\caption{
This graph illustrates the relationship between SMBH accretion efficiency and mass for the flux- and volume-limited QUOTAS+QuasarNET+DL11 dataset. 
In the plot, the solid green lines represent $L'_{max}$, $L'_{min}$, $\lambda'_{max}$, and $\lambda'_{min}$ for the QUOTAS+QuasarNET+DL11 dataset; where the solid grey lines represent $L_{max}$, $L_{min}$, $\lambda_{max}$, and $\lambda_{min}$ for the DL11 dataset.
As depicted in the plot, there is a positive correlation between SMBH mass and accretion efficiency. This indicates that higher SMBH masses are associated with greater accretion efficiency.
}
\label{fig_ep_m_fit_qn_d11}
\end{figure}

In Fig. \ref{fig_ep_m_fit_qn_d11}, the relationship between SMBHs accretion efficiency and mass is plotted using the flux- and volume-limited QUOTAS+QuasarNET+DL11. The original data of the QUOTAS+QuasarNET dataset contained about 37,648 quasars, and about 75 final quasars had the required information, including the optical luminosity, the bolometric luminosity, the Eddington ratio, redshift, and the SMBH mass of each object after the data pre-processing. 

The QUOTAS+QuasarNET dataset is designed for quasars with high redshift. It is indicated that adding the DL11 dataset of the 79 PG quasars to the flux- and volume-limited QUOTAS+QuasarNET data collection and conducting a comparison between the two datasets could be an exciting discovery. It has been noted that after removing the nonphysical and duplicated data, the combined dataset of flux- and volume-limited QUOTAS+QuasarNET and the PG quasars consist of approximately 154 quasars.

As shown in the plot, there is a positive correlation between SMBH mass and accretion efficiency. This means that the more mass of SMBHs there is, the more efficient the accretion is.
Eq. (\ref{lambda_edd}) has been used to calculate the maximum and minimum Eddington ratio.
Bolometric luminosity reaches its maximum, Eddington luminosity reaches its minimum, and the Eddington ratio reaches its maximum, and vice versa.
This way, the related lines to $\lambda_{min}$ and $\lambda_{max}$ can be plotted.

Lastly, the maximum and minimum values of bolometric luminosity are gathered to calculate the values for accretion efficiency using Eq. (\ref{epsilon_opt}). This equation provides a way to estimate the efficiency of accretion based on the luminosity produced by the accretion process. Insights into the physical processes that govern SMBH growth and accretion are gained by using these equations. A specific bond is created by the lines evident in the graphs.

It is important to note that most of the data used in Fig. \ref{fig_ep_m_fit_qn_d11} are in the valid range of estimation for the standard accretion disk efficiency of $0.038 < \epsilon < 0.42$ \cite{zhang2020extracting}. Notably, none of the calculated accretion efficiencies for the QUOTAS+QuasarNET dataset exceed the upper constraint of the estimated range.
It is evident that due to the low redshift data of DL11, the accretion efficiency is found in the lower SMBH mass range.
An intriguing finding is that the average range of accretion efficiency decreases when the SMBH mass range is lowered. The accretion efficiency is projected to decrease for SMBHs with lower mass in low redshift, just as it rises with SMBHs with higher mass of the high redshift data. Thus, the idea can be put forth that there is a correlation between SMBH accretion efficiency and mass independent of redshift.

\section{Accretion efficiency history} \label{result}

In the last Section, a direct correlation between the SMBH accretion efficiency and mass was found, independent of redshift.
Now, it must be checked whether redshift is directly correlated with accretion efficiency and, if so, to what extent.
This Section explores the possibility of a correlation between redshift and accretion efficiency.
As a result, a color bar illustrates the central SMBH mass, and the overall plots show the accretion efficiency in terms of redshift.

\begin{figure}[h!]
 \centering
  \includegraphics[width=0.5\textwidth]{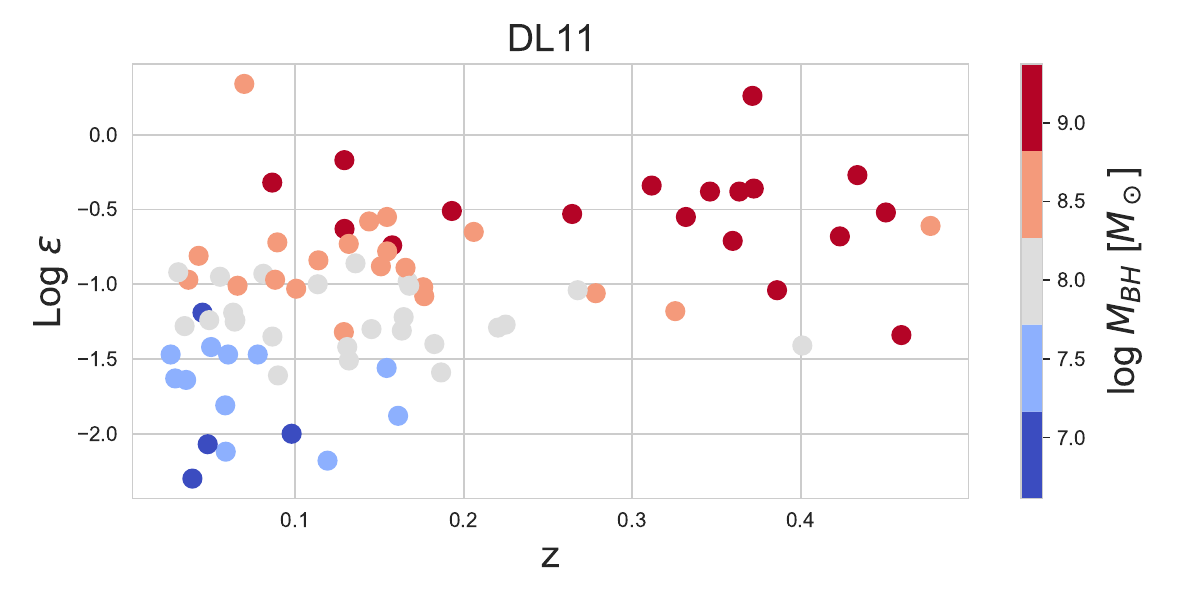}
\caption{
This plot shows the relationship between accretion efficiency, redshift, and SMBH mass of the DL11 low redshift PG quasars. The visualization reveals an intriguing correlation between increasing redshift and accretion efficiency. 
As redshift values rise, a corresponding increase in accretion efficiency is expected.}\label{fig_ep_m_col_11}
\end{figure}

\begin{figure}[h!]
 \centering
  \includegraphics[width=0.5\textwidth]{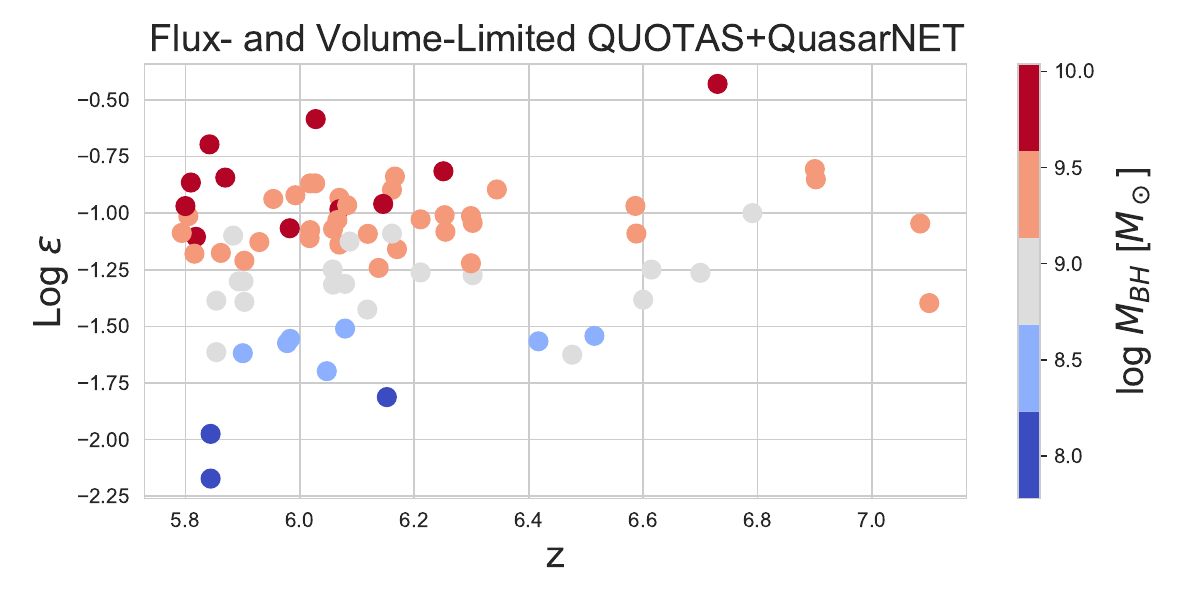}
\caption{
This plot illustrates the relationship between accretion efficiency, redshift, and SMBH mass using the flux- and volume-limited samples from the QUOTAS+QuasarNET dataset. At redshifts exceeding 5 ($z > 5$), the correlation between SMBH mass and redshift is apparent, showing a decrease in SMBH mass as redshift increases. Consequently, consistent results are obtained.
}\label{fig_ep_m_col_qn_vol}
\end{figure}

As seen in Fig. \ref{fig_ep_m_col_11}, it is illustrated that an interesting correlation between the growth of redshift and the increase in accretion efficiency is present for low redshift SMBHs. 
It is clear that in redshifts lower than $z \sim 0.5$, the mass of the observed SMBHs decreases. 
According to the flux- and volume-limited QUOTAS+QuasarNET dataset, which has a redshift range of $5.844 \lesssim z \lesssim 7.1$, the population of higher mass SMBHs increases with the decrease of redshift as shown in Fig. \ref{fig_ep_m_col_qn_vol}.
Furthermore, with decreasing the redshift, SMBHs mass in the order of $10^7 M_\odot < M_{BH} < 10^9 M_\odot$ can be observed as well.

What can be understood from the said graphs is that in low redshifts, the mass of the central SMBHs of quasars in the local universe tends to decrease as $z \sim 0.025$. Therefore, SMBHs with closer distance have less mass compared to farther SMBHs.

\begin{figure}[h!]
 \centering
  \includegraphics[width=0.5\textwidth]{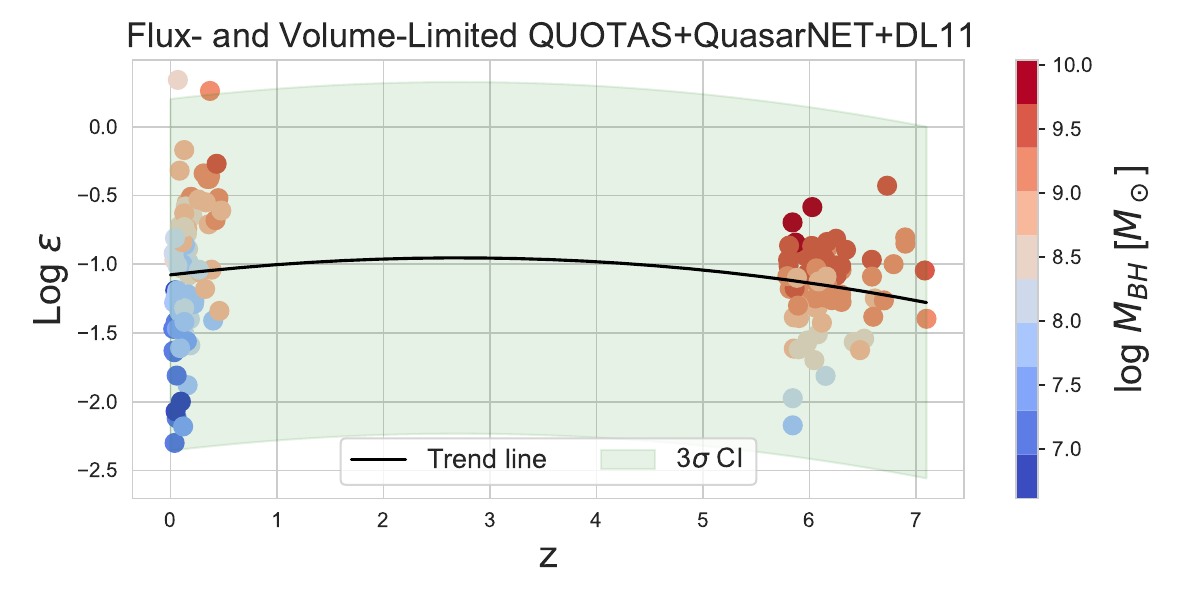}
\caption{
This graph illustrates the relationship between accretion efficiency, redshift, and SMBH mass using the flux- and volume-limited samples from the QUOTAS+QuasarNET+DL11 datasets, along with a trend line indicating the peak of accretion efficiency. A regression fit has been applied to the flux- and volume-limited datasets, clearly showing the peaks of the data, with observed values around $z \sim 2.708$.
}\label{fig_ep_m_col_vol}
\end{figure}

The situation for the high redshift data is inversely correlated. Therefore, investigating the two datasets of DL11 and QUOTAS+QuasarNET together, the QUOTAS+QuasarNET+DL11 datasets can anticipate a peak where the two trends meet, as can be seen in Fig. \ref{fig_ep_m_col_vol}.

To understand how accretion efficiency behaves over time, it is necessary to compare new and corrected data that have been subjected to flux- and volume-limitations. These restrictions are prominent to ensure that the data is as precise and reliable as possible. In Sec. \ref{fluxlim}, a deeper delve is made into why these limitations are deemed essential in the context of this study.
Despite a more than 50\% reduction in the abundance of the data, accretion efficiency shows growth as a function of growing mass. Therefore, the correlation between SMBHs accretion efficiency and SMBHs mass is shown independent of flux- and volume-limitations.

It should be mentioned again that since flux- and volume-limit corrections have little to no effect on data with low redshifts, the plot of flux- and volume-limited for the DL11 dataset is identical to Fig. \ref{fig_ep_m_col_11}.

Visualizing the results is also essential to understand the data better. 
In Fig. \ref{fig_ep_m_col_vol}, a regression fit has been applied to the flux- and volume-limited data. This results in a clear indication of the peaks of the data, with values of approximately $z \sim 2.708$ being observed.

As mentioned before, the QUOTAS+QuasarNET dataset is the high redshift data, which after flux- and volume-limiting is finalized to the range of $5.794 \leq z \leq 7.1$, and the DL11 dataset containing the PG quasars is the low redshift data which is in the range of $0.025 \leq z \leq 0.5$; thus, the model is done utilizing the mentioned datasets. 
Therefore, it is essential to determine whether the model incorporates results from quasars within a redshift range for which no data was used in its construction.

Furthermore, as mentioned in Section \ref{QUOTAS}, a validation dataset should be considered for the data with which the main regression was done. Hence, a total of 103 data have been obtained from multiple observatories.

\begin{figure}[h!]
 \centering
  \includegraphics[width=0.45\textwidth]{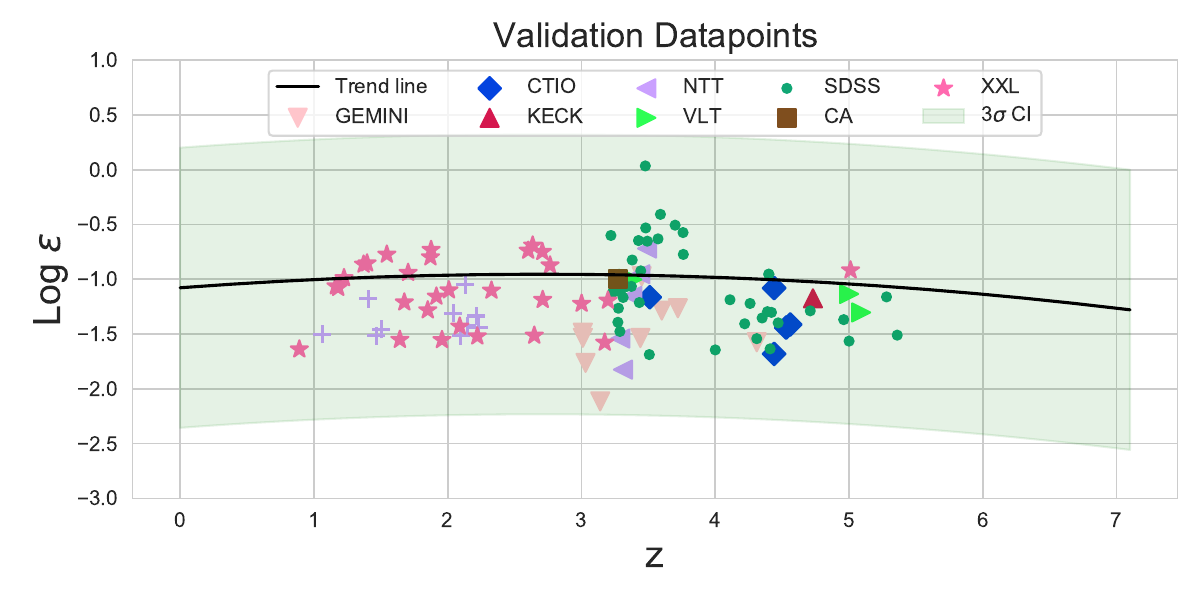}
\caption{
In this graph, the presented trend line based on the QUOTAS+QuasarNET+DL11 data along the $3\sigma$ confidence interval bond, in order to investigate its performance in the redshift range besides the original dataset is illustrated.
The validation data points include the data from the GEMINI observatory shown with the bright pink triangle-down marker; the data from the XXL survey shown with the pink star marker; the data from the SDSS (DR7 and DR14) shown with the green point marker; the data from the VLT shown with the bright green triangle-right marker; the data from the NTT shown with the bright purple triangle-left marker; the data from the CA Observatory shown with the brown square marker; the data from the CTIO shown with the blue diamond marker; and the data from the KECK Observatory shown with the maroon triangle-up marker. It seems as though the entirety of the validation points are present inside the bond of the presented model.
}\label{valid_fig}
\end{figure}

Additionally, Fig. \ref{valid_fig} illustrates the $3\sigma$ confidence interval bond represented in this work. In this plot, the data from different observatories is shown in different colors. As it is clear from Fig. \ref{valid_fig}, all of the data are located in the suggested bond, which is a clear depiction of how the results concluded from this work in order to describe the data in the range of $0.9 \leq z \leq 5.6$, works satisfactory.
The exact details of the mentioned data are in Table \ref{tab:val_dat}.

The peak observed in Fig. \ref{fig_ep_m_col_vol} can serve as a reminder for the redshift range, $z_{SFR} \sim 1-3$, in which the peak of Star Formation Rate (SFR) occurs.
SFR shows the amount of mass produced per year in a galaxy. Using the low-mass X-ray binaries (LMXBs) data, Ref. \cite{white1998low} has reported the SFR peak to be $z \sim 1.5$.
Moreover, multiple works, such as Ref. \cite{atek2014hubble,zolotov2015compaction,driver2018gama} show that the peak of SFR is at $z \sim 2$, about 3.5 billion years after the Big Bang.
To go further into detail of the redshift range of the peak of SFR, Ref. \cite{hopkins2006normalization} produced an extensive collection of observations of luminosity density, covering a wide range of redshifts up to $z = 6$. Their findings indicate a significant increase in the SFR at low redshifts, reaching its highest point at around $z \simeq 2$ and declining at higher redshifts \cite{vangioni2015impact}.
Ref. \cite{cucciati2012star} reports the Star Formation Rate Density (SFRD) peak to be $z \sim 2$, using the VIMOS-VLT Deep Survey (VVDS).
In Ref. \cite{jo2021star}, for the 1752 objects of the Spitzer COSMOS Legacy survey, the SFRD peak is approximated between $z \sim 2.2 - 2.4$. 
Moreso, in Ref. \cite{schinnerer2016gas,scoville2017evolution,tacconi2018phibss}, using the data from Main Sequence (MS) galaxies, the peak of SFR is located in $z \sim 1.5 - 3$.
Additionally, according to Ref. \cite{reddy2007multi,madau2014cosmic,hodge2020high}, the redshift in which SFR is at its maximum is thought to be around $z \sim 1-3$.
If the target data is increased and more accurate surveys are used, with the condition that they will include all the necessary parameters for calculating the accretion efficiency, the final result will be more precisely finalized.
By analyzing this data, a better understanding of how accretion efficiency behaves over time can be obtained, and more accurate predictions about future trends can be made.

By fitting a trend line to the corrected flux- and volume-limited QUOTAS+QuasarNET+DL11 dataset, a new equation is attained,
\begin{equation} \label{fit_ep_tot_vol}
log(\epsilon_{3\sigma}) = -0.01685z^2 + 0.09128z -1.07776 \pm 1.2789.
\end{equation}
As mentioned before, the peak approaches the SFR peak much more closely.
Based on Fig. \ref{fig_ep_m_col_vol}, and the observed peak, the accretion efficiency should be part of the parameters where the downsizing problem must be investigated for.
The term "downsizing" was first introduced by Ref. \cite{cowie1996new}, where the "downsizing problem" has been established as lower-mass galaxies continuing star formation for long periods, whereas the most massive galaxies created the vast bulk of their stars during the early universe \cite{firmani2010can}. 

The proportion of galaxies exhibiting very low star formation appears to rise as a transition occurs to higher masses. Meanwhile, the median SFR decreases within the high mass range. 
This trend is less noticeable at lower masses, as galaxies are experiencing star formation in specific time ranges and remain highly active overall \cite{villar2011star}.

Hierarchical theories of galaxy formation, on the other hand, proposed that smaller galaxies would merge into larger ones over time, and the most massive galaxies should have had more time to form.
Next, it was found that stellar populations, average SFRs, stellar masses, and dark matter halos mass are all shrinking as a function of cosmic time \cite{de2007hierarchical}. 

Moreover, in more massive systems at $z > 1-2$, the peak of cosmic star formation activity and spatial density of star-forming galaxies have occurred earlier than in lower mass systems at $z < 1$ \cite{firmani2010galaxy,conselice2018halo,somerville2018relationship}.

This suggests that as the universe expanded, the most intensive star formation events occurred in ever less massive dark matter halos.
The process of downsizing has created challenges for scientific theories that attempt to explain the rapid and early development of the largest galaxies, as well as the period during which their star formation has been halted, leading to the formation of inactive galaxies that populate the universe today \cite{mutch2013simplest}.

Focusing on the SMBHs, if SMBHs are thought to originate from early Ultracompact Dwarfs (UCD) or hyper-massive star-burst clusters, then SMBHs cannot develop for spheroid masses below a certain threshold, leaving only the accumulated nuclear cluster. This indicates that SMBH creation is a shrinking process, with lower-mass galaxies creating their SMBHs on different timelines than higher-mass galaxies \cite{kroupa2020very}. 

In conclusion, downsizing is the general tendency towards lower mass galaxies having a shorter duration for the creation of SMBHs than higher mass galaxies \cite{fontanot2009many}.
Better said, the downsizing problem is the discrepancy between the masses of stars measured directly and those inferred from their luminosity.
Thus, to understand how SMBHs and their host galaxies have evolved, the downsizing problem can be recognized \cite{li2012cosmological,izumi2018supermassive}.
It is anticipated that the same peak that can be seen in the SFR in terms of redshift plots \cite{neistein2006natural,fontanot2009many}, which depends on the mass of the host structures, can appear in the plot of the SMBHs mass in terms of redshift.
This is expected because the mass of the host structures and the mass of their central SMBHs are directly correlated.

Thus, future works could delve deeper into this concept, focusing on the downsizing process and considering accretion efficiency.

\section{Conclusion}
In this paper, we have employed datasets containing SMBHs in the center of quasars, using the information of QUOTAS+QuasarNET+DL11 datasets.
The QUOTAS+QuasarNET datasets used by projects, such as the QUOTAS and QuasarNET datasets, encompass 37648 high redshift objects, and the DL11 dataset includes 80 PG low redshift quasars.
These datasets include essential parameters, such as SMBH mass, bolometric luminosity, optical luminosity, and Eddington luminosity.

Using these parameters, numerous models have been developed to estimate the accretion efficiency for SMBHs, and various previous calculations and simulations have been conducted. 
However, whether we can accurately make a model to find the dependency of accretion efficiency on redshift and SMBH mass has been questioned.
Different methods have been proposed for calculating this parameter, and in this work, we have opted for an approach based on optical luminosity because other methods yield values beyond the feasible range ($0.038 < \epsilon < 0.42$) for accretion efficiency.

Since there is a need for a comprehensive model to calculate the accretion efficiency in all redshifts, the DL11 low redshift ($z < 0.5$) dataset has been used in addition to the QUOTAS+QuasarNET.
By fitting the spectra of individual AGNs with the standard thin accretion disk model, it is possible to deduce the absolute accretion efficiency. This dataset provides a robust estimate of the bolometric luminosity as it includes a wide variety of bands, from optical to far-UV and X-rays. These luminosities allow one to ascertain the accretion efficiency of each source.

As a part of the data pre-processing, we have excluded AGNs with RIAF and slim disks. Additionally, in a subsequent data refinement step, we have applied the flux- and volume-limiting corrections to be able to work on the dataset through redshift.
The primary focus has been on investigating the evolution of the accretion efficiency over redshift or cosmic time to determine if it follows a discernible pattern. 

Given the research conducted by Ref. \cite{tabasi2023modeling} on SMBHs mass, we were intrigued by the possibility of modeling based on redshift. In the mentioned work, they found a peak for SMBH mass evolution through redshift; therefore, we were curious to see whether the accretion efficiency had the same peak or not.

In the QUOTAS+QuasarNET dataset, there is a visible pattern where accretion efficiency rises as the redshift decreases, along with an increase in mass growth.
On the other hand, in the DL11 dataset, as redshift increased, both accretion efficiency and mass exhibited growth, leading us to identify a redshift peak between the trends of these two datasets.
Subsequently, we have recognized the optimal fit for the flux- and volume-limited QUOTAS+QuasarNET+DL11 dataset.
The calculations of this paper begin by examining the correlation between the mass of SMBHs and the accretion efficiency.
It is evident from Fig. \ref{fig_ep_m_fit_qn_d11} that SMBHs with a higher mass also have higher accretion efficiency.
Nonetheless, the fact that the central SMBHs of quasars are heavier at higher redshifts is evident. Then, an examination is conducted into the correlation between accretion efficiency and redshift. The peak of the accretion efficiency factor is observed with the information of 79 quasars from the DL11 dataset and 75 quasars from the QUOTAS+QuasarNET dataset, after flux- and volume-limiting corrections, as illustrated in Fig. \ref{fig_ep_m_col_vol}.

The redshift peak for the flux- and volume-limited dataset occurred at approximately $ z \sim 2.708$. This peak can be a recollection of the SFR peak, which is between $1 < z < 3$, prompting us to explore potential similarities between this peak and the SFR peak, as well as investigate whether the accretion efficiency could mirror the SFR in AGNs. The peak we have found is closely aligned with existing introduced values by different works, as mentioned in Sec. \ref{result}.

\raggedbottom
To check if the model includes results from quasars in the mid-range redshift of the initial datasets, a separate dataset has been utilized to validate the main regression. 103 data points from various observatories have been used, including the GEMINI observatory, specifically the Gemini North telescope in Hawaii, the Very Large Telescope (VLT) in Chile, the New Technology Telescope (NTT) in Chile, the Calar Alto (CA) Observatory in Spain, the Cerro Tololo Inter-American Observatory (CTIO) in Chile, the W. M. Keck (KECK) Observatory in Hawaii, the XXL survey, and the SDSS (DR7 and DR14).
Fig. \ref{valid_fig} represents data from different observatories in distinct colors, as well as the plot displaying the $3\sigma$ confidence interval boundary. As evident from the plot, all the data falls within the suggested boundary, indicating satisfactory results for describing data within $0.9 \leq z \leq 5.6$.

It is worth noting that Ref. \cite{tabasi2023modeling} has created a model for SMBHs mass in terms of redshift, and the observed peak has been approximately $z \sim 4.72$, which is comparable to the peak associated with different downsizing plots. They have also stated that they have not considered SFR in their work. 
In contrast, our work has used the calculations related to the accretion efficiency for the mentioned result, and it seems that the final peak has automatically considered the SFR, where it has been implemented internally.

Future works can investigate the relationship between the mass of central SMBHs of quasars, SFR, and accretion efficiency. Moreover, a comprehensive and complete model can be achieved by incorporating data from additional surveys, utilizing a larger dataset across all redshifts, and comparing the final models with the newly added data from the JWST dataset.
In conclusion, it is critical to underscore the importance of priming any constructed model on empirical evidence; relying on simulation, given the present state of knowledge, would be impractical because an accretion model should be chosen between various models for simulations at the beginning.
Meanwhile, none of the models are perfect for all quasars in different redshifts.
Thus, more accurate observations are needed to estimate the accretion efficiency reliably.
Therefore, the physics behind the accretion of SMBHs lacks a flawless model, which can be improved with a significant increase in data.

\section*{Data Availability}
The code behind the data pre-processing of this article is publicly available on \href{https://github.com/artakh10/Flux-and-Volume-limit}{GitHub}.

We have used the QUOTAS dataset \cite{natarajan2023quotas} in this work, which supports the findings of this study and is publicly available on the Google Kaggle platform and can be found at \href{https://www.kaggle.com/data sets/quotasplatform/quotas}{https://www.kaggle.com/data sets/quotasplatform/quotas}.

\bibliography{ref}

\begin{thebibliography}{245}%
\makeatletter
\providecommand \@ifxundefined [1]{%
 \@ifx{#1\undefined}
}%
\providecommand \@ifnum [1]{%
 \ifnum #1\expandafter \@firstoftwo
 \else \expandafter \@secondoftwo
 \fi
}%
\providecommand \@ifx [1]{%
 \ifx #1\expandafter \@firstoftwo
 \else \expandafter \@secondoftwo
 \fi
}%
\providecommand \natexlab [1]{#1}%
\providecommand \enquote  [1]{``#1''}%
\providecommand \bibnamefont  [1]{#1}%
\providecommand \bibfnamefont [1]{#1}%
\providecommand \citenamefont [1]{#1}%
\providecommand \href@noop [0]{\@secondoftwo}%
\providecommand \href [0]{\begingroup \@sanitize@url \@href}%
\providecommand \@href[1]{\@@startlink{#1}\@@href}%
\providecommand \@@href[1]{\endgroup#1\@@endlink}%
\providecommand \@sanitize@url [0]{\catcode `\\12\catcode `\$12\catcode `\&12\catcode `\#12\catcode `\^12\catcode `\_12\catcode `\%12\relax}%
\providecommand \@@startlink[1]{}%
\providecommand \@@endlink[0]{}%
\providecommand \url  [0]{\begingroup\@sanitize@url \@url }%
\providecommand \@url [1]{\endgroup\@href {#1}{\urlprefix }}%
\providecommand \urlprefix  [0]{URL }%
\providecommand \Eprint [0]{\href }%
\providecommand \doibase [0]{https://doi.org/}%
\providecommand \selectlanguage [0]{\@gobble}%
\providecommand \bibinfo  [0]{\@secondoftwo}%
\providecommand \bibfield  [0]{\@secondoftwo}%
\providecommand \translation [1]{[#1]}%
\providecommand \BibitemOpen [0]{}%
\providecommand \bibitemStop [0]{}%
\providecommand \bibitemNoStop [0]{.\EOS\space}%
\providecommand \EOS [0]{\spacefactor3000\relax}%
\providecommand \BibitemShut  [1]{\csname bibitem#1\endcsname}%
\let\auto@bib@innerbib\@empty
%</preamble>
\bibitem [{\citenamefont {Lynden-Bell}(1969)}]{lynden1969galactic}%
  \BibitemOpen
  \bibfield  {author} {\bibinfo {author} {\bibfnamefont {D.}~\bibnamefont {Lynden-Bell}},\ }\bibfield  {title} {\bibinfo {title} {Galactic nuclei as collapsed old quasars.},\ }\href@noop {} {\bibfield  {journal} {\bibinfo  {journal} {Nature}\ }\textbf {\bibinfo {volume} {223}} (\bibinfo {year} {1969})}\BibitemShut {NoStop}%
\bibitem [{\citenamefont {Rees}(1984)}]{rees1984black}%
  \BibitemOpen
  \bibfield  {author} {\bibinfo {author} {\bibfnamefont {M.~J.}\ \bibnamefont {Rees}},\ }\bibfield  {title} {\bibinfo {title} {Black hole models for active galactic nuclei},\ }\href@noop {} {\bibfield  {journal} {\bibinfo  {journal} {IN: Annual review of astronomy and astrophysics. Volume 22. Palo Alto, CA, Annual Reviews, Inc., 1984, p. 471-506.}\ }\textbf {\bibinfo {volume} {22}},\ \bibinfo {pages} {471} (\bibinfo {year} {1984})}\BibitemShut {NoStop}%
\bibitem [{\citenamefont {Wang}\ \emph {et~al.}(2007)\citenamefont {Wang}, \citenamefont {Chen}, \citenamefont {Yan},\ and\ \citenamefont {Hu}}]{wang2007early}%
  \BibitemOpen
  \bibfield  {author} {\bibinfo {author} {\bibfnamefont {J.-M.}\ \bibnamefont {Wang}}, \bibinfo {author} {\bibfnamefont {Y.-M.}\ \bibnamefont {Chen}}, \bibinfo {author} {\bibfnamefont {C.-S.}\ \bibnamefont {Yan}},\ and\ \bibinfo {author} {\bibfnamefont {C.}~\bibnamefont {Hu}},\ }\bibfield  {title} {\bibinfo {title} {Early growth of massive black holes in quasars},\ }\href@noop {} {\bibfield  {journal} {\bibinfo  {journal} {The Astrophysical Journal}\ }\textbf {\bibinfo {volume} {673}},\ \bibinfo {pages} {L9} (\bibinfo {year} {2007})}\BibitemShut {NoStop}%
\bibitem [{\citenamefont {Treister}\ \emph {et~al.}(2010)\citenamefont {Treister}, \citenamefont {Natarajan}, \citenamefont {Sanders}, \citenamefont {Urry}, \citenamefont {Schawinski},\ and\ \citenamefont {Kartaltepe}}]{treister2010major}%
  \BibitemOpen
  \bibfield  {author} {\bibinfo {author} {\bibfnamefont {E.}~\bibnamefont {Treister}}, \bibinfo {author} {\bibfnamefont {P.}~\bibnamefont {Natarajan}}, \bibinfo {author} {\bibfnamefont {D.~B.}\ \bibnamefont {Sanders}}, \bibinfo {author} {\bibfnamefont {C.~M.}\ \bibnamefont {Urry}}, \bibinfo {author} {\bibfnamefont {K.}~\bibnamefont {Schawinski}},\ and\ \bibinfo {author} {\bibfnamefont {J.}~\bibnamefont {Kartaltepe}},\ }\bibfield  {title} {\bibinfo {title} {Major galaxy mergers and the growth of supermassive black holes in quasars},\ }\href@noop {} {\bibfield  {journal} {\bibinfo  {journal} {Science}\ }\textbf {\bibinfo {volume} {328}},\ \bibinfo {pages} {600} (\bibinfo {year} {2010})}\BibitemShut {NoStop}%
\bibitem [{\citenamefont {Melia}(2019)}]{melia2019cosmological}%
  \BibitemOpen
  \bibfield  {author} {\bibinfo {author} {\bibfnamefont {F.}~\bibnamefont {Melia}},\ }\bibfield  {title} {\bibinfo {title} {Cosmological test using the hubble diagram of high-z quasars},\ }\href@noop {} {\bibfield  {journal} {\bibinfo  {journal} {Monthly Notices of the Royal Astronomical Society}\ }\textbf {\bibinfo {volume} {489}},\ \bibinfo {pages} {517} (\bibinfo {year} {2019})}\BibitemShut {NoStop}%
\bibitem [{\citenamefont {Bromley}\ \emph {et~al.}(2004)\citenamefont {Bromley}, \citenamefont {Somerville},\ and\ \citenamefont {Fabian}}]{bromley2004high}%
  \BibitemOpen
  \bibfield  {author} {\bibinfo {author} {\bibfnamefont {J.~M.}\ \bibnamefont {Bromley}}, \bibinfo {author} {\bibfnamefont {R.}~\bibnamefont {Somerville}},\ and\ \bibinfo {author} {\bibfnamefont {A.}~\bibnamefont {Fabian}},\ }\bibfield  {title} {\bibinfo {title} {High-redshift quasars and the supermassive black hole mass budget: constraints on quasar formation models},\ }\href@noop {} {\bibfield  {journal} {\bibinfo  {journal} {Monthly Notices of the Royal Astronomical Society}\ }\textbf {\bibinfo {volume} {350}},\ \bibinfo {pages} {456} (\bibinfo {year} {2004})}\BibitemShut {NoStop}%
\bibitem [{\citenamefont {Koz{\l}owski}\ \emph {et~al.}(2009)\citenamefont {Koz{\l}owski}, \citenamefont {Kochanek}, \citenamefont {Udalski}, \citenamefont {Soszy{\'n}ski}, \citenamefont {Szyma{\'n}ski}, \citenamefont {Kubiak}, \citenamefont {Pietrzy{\'n}ski}, \citenamefont {Szewczyk}, \citenamefont {Ulaczyk}, \citenamefont {Poleski} \emph {et~al.}}]{kozlowski2009quantifying}%
  \BibitemOpen
  \bibfield  {author} {\bibinfo {author} {\bibfnamefont {S.}~\bibnamefont {Koz{\l}owski}}, \bibinfo {author} {\bibfnamefont {C.~S.}\ \bibnamefont {Kochanek}}, \bibinfo {author} {\bibfnamefont {A.}~\bibnamefont {Udalski}}, \bibinfo {author} {\bibfnamefont {I.}~\bibnamefont {Soszy{\'n}ski}}, \bibinfo {author} {\bibfnamefont {M.}~\bibnamefont {Szyma{\'n}ski}}, \bibinfo {author} {\bibfnamefont {M.}~\bibnamefont {Kubiak}}, \bibinfo {author} {\bibfnamefont {G.}~\bibnamefont {Pietrzy{\'n}ski}}, \bibinfo {author} {\bibfnamefont {O.}~\bibnamefont {Szewczyk}}, \bibinfo {author} {\bibfnamefont {K.}~\bibnamefont {Ulaczyk}}, \bibinfo {author} {\bibfnamefont {R.}~\bibnamefont {Poleski}}, \emph {et~al.},\ }\bibfield  {title} {\bibinfo {title} {Quantifying quasar variability as part of a general approach to classifying continuously varying sources},\ }\href@noop {} {\bibfield  {journal} {\bibinfo  {journal} {The Astrophysical Journal}\ }\textbf {\bibinfo {volume} {708}},\ \bibinfo {pages} {927} (\bibinfo {year}
  {2009})}\BibitemShut {NoStop}%
\bibitem [{\citenamefont {Cherepashchuk}(2016)}]{cherepashchuk2016observing}%
  \BibitemOpen
  \bibfield  {author} {\bibinfo {author} {\bibfnamefont {A.~M.}\ \bibnamefont {Cherepashchuk}},\ }\bibfield  {title} {\bibinfo {title} {Observing stellar-mass and supermassive black holes},\ }\href@noop {} {\bibfield  {journal} {\bibinfo  {journal} {Physics-Uspekhi}\ }\textbf {\bibinfo {volume} {59}},\ \bibinfo {pages} {702} (\bibinfo {year} {2016})}\BibitemShut {NoStop}%
\bibitem [{\citenamefont {Milgrom}(1983)}]{milgrom1983modification}%
  \BibitemOpen
  \bibfield  {author} {\bibinfo {author} {\bibfnamefont {M.}~\bibnamefont {Milgrom}},\ }\bibfield  {title} {\bibinfo {title} {A modification of the newtonian dynamics as a possible alternative to the hidden mass hypothesis},\ }\href@noop {} {\bibfield  {journal} {\bibinfo  {journal} {Astrophysical Journal, Part 1 (ISSN 0004-637X), vol. 270, July 15, 1983, p. 365-370. Research supported by the US-Israel Binational Science Foundation.}\ }\textbf {\bibinfo {volume} {270}},\ \bibinfo {pages} {365} (\bibinfo {year} {1983})}\BibitemShut {NoStop}%
\bibitem [{\citenamefont {Ghez}\ \emph {et~al.}(2005)\citenamefont {Ghez}, \citenamefont {Salim}, \citenamefont {Hornstein}, \citenamefont {Tanner}, \citenamefont {Lu}, \citenamefont {Morris}, \citenamefont {Becklin},\ and\ \citenamefont {Duch{\^e}ne}}]{ghez2005stellar}%
  \BibitemOpen
  \bibfield  {author} {\bibinfo {author} {\bibfnamefont {A.}~\bibnamefont {Ghez}}, \bibinfo {author} {\bibfnamefont {S.}~\bibnamefont {Salim}}, \bibinfo {author} {\bibfnamefont {S.~D.}\ \bibnamefont {Hornstein}}, \bibinfo {author} {\bibfnamefont {A.}~\bibnamefont {Tanner}}, \bibinfo {author} {\bibfnamefont {J.}~\bibnamefont {Lu}}, \bibinfo {author} {\bibfnamefont {M.}~\bibnamefont {Morris}}, \bibinfo {author} {\bibfnamefont {E.}~\bibnamefont {Becklin}},\ and\ \bibinfo {author} {\bibfnamefont {G.}~\bibnamefont {Duch{\^e}ne}},\ }\bibfield  {title} {\bibinfo {title} {Stellar orbits around the galactic center black hole},\ }\href@noop {} {\bibfield  {journal} {\bibinfo  {journal} {The Astrophysical Journal}\ }\textbf {\bibinfo {volume} {620}},\ \bibinfo {pages} {744} (\bibinfo {year} {2005})}\BibitemShut {NoStop}%
\bibitem [{\citenamefont {Li}\ \emph {et~al.}(2023{\natexlab{a}})\citenamefont {Li}, \citenamefont {Zhong}, \citenamefont {Berczik}, \citenamefont {Spurzem}, \citenamefont {Chen},\ and\ \citenamefont {Liu}}]{li2023tracing}%
  \BibitemOpen
  \bibfield  {author} {\bibinfo {author} {\bibfnamefont {S.}~\bibnamefont {Li}}, \bibinfo {author} {\bibfnamefont {S.}~\bibnamefont {Zhong}}, \bibinfo {author} {\bibfnamefont {P.}~\bibnamefont {Berczik}}, \bibinfo {author} {\bibfnamefont {R.}~\bibnamefont {Spurzem}}, \bibinfo {author} {\bibfnamefont {X.}~\bibnamefont {Chen}},\ and\ \bibinfo {author} {\bibfnamefont {F.}~\bibnamefont {Liu}},\ }\bibfield  {title} {\bibinfo {title} {Tracing the evolution of smbhs and stellar objects in galaxy mergers: A multi-mass direct n-body model},\ }\href@noop {} {\bibfield  {journal} {\bibinfo  {journal} {The Astrophysical Journal}\ }\textbf {\bibinfo {volume} {944}},\ \bibinfo {pages} {109} (\bibinfo {year} {2023}{\natexlab{a}})}\BibitemShut {NoStop}%
\bibitem [{\citenamefont {Davis}\ \emph {et~al.}(2018)\citenamefont {Davis}, \citenamefont {Graham},\ and\ \citenamefont {Cameron}}]{davis2018black}%
  \BibitemOpen
  \bibfield  {author} {\bibinfo {author} {\bibfnamefont {B.~L.}\ \bibnamefont {Davis}}, \bibinfo {author} {\bibfnamefont {A.~W.}\ \bibnamefont {Graham}},\ and\ \bibinfo {author} {\bibfnamefont {E.}~\bibnamefont {Cameron}},\ }\bibfield  {title} {\bibinfo {title} {Black hole mass scaling relations for spiral galaxies. ii. mbh--m*, tot and mbh--m*, disk},\ }\href@noop {} {\bibfield  {journal} {\bibinfo  {journal} {The Astrophysical Journal}\ }\textbf {\bibinfo {volume} {869}},\ \bibinfo {pages} {113} (\bibinfo {year} {2018})}\BibitemShut {NoStop}%
\bibitem [{\citenamefont {Davis}\ \emph {et~al.}(2019)\citenamefont {Davis}, \citenamefont {Graham},\ and\ \citenamefont {Cameron}}]{davis2019black}%
  \BibitemOpen
  \bibfield  {author} {\bibinfo {author} {\bibfnamefont {B.~L.}\ \bibnamefont {Davis}}, \bibinfo {author} {\bibfnamefont {A.~W.}\ \bibnamefont {Graham}},\ and\ \bibinfo {author} {\bibfnamefont {E.}~\bibnamefont {Cameron}},\ }\bibfield  {title} {\bibinfo {title} {Black hole mass scaling relations for spiral galaxies. i. mbh--m*, sph},\ }\href@noop {} {\bibfield  {journal} {\bibinfo  {journal} {The Astrophysical Journal}\ }\textbf {\bibinfo {volume} {873}},\ \bibinfo {pages} {85} (\bibinfo {year} {2019})}\BibitemShut {NoStop}%
\bibitem [{\citenamefont {Osmer}(1982)}]{osmer1982quasars}%
  \BibitemOpen
  \bibfield  {author} {\bibinfo {author} {\bibfnamefont {P.~S.}\ \bibnamefont {Osmer}},\ }\bibfield  {title} {\bibinfo {title} {Quasars as probes of the distant and early universe},\ }\href@noop {} {\bibfield  {journal} {\bibinfo  {journal} {Scientific American}\ }\textbf {\bibinfo {volume} {246}},\ \bibinfo {pages} {126} (\bibinfo {year} {1982})}\BibitemShut {NoStop}%
\bibitem [{\citenamefont {Fan}\ \emph {et~al.}(1999)\citenamefont {Fan}, \citenamefont {Strauss}, \citenamefont {Schneider}, \citenamefont {Gunn}, \citenamefont {Lupton}, \citenamefont {Yanny}, \citenamefont {Anderson}, \citenamefont {Anderson~Jr}, \citenamefont {Annis}, \citenamefont {Bahcall} \emph {et~al.}}]{fan1999high}%
  \BibitemOpen
  \bibfield  {author} {\bibinfo {author} {\bibfnamefont {X.}~\bibnamefont {Fan}}, \bibinfo {author} {\bibfnamefont {M.~A.}\ \bibnamefont {Strauss}}, \bibinfo {author} {\bibfnamefont {D.~P.}\ \bibnamefont {Schneider}}, \bibinfo {author} {\bibfnamefont {J.~E.}\ \bibnamefont {Gunn}}, \bibinfo {author} {\bibfnamefont {R.~H.}\ \bibnamefont {Lupton}}, \bibinfo {author} {\bibfnamefont {B.}~\bibnamefont {Yanny}}, \bibinfo {author} {\bibfnamefont {S.~F.}\ \bibnamefont {Anderson}}, \bibinfo {author} {\bibfnamefont {J.~E.}\ \bibnamefont {Anderson~Jr}}, \bibinfo {author} {\bibfnamefont {J.}~\bibnamefont {Annis}}, \bibinfo {author} {\bibfnamefont {N.~A.}\ \bibnamefont {Bahcall}}, \emph {et~al.},\ }\bibfield  {title} {\bibinfo {title} {High-redshift quasars found in sloan digital sky survey commissioning data},\ }\href@noop {} {\bibfield  {journal} {\bibinfo  {journal} {The Astronomical Journal}\ }\textbf {\bibinfo {volume} {118}},\ \bibinfo {pages} {1} (\bibinfo {year} {1999})}\BibitemShut {NoStop}%
\bibitem [{\citenamefont {McLure}\ and\ \citenamefont {Jarvis}(2002)}]{mclure2002measuring}%
  \BibitemOpen
  \bibfield  {author} {\bibinfo {author} {\bibfnamefont {R.~J.}\ \bibnamefont {McLure}}\ and\ \bibinfo {author} {\bibfnamefont {M.~J.}\ \bibnamefont {Jarvis}},\ }\bibfield  {title} {\bibinfo {title} {Measuring the black hole masses of high-redshift quasars},\ }\href@noop {} {\bibfield  {journal} {\bibinfo  {journal} {Monthly Notices of the Royal Astronomical Society}\ }\textbf {\bibinfo {volume} {337}},\ \bibinfo {pages} {109} (\bibinfo {year} {2002})}\BibitemShut {NoStop}%
\bibitem [{\citenamefont {Shakura}\ and\ \citenamefont {Sunyaev}(1973)}]{shakura1973black}%
  \BibitemOpen
  \bibfield  {author} {\bibinfo {author} {\bibfnamefont {N.~I.}\ \bibnamefont {Shakura}}\ and\ \bibinfo {author} {\bibfnamefont {R.~A.}\ \bibnamefont {Sunyaev}},\ }\bibfield  {title} {\bibinfo {title} {Black holes in binary systems. observational appearance.},\ }\href@noop {} {\bibfield  {journal} {\bibinfo  {journal} {Astronomy and Astrophysics, Vol. 24, p. 337-355}\ }\textbf {\bibinfo {volume} {24}},\ \bibinfo {pages} {337} (\bibinfo {year} {1973})}\BibitemShut {NoStop}%
\bibitem [{\citenamefont {Ding}\ \emph {et~al.}(2022)\citenamefont {Ding}, \citenamefont {Li}, \citenamefont {Ho},\ and\ \citenamefont {Ricci}}]{ding2022accretion}%
  \BibitemOpen
  \bibfield  {author} {\bibinfo {author} {\bibfnamefont {Y.}~\bibnamefont {Ding}}, \bibinfo {author} {\bibfnamefont {R.}~\bibnamefont {Li}}, \bibinfo {author} {\bibfnamefont {L.~C.}\ \bibnamefont {Ho}},\ and\ \bibinfo {author} {\bibfnamefont {C.}~\bibnamefont {Ricci}},\ }\bibfield  {title} {\bibinfo {title} {Accretion disk outflow during the x-ray flare of the super-eddington active nucleus of i zwicky 1},\ }\href@noop {} {\bibfield  {journal} {\bibinfo  {journal} {The Astrophysical Journal}\ }\textbf {\bibinfo {volume} {931}},\ \bibinfo {pages} {77} (\bibinfo {year} {2022})}\BibitemShut {NoStop}%
\bibitem [{\citenamefont {Kaspi}\ \emph {et~al.}(2000)\citenamefont {Kaspi}, \citenamefont {Smith}, \citenamefont {Netzer}, \citenamefont {Maoz}, \citenamefont {Jannuzi},\ and\ \citenamefont {Giveon}}]{kaspi2000reverberation}%
  \BibitemOpen
  \bibfield  {author} {\bibinfo {author} {\bibfnamefont {S.}~\bibnamefont {Kaspi}}, \bibinfo {author} {\bibfnamefont {P.~S.}\ \bibnamefont {Smith}}, \bibinfo {author} {\bibfnamefont {H.}~\bibnamefont {Netzer}}, \bibinfo {author} {\bibfnamefont {D.}~\bibnamefont {Maoz}}, \bibinfo {author} {\bibfnamefont {B.~T.}\ \bibnamefont {Jannuzi}},\ and\ \bibinfo {author} {\bibfnamefont {U.}~\bibnamefont {Giveon}},\ }\bibfield  {title} {\bibinfo {title} {Reverberation measurements for 17 quasars and the size-mass-luminosity relations in active galactic nuclei},\ }\href@noop {} {\bibfield  {journal} {\bibinfo  {journal} {The Astrophysical Journal}\ }\textbf {\bibinfo {volume} {533}},\ \bibinfo {pages} {631} (\bibinfo {year} {2000})}\BibitemShut {NoStop}%
\bibitem [{\citenamefont {Fan}\ \emph {et~al.}(2003)\citenamefont {Fan}, \citenamefont {Strauss}, \citenamefont {Schneider}, \citenamefont {Becker}, \citenamefont {White}, \citenamefont {Haiman}, \citenamefont {Gregg}, \citenamefont {Pentericci}, \citenamefont {Grebel}, \citenamefont {Narayanan} \emph {et~al.}}]{fan2003survey}%
  \BibitemOpen
  \bibfield  {author} {\bibinfo {author} {\bibfnamefont {X.}~\bibnamefont {Fan}}, \bibinfo {author} {\bibfnamefont {M.~A.}\ \bibnamefont {Strauss}}, \bibinfo {author} {\bibfnamefont {D.~P.}\ \bibnamefont {Schneider}}, \bibinfo {author} {\bibfnamefont {R.~H.}\ \bibnamefont {Becker}}, \bibinfo {author} {\bibfnamefont {R.~L.}\ \bibnamefont {White}}, \bibinfo {author} {\bibfnamefont {Z.}~\bibnamefont {Haiman}}, \bibinfo {author} {\bibfnamefont {M.}~\bibnamefont {Gregg}}, \bibinfo {author} {\bibfnamefont {L.}~\bibnamefont {Pentericci}}, \bibinfo {author} {\bibfnamefont {E.~K.}\ \bibnamefont {Grebel}}, \bibinfo {author} {\bibfnamefont {V.~K.}\ \bibnamefont {Narayanan}}, \emph {et~al.},\ }\bibfield  {title} {\bibinfo {title} {A survey of z> 5.7 quasars in the sloan digital sky survey. ii. discovery of three additional quasars at z> 6},\ }\href@noop {} {\bibfield  {journal} {\bibinfo  {journal} {The Astronomical Journal}\ }\textbf {\bibinfo {volume} {125}},\ \bibinfo {pages} {1649} (\bibinfo {year}
  {2003})}\BibitemShut {NoStop}%
\bibitem [{\citenamefont {Shen}\ \emph {et~al.}(2019{\natexlab{a}})\citenamefont {Shen}, \citenamefont {Wu}, \citenamefont {Jiang}, \citenamefont {Ba{\~n}ados}, \citenamefont {Fan}, \citenamefont {Ho}, \citenamefont {Riechers}, \citenamefont {Strauss}, \citenamefont {Venemans}, \citenamefont {Vestergaard} \emph {et~al.}}]{shen2019gemini}%
  \BibitemOpen
  \bibfield  {author} {\bibinfo {author} {\bibfnamefont {Y.}~\bibnamefont {Shen}}, \bibinfo {author} {\bibfnamefont {J.}~\bibnamefont {Wu}}, \bibinfo {author} {\bibfnamefont {L.}~\bibnamefont {Jiang}}, \bibinfo {author} {\bibfnamefont {E.}~\bibnamefont {Ba{\~n}ados}}, \bibinfo {author} {\bibfnamefont {X.}~\bibnamefont {Fan}}, \bibinfo {author} {\bibfnamefont {L.~C.}\ \bibnamefont {Ho}}, \bibinfo {author} {\bibfnamefont {D.~A.}\ \bibnamefont {Riechers}}, \bibinfo {author} {\bibfnamefont {M.~A.}\ \bibnamefont {Strauss}}, \bibinfo {author} {\bibfnamefont {B.}~\bibnamefont {Venemans}}, \bibinfo {author} {\bibfnamefont {M.}~\bibnamefont {Vestergaard}}, \emph {et~al.},\ }\bibfield  {title} {\bibinfo {title} {Gemini gnirs near-infrared spectroscopy of 50 quasars at $z \geq 5.7$},\ }\href@noop {} {\bibfield  {journal} {\bibinfo  {journal} {The Astrophysical Journal}\ }\textbf {\bibinfo {volume} {873}},\ \bibinfo {pages} {35} (\bibinfo {year} {2019}{\natexlab{a}})}\BibitemShut {NoStop}%
\bibitem [{\citenamefont {Kawaguchi}\ \emph {et~al.}(2004)\citenamefont {Kawaguchi}, \citenamefont {Aoki}, \citenamefont {Ohta},\ and\ \citenamefont {Collin}}]{kawaguchi2004growth}%
  \BibitemOpen
  \bibfield  {author} {\bibinfo {author} {\bibfnamefont {T.}~\bibnamefont {Kawaguchi}}, \bibinfo {author} {\bibfnamefont {K.}~\bibnamefont {Aoki}}, \bibinfo {author} {\bibfnamefont {K.}~\bibnamefont {Ohta}},\ and\ \bibinfo {author} {\bibfnamefont {S.}~\bibnamefont {Collin}},\ }\bibfield  {title} {\bibinfo {title} {Growth of massive black holes by super-eddington accretion},\ }\href@noop {} {\bibfield  {journal} {\bibinfo  {journal} {Astronomy \& Astrophysics}\ }\textbf {\bibinfo {volume} {420}},\ \bibinfo {pages} {L23} (\bibinfo {year} {2004})}\BibitemShut {NoStop}%
\bibitem [{\citenamefont {Kawasaki}\ \emph {et~al.}(2012)\citenamefont {Kawasaki}, \citenamefont {Kusenko},\ and\ \citenamefont {Yanagida}}]{kawasaki2012primordial}%
  \BibitemOpen
  \bibfield  {author} {\bibinfo {author} {\bibfnamefont {M.}~\bibnamefont {Kawasaki}}, \bibinfo {author} {\bibfnamefont {A.}~\bibnamefont {Kusenko}},\ and\ \bibinfo {author} {\bibfnamefont {T.~T.}\ \bibnamefont {Yanagida}},\ }\bibfield  {title} {\bibinfo {title} {Primordial seeds of supermassive black holes},\ }\href@noop {} {\bibfield  {journal} {\bibinfo  {journal} {Physics Letters B}\ }\textbf {\bibinfo {volume} {711}},\ \bibinfo {pages} {1} (\bibinfo {year} {2012})}\BibitemShut {NoStop}%
\bibitem [{\citenamefont {Pohlen}\ and\ \citenamefont {Trujillo}(2006)}]{pohlen2006structure}%
  \BibitemOpen
  \bibfield  {author} {\bibinfo {author} {\bibfnamefont {M.}~\bibnamefont {Pohlen}}\ and\ \bibinfo {author} {\bibfnamefont {I.}~\bibnamefont {Trujillo}},\ }\bibfield  {title} {\bibinfo {title} {The structure of galactic disks-studying late-type spiral galaxies using sdss},\ }\href@noop {} {\bibfield  {journal} {\bibinfo  {journal} {Astronomy \& Astrophysics}\ }\textbf {\bibinfo {volume} {454}},\ \bibinfo {pages} {759} (\bibinfo {year} {2006})}\BibitemShut {NoStop}%
\bibitem [{\citenamefont {Percival}\ \emph {et~al.}(2007)\citenamefont {Percival}, \citenamefont {Nichol}, \citenamefont {Eisenstein}, \citenamefont {Weinberg}, \citenamefont {Fukugita}, \citenamefont {Pope}, \citenamefont {Schneider}, \citenamefont {Szalay}, \citenamefont {Vogeley}, \citenamefont {Zehavi} \emph {et~al.}}]{percival2007measuring}%
  \BibitemOpen
  \bibfield  {author} {\bibinfo {author} {\bibfnamefont {W.~J.}\ \bibnamefont {Percival}}, \bibinfo {author} {\bibfnamefont {R.~C.}\ \bibnamefont {Nichol}}, \bibinfo {author} {\bibfnamefont {D.~J.}\ \bibnamefont {Eisenstein}}, \bibinfo {author} {\bibfnamefont {D.~H.}\ \bibnamefont {Weinberg}}, \bibinfo {author} {\bibfnamefont {M.}~\bibnamefont {Fukugita}}, \bibinfo {author} {\bibfnamefont {A.~C.}\ \bibnamefont {Pope}}, \bibinfo {author} {\bibfnamefont {D.~P.}\ \bibnamefont {Schneider}}, \bibinfo {author} {\bibfnamefont {A.~S.}\ \bibnamefont {Szalay}}, \bibinfo {author} {\bibfnamefont {M.~S.}\ \bibnamefont {Vogeley}}, \bibinfo {author} {\bibfnamefont {I.}~\bibnamefont {Zehavi}}, \emph {et~al.},\ }\bibfield  {title} {\bibinfo {title} {Measuring the matter density using baryon oscillations in the sdss},\ }\href@noop {} {\bibfield  {journal} {\bibinfo  {journal} {The Astrophysical Journal}\ }\textbf {\bibinfo {volume} {657}},\ \bibinfo {pages} {51} (\bibinfo {year} {2007})}\BibitemShut {NoStop}%
\bibitem [{\citenamefont {Lyke}\ \emph {et~al.}(2020)\citenamefont {Lyke}, \citenamefont {Higley}, \citenamefont {McLane}, \citenamefont {Schurhammer}, \citenamefont {Myers}, \citenamefont {Ross}, \citenamefont {Dawson}, \citenamefont {Chabanier}, \citenamefont {Martini}, \citenamefont {Des~Bourboux} \emph {et~al.}}]{lyke2020sloan}%
  \BibitemOpen
  \bibfield  {author} {\bibinfo {author} {\bibfnamefont {B.~W.}\ \bibnamefont {Lyke}}, \bibinfo {author} {\bibfnamefont {A.~N.}\ \bibnamefont {Higley}}, \bibinfo {author} {\bibfnamefont {J.}~\bibnamefont {McLane}}, \bibinfo {author} {\bibfnamefont {D.~P.}\ \bibnamefont {Schurhammer}}, \bibinfo {author} {\bibfnamefont {A.~D.}\ \bibnamefont {Myers}}, \bibinfo {author} {\bibfnamefont {A.~J.}\ \bibnamefont {Ross}}, \bibinfo {author} {\bibfnamefont {K.}~\bibnamefont {Dawson}}, \bibinfo {author} {\bibfnamefont {S.}~\bibnamefont {Chabanier}}, \bibinfo {author} {\bibfnamefont {P.}~\bibnamefont {Martini}}, \bibinfo {author} {\bibfnamefont {H.~D.~M.}\ \bibnamefont {Des~Bourboux}}, \emph {et~al.},\ }\bibfield  {title} {\bibinfo {title} {The sloan digital sky survey quasar catalog: Sixteenth data release},\ }\href@noop {} {\bibfield  {journal} {\bibinfo  {journal} {The Astrophysical Journal Supplement Series}\ }\textbf {\bibinfo {volume} {250}},\ \bibinfo {pages} {8} (\bibinfo {year} {2020})}\BibitemShut {NoStop}%
\bibitem [{\citenamefont {Abazajian}\ \emph {et~al.}(2009)\citenamefont {Abazajian}, \citenamefont {Adelman-McCarthy}, \citenamefont {Ag{\"u}eros}, \citenamefont {Allam}, \citenamefont {Prieto}, \citenamefont {An}, \citenamefont {Anderson}, \citenamefont {Anderson}, \citenamefont {Annis}, \citenamefont {Bahcall} \emph {et~al.}}]{abazajian2009seventh}%
  \BibitemOpen
  \bibfield  {author} {\bibinfo {author} {\bibfnamefont {K.~N.}\ \bibnamefont {Abazajian}}, \bibinfo {author} {\bibfnamefont {J.~K.}\ \bibnamefont {Adelman-McCarthy}}, \bibinfo {author} {\bibfnamefont {M.~A.}\ \bibnamefont {Ag{\"u}eros}}, \bibinfo {author} {\bibfnamefont {S.~S.}\ \bibnamefont {Allam}}, \bibinfo {author} {\bibfnamefont {C.~A.}\ \bibnamefont {Prieto}}, \bibinfo {author} {\bibfnamefont {D.}~\bibnamefont {An}}, \bibinfo {author} {\bibfnamefont {K.~S.}\ \bibnamefont {Anderson}}, \bibinfo {author} {\bibfnamefont {S.~F.}\ \bibnamefont {Anderson}}, \bibinfo {author} {\bibfnamefont {J.}~\bibnamefont {Annis}}, \bibinfo {author} {\bibfnamefont {N.~A.}\ \bibnamefont {Bahcall}}, \emph {et~al.},\ }\bibfield  {title} {\bibinfo {title} {The seventh data release of the sloan digital sky survey},\ }\href@noop {} {\bibfield  {journal} {\bibinfo  {journal} {The Astrophysical Journal Supplement Series}\ }\textbf {\bibinfo {volume} {182}},\ \bibinfo {pages} {543} (\bibinfo {year} {2009})}\BibitemShut {NoStop}%
\bibitem [{\citenamefont {Richards}\ \emph {et~al.}(2006)\citenamefont {Richards}, \citenamefont {Strauss}, \citenamefont {Fan}, \citenamefont {Hall}, \citenamefont {Jester}, \citenamefont {Schneider}, \citenamefont {Berk}, \citenamefont {Stoughton}, \citenamefont {Anderson}, \citenamefont {Brunner} \emph {et~al.}}]{richards2006sloan}%
  \BibitemOpen
  \bibfield  {author} {\bibinfo {author} {\bibfnamefont {G.~T.}\ \bibnamefont {Richards}}, \bibinfo {author} {\bibfnamefont {M.~A.}\ \bibnamefont {Strauss}}, \bibinfo {author} {\bibfnamefont {X.}~\bibnamefont {Fan}}, \bibinfo {author} {\bibfnamefont {P.~B.}\ \bibnamefont {Hall}}, \bibinfo {author} {\bibfnamefont {S.}~\bibnamefont {Jester}}, \bibinfo {author} {\bibfnamefont {D.~P.}\ \bibnamefont {Schneider}}, \bibinfo {author} {\bibfnamefont {D.~E.~V.}\ \bibnamefont {Berk}}, \bibinfo {author} {\bibfnamefont {C.}~\bibnamefont {Stoughton}}, \bibinfo {author} {\bibfnamefont {S.~F.}\ \bibnamefont {Anderson}}, \bibinfo {author} {\bibfnamefont {R.~J.}\ \bibnamefont {Brunner}}, \emph {et~al.},\ }\bibfield  {title} {\bibinfo {title} {The sloan digital sky survey quasar survey: Quasar luminosity function from data release 3},\ }\href@noop {} {\bibfield  {journal} {\bibinfo  {journal} {The Astronomical Journal}\ }\textbf {\bibinfo {volume} {131}},\ \bibinfo {pages} {2766} (\bibinfo {year} {2006})}\BibitemShut {NoStop}%
\bibitem [{\citenamefont {Abolfathi}\ \emph {et~al.}(2018)\citenamefont {Abolfathi}, \citenamefont {Aguado}, \citenamefont {Aguilar}, \citenamefont {Prieto}, \citenamefont {Almeida}, \citenamefont {Ananna}, \citenamefont {Anders}, \citenamefont {Anderson}, \citenamefont {Andrews}, \citenamefont {Anguiano} \emph {et~al.}}]{abolfathi2018fourteenth}%
  \BibitemOpen
  \bibfield  {author} {\bibinfo {author} {\bibfnamefont {B.}~\bibnamefont {Abolfathi}}, \bibinfo {author} {\bibfnamefont {D.}~\bibnamefont {Aguado}}, \bibinfo {author} {\bibfnamefont {G.}~\bibnamefont {Aguilar}}, \bibinfo {author} {\bibfnamefont {C.~A.}\ \bibnamefont {Prieto}}, \bibinfo {author} {\bibfnamefont {A.}~\bibnamefont {Almeida}}, \bibinfo {author} {\bibfnamefont {T.~T.}\ \bibnamefont {Ananna}}, \bibinfo {author} {\bibfnamefont {F.}~\bibnamefont {Anders}}, \bibinfo {author} {\bibfnamefont {S.~F.}\ \bibnamefont {Anderson}}, \bibinfo {author} {\bibfnamefont {B.~H.}\ \bibnamefont {Andrews}}, \bibinfo {author} {\bibfnamefont {B.}~\bibnamefont {Anguiano}}, \emph {et~al.},\ }\bibfield  {title} {\bibinfo {title} {The fourteenth data release of the sloan digital sky survey: First spectroscopic data from the extended baryon oscillation spectroscopic survey and from the second phase of the apache point observatory galactic evolution experiment},\ }\href@noop {} {\bibfield  {journal} {\bibinfo  {journal} {The
  Astrophysical Journal Supplement Series}\ }\textbf {\bibinfo {volume} {235}},\ \bibinfo {pages} {42} (\bibinfo {year} {2018})}\BibitemShut {NoStop}%
\bibitem [{\citenamefont {Beutler}\ \emph {et~al.}(2017)\citenamefont {Beutler}, \citenamefont {Seo}, \citenamefont {Saito}, \citenamefont {Chuang}, \citenamefont {Cuesta}, \citenamefont {Eisenstein}, \citenamefont {Gil-Mar{\'\i}n}, \citenamefont {Grieb}, \citenamefont {Hand}, \citenamefont {Kitaura} \emph {et~al.}}]{beutler2017clustering}%
  \BibitemOpen
  \bibfield  {author} {\bibinfo {author} {\bibfnamefont {F.}~\bibnamefont {Beutler}}, \bibinfo {author} {\bibfnamefont {H.-J.}\ \bibnamefont {Seo}}, \bibinfo {author} {\bibfnamefont {S.}~\bibnamefont {Saito}}, \bibinfo {author} {\bibfnamefont {C.-H.}\ \bibnamefont {Chuang}}, \bibinfo {author} {\bibfnamefont {A.~J.}\ \bibnamefont {Cuesta}}, \bibinfo {author} {\bibfnamefont {D.~J.}\ \bibnamefont {Eisenstein}}, \bibinfo {author} {\bibfnamefont {H.}~\bibnamefont {Gil-Mar{\'\i}n}}, \bibinfo {author} {\bibfnamefont {J.~N.}\ \bibnamefont {Grieb}}, \bibinfo {author} {\bibfnamefont {N.}~\bibnamefont {Hand}}, \bibinfo {author} {\bibfnamefont {F.-S.}\ \bibnamefont {Kitaura}}, \emph {et~al.},\ }\bibfield  {title} {\bibinfo {title} {The clustering of galaxies in the completed sdss-iii baryon oscillation spectroscopic survey: anisotropic galaxy clustering in fourier space},\ }\href@noop {} {\bibfield  {journal} {\bibinfo  {journal} {Monthly Notices of the Royal Astronomical Society}\ }\textbf {\bibinfo {volume} {466}},\
  \bibinfo {pages} {2242} (\bibinfo {year} {2017})}\BibitemShut {NoStop}%
\bibitem [{\citenamefont {Zhao}\ \emph {et~al.}(2019)\citenamefont {Zhao}, \citenamefont {Wang}, \citenamefont {Saito}, \citenamefont {Gil-Mar{\'\i}n}, \citenamefont {Percival}, \citenamefont {Wang}, \citenamefont {Chuang}, \citenamefont {Ruggeri}, \citenamefont {Mueller}, \citenamefont {Zhu} \emph {et~al.}}]{zhao2019clustering}%
  \BibitemOpen
  \bibfield  {author} {\bibinfo {author} {\bibfnamefont {G.-B.}\ \bibnamefont {Zhao}}, \bibinfo {author} {\bibfnamefont {Y.}~\bibnamefont {Wang}}, \bibinfo {author} {\bibfnamefont {S.}~\bibnamefont {Saito}}, \bibinfo {author} {\bibfnamefont {H.}~\bibnamefont {Gil-Mar{\'\i}n}}, \bibinfo {author} {\bibfnamefont {W.~J.}\ \bibnamefont {Percival}}, \bibinfo {author} {\bibfnamefont {D.}~\bibnamefont {Wang}}, \bibinfo {author} {\bibfnamefont {C.-H.}\ \bibnamefont {Chuang}}, \bibinfo {author} {\bibfnamefont {R.}~\bibnamefont {Ruggeri}}, \bibinfo {author} {\bibfnamefont {E.-M.}\ \bibnamefont {Mueller}}, \bibinfo {author} {\bibfnamefont {F.}~\bibnamefont {Zhu}}, \emph {et~al.},\ }\bibfield  {title} {\bibinfo {title} {The clustering of the sdss-iv extended baryon oscillation spectroscopic survey dr14 quasar sample: a tomographic measurement of cosmic structure growth and expansion rate based on optimal redshift weights},\ }\href@noop {} {\bibfield  {journal} {\bibinfo  {journal} {Monthly Notices of the Royal
  Astronomical Society}\ }\textbf {\bibinfo {volume} {482}},\ \bibinfo {pages} {3497} (\bibinfo {year} {2019})}\BibitemShut {NoStop}%
\bibitem [{\citenamefont {Alam}\ \emph {et~al.}(2021)\citenamefont {Alam}, \citenamefont {Aubert}, \citenamefont {Avila}, \citenamefont {Balland}, \citenamefont {Bautista}, \citenamefont {Bershady}, \citenamefont {Bizyaev}, \citenamefont {Blanton}, \citenamefont {Bolton}, \citenamefont {Bovy} \emph {et~al.}}]{alam2021completed}%
  \BibitemOpen
  \bibfield  {author} {\bibinfo {author} {\bibfnamefont {S.}~\bibnamefont {Alam}}, \bibinfo {author} {\bibfnamefont {M.}~\bibnamefont {Aubert}}, \bibinfo {author} {\bibfnamefont {S.}~\bibnamefont {Avila}}, \bibinfo {author} {\bibfnamefont {C.}~\bibnamefont {Balland}}, \bibinfo {author} {\bibfnamefont {J.~E.}\ \bibnamefont {Bautista}}, \bibinfo {author} {\bibfnamefont {M.~A.}\ \bibnamefont {Bershady}}, \bibinfo {author} {\bibfnamefont {D.}~\bibnamefont {Bizyaev}}, \bibinfo {author} {\bibfnamefont {M.~R.}\ \bibnamefont {Blanton}}, \bibinfo {author} {\bibfnamefont {A.~S.}\ \bibnamefont {Bolton}}, \bibinfo {author} {\bibfnamefont {J.}~\bibnamefont {Bovy}}, \emph {et~al.},\ }\bibfield  {title} {\bibinfo {title} {Completed sdss-iv extended baryon oscillation spectroscopic survey: Cosmological implications from two decades of spectroscopic surveys at the apache point observatory},\ }\href@noop {} {\bibfield  {journal} {\bibinfo  {journal} {Physical Review D}\ }\textbf {\bibinfo {volume} {103}},\ \bibinfo {pages}
  {083533} (\bibinfo {year} {2021})}\BibitemShut {NoStop}%
\bibitem [{\citenamefont {Xu}\ \emph {et~al.}(2023)\citenamefont {Xu}, \citenamefont {Jing}, \citenamefont {Zhao},\ and\ \citenamefont {Cuesta}}]{xu2023evidence}%
  \BibitemOpen
  \bibfield  {author} {\bibinfo {author} {\bibfnamefont {K.}~\bibnamefont {Xu}}, \bibinfo {author} {\bibfnamefont {Y.}~\bibnamefont {Jing}}, \bibinfo {author} {\bibfnamefont {G.-B.}\ \bibnamefont {Zhao}},\ and\ \bibinfo {author} {\bibfnamefont {A.~J.}\ \bibnamefont {Cuesta}},\ }\bibfield  {title} {\bibinfo {title} {Evidence for baryon acoustic oscillations from galaxy--ellipticity correlations},\ }\href@noop {} {\bibfield  {journal} {\bibinfo  {journal} {Nature Astronomy}\ }\textbf {\bibinfo {volume} {7}},\ \bibinfo {pages} {1259} (\bibinfo {year} {2023})}\BibitemShut {NoStop}%
\bibitem [{\citenamefont {Eisenstein}\ \emph {et~al.}(2005)\citenamefont {Eisenstein}, \citenamefont {Zehavi}, \citenamefont {Hogg}, \citenamefont {Scoccimarro}, \citenamefont {Blanton}, \citenamefont {Nichol}, \citenamefont {Scranton}, \citenamefont {Seo}, \citenamefont {Tegmark}, \citenamefont {Zheng} \emph {et~al.}}]{eisenstein2005detection}%
  \BibitemOpen
  \bibfield  {author} {\bibinfo {author} {\bibfnamefont {D.~J.}\ \bibnamefont {Eisenstein}}, \bibinfo {author} {\bibfnamefont {I.}~\bibnamefont {Zehavi}}, \bibinfo {author} {\bibfnamefont {D.~W.}\ \bibnamefont {Hogg}}, \bibinfo {author} {\bibfnamefont {R.}~\bibnamefont {Scoccimarro}}, \bibinfo {author} {\bibfnamefont {M.~R.}\ \bibnamefont {Blanton}}, \bibinfo {author} {\bibfnamefont {R.~C.}\ \bibnamefont {Nichol}}, \bibinfo {author} {\bibfnamefont {R.}~\bibnamefont {Scranton}}, \bibinfo {author} {\bibfnamefont {H.-J.}\ \bibnamefont {Seo}}, \bibinfo {author} {\bibfnamefont {M.}~\bibnamefont {Tegmark}}, \bibinfo {author} {\bibfnamefont {Z.}~\bibnamefont {Zheng}}, \emph {et~al.},\ }\bibfield  {title} {\bibinfo {title} {Detection of the baryon acoustic peak in the large-scale correlation function of sdss luminous red galaxies},\ }\href@noop {} {\bibfield  {journal} {\bibinfo  {journal} {The Astrophysical Journal}\ }\textbf {\bibinfo {volume} {633}},\ \bibinfo {pages} {560} (\bibinfo {year} {2005})}\BibitemShut
  {NoStop}%
\bibitem [{\citenamefont {Schlegel}\ \emph {et~al.}(2009)\citenamefont {Schlegel}, \citenamefont {White},\ and\ \citenamefont {Eisenstein}}]{schlegel2009baryon}%
  \BibitemOpen
  \bibfield  {author} {\bibinfo {author} {\bibfnamefont {D.}~\bibnamefont {Schlegel}}, \bibinfo {author} {\bibfnamefont {M.}~\bibnamefont {White}},\ and\ \bibinfo {author} {\bibfnamefont {D.}~\bibnamefont {Eisenstein}},\ }\bibfield  {title} {\bibinfo {title} {The baryon oscillation spectroscopic survey: Precision measurements of the absolute cosmic distance scale},\ }\href@noop {} {\bibfield  {journal} {\bibinfo  {journal} {arXiv preprint arXiv:0902.4680}\ } (\bibinfo {year} {2009})}\BibitemShut {NoStop}%
\bibitem [{\citenamefont {Aubourg}\ \emph {et~al.}(2015)\citenamefont {Aubourg}, \citenamefont {Bailey}, \citenamefont {Bautista}, \citenamefont {Beutler}, \citenamefont {Bhardwaj}, \citenamefont {Bizyaev}, \citenamefont {Blanton}, \citenamefont {Blomqvist}, \citenamefont {Bolton}, \citenamefont {Bovy} \emph {et~al.}}]{aubourg2015cosmological}%
  \BibitemOpen
  \bibfield  {author} {\bibinfo {author} {\bibfnamefont {{\'E}.}~\bibnamefont {Aubourg}}, \bibinfo {author} {\bibfnamefont {S.}~\bibnamefont {Bailey}}, \bibinfo {author} {\bibfnamefont {J.~E.}\ \bibnamefont {Bautista}}, \bibinfo {author} {\bibfnamefont {F.}~\bibnamefont {Beutler}}, \bibinfo {author} {\bibfnamefont {V.}~\bibnamefont {Bhardwaj}}, \bibinfo {author} {\bibfnamefont {D.}~\bibnamefont {Bizyaev}}, \bibinfo {author} {\bibfnamefont {M.}~\bibnamefont {Blanton}}, \bibinfo {author} {\bibfnamefont {M.}~\bibnamefont {Blomqvist}}, \bibinfo {author} {\bibfnamefont {A.~S.}\ \bibnamefont {Bolton}}, \bibinfo {author} {\bibfnamefont {J.}~\bibnamefont {Bovy}}, \emph {et~al.},\ }\bibfield  {title} {\bibinfo {title} {Cosmological implications of baryon acoustic oscillation measurements},\ }\href@noop {} {\bibfield  {journal} {\bibinfo  {journal} {Physical Review D}\ }\textbf {\bibinfo {volume} {92}},\ \bibinfo {pages} {123516} (\bibinfo {year} {2015})}\BibitemShut {NoStop}%
\bibitem [{\citenamefont {Wang}\ \emph {et~al.}(2017)\citenamefont {Wang}, \citenamefont {Xu},\ and\ \citenamefont {Zhao}}]{wang2017measurement}%
  \BibitemOpen
  \bibfield  {author} {\bibinfo {author} {\bibfnamefont {Y.}~\bibnamefont {Wang}}, \bibinfo {author} {\bibfnamefont {L.}~\bibnamefont {Xu}},\ and\ \bibinfo {author} {\bibfnamefont {G.-B.}\ \bibnamefont {Zhao}},\ }\bibfield  {title} {\bibinfo {title} {A measurement of the hubble constant using galaxy redshift surveys},\ }\href@noop {} {\bibfield  {journal} {\bibinfo  {journal} {The Astrophysical Journal}\ }\textbf {\bibinfo {volume} {849}},\ \bibinfo {pages} {84} (\bibinfo {year} {2017})}\BibitemShut {NoStop}%
\bibitem [{\citenamefont {Bondi}\ and\ \citenamefont {Hoyle}(1944)}]{bondi1944mechanism}%
  \BibitemOpen
  \bibfield  {author} {\bibinfo {author} {\bibfnamefont {H.}~\bibnamefont {Bondi}}\ and\ \bibinfo {author} {\bibfnamefont {F.}~\bibnamefont {Hoyle}},\ }\bibfield  {title} {\bibinfo {title} {On the mechanism of accretion by stars},\ }\href@noop {} {\bibfield  {journal} {\bibinfo  {journal} {Monthly Notices of the Royal Astronomical Society}\ }\textbf {\bibinfo {volume} {104}},\ \bibinfo {pages} {273} (\bibinfo {year} {1944})}\BibitemShut {NoStop}%
\bibitem [{\citenamefont {Bondi}(1952)}]{bondi1952spherically}%
  \BibitemOpen
  \bibfield  {author} {\bibinfo {author} {\bibfnamefont {H.}~\bibnamefont {Bondi}},\ }\bibfield  {title} {\bibinfo {title} {On spherically symmetrical accretion},\ }\href@noop {} {\bibfield  {journal} {\bibinfo  {journal} {Monthly Notices of the Royal Astronomical Society}\ }\textbf {\bibinfo {volume} {112}},\ \bibinfo {pages} {195} (\bibinfo {year} {1952})}\BibitemShut {NoStop}%
\bibitem [{\citenamefont {Edgar}(2004)}]{edgar2004review}%
  \BibitemOpen
  \bibfield  {author} {\bibinfo {author} {\bibfnamefont {R.}~\bibnamefont {Edgar}},\ }\bibfield  {title} {\bibinfo {title} {A review of bondi--hoyle--lyttleton accretion},\ }\href@noop {} {\bibfield  {journal} {\bibinfo  {journal} {New Astronomy Reviews}\ }\textbf {\bibinfo {volume} {48}},\ \bibinfo {pages} {843} (\bibinfo {year} {2004})}\BibitemShut {NoStop}%
\bibitem [{\citenamefont {Gregoris}(2023)}]{gregoris2023black}%
  \BibitemOpen
  \bibfield  {author} {\bibinfo {author} {\bibfnamefont {D.}~\bibnamefont {Gregoris}},\ }\bibfield  {title} {\bibinfo {title} {Black hole evolution in the bondi--hoyle--lyttleton accretion model},\ }\href@noop {} {\bibfield  {journal} {\bibinfo  {journal} {General Relativity and Gravitation}\ }\textbf {\bibinfo {volume} {55}},\ \bibinfo {pages} {97} (\bibinfo {year} {2023})}\BibitemShut {NoStop}%
\bibitem [{\citenamefont {Xu}\ and\ \citenamefont {Stone}(2019)}]{xu2019bondi}%
  \BibitemOpen
  \bibfield  {author} {\bibinfo {author} {\bibfnamefont {W.}~\bibnamefont {Xu}}\ and\ \bibinfo {author} {\bibfnamefont {J.~M.}\ \bibnamefont {Stone}},\ }\bibfield  {title} {\bibinfo {title} {Bondi--hoyle--lyttleton accretion in supergiant x-ray binaries: stability and disc formation},\ }\href@noop {} {\bibfield  {journal} {\bibinfo  {journal} {Monthly Notices of the Royal Astronomical Society}\ }\textbf {\bibinfo {volume} {488}},\ \bibinfo {pages} {5162} (\bibinfo {year} {2019})}\BibitemShut {NoStop}%
\bibitem [{\citenamefont {Tejeda}\ and\ \citenamefont {Aguayo-Ortiz}(2019)}]{tejeda2019relativistic}%
  \BibitemOpen
  \bibfield  {author} {\bibinfo {author} {\bibfnamefont {E.}~\bibnamefont {Tejeda}}\ and\ \bibinfo {author} {\bibfnamefont {A.}~\bibnamefont {Aguayo-Ortiz}},\ }\bibfield  {title} {\bibinfo {title} {Relativistic wind accretion on to a schwarzschild black hole},\ }\href@noop {} {\bibfield  {journal} {\bibinfo  {journal} {Monthly Notices of the Royal Astronomical Society}\ }\textbf {\bibinfo {volume} {487}},\ \bibinfo {pages} {3607} (\bibinfo {year} {2019})}\BibitemShut {NoStop}%
\bibitem [{\citenamefont {Cruz-Osorio}\ \emph {et~al.}(2023)\citenamefont {Cruz-Osorio}, \citenamefont {Rezzolla}, \citenamefont {Lora-Clavijo}, \citenamefont {Font}, \citenamefont {Herdeiro},\ and\ \citenamefont {Radu}}]{cruz2023bondi}%
  \BibitemOpen
  \bibfield  {author} {\bibinfo {author} {\bibfnamefont {A.}~\bibnamefont {Cruz-Osorio}}, \bibinfo {author} {\bibfnamefont {L.}~\bibnamefont {Rezzolla}}, \bibinfo {author} {\bibfnamefont {F.~D.}\ \bibnamefont {Lora-Clavijo}}, \bibinfo {author} {\bibfnamefont {J.~A.}\ \bibnamefont {Font}}, \bibinfo {author} {\bibfnamefont {C.}~\bibnamefont {Herdeiro}},\ and\ \bibinfo {author} {\bibfnamefont {E.}~\bibnamefont {Radu}},\ }\bibfield  {title} {\bibinfo {title} {Bondi-hoyle-lyttleton accretion onto a rotating black hole with ultralight scalar hair},\ }\href@noop {} {\bibfield  {journal} {\bibinfo  {journal} {Journal of Cosmology and Astroparticle Physics}\ }\textbf {\bibinfo {volume} {2023}}\bibinfo  {number} { (08)},\ \bibinfo {pages} {057}}\BibitemShut {NoStop}%
\bibitem [{\citenamefont {Font}\ \emph {et~al.}(1999)\citenamefont {Font}, \citenamefont {Ib{\'a}{\~n}ez},\ and\ \citenamefont {Papadopoulos}}]{font1999non}%
  \BibitemOpen
\bibfield  {number} {  }\bibfield  {author} {\bibinfo {author} {\bibfnamefont {J.~A.}\ \bibnamefont {Font}}, \bibinfo {author} {\bibfnamefont {J.~M.}\ \bibnamefont {Ib{\'a}{\~n}ez}},\ and\ \bibinfo {author} {\bibfnamefont {P.}~\bibnamefont {Papadopoulos}},\ }\bibfield  {title} {\bibinfo {title} {Non-axisymmetric relativistic bondi-hoyle accretion on to a kerr black hole},\ }\href@noop {} {\bibfield  {journal} {\bibinfo  {journal} {Monthly Notices of the Royal Astronomical Society}\ }\textbf {\bibinfo {volume} {305}},\ \bibinfo {pages} {920} (\bibinfo {year} {1999})}\BibitemShut {NoStop}%
\bibitem [{\citenamefont {Li}\ \emph {et~al.}(2023{\natexlab{b}})\citenamefont {Li}, \citenamefont {Liu},\ and\ \citenamefont {Zhai}}]{li2023accretion}%
  \BibitemOpen
  \bibfield  {author} {\bibinfo {author} {\bibfnamefont {P.}~\bibnamefont {Li}}, \bibinfo {author} {\bibfnamefont {Y.-q.}\ \bibnamefont {Liu}},\ and\ \bibinfo {author} {\bibfnamefont {X.-h.}\ \bibnamefont {Zhai}},\ }\bibfield  {title} {\bibinfo {title} {Accretion of the relativistic vlasov gas onto a kerr black hole},\ }\href@noop {} {\bibfield  {journal} {\bibinfo  {journal} {Physical Review D}\ }\textbf {\bibinfo {volume} {108}},\ \bibinfo {pages} {124022} (\bibinfo {year} {2023}{\natexlab{b}})}\BibitemShut {NoStop}%
\bibitem [{\citenamefont {Lora-Clavijo}\ and\ \citenamefont {Guzm{\'a}n}(2013)}]{lora2013axisymmetric}%
  \BibitemOpen
  \bibfield  {author} {\bibinfo {author} {\bibfnamefont {F.}~\bibnamefont {Lora-Clavijo}}\ and\ \bibinfo {author} {\bibfnamefont {F.}~\bibnamefont {Guzm{\'a}n}},\ }\bibfield  {title} {\bibinfo {title} {Axisymmetric bondi--hoyle accretion on to a schwarzschild black hole: shock cone vibrations},\ }\href@noop {} {\bibfield  {journal} {\bibinfo  {journal} {Monthly Notices of the Royal Astronomical Society}\ }\textbf {\bibinfo {volume} {429}},\ \bibinfo {pages} {3144} (\bibinfo {year} {2013})}\BibitemShut {NoStop}%
\bibitem [{\citenamefont {Lora-Clavijo}\ \emph {et~al.}(2015)\citenamefont {Lora-Clavijo}, \citenamefont {Cruz-Osorio},\ and\ \citenamefont {M{\'e}ndez}}]{lora2015relativistic}%
  \BibitemOpen
  \bibfield  {author} {\bibinfo {author} {\bibfnamefont {F.}~\bibnamefont {Lora-Clavijo}}, \bibinfo {author} {\bibfnamefont {A.}~\bibnamefont {Cruz-Osorio}},\ and\ \bibinfo {author} {\bibfnamefont {E.~M.}\ \bibnamefont {M{\'e}ndez}},\ }\bibfield  {title} {\bibinfo {title} {Relativistic bondi--hoyle--lyttleton accretion onto a rotating black hole: density gradients},\ }\href@noop {} {\bibfield  {journal} {\bibinfo  {journal} {The Astrophysical Journal Supplement Series}\ }\textbf {\bibinfo {volume} {219}},\ \bibinfo {pages} {30} (\bibinfo {year} {2015})}\BibitemShut {NoStop}%
\bibitem [{\citenamefont {Blakely}\ and\ \citenamefont {Nikiforakis}(2015)}]{blakely2015relativistic}%
  \BibitemOpen
  \bibfield  {author} {\bibinfo {author} {\bibfnamefont {P.}~\bibnamefont {Blakely}}\ and\ \bibinfo {author} {\bibfnamefont {N.}~\bibnamefont {Nikiforakis}},\ }\bibfield  {title} {\bibinfo {title} {Relativistic bondi-hoyle-lyttleton accretion: A parametric study},\ }\href@noop {} {\bibfield  {journal} {\bibinfo  {journal} {Astronomy \& Astrophysics}\ }\textbf {\bibinfo {volume} {583}},\ \bibinfo {pages} {A90} (\bibinfo {year} {2015})}\BibitemShut {NoStop}%
\bibitem [{\citenamefont {Ruffert}(1996)}]{ruffert1996non}%
  \BibitemOpen
  \bibfield  {author} {\bibinfo {author} {\bibfnamefont {M.}~\bibnamefont {Ruffert}},\ }\bibfield  {title} {\bibinfo {title} {Non-axisymmetric wind-accretion simulations i. velocity gradients of 3\% and 20\% over one accretion radius},\ }\href@noop {} {\bibfield  {journal} {\bibinfo  {journal} {arXiv preprint astro-ph/9605072}\ } (\bibinfo {year} {1996})}\BibitemShut {NoStop}%
\bibitem [{\citenamefont {Fukue}\ and\ \citenamefont {Ioroi}(1999)}]{fukue1999hoyle}%
  \BibitemOpen
  \bibfield  {author} {\bibinfo {author} {\bibfnamefont {J.}~\bibnamefont {Fukue}}\ and\ \bibinfo {author} {\bibfnamefont {M.}~\bibnamefont {Ioroi}},\ }\bibfield  {title} {\bibinfo {title} {Hoyle-lyttleton accretion onto accretion disks},\ }\href@noop {} {\bibfield  {journal} {\bibinfo  {journal} {Publications of the Astronomical Society of Japan}\ }\textbf {\bibinfo {volume} {51}},\ \bibinfo {pages} {151} (\bibinfo {year} {1999})}\BibitemShut {NoStop}%
\bibitem [{\citenamefont {Fukue}(2001)}]{fukue2001bondi}%
  \BibitemOpen
  \bibfield  {author} {\bibinfo {author} {\bibfnamefont {J.}~\bibnamefont {Fukue}},\ }\bibfield  {title} {\bibinfo {title} {Bondi accretion onto a luminous object},\ }\href@noop {} {\bibfield  {journal} {\bibinfo  {journal} {Publications of the Astronomical Society of Japan}\ }\textbf {\bibinfo {volume} {53}},\ \bibinfo {pages} {687} (\bibinfo {year} {2001})}\BibitemShut {NoStop}%
\bibitem [{\citenamefont {Scicluna}\ \emph {et~al.}(2014)\citenamefont {Scicluna}, \citenamefont {Rosotti}, \citenamefont {Dale},\ and\ \citenamefont {Testi}}]{scicluna2014old}%
  \BibitemOpen
  \bibfield  {author} {\bibinfo {author} {\bibfnamefont {P.}~\bibnamefont {Scicluna}}, \bibinfo {author} {\bibfnamefont {G.}~\bibnamefont {Rosotti}}, \bibinfo {author} {\bibfnamefont {J.}~\bibnamefont {Dale}},\ and\ \bibinfo {author} {\bibfnamefont {L.}~\bibnamefont {Testi}},\ }\bibfield  {title} {\bibinfo {title} {Old pre-main-sequence stars-disc reformation by bondi-hoyle accretion},\ }\href@noop {} {\bibfield  {journal} {\bibinfo  {journal} {Astronomy \& Astrophysics}\ }\textbf {\bibinfo {volume} {566}},\ \bibinfo {pages} {L3} (\bibinfo {year} {2014})}\BibitemShut {NoStop}%
\bibitem [{\citenamefont {Comerford}\ \emph {et~al.}(2019)\citenamefont {Comerford}, \citenamefont {Izzard}, \citenamefont {Booth},\ and\ \citenamefont {Rosotti}}]{comerford2019bondi}%
  \BibitemOpen
  \bibfield  {author} {\bibinfo {author} {\bibfnamefont {T.}~\bibnamefont {Comerford}}, \bibinfo {author} {\bibfnamefont {R.}~\bibnamefont {Izzard}}, \bibinfo {author} {\bibfnamefont {R.}~\bibnamefont {Booth}},\ and\ \bibinfo {author} {\bibfnamefont {G.}~\bibnamefont {Rosotti}},\ }\bibfield  {title} {\bibinfo {title} {Bondi--hoyle--lyttleton accretion by binary stars},\ }\href@noop {} {\bibfield  {journal} {\bibinfo  {journal} {Monthly Notices of the Royal Astronomical Society}\ }\textbf {\bibinfo {volume} {490}},\ \bibinfo {pages} {5196} (\bibinfo {year} {2019})}\BibitemShut {NoStop}%
\bibitem [{\citenamefont {Torres}(2010)}]{torres2010use}%
  \BibitemOpen
  \bibfield  {author} {\bibinfo {author} {\bibfnamefont {G.}~\bibnamefont {Torres}},\ }\bibfield  {title} {\bibinfo {title} {On the use of empirical bolometric corrections for stars},\ }\href@noop {} {\bibfield  {journal} {\bibinfo  {journal} {The Astronomical Journal}\ }\textbf {\bibinfo {volume} {140}},\ \bibinfo {pages} {1158} (\bibinfo {year} {2010})}\BibitemShut {NoStop}%
\bibitem [{\citenamefont {Wielgus}\ \emph {et~al.}(2015)\citenamefont {Wielgus}, \citenamefont {Klu{\'z}niak}, \citenamefont {Sadowski}, \citenamefont {Narayan},\ and\ \citenamefont {Abramowicz}}]{wielgus2015stable}%
  \BibitemOpen
  \bibfield  {author} {\bibinfo {author} {\bibfnamefont {M.}~\bibnamefont {Wielgus}}, \bibinfo {author} {\bibfnamefont {W.}~\bibnamefont {Klu{\'z}niak}}, \bibinfo {author} {\bibfnamefont {A.}~\bibnamefont {Sadowski}}, \bibinfo {author} {\bibfnamefont {R.}~\bibnamefont {Narayan}},\ and\ \bibinfo {author} {\bibfnamefont {M.}~\bibnamefont {Abramowicz}},\ }\bibfield  {title} {\bibinfo {title} {Stable, levitating, optically thin atmospheres of eddington-luminosity neutron stars},\ }\href@noop {} {\bibfield  {journal} {\bibinfo  {journal} {Monthly Notices of the Royal Astronomical Society}\ }\textbf {\bibinfo {volume} {454}},\ \bibinfo {pages} {3766} (\bibinfo {year} {2015})}\BibitemShut {NoStop}%
\bibitem [{\citenamefont {Rybicki}\ and\ \citenamefont {Lightman}(1991)}]{rybicki1991radiative}%
  \BibitemOpen
  \bibfield  {author} {\bibinfo {author} {\bibfnamefont {G.~B.}\ \bibnamefont {Rybicki}}\ and\ \bibinfo {author} {\bibfnamefont {A.~P.}\ \bibnamefont {Lightman}},\ }\href@noop {} {\emph {\bibinfo {title} {Radiative processes in astrophysics}}}\ (\bibinfo  {publisher} {John Wiley \& Sons},\ \bibinfo {year} {1991})\BibitemShut {NoStop}%
\bibitem [{\citenamefont {Zhang}\ and\ \citenamefont {Lu}(2017)}]{zhang2017mean}%
  \BibitemOpen
  \bibfield  {author} {\bibinfo {author} {\bibfnamefont {X.}~\bibnamefont {Zhang}}\ and\ \bibinfo {author} {\bibfnamefont {Y.}~\bibnamefont {Lu}},\ }\bibfield  {title} {\bibinfo {title} {On the mean radiative efficiency of accreting massive black holes in agns and qsos},\ }\href@noop {} {\bibfield  {journal} {\bibinfo  {journal} {Science China Physics, Mechanics \& Astronomy}\ }\textbf {\bibinfo {volume} {60}},\ \bibinfo {pages} {1} (\bibinfo {year} {2017})}\BibitemShut {NoStop}%
\bibitem [{\citenamefont {Eddington}(1988)}]{eddington1988internal}%
  \BibitemOpen
  \bibfield  {author} {\bibinfo {author} {\bibfnamefont {A.~S.}\ \bibnamefont {Eddington}},\ }\href@noop {} {\emph {\bibinfo {title} {The internal constitution of the stars}}}\ (\bibinfo  {publisher} {Cambridge University Press},\ \bibinfo {year} {1988})\BibitemShut {NoStop}%
\bibitem [{\citenamefont {Bardeen}\ \emph {et~al.}(1972)\citenamefont {Bardeen}, \citenamefont {Press},\ and\ \citenamefont {Teukolsky}}]{bardeen1972rotating}%
  \BibitemOpen
  \bibfield  {author} {\bibinfo {author} {\bibfnamefont {J.~M.}\ \bibnamefont {Bardeen}}, \bibinfo {author} {\bibfnamefont {W.~H.}\ \bibnamefont {Press}},\ and\ \bibinfo {author} {\bibfnamefont {S.~A.}\ \bibnamefont {Teukolsky}},\ }\bibfield  {title} {\bibinfo {title} {Rotating black holes: locally nonrotating frames, energy extraction, and scalar synchrotron radiation},\ }\href@noop {} {\bibfield  {journal} {\bibinfo  {journal} {Astrophysical Journal, Vol. 178, pp. 347-370 (1972)}\ }\textbf {\bibinfo {volume} {178}},\ \bibinfo {pages} {347} (\bibinfo {year} {1972})}\BibitemShut {NoStop}%
\bibitem [{\citenamefont {Novikov}\ and\ \citenamefont {Thorne}(1973)}]{novikov1973astrophysics}%
  \BibitemOpen
  \bibfield  {author} {\bibinfo {author} {\bibfnamefont {I.~D.}\ \bibnamefont {Novikov}}\ and\ \bibinfo {author} {\bibfnamefont {K.~S.}\ \bibnamefont {Thorne}},\ }\bibfield  {title} {\bibinfo {title} {Astrophysics of black holes},\ }\href@noop {} {\bibfield  {journal} {\bibinfo  {journal} {Black holes (Les astres occlus)}\ }\textbf {\bibinfo {volume} {1}},\ \bibinfo {pages} {343} (\bibinfo {year} {1973})}\BibitemShut {NoStop}%
\bibitem [{\citenamefont {Abramowicz}\ \emph {et~al.}(1988)\citenamefont {Abramowicz}, \citenamefont {Czerny}, \citenamefont {Lasota},\ and\ \citenamefont {Szuszkiewicz}}]{abramowicz1988slim}%
  \BibitemOpen
  \bibfield  {author} {\bibinfo {author} {\bibfnamefont {M.}~\bibnamefont {Abramowicz}}, \bibinfo {author} {\bibfnamefont {B.}~\bibnamefont {Czerny}}, \bibinfo {author} {\bibfnamefont {J.}~\bibnamefont {Lasota}},\ and\ \bibinfo {author} {\bibfnamefont {E.}~\bibnamefont {Szuszkiewicz}},\ }\bibfield  {title} {\bibinfo {title} {Slim accretion disks},\ }\href@noop {} {\bibfield  {journal} {\bibinfo  {journal} {Astrophysical Journal, Part 1 (ISSN 0004-637X), vol. 332, Sept. 15, 1988, p. 646-658. Research supported by Observatoire de Paris and NASA.}\ }\textbf {\bibinfo {volume} {332}},\ \bibinfo {pages} {646} (\bibinfo {year} {1988})}\BibitemShut {NoStop}%
\bibitem [{\citenamefont {Heinzeller}\ and\ \citenamefont {Duschl}(2007)}]{heinzeller2007eddington}%
  \BibitemOpen
  \bibfield  {author} {\bibinfo {author} {\bibfnamefont {D.}~\bibnamefont {Heinzeller}}\ and\ \bibinfo {author} {\bibfnamefont {W.}~\bibnamefont {Duschl}},\ }\bibfield  {title} {\bibinfo {title} {On the eddington limit in accretion discs},\ }\href@noop {} {\bibfield  {journal} {\bibinfo  {journal} {Monthly Notices of the Royal Astronomical Society}\ }\textbf {\bibinfo {volume} {374}},\ \bibinfo {pages} {1146} (\bibinfo {year} {2007})}\BibitemShut {NoStop}%
\bibitem [{\citenamefont {Piotrovich}\ \emph {et~al.}(2023{\natexlab{a}})\citenamefont {Piotrovich}, \citenamefont {Buliga},\ and\ \citenamefont {Natsvlishvili}}]{piotrovich2023estimate}%
  \BibitemOpen
  \bibfield  {author} {\bibinfo {author} {\bibfnamefont {M.}~\bibnamefont {Piotrovich}}, \bibinfo {author} {\bibfnamefont {S.}~\bibnamefont {Buliga}},\ and\ \bibinfo {author} {\bibfnamefont {T.}~\bibnamefont {Natsvlishvili}},\ }\bibfield  {title} {\bibinfo {title} {Estimate of smbh spin for narrow-line seyfert 1 galaxies},\ }\href@noop {} {\bibfield  {journal} {\bibinfo  {journal} {Universe}\ }\textbf {\bibinfo {volume} {9}},\ \bibinfo {pages} {175} (\bibinfo {year} {2023}{\natexlab{a}})}\BibitemShut {NoStop}%
\bibitem [{\citenamefont {Robson}(1996)}]{robson1996active}%
  \BibitemOpen
  \bibfield  {author} {\bibinfo {author} {\bibfnamefont {I.}~\bibnamefont {Robson}},\ }\bibfield  {title} {\bibinfo {title} {Active galactic nuclei},\ }\href@noop {} {\bibfield  {journal} {\bibinfo  {journal} {Wiley-Praxis series in astronomy and astrophysics}\ } (\bibinfo {year} {1996})}\BibitemShut {NoStop}%
\bibitem [{\citenamefont {Netzer}(2015)}]{netzer2015revisiting}%
  \BibitemOpen
  \bibfield  {author} {\bibinfo {author} {\bibfnamefont {H.}~\bibnamefont {Netzer}},\ }\bibfield  {title} {\bibinfo {title} {Revisiting the unified model of active galactic nuclei},\ }\href@noop {} {\bibfield  {journal} {\bibinfo  {journal} {Annual Review of Astronomy and Astrophysics}\ }\textbf {\bibinfo {volume} {53}},\ \bibinfo {pages} {365} (\bibinfo {year} {2015})}\BibitemShut {NoStop}%
\bibitem [{\citenamefont {Du}\ \emph {et~al.}(2014)\citenamefont {Du}, \citenamefont {Hu}, \citenamefont {Lu}, \citenamefont {Wang}, \citenamefont {Qiu}, \citenamefont {Li}, \citenamefont {Bai}, \citenamefont {Kaspi}, \citenamefont {Netzer}, \citenamefont {Wang} \emph {et~al.}}]{du2014supermassive}%
  \BibitemOpen
  \bibfield  {author} {\bibinfo {author} {\bibfnamefont {P.}~\bibnamefont {Du}}, \bibinfo {author} {\bibfnamefont {C.}~\bibnamefont {Hu}}, \bibinfo {author} {\bibfnamefont {K.-X.}\ \bibnamefont {Lu}}, \bibinfo {author} {\bibfnamefont {F.}~\bibnamefont {Wang}}, \bibinfo {author} {\bibfnamefont {J.}~\bibnamefont {Qiu}}, \bibinfo {author} {\bibfnamefont {Y.-R.}\ \bibnamefont {Li}}, \bibinfo {author} {\bibfnamefont {J.-M.}\ \bibnamefont {Bai}}, \bibinfo {author} {\bibfnamefont {S.}~\bibnamefont {Kaspi}}, \bibinfo {author} {\bibfnamefont {H.}~\bibnamefont {Netzer}}, \bibinfo {author} {\bibfnamefont {J.-M.}\ \bibnamefont {Wang}}, \emph {et~al.},\ }\bibfield  {title} {\bibinfo {title} {Supermassive black holes with high accretion rates in active galactic nuclei. i. first results from a new reverberation mapping campaign},\ }\href@noop {} {\bibfield  {journal} {\bibinfo  {journal} {The Astrophysical Journal}\ }\textbf {\bibinfo {volume} {782}},\ \bibinfo {pages} {45} (\bibinfo {year} {2014})}\BibitemShut {NoStop}%
\bibitem [{\citenamefont {Czerny}\ \emph {et~al.}(2011)\citenamefont {Czerny}, \citenamefont {Hryniewicz}, \citenamefont {Niko{\l}ajuk},\ and\ \citenamefont {Sadowski}}]{czerny2011constraints}%
  \BibitemOpen
  \bibfield  {author} {\bibinfo {author} {\bibfnamefont {B.}~\bibnamefont {Czerny}}, \bibinfo {author} {\bibfnamefont {K.}~\bibnamefont {Hryniewicz}}, \bibinfo {author} {\bibfnamefont {M.}~\bibnamefont {Niko{\l}ajuk}},\ and\ \bibinfo {author} {\bibfnamefont {A.}~\bibnamefont {Sadowski}},\ }\bibfield  {title} {\bibinfo {title} {Constraints on the black hole spin in the quasar sdss j094533. 99+ 100950.1},\ }\href@noop {} {\bibfield  {journal} {\bibinfo  {journal} {Monthly Notices of the Royal Astronomical Society}\ }\textbf {\bibinfo {volume} {415}},\ \bibinfo {pages} {2942} (\bibinfo {year} {2011})}\BibitemShut {NoStop}%
\bibitem [{\citenamefont {Davis}\ and\ \citenamefont {Laor}(2011)}]{davis2011radiative}%
  \BibitemOpen
  \bibfield  {author} {\bibinfo {author} {\bibfnamefont {S.~W.}\ \bibnamefont {Davis}}\ and\ \bibinfo {author} {\bibfnamefont {A.}~\bibnamefont {Laor}},\ }\bibfield  {title} {\bibinfo {title} {The radiative efficiency of accretion flows in individual active galactic nuclei},\ }\href@noop {} {\bibfield  {journal} {\bibinfo  {journal} {The Astrophysical Journal}\ }\textbf {\bibinfo {volume} {728}},\ \bibinfo {pages} {98} (\bibinfo {year} {2011})}\BibitemShut {NoStop}%
\bibitem [{\citenamefont {Raimundo}\ \emph {et~al.}(2012)\citenamefont {Raimundo}, \citenamefont {Fabian}, \citenamefont {Vasudevan}, \citenamefont {Gandhi},\ and\ \citenamefont {Wu}}]{raimundo2012can}%
  \BibitemOpen
  \bibfield  {author} {\bibinfo {author} {\bibfnamefont {S.}~\bibnamefont {Raimundo}}, \bibinfo {author} {\bibfnamefont {A.}~\bibnamefont {Fabian}}, \bibinfo {author} {\bibfnamefont {R.}~\bibnamefont {Vasudevan}}, \bibinfo {author} {\bibfnamefont {P.}~\bibnamefont {Gandhi}},\ and\ \bibinfo {author} {\bibfnamefont {J.}~\bibnamefont {Wu}},\ }\bibfield  {title} {\bibinfo {title} {Can we measure the accretion efficiency of active galactic nuclei?},\ }\href@noop {} {\bibfield  {journal} {\bibinfo  {journal} {Monthly Notices of the Royal Astronomical Society}\ }\textbf {\bibinfo {volume} {419}},\ \bibinfo {pages} {2529} (\bibinfo {year} {2012})}\BibitemShut {NoStop}%
\bibitem [{\citenamefont {Trakhtenbrot}(2014)}]{trakhtenbrot2014most}%
  \BibitemOpen
  \bibfield  {author} {\bibinfo {author} {\bibfnamefont {B.}~\bibnamefont {Trakhtenbrot}},\ }\bibfield  {title} {\bibinfo {title} {The most massive active black holes at z~ 1.5--3.5 have high spins and radiative efficiencies},\ }\href@noop {} {\bibfield  {journal} {\bibinfo  {journal} {The Astrophysical Journal Letters}\ }\textbf {\bibinfo {volume} {789}},\ \bibinfo {pages} {L9} (\bibinfo {year} {2014})}\BibitemShut {NoStop}%
\bibitem [{\citenamefont {Lawther}\ \emph {et~al.}(2017)\citenamefont {Lawther}, \citenamefont {Vestergaard}, \citenamefont {Raimundo},\ and\ \citenamefont {Grupe}}]{lawther2017catalogue}%
  \BibitemOpen
  \bibfield  {author} {\bibinfo {author} {\bibfnamefont {D.}~\bibnamefont {Lawther}}, \bibinfo {author} {\bibfnamefont {M.}~\bibnamefont {Vestergaard}}, \bibinfo {author} {\bibfnamefont {S.}~\bibnamefont {Raimundo}},\ and\ \bibinfo {author} {\bibfnamefont {D.}~\bibnamefont {Grupe}},\ }\bibfield  {title} {\bibinfo {title} {A catalogue of optical to x-ray spectral energy distributions of $z \approx 2$ quasars observed with swift--i. first results},\ }\href@noop {} {\bibfield  {journal} {\bibinfo  {journal} {Monthly Notices of the Royal Astronomical Society}\ }\textbf {\bibinfo {volume} {467}},\ \bibinfo {pages} {4674} (\bibinfo {year} {2017})}\BibitemShut {NoStop}%
\bibitem [{\citenamefont {Schmidt}\ and\ \citenamefont {Green}(1983)}]{schmidt1983quasar}%
  \BibitemOpen
  \bibfield  {author} {\bibinfo {author} {\bibfnamefont {M.}~\bibnamefont {Schmidt}}\ and\ \bibinfo {author} {\bibfnamefont {R.~F.}\ \bibnamefont {Green}},\ }\bibfield  {title} {\bibinfo {title} {Quasar evolution derived from the palomar bright quasar survey and other complete quasar surveys},\ }\href@noop {} {\bibfield  {journal} {\bibinfo  {journal} {Astrophysical Journal, Part 1 (ISSN 0004-637X), vol. 269, June 15, 1983, p. 352-374.}\ }\textbf {\bibinfo {volume} {269}},\ \bibinfo {pages} {352} (\bibinfo {year} {1983})}\BibitemShut {NoStop}%
\bibitem [{\citenamefont {Plotkin}\ \emph {et~al.}(2016)\citenamefont {Plotkin}, \citenamefont {Gallo}, \citenamefont {Haardt}, \citenamefont {Miller}, \citenamefont {Wood}, \citenamefont {Reines}, \citenamefont {Wu},\ and\ \citenamefont {Greene}}]{plotkin2016x}%
  \BibitemOpen
  \bibfield  {author} {\bibinfo {author} {\bibfnamefont {R.~M.}\ \bibnamefont {Plotkin}}, \bibinfo {author} {\bibfnamefont {E.}~\bibnamefont {Gallo}}, \bibinfo {author} {\bibfnamefont {F.}~\bibnamefont {Haardt}}, \bibinfo {author} {\bibfnamefont {B.~P.}\ \bibnamefont {Miller}}, \bibinfo {author} {\bibfnamefont {C.~J.}\ \bibnamefont {Wood}}, \bibinfo {author} {\bibfnamefont {A.~E.}\ \bibnamefont {Reines}}, \bibinfo {author} {\bibfnamefont {J.}~\bibnamefont {Wu}},\ and\ \bibinfo {author} {\bibfnamefont {J.~E.}\ \bibnamefont {Greene}},\ }\bibfield  {title} {\bibinfo {title} {The x-ray properties of million solar mass black holes},\ }\href@noop {} {\bibfield  {journal} {\bibinfo  {journal} {The Astrophysical Journal}\ }\textbf {\bibinfo {volume} {825}},\ \bibinfo {pages} {139} (\bibinfo {year} {2016})}\BibitemShut {NoStop}%
\bibitem [{\citenamefont {Shields}(1978)}]{shields1978thermal}%
  \BibitemOpen
  \bibfield  {author} {\bibinfo {author} {\bibfnamefont {G.}~\bibnamefont {Shields}},\ }\bibfield  {title} {\bibinfo {title} {Thermal continuum from accretion disks in quasars},\ }\href@noop {} {\bibfield  {journal} {\bibinfo  {journal} {Nature}\ }\textbf {\bibinfo {volume} {272}},\ \bibinfo {pages} {706} (\bibinfo {year} {1978})}\BibitemShut {NoStop}%
\bibitem [{\citenamefont {Shang}\ \emph {et~al.}(2005)\citenamefont {Shang}, \citenamefont {Brotherton}, \citenamefont {Green}, \citenamefont {Kriss}, \citenamefont {Scott}, \citenamefont {Quijano}, \citenamefont {Blaes}, \citenamefont {Hubeny}, \citenamefont {Hutchings}, \citenamefont {Kaiser} \emph {et~al.}}]{shang2005quasars}%
  \BibitemOpen
  \bibfield  {author} {\bibinfo {author} {\bibfnamefont {Z.}~\bibnamefont {Shang}}, \bibinfo {author} {\bibfnamefont {M.~S.}\ \bibnamefont {Brotherton}}, \bibinfo {author} {\bibfnamefont {R.~F.}\ \bibnamefont {Green}}, \bibinfo {author} {\bibfnamefont {G.~A.}\ \bibnamefont {Kriss}}, \bibinfo {author} {\bibfnamefont {J.}~\bibnamefont {Scott}}, \bibinfo {author} {\bibfnamefont {J.~K.}\ \bibnamefont {Quijano}}, \bibinfo {author} {\bibfnamefont {O.}~\bibnamefont {Blaes}}, \bibinfo {author} {\bibfnamefont {I.}~\bibnamefont {Hubeny}}, \bibinfo {author} {\bibfnamefont {J.}~\bibnamefont {Hutchings}}, \bibinfo {author} {\bibfnamefont {M.~E.}\ \bibnamefont {Kaiser}}, \emph {et~al.},\ }\bibfield  {title} {\bibinfo {title} {Quasars and the big blue bump},\ }\href@noop {} {\bibfield  {journal} {\bibinfo  {journal} {The Astrophysical Journal}\ }\textbf {\bibinfo {volume} {619}},\ \bibinfo {pages} {41} (\bibinfo {year} {2005})}\BibitemShut {NoStop}%
\bibitem [{\citenamefont {Liu}\ \emph {et~al.}(2008)\citenamefont {Liu}, \citenamefont {Bai}, \citenamefont {Zhao},\ and\ \citenamefont {Ma}}]{liu2008tests}%
  \BibitemOpen
  \bibfield  {author} {\bibinfo {author} {\bibfnamefont {H.}~\bibnamefont {Liu}}, \bibinfo {author} {\bibfnamefont {J.}~\bibnamefont {Bai}}, \bibinfo {author} {\bibfnamefont {X.}~\bibnamefont {Zhao}},\ and\ \bibinfo {author} {\bibfnamefont {L.}~\bibnamefont {Ma}},\ }\bibfield  {title} {\bibinfo {title} {Tests for standard accretion disk models by variability in active galactic nuclei},\ }\href@noop {} {\bibfield  {journal} {\bibinfo  {journal} {The Astrophysical Journal}\ }\textbf {\bibinfo {volume} {677}},\ \bibinfo {pages} {884} (\bibinfo {year} {2008})}\BibitemShut {NoStop}%
\bibitem [{\citenamefont {Inoue}(2010)}]{inoue2010lyman}%
  \BibitemOpen
  \bibfield  {author} {\bibinfo {author} {\bibfnamefont {A.~K.}\ \bibnamefont {Inoue}},\ }\bibfield  {title} {\bibinfo {title} {Lyman ‘bump’galaxies--i. spectral energy distribution of galaxies with an escape of nebular lyman continuum},\ }\href@noop {} {\bibfield  {journal} {\bibinfo  {journal} {Monthly Notices of the Royal Astronomical Society}\ }\textbf {\bibinfo {volume} {401}},\ \bibinfo {pages} {1325} (\bibinfo {year} {2010})}\BibitemShut {NoStop}%
\bibitem [{\citenamefont {Davidson}(1976)}]{davidson1976some}%
  \BibitemOpen
  \bibfield  {author} {\bibinfo {author} {\bibfnamefont {K.}~\bibnamefont {Davidson}},\ }\bibfield  {title} {\bibinfo {title} {Some remarks concerning lyman-continuum emission in quasar spectra},\ }\href@noop {} {\bibfield  {journal} {\bibinfo  {journal} {Astrophysical Journal, vol. 207, Aug. 1, 1976, pt. 1, p. 710-712. Research supported by the Alfred P. Sloan Foundation}\ }\textbf {\bibinfo {volume} {207}},\ \bibinfo {pages} {710} (\bibinfo {year} {1976})}\BibitemShut {NoStop}%
\bibitem [{\citenamefont {Shukla}\ \emph {et~al.}(2016)\citenamefont {Shukla}, \citenamefont {Mellema}, \citenamefont {Iliev},\ and\ \citenamefont {Shapiro}}]{shukla2016effects}%
  \BibitemOpen
  \bibfield  {author} {\bibinfo {author} {\bibfnamefont {H.}~\bibnamefont {Shukla}}, \bibinfo {author} {\bibfnamefont {G.}~\bibnamefont {Mellema}}, \bibinfo {author} {\bibfnamefont {I.~T.}\ \bibnamefont {Iliev}},\ and\ \bibinfo {author} {\bibfnamefont {P.~R.}\ \bibnamefont {Shapiro}},\ }\bibfield  {title} {\bibinfo {title} {The effects of lyman-limit systems on the evolution and observability of the epoch of reionization},\ }\href@noop {} {\bibfield  {journal} {\bibinfo  {journal} {Monthly Notices of the Royal Astronomical Society}\ }\textbf {\bibinfo {volume} {458}},\ \bibinfo {pages} {135} (\bibinfo {year} {2016})}\BibitemShut {NoStop}%
\bibitem [{\citenamefont {Sellwood}\ and\ \citenamefont {Masters}(2022)}]{sellwood2022spirals}%
  \BibitemOpen
  \bibfield  {author} {\bibinfo {author} {\bibfnamefont {J.~A.}\ \bibnamefont {Sellwood}}\ and\ \bibinfo {author} {\bibfnamefont {K.~L.}\ \bibnamefont {Masters}},\ }\bibfield  {title} {\bibinfo {title} {Spirals in galaxies},\ }\href@noop {} {\bibfield  {journal} {\bibinfo  {journal} {Annual Review of Astronomy and Astrophysics}\ }\textbf {\bibinfo {volume} {60}},\ \bibinfo {pages} {73} (\bibinfo {year} {2022})}\BibitemShut {NoStop}%
\bibitem [{\citenamefont {Chakrabarti}(1996)}]{chakrabarti1996accretion}%
  \BibitemOpen
  \bibfield  {author} {\bibinfo {author} {\bibfnamefont {S.~K.}\ \bibnamefont {Chakrabarti}},\ }\bibfield  {title} {\bibinfo {title} {Accretion processes on a black hole},\ }\href@noop {} {\bibfield  {journal} {\bibinfo  {journal} {Physics Reports}\ }\textbf {\bibinfo {volume} {266}},\ \bibinfo {pages} {229} (\bibinfo {year} {1996})}\BibitemShut {NoStop}%
\bibitem [{\citenamefont {Bu}\ and\ \citenamefont {Yang}(2019)}]{bu2019real}%
  \BibitemOpen
  \bibfield  {author} {\bibinfo {author} {\bibfnamefont {D.-F.}\ \bibnamefont {Bu}}\ and\ \bibinfo {author} {\bibfnamefont {X.-H.}\ \bibnamefont {Yang}},\ }\bibfield  {title} {\bibinfo {title} {What is the real accretion rate on to a black hole for low-angular-momentum accretion?},\ }\href@noop {} {\bibfield  {journal} {\bibinfo  {journal} {Monthly Notices of the Royal Astronomical Society}\ }\textbf {\bibinfo {volume} {484}},\ \bibinfo {pages} {1724} (\bibinfo {year} {2019})}\BibitemShut {NoStop}%
\bibitem [{\citenamefont {Bu}\ and\ \citenamefont {Zhang}(2023)}]{bu2023black}%
  \BibitemOpen
  \bibfield  {author} {\bibinfo {author} {\bibfnamefont {Q.}~\bibnamefont {Bu}}\ and\ \bibinfo {author} {\bibfnamefont {S.}~\bibnamefont {Zhang}},\ }\bibfield  {title} {\bibinfo {title} {Black holes: accretion processes in x-ray binaries},\ }\href@noop {} {\bibfield  {journal} {\bibinfo  {journal} {arXiv preprint arXiv:2310.20637}\ } (\bibinfo {year} {2023})}\BibitemShut {NoStop}%
\bibitem [{\citenamefont {He}\ \emph {et~al.}(2022)\citenamefont {He}, \citenamefont {Cai},\ and\ \citenamefont {Yang}}]{he2022thin}%
  \BibitemOpen
  \bibfield  {author} {\bibinfo {author} {\bibfnamefont {T.-Y.}\ \bibnamefont {He}}, \bibinfo {author} {\bibfnamefont {Z.}~\bibnamefont {Cai}},\ and\ \bibinfo {author} {\bibfnamefont {R.-J.}\ \bibnamefont {Yang}},\ }\bibfield  {title} {\bibinfo {title} {Thin accretion disks around a black hole in einstein-aether-scalar theory},\ }\href@noop {} {\bibfield  {journal} {\bibinfo  {journal} {The European Physical Journal C}\ }\textbf {\bibinfo {volume} {82}},\ \bibinfo {pages} {1067} (\bibinfo {year} {2022})}\BibitemShut {NoStop}%
\bibitem [{\citenamefont {Trump}\ \emph {et~al.}(2011)\citenamefont {Trump}, \citenamefont {Impey}, \citenamefont {Kelly}, \citenamefont {Civano}, \citenamefont {Gabor}, \citenamefont {Diamond-Stanic}, \citenamefont {Merloni}, \citenamefont {Urry}, \citenamefont {Hao}, \citenamefont {Jahnke} \emph {et~al.}}]{trump2011accretion}%
  \BibitemOpen
  \bibfield  {author} {\bibinfo {author} {\bibfnamefont {J.~R.}\ \bibnamefont {Trump}}, \bibinfo {author} {\bibfnamefont {C.~D.}\ \bibnamefont {Impey}}, \bibinfo {author} {\bibfnamefont {B.~C.}\ \bibnamefont {Kelly}}, \bibinfo {author} {\bibfnamefont {F.}~\bibnamefont {Civano}}, \bibinfo {author} {\bibfnamefont {J.~M.}\ \bibnamefont {Gabor}}, \bibinfo {author} {\bibfnamefont {A.~M.}\ \bibnamefont {Diamond-Stanic}}, \bibinfo {author} {\bibfnamefont {A.}~\bibnamefont {Merloni}}, \bibinfo {author} {\bibfnamefont {C.~M.}\ \bibnamefont {Urry}}, \bibinfo {author} {\bibfnamefont {H.}~\bibnamefont {Hao}}, \bibinfo {author} {\bibfnamefont {K.}~\bibnamefont {Jahnke}}, \emph {et~al.},\ }\bibfield  {title} {\bibinfo {title} {Accretion rate and the physical nature of unobscured active galaxiesbased on observations with the xmm-newton satellite, an esa science mission with instruments and contributions directly funded by esa member states and nasa; the magellan telescope, operated by the carnegie observatories; the eso
  very large telescope; and the mmt observatory, a joint facility of the university of arizona and the smithsonian institution; the subaru telescope, operated by the national astronomical observatory of japan; and the nasa/esa hubble space telescope, operated at the space telescope science institute, which is operated by aura inc., under nasa contract nas 5-26555.},\ }\href@noop {} {\bibfield  {journal} {\bibinfo  {journal} {Astrophysical Journal}\ }\textbf {\bibinfo {volume} {733}},\ \bibinfo {pages} {60} (\bibinfo {year} {2011})}\BibitemShut {NoStop}%
\bibitem [{\citenamefont {Kollmeier}\ \emph {et~al.}(2006)\citenamefont {Kollmeier}, \citenamefont {Onken}, \citenamefont {Kochanek}, \citenamefont {Gould}, \citenamefont {Weinberg}, \citenamefont {Dietrich}, \citenamefont {Cool}, \citenamefont {Dey}, \citenamefont {Eisenstein}, \citenamefont {Jannuzi} \emph {et~al.}}]{kollmeier2006black}%
  \BibitemOpen
  \bibfield  {author} {\bibinfo {author} {\bibfnamefont {J.~A.}\ \bibnamefont {Kollmeier}}, \bibinfo {author} {\bibfnamefont {C.~A.}\ \bibnamefont {Onken}}, \bibinfo {author} {\bibfnamefont {C.~S.}\ \bibnamefont {Kochanek}}, \bibinfo {author} {\bibfnamefont {A.}~\bibnamefont {Gould}}, \bibinfo {author} {\bibfnamefont {D.~H.}\ \bibnamefont {Weinberg}}, \bibinfo {author} {\bibfnamefont {M.}~\bibnamefont {Dietrich}}, \bibinfo {author} {\bibfnamefont {R.}~\bibnamefont {Cool}}, \bibinfo {author} {\bibfnamefont {A.}~\bibnamefont {Dey}}, \bibinfo {author} {\bibfnamefont {D.~J.}\ \bibnamefont {Eisenstein}}, \bibinfo {author} {\bibfnamefont {B.~T.}\ \bibnamefont {Jannuzi}}, \emph {et~al.},\ }\bibfield  {title} {\bibinfo {title} {Black hole masses and eddington ratios at 0.3< z< 4},\ }\href@noop {} {\bibfield  {journal} {\bibinfo  {journal} {The Astrophysical Journal}\ }\textbf {\bibinfo {volume} {648}},\ \bibinfo {pages} {128} (\bibinfo {year} {2006})}\BibitemShut {NoStop}%
\bibitem [{\citenamefont {Trump}\ \emph {et~al.}(2009)\citenamefont {Trump}, \citenamefont {Impey}, \citenamefont {Kelly}, \citenamefont {Elvis}, \citenamefont {Merloni}, \citenamefont {Bongiorno}, \citenamefont {Gabor}, \citenamefont {Hao}, \citenamefont {McCarthy}, \citenamefont {Huchra} \emph {et~al.}}]{trump2009observational}%
  \BibitemOpen
  \bibfield  {author} {\bibinfo {author} {\bibfnamefont {J.~R.}\ \bibnamefont {Trump}}, \bibinfo {author} {\bibfnamefont {C.~D.}\ \bibnamefont {Impey}}, \bibinfo {author} {\bibfnamefont {B.~C.}\ \bibnamefont {Kelly}}, \bibinfo {author} {\bibfnamefont {M.}~\bibnamefont {Elvis}}, \bibinfo {author} {\bibfnamefont {A.}~\bibnamefont {Merloni}}, \bibinfo {author} {\bibfnamefont {A.}~\bibnamefont {Bongiorno}}, \bibinfo {author} {\bibfnamefont {J.}~\bibnamefont {Gabor}}, \bibinfo {author} {\bibfnamefont {H.}~\bibnamefont {Hao}}, \bibinfo {author} {\bibfnamefont {P.~J.}\ \bibnamefont {McCarthy}}, \bibinfo {author} {\bibfnamefont {J.~P.}\ \bibnamefont {Huchra}}, \emph {et~al.},\ }\bibfield  {title} {\bibinfo {title} {Observational limits on type 1 active galactic nucleus accretion rate in cosmos},\ }\href@noop {} {\bibfield  {journal} {\bibinfo  {journal} {The Astrophysical Journal}\ }\textbf {\bibinfo {volume} {700}},\ \bibinfo {pages} {49} (\bibinfo {year} {2009})}\BibitemShut {NoStop}%
\bibitem [{\citenamefont {Gaskell}\ and\ \citenamefont {Goosmann}(2013)}]{gaskell2013line}%
  \BibitemOpen
  \bibfield  {author} {\bibinfo {author} {\bibfnamefont {C.~M.}\ \bibnamefont {Gaskell}}\ and\ \bibinfo {author} {\bibfnamefont {R.~W.}\ \bibnamefont {Goosmann}},\ }\bibfield  {title} {\bibinfo {title} {Line shifts, broad-line region inflow, and the feeding of active galactic nuclei},\ }\href@noop {} {\bibfield  {journal} {\bibinfo  {journal} {The Astrophysical Journal}\ }\textbf {\bibinfo {volume} {769}},\ \bibinfo {pages} {30} (\bibinfo {year} {2013})}\BibitemShut {NoStop}%
\bibitem [{\citenamefont {Neri-Larios}\ \emph {et~al.}(2011)\citenamefont {Neri-Larios}, \citenamefont {Coziol}, \citenamefont {Torres-Papaqui}, \citenamefont {Andernach}, \citenamefont {Islas-Islas}, \citenamefont {Plauchu-Frayn},\ and\ \citenamefont {Ortega-Minakata}}]{neri2011narrow}%
  \BibitemOpen
  \bibfield  {author} {\bibinfo {author} {\bibfnamefont {D.}~\bibnamefont {Neri-Larios}}, \bibinfo {author} {\bibfnamefont {R.}~\bibnamefont {Coziol}}, \bibinfo {author} {\bibfnamefont {J.}~\bibnamefont {Torres-Papaqui}}, \bibinfo {author} {\bibfnamefont {H.}~\bibnamefont {Andernach}}, \bibinfo {author} {\bibfnamefont {J.}~\bibnamefont {Islas-Islas}}, \bibinfo {author} {\bibfnamefont {I.}~\bibnamefont {Plauchu-Frayn}},\ and\ \bibinfo {author} {\bibfnamefont {R.}~\bibnamefont {Ortega-Minakata}},\ }\bibfield  {title} {\bibinfo {title} {Narrow-line agns: confirming the relationship between metallicity and accretion rate},\ }\href@noop {} {\bibfield  {journal} {\bibinfo  {journal} {arXiv preprint arXiv:1106.1561}\ } (\bibinfo {year} {2011})}\BibitemShut {NoStop}%
\bibitem [{\citenamefont {Begelman}\ \emph {et~al.}(1984)\citenamefont {Begelman}, \citenamefont {Blandford},\ and\ \citenamefont {Rees}}]{begelman1984theory}%
  \BibitemOpen
  \bibfield  {author} {\bibinfo {author} {\bibfnamefont {M.~C.}\ \bibnamefont {Begelman}}, \bibinfo {author} {\bibfnamefont {R.~D.}\ \bibnamefont {Blandford}},\ and\ \bibinfo {author} {\bibfnamefont {M.~J.}\ \bibnamefont {Rees}},\ }\bibfield  {title} {\bibinfo {title} {Theory of extragalactic radio sources},\ }\href@noop {} {\bibfield  {journal} {\bibinfo  {journal} {Reviews of Modern Physics}\ }\textbf {\bibinfo {volume} {56}},\ \bibinfo {pages} {255} (\bibinfo {year} {1984})}\BibitemShut {NoStop}%
\bibitem [{\citenamefont {Narayan}\ \emph {et~al.}(1995)\citenamefont {Narayan}, \citenamefont {Yi},\ and\ \citenamefont {Mahadevan}}]{narayan1995explaining}%
  \BibitemOpen
  \bibfield  {author} {\bibinfo {author} {\bibfnamefont {R.}~\bibnamefont {Narayan}}, \bibinfo {author} {\bibfnamefont {I.}~\bibnamefont {Yi}},\ and\ \bibinfo {author} {\bibfnamefont {R.}~\bibnamefont {Mahadevan}},\ }\bibfield  {title} {\bibinfo {title} {Explaining the spectrum of sagittarius a* with a model of an accreting black hole},\ }\href@noop {} {\bibfield  {journal} {\bibinfo  {journal} {Nature}\ }\textbf {\bibinfo {volume} {374}},\ \bibinfo {pages} {623} (\bibinfo {year} {1995})}\BibitemShut {NoStop}%
\bibitem [{\citenamefont {Yuan}\ and\ \citenamefont {Narayan}(2004)}]{yuan2004nature}%
  \BibitemOpen
  \bibfield  {author} {\bibinfo {author} {\bibfnamefont {F.}~\bibnamefont {Yuan}}\ and\ \bibinfo {author} {\bibfnamefont {R.}~\bibnamefont {Narayan}},\ }\bibfield  {title} {\bibinfo {title} {On the nature of x-ray-bright, optically normal galaxies},\ }\href@noop {} {\bibfield  {journal} {\bibinfo  {journal} {The Astrophysical Journal}\ }\textbf {\bibinfo {volume} {612}},\ \bibinfo {pages} {724} (\bibinfo {year} {2004})}\BibitemShut {NoStop}%
\bibitem [{\citenamefont {Narayan}\ and\ \citenamefont {McClintock}(2008)}]{narayan2008advection}%
  \BibitemOpen
  \bibfield  {author} {\bibinfo {author} {\bibfnamefont {R.}~\bibnamefont {Narayan}}\ and\ \bibinfo {author} {\bibfnamefont {J.~E.}\ \bibnamefont {McClintock}},\ }\bibfield  {title} {\bibinfo {title} {Advection-dominated accretion and the black hole event horizon},\ }\href@noop {} {\bibfield  {journal} {\bibinfo  {journal} {New Astronomy Reviews}\ }\textbf {\bibinfo {volume} {51}},\ \bibinfo {pages} {733} (\bibinfo {year} {2008})}\BibitemShut {NoStop}%
\bibitem [{\citenamefont {Ichimaru}(1977)}]{ichimaru1977bimodal}%
  \BibitemOpen
  \bibfield  {author} {\bibinfo {author} {\bibfnamefont {S.}~\bibnamefont {Ichimaru}},\ }\bibfield  {title} {\bibinfo {title} {Bimodal behavior of accretion disks-theory and application to cygnus x-1 transitions},\ }\href@noop {} {\bibfield  {journal} {\bibinfo  {journal} {The Astrophysical Journal}\ }\textbf {\bibinfo {volume} {214}},\ \bibinfo {pages} {840} (\bibinfo {year} {1977})}\BibitemShut {NoStop}%
\bibitem [{\citenamefont {Rees}\ \emph {et~al.}(1982)\citenamefont {Rees}, \citenamefont {Begelman}, \citenamefont {Blandford},\ and\ \citenamefont {Phinney}}]{rees1982ion}%
  \BibitemOpen
  \bibfield  {author} {\bibinfo {author} {\bibfnamefont {M.}~\bibnamefont {Rees}}, \bibinfo {author} {\bibfnamefont {M.}~\bibnamefont {Begelman}}, \bibinfo {author} {\bibfnamefont {R.}~\bibnamefont {Blandford}},\ and\ \bibinfo {author} {\bibfnamefont {E.}~\bibnamefont {Phinney}},\ }\bibfield  {title} {\bibinfo {title} {Ion-supported tori and the origin of radio jets},\ }\href@noop {} {\bibfield  {journal} {\bibinfo  {journal} {Nature}\ }\textbf {\bibinfo {volume} {295}},\ \bibinfo {pages} {17} (\bibinfo {year} {1982})}\BibitemShut {NoStop}%
\bibitem [{\citenamefont {Narayan}\ and\ \citenamefont {Yi}(1994)}]{narayan1994advection}%
  \BibitemOpen
  \bibfield  {author} {\bibinfo {author} {\bibfnamefont {R.}~\bibnamefont {Narayan}}\ and\ \bibinfo {author} {\bibfnamefont {I.}~\bibnamefont {Yi}},\ }\bibfield  {title} {\bibinfo {title} {Advection-dominated accretion: A self-similar solution},\ }\href@noop {} {\bibfield  {journal} {\bibinfo  {journal} {arXiv preprint astro-ph/9403052}\ } (\bibinfo {year} {1994})}\BibitemShut {NoStop}%
\bibitem [{\citenamefont {Quataert}(2003)}]{quataert2003radiatively}%
  \BibitemOpen
  \bibfield  {author} {\bibinfo {author} {\bibfnamefont {E.}~\bibnamefont {Quataert}},\ }\bibfield  {title} {\bibinfo {title} {Radiatively inefficient accretion flow models of sgr a},\ }\href@noop {} {\bibfield  {journal} {\bibinfo  {journal} {Astronomische Nachrichten: Astronomical Notes}\ }\textbf {\bibinfo {volume} {324}},\ \bibinfo {pages} {435} (\bibinfo {year} {2003})}\BibitemShut {NoStop}%
\bibitem [{\citenamefont {Katz}(1977)}]{katz1977x}%
  \BibitemOpen
  \bibfield  {author} {\bibinfo {author} {\bibfnamefont {J.}~\bibnamefont {Katz}},\ }\bibfield  {title} {\bibinfo {title} {X-rays from spherical accretion onto degenerate dwarfs},\ }\href@noop {} {\bibfield  {journal} {\bibinfo  {journal} {Astrophysical Journal, Part 1, vol. 215, July 1, 1977, p. 265-275.}\ }\textbf {\bibinfo {volume} {215}},\ \bibinfo {pages} {265} (\bibinfo {year} {1977})}\BibitemShut {NoStop}%
\bibitem [{\citenamefont {Begelman}(1978)}]{begelman1978black}%
  \BibitemOpen
  \bibfield  {author} {\bibinfo {author} {\bibfnamefont {M.~C.}\ \bibnamefont {Begelman}},\ }\bibfield  {title} {\bibinfo {title} {Black holes in radiation-dominated gas: an analogue of the bondi accretion problem},\ }\href@noop {} {\bibfield  {journal} {\bibinfo  {journal} {Monthly Notices of the Royal Astronomical Society}\ }\textbf {\bibinfo {volume} {184}},\ \bibinfo {pages} {53} (\bibinfo {year} {1978})}\BibitemShut {NoStop}%
\bibitem [{\citenamefont {Lipunova}\ \emph {et~al.}(2018)\citenamefont {Lipunova}, \citenamefont {Malanchev},\ and\ \citenamefont {Shakura}}]{lipunova2018standard}%
  \BibitemOpen
  \bibfield  {author} {\bibinfo {author} {\bibfnamefont {G.}~\bibnamefont {Lipunova}}, \bibinfo {author} {\bibfnamefont {K.}~\bibnamefont {Malanchev}},\ and\ \bibinfo {author} {\bibfnamefont {N.}~\bibnamefont {Shakura}},\ }\bibfield  {title} {\bibinfo {title} {The standard model of disc accretion},\ }\href@noop {} {\bibfield  {journal} {\bibinfo  {journal} {Accretion Flows in Astrophysics}\ ,\ \bibinfo {pages} {1}} (\bibinfo {year} {2018})}\BibitemShut {NoStop}%
\bibitem [{\citenamefont {Liu}\ and\ \citenamefont {Qiao}(2022)}]{liu2022accretion}%
  \BibitemOpen
  \bibfield  {author} {\bibinfo {author} {\bibfnamefont {B.}~\bibnamefont {Liu}}\ and\ \bibinfo {author} {\bibfnamefont {E.}~\bibnamefont {Qiao}},\ }\bibfield  {title} {\bibinfo {title} {Accretion around black holes: The geometry and spectra},\ }\href@noop {} {\bibfield  {journal} {\bibinfo  {journal} {Iscience}\ }\textbf {\bibinfo {volume} {25}} (\bibinfo {year} {2022})}\BibitemShut {NoStop}%
\bibitem [{\citenamefont {Yuan}\ and\ \citenamefont {Narayan}(2014)}]{yuan2014hot}%
  \BibitemOpen
  \bibfield  {author} {\bibinfo {author} {\bibfnamefont {F.}~\bibnamefont {Yuan}}\ and\ \bibinfo {author} {\bibfnamefont {R.}~\bibnamefont {Narayan}},\ }\bibfield  {title} {\bibinfo {title} {Hot accretion flows around black holes},\ }\href@noop {} {\bibfield  {journal} {\bibinfo  {journal} {Annual Review of Astronomy and Astrophysics}\ }\textbf {\bibinfo {volume} {52}},\ \bibinfo {pages} {529} (\bibinfo {year} {2014})}\BibitemShut {NoStop}%
\bibitem [{\citenamefont {Bian}\ and\ \citenamefont {Zhao}(2003)}]{bian2003accretion}%
  \BibitemOpen
  \bibfield  {author} {\bibinfo {author} {\bibfnamefont {W.-H.}\ \bibnamefont {Bian}}\ and\ \bibinfo {author} {\bibfnamefont {Y.-H.}\ \bibnamefont {Zhao}},\ }\bibfield  {title} {\bibinfo {title} {Accretion rates and the accretion efficiency in agns},\ }\href@noop {} {\bibfield  {journal} {\bibinfo  {journal} {Publications of the Astronomical Society of Japan}\ }\textbf {\bibinfo {volume} {55}},\ \bibinfo {pages} {599} (\bibinfo {year} {2003})}\BibitemShut {NoStop}%
\bibitem [{\citenamefont {Zhou}\ \emph {et~al.}(2017)\citenamefont {Zhou}, \citenamefont {Wang}, \citenamefont {Yuan}, \citenamefont {Lu}, \citenamefont {Dong}, \citenamefont {Wang},\ and\ \citenamefont {Lu}}]{zhou2017vizier}%
  \BibitemOpen
  \bibfield  {author} {\bibinfo {author} {\bibfnamefont {H.}~\bibnamefont {Zhou}}, \bibinfo {author} {\bibfnamefont {T.}~\bibnamefont {Wang}}, \bibinfo {author} {\bibfnamefont {W.}~\bibnamefont {Yuan}}, \bibinfo {author} {\bibfnamefont {H.}~\bibnamefont {Lu}}, \bibinfo {author} {\bibfnamefont {X.}~\bibnamefont {Dong}}, \bibinfo {author} {\bibfnamefont {J.}~\bibnamefont {Wang}},\ and\ \bibinfo {author} {\bibfnamefont {Y.}~\bibnamefont {Lu}},\ }\bibfield  {title} {\bibinfo {title} {Vizier online data catalog: Narrow line seyfert 1 galaxies from sdss-dr3 (zhou+, 2006)},\ }\href@noop {} {\bibfield  {journal} {\bibinfo  {journal} {VizieR Online Data Catalog}\ ,\ \bibinfo {pages} {J}} (\bibinfo {year} {2017})}\BibitemShut {NoStop}%
\bibitem [{\citenamefont {Czerny}(2019)}]{czerny2019slim}%
  \BibitemOpen
  \bibfield  {author} {\bibinfo {author} {\bibfnamefont {B.}~\bibnamefont {Czerny}},\ }\bibfield  {title} {\bibinfo {title} {Slim accretion disks: theory and observational consequences},\ }\href@noop {} {\bibfield  {journal} {\bibinfo  {journal} {Universe}\ }\textbf {\bibinfo {volume} {5}},\ \bibinfo {pages} {131} (\bibinfo {year} {2019})}\BibitemShut {NoStop}%
\bibitem [{\citenamefont {Lynden-Bell}\ and\ \citenamefont {Pringle}(1974)}]{lynden1974evolution}%
  \BibitemOpen
  \bibfield  {author} {\bibinfo {author} {\bibfnamefont {D.}~\bibnamefont {Lynden-Bell}}\ and\ \bibinfo {author} {\bibfnamefont {J.~E.}\ \bibnamefont {Pringle}},\ }\bibfield  {title} {\bibinfo {title} {The evolution of viscous discs and the origin of the nebular variables},\ }\href@noop {} {\bibfield  {journal} {\bibinfo  {journal} {Monthly Notices of the Royal Astronomical Society}\ }\textbf {\bibinfo {volume} {168}},\ \bibinfo {pages} {603} (\bibinfo {year} {1974})}\BibitemShut {NoStop}%
\bibitem [{\citenamefont {Burke}\ \emph {et~al.}(2021)\citenamefont {Burke}, \citenamefont {Shen}, \citenamefont {Blaes}, \citenamefont {Gammie}, \citenamefont {Horne}, \citenamefont {Jiang}, \citenamefont {Liu}, \citenamefont {McHardy}, \citenamefont {Morgan}, \citenamefont {Scaringi} \emph {et~al.}}]{burke2021characteristic}%
  \BibitemOpen
  \bibfield  {author} {\bibinfo {author} {\bibfnamefont {C.~J.}\ \bibnamefont {Burke}}, \bibinfo {author} {\bibfnamefont {Y.}~\bibnamefont {Shen}}, \bibinfo {author} {\bibfnamefont {O.}~\bibnamefont {Blaes}}, \bibinfo {author} {\bibfnamefont {C.~F.}\ \bibnamefont {Gammie}}, \bibinfo {author} {\bibfnamefont {K.}~\bibnamefont {Horne}}, \bibinfo {author} {\bibfnamefont {Y.-F.}\ \bibnamefont {Jiang}}, \bibinfo {author} {\bibfnamefont {X.}~\bibnamefont {Liu}}, \bibinfo {author} {\bibfnamefont {I.~M.}\ \bibnamefont {McHardy}}, \bibinfo {author} {\bibfnamefont {C.~W.}\ \bibnamefont {Morgan}}, \bibinfo {author} {\bibfnamefont {S.}~\bibnamefont {Scaringi}}, \emph {et~al.},\ }\bibfield  {title} {\bibinfo {title} {A characteristic optical variability time scale in astrophysical accretion disks},\ }\href@noop {} {\bibfield  {journal} {\bibinfo  {journal} {Science}\ }\textbf {\bibinfo {volume} {373}},\ \bibinfo {pages} {789} (\bibinfo {year} {2021})}\BibitemShut {NoStop}%
\bibitem [{\citenamefont {Hanawa}(1989)}]{hanawa1989x}%
  \BibitemOpen
  \bibfield  {author} {\bibinfo {author} {\bibfnamefont {T.}~\bibnamefont {Hanawa}},\ }\bibfield  {title} {\bibinfo {title} {X-ray emission from accretion disks in low-mass x-ray binaries},\ }\href@noop {} {\bibfield  {journal} {\bibinfo  {journal} {Astrophysical Journal, Part 1 (ISSN 0004-637X), vol. 341, June 15, 1989, p. 948-954.}\ }\textbf {\bibinfo {volume} {341}},\ \bibinfo {pages} {948} (\bibinfo {year} {1989})}\BibitemShut {NoStop}%
\bibitem [{\citenamefont {Ebisawa}\ \emph {et~al.}(1991)\citenamefont {Ebisawa}, \citenamefont {Mitsuda},\ and\ \citenamefont {Hanawa}}]{ebisawa1991application}%
  \BibitemOpen
  \bibfield  {author} {\bibinfo {author} {\bibfnamefont {K.}~\bibnamefont {Ebisawa}}, \bibinfo {author} {\bibfnamefont {K.}~\bibnamefont {Mitsuda}},\ and\ \bibinfo {author} {\bibfnamefont {T.}~\bibnamefont {Hanawa}},\ }\bibfield  {title} {\bibinfo {title} {Application of a general relativistic accretion disk model to lmc x-1, lmc x-3, x1608-522, and x1636-536},\ }\href@noop {} {\bibfield  {journal} {\bibinfo  {journal} {Astrophysical Journal, Part 1 (ISSN 0004-637X), vol. 367, Jan. 20, 1991, p. 213-220.}\ }\textbf {\bibinfo {volume} {367}},\ \bibinfo {pages} {213} (\bibinfo {year} {1991})}\BibitemShut {NoStop}%
\bibitem [{\citenamefont {Li}\ \emph {et~al.}(2005)\citenamefont {Li}, \citenamefont {Zimmerman}, \citenamefont {Narayan},\ and\ \citenamefont {McClintock}}]{li2005multitemperature}%
  \BibitemOpen
  \bibfield  {author} {\bibinfo {author} {\bibfnamefont {L.-X.}\ \bibnamefont {Li}}, \bibinfo {author} {\bibfnamefont {E.~R.}\ \bibnamefont {Zimmerman}}, \bibinfo {author} {\bibfnamefont {R.}~\bibnamefont {Narayan}},\ and\ \bibinfo {author} {\bibfnamefont {J.~E.}\ \bibnamefont {McClintock}},\ }\bibfield  {title} {\bibinfo {title} {Multitemperature blackbody spectrum of a thin accretion disk around a kerr black hole: model computations and comparison with observations},\ }\href@noop {} {\bibfield  {journal} {\bibinfo  {journal} {The Astrophysical Journal Supplement Series}\ }\textbf {\bibinfo {volume} {157}},\ \bibinfo {pages} {335} (\bibinfo {year} {2005})}\BibitemShut {NoStop}%
\bibitem [{\citenamefont {Zimmerman}\ \emph {et~al.}(2005)\citenamefont {Zimmerman}, \citenamefont {Narayan}, \citenamefont {McClintock},\ and\ \citenamefont {Miller}}]{zimmerman2005multitemperature}%
  \BibitemOpen
  \bibfield  {author} {\bibinfo {author} {\bibfnamefont {E.}~\bibnamefont {Zimmerman}}, \bibinfo {author} {\bibfnamefont {R.}~\bibnamefont {Narayan}}, \bibinfo {author} {\bibfnamefont {J.}~\bibnamefont {McClintock}},\ and\ \bibinfo {author} {\bibfnamefont {J.}~\bibnamefont {Miller}},\ }\bibfield  {title} {\bibinfo {title} {Multitemperature blackbody spectra of thin accretion disks with and without a zero-torque inner boundary condition},\ }\href@noop {} {\bibfield  {journal} {\bibinfo  {journal} {The Astrophysical Journal}\ }\textbf {\bibinfo {volume} {618}},\ \bibinfo {pages} {832} (\bibinfo {year} {2005})}\BibitemShut {NoStop}%
\bibitem [{\citenamefont {Pereyra}\ \emph {et~al.}(2006)\citenamefont {Pereyra}, \citenamefont {Berk}, \citenamefont {Turnshek}, \citenamefont {Hillier}, \citenamefont {Wilhite}, \citenamefont {Kron}, \citenamefont {Schneider},\ and\ \citenamefont {Brinkmann}}]{pereyra2006characteristic}%
  \BibitemOpen
  \bibfield  {author} {\bibinfo {author} {\bibfnamefont {N.~A.}\ \bibnamefont {Pereyra}}, \bibinfo {author} {\bibfnamefont {D.~E.~V.}\ \bibnamefont {Berk}}, \bibinfo {author} {\bibfnamefont {D.~A.}\ \bibnamefont {Turnshek}}, \bibinfo {author} {\bibfnamefont {D.~J.}\ \bibnamefont {Hillier}}, \bibinfo {author} {\bibfnamefont {B.~C.}\ \bibnamefont {Wilhite}}, \bibinfo {author} {\bibfnamefont {R.~G.}\ \bibnamefont {Kron}}, \bibinfo {author} {\bibfnamefont {D.~P.}\ \bibnamefont {Schneider}},\ and\ \bibinfo {author} {\bibfnamefont {J.}~\bibnamefont {Brinkmann}},\ }\bibfield  {title} {\bibinfo {title} {Characteristic qso accretion disk temperatures from spectroscopic continuum variability},\ }\href@noop {} {\bibfield  {journal} {\bibinfo  {journal} {The Astrophysical Journal}\ }\textbf {\bibinfo {volume} {642}},\ \bibinfo {pages} {87} (\bibinfo {year} {2006})}\BibitemShut {NoStop}%
\bibitem [{\citenamefont {Calvet}\ \emph {et~al.}(1999)\citenamefont {Calvet}, \citenamefont {Hartmann},\ and\ \citenamefont {Strom}}]{calvet1999evolution}%
  \BibitemOpen
  \bibfield  {author} {\bibinfo {author} {\bibfnamefont {N.}~\bibnamefont {Calvet}}, \bibinfo {author} {\bibfnamefont {L.}~\bibnamefont {Hartmann}},\ and\ \bibinfo {author} {\bibfnamefont {S.~E.}\ \bibnamefont {Strom}},\ }\bibfield  {title} {\bibinfo {title} {Evolution of disk accretion},\ }\href@noop {} {\bibfield  {journal} {\bibinfo  {journal} {arXiv preprint astro-ph/9902335}\ } (\bibinfo {year} {1999})}\BibitemShut {NoStop}%
\bibitem [{\citenamefont {Caroline}\ and\ \citenamefont {Terquem}(2007)}]{caroline2007theory}%
  \BibitemOpen
  \bibfield  {author} {\bibinfo {author} {\bibfnamefont {E.}~\bibnamefont {Caroline}}\ and\ \bibinfo {author} {\bibfnamefont {L.}~\bibnamefont {Terquem}},\ }\bibfield  {title} {\bibinfo {title} {Theory and models of standard accretion disks},\ }in\ \href@noop {} {\emph {\bibinfo {booktitle} {Jets from Young Stars: Models and Constraints}}}\ (\bibinfo  {publisher} {Springer},\ \bibinfo {year} {2007})\ pp.\ \bibinfo {pages} {103--115}\BibitemShut {NoStop}%
\bibitem [{\citenamefont {Lorenzin}\ and\ \citenamefont {Zampieri}(2009)}]{lorenzin2009comparative}%
  \BibitemOpen
  \bibfield  {author} {\bibinfo {author} {\bibfnamefont {A.}~\bibnamefont {Lorenzin}}\ and\ \bibinfo {author} {\bibfnamefont {L.}~\bibnamefont {Zampieri}},\ }\bibfield  {title} {\bibinfo {title} {A comparative analysis of standard accretion discs spectra: an application to ultraluminous x-ray sources},\ }\href@noop {} {\bibfield  {journal} {\bibinfo  {journal} {Monthly Notices of the Royal Astronomical Society}\ }\textbf {\bibinfo {volume} {394}},\ \bibinfo {pages} {1588} (\bibinfo {year} {2009})}\BibitemShut {NoStop}%
\bibitem [{\citenamefont {Armijo}(2012)}]{armijo2012accretion}%
  \BibitemOpen
  \bibfield  {author} {\bibinfo {author} {\bibfnamefont {M.~M.}\ \bibnamefont {Armijo}},\ }\href@noop {} {\emph {\bibinfo {title} {Accretion disk theory}}},\ \bibinfo {type} {Tech. Rep.}\ (\bibinfo {year} {2012})\BibitemShut {NoStop}%
\bibitem [{\citenamefont {Beloborodov}(2001)}]{beloborodov2001accretion}%
  \BibitemOpen
  \bibfield  {author} {\bibinfo {author} {\bibfnamefont {A.~M.}\ \bibnamefont {Beloborodov}},\ }\bibfield  {title} {\bibinfo {title} {Accretion disk models of luminous black holes},\ }\href@noop {} {\bibfield  {journal} {\bibinfo  {journal} {Advances in Space Research}\ }\textbf {\bibinfo {volume} {28}},\ \bibinfo {pages} {411} (\bibinfo {year} {2001})}\BibitemShut {NoStop}%
\bibitem [{\citenamefont {Khesali}\ and\ \citenamefont {Khosravi}(2013)}]{khesali2013local}%
  \BibitemOpen
  \bibfield  {author} {\bibinfo {author} {\bibfnamefont {A.}~\bibnamefont {Khesali}}\ and\ \bibinfo {author} {\bibfnamefont {A.}~\bibnamefont {Khosravi}},\ }\bibfield  {title} {\bibinfo {title} {The local stability of accretion disk models with considering the role of various viscosity and cooling mechanisms},\ }\href@noop {} {\bibfield  {journal} {\bibinfo  {journal} {Astrophysics and Space Science}\ }\textbf {\bibinfo {volume} {348}},\ \bibinfo {pages} {143} (\bibinfo {year} {2013})}\BibitemShut {NoStop}%
\bibitem [{\citenamefont {Smith}(2021)}]{smith2021timing}%
  \BibitemOpen
  \bibfield  {author} {\bibinfo {author} {\bibfnamefont {E.~A.}\ \bibnamefont {Smith}},\ }\bibfield  {title} {\bibinfo {title} {Timing properties of active galactic nuclei},\ }\href@noop {} {\  (\bibinfo {year} {2021})}\BibitemShut {NoStop}%
\bibitem [{\citenamefont {Shapiro}\ \emph {et~al.}(1976)\citenamefont {Shapiro}, \citenamefont {Lightman},\ and\ \citenamefont {Eardley}}]{shapiro1976two}%
  \BibitemOpen
  \bibfield  {author} {\bibinfo {author} {\bibfnamefont {S.~L.}\ \bibnamefont {Shapiro}}, \bibinfo {author} {\bibfnamefont {A.~P.}\ \bibnamefont {Lightman}},\ and\ \bibinfo {author} {\bibfnamefont {D.~M.}\ \bibnamefont {Eardley}},\ }\bibfield  {title} {\bibinfo {title} {A two-temperature accretion disk model for cygnus x-1-structure and spectrum},\ }\href@noop {} {\bibfield  {journal} {\bibinfo  {journal} {Astrophysical Journal, vol. 204, Feb. 15, 1976, pt. 1, p. 187-199.}\ }\textbf {\bibinfo {volume} {204}},\ \bibinfo {pages} {187} (\bibinfo {year} {1976})}\BibitemShut {NoStop}%
\bibitem [{\citenamefont {Kokubo}(2018)}]{kokubo2018constraints}%
  \BibitemOpen
  \bibfield  {author} {\bibinfo {author} {\bibfnamefont {M.}~\bibnamefont {Kokubo}},\ }\bibfield  {title} {\bibinfo {title} {Constraints on accretion disk size in the massive type 1 quasar pg 2308+ 098 from optical continuum reverberation lags},\ }\href@noop {} {\bibfield  {journal} {\bibinfo  {journal} {Publications of the Astronomical Society of Japan}\ }\textbf {\bibinfo {volume} {70}},\ \bibinfo {pages} {97} (\bibinfo {year} {2018})}\BibitemShut {NoStop}%
\bibitem [{\citenamefont {Yoneyama}\ and\ \citenamefont {Dotani}(2023)}]{yoneyama2023x}%
  \BibitemOpen
  \bibfield  {author} {\bibinfo {author} {\bibfnamefont {T.}~\bibnamefont {Yoneyama}}\ and\ \bibinfo {author} {\bibfnamefont {T.}~\bibnamefont {Dotani}},\ }\bibfield  {title} {\bibinfo {title} {X-ray spectroscopy of the accretion disk corona source 2s 0921- 630 with suzaku archival data},\ }\href@noop {} {\bibfield  {journal} {\bibinfo  {journal} {Publications of the Astronomical Society of Japan}\ }\textbf {\bibinfo {volume} {75}},\ \bibinfo {pages} {30} (\bibinfo {year} {2023})}\BibitemShut {NoStop}%
\bibitem [{\citenamefont {Montesinos}(2012)}]{montesinos2012accretion}%
  \BibitemOpen
  \bibfield  {author} {\bibinfo {author} {\bibfnamefont {M.}~\bibnamefont {Montesinos}},\ }\bibfield  {title} {\bibinfo {title} {Accretion disk theory},\ }\href@noop {} {\bibfield  {journal} {\bibinfo  {journal} {arXiv preprint arXiv:1203.6851}\ } (\bibinfo {year} {2012})}\BibitemShut {NoStop}%
\bibitem [{\citenamefont {Cannizzo}(1993)}]{cannizzo1993accretion}%
  \BibitemOpen
  \bibfield  {author} {\bibinfo {author} {\bibfnamefont {J.~K.}\ \bibnamefont {Cannizzo}},\ }\bibfield  {title} {\bibinfo {title} {The accretion disk limit cycle model: toward an understanding of the long-term behavior of ss cygni},\ }\href@noop {} {\bibfield  {journal} {\bibinfo  {journal} {Astrophysical Journal v. 419, p. 318}\ }\textbf {\bibinfo {volume} {419}},\ \bibinfo {pages} {318} (\bibinfo {year} {1993})}\BibitemShut {NoStop}%
\bibitem [{\citenamefont {Sun}\ and\ \citenamefont {Malkan}(1986)}]{sun1986new}%
  \BibitemOpen
  \bibfield  {author} {\bibinfo {author} {\bibfnamefont {W.-H.}\ \bibnamefont {Sun}}\ and\ \bibinfo {author} {\bibfnamefont {M.~A.}\ \bibnamefont {Malkan}},\ }\bibfield  {title} {\bibinfo {title} {New accretion disk modeling of active galactic nuclei and quasars},\ }in\ \href@noop {} {\emph {\bibinfo {booktitle} {New Insights in Astrophysics. Eight Years of UV Astronomy with IUE}}},\ Vol.\ \bibinfo {volume} {263}\ (\bibinfo {year} {1986})\BibitemShut {NoStop}%
\bibitem [{\citenamefont {Schneider}\ \emph {et~al.}(2010)\citenamefont {Schneider}, \citenamefont {Richards}, \citenamefont {Hall}, \citenamefont {Strauss}, \citenamefont {Anderson}, \citenamefont {Boroson}, \citenamefont {Ross}, \citenamefont {Shen}, \citenamefont {Brandt}, \citenamefont {Fan} \emph {et~al.}}]{schneider2010sloan}%
  \BibitemOpen
  \bibfield  {author} {\bibinfo {author} {\bibfnamefont {D.~P.}\ \bibnamefont {Schneider}}, \bibinfo {author} {\bibfnamefont {G.~T.}\ \bibnamefont {Richards}}, \bibinfo {author} {\bibfnamefont {P.~B.}\ \bibnamefont {Hall}}, \bibinfo {author} {\bibfnamefont {M.~A.}\ \bibnamefont {Strauss}}, \bibinfo {author} {\bibfnamefont {S.~F.}\ \bibnamefont {Anderson}}, \bibinfo {author} {\bibfnamefont {T.~A.}\ \bibnamefont {Boroson}}, \bibinfo {author} {\bibfnamefont {N.~P.}\ \bibnamefont {Ross}}, \bibinfo {author} {\bibfnamefont {Y.}~\bibnamefont {Shen}}, \bibinfo {author} {\bibfnamefont {W.}~\bibnamefont {Brandt}}, \bibinfo {author} {\bibfnamefont {X.}~\bibnamefont {Fan}}, \emph {et~al.},\ }\bibfield  {title} {\bibinfo {title} {The sloan digital sky survey quasar catalog. v. seventh data release},\ }\href@noop {} {\bibfield  {journal} {\bibinfo  {journal} {The Astronomical Journal}\ }\textbf {\bibinfo {volume} {139}},\ \bibinfo {pages} {2360} (\bibinfo {year} {2010})}\BibitemShut {NoStop}%
\bibitem [{\citenamefont {Natarajan}\ \emph {et~al.}(2023)\citenamefont {Natarajan}, \citenamefont {Tang}, \citenamefont {McGibbon}, \citenamefont {Khochfar}, \citenamefont {Nord}, \citenamefont {Sigurdsson}, \citenamefont {Tricot}, \citenamefont {Cappelluti}, \citenamefont {George},\ and\ \citenamefont {Hidary}}]{natarajan2023quotas}%
  \BibitemOpen
  \bibfield  {author} {\bibinfo {author} {\bibfnamefont {P.}~\bibnamefont {Natarajan}}, \bibinfo {author} {\bibfnamefont {K.~S.}\ \bibnamefont {Tang}}, \bibinfo {author} {\bibfnamefont {R.}~\bibnamefont {McGibbon}}, \bibinfo {author} {\bibfnamefont {S.}~\bibnamefont {Khochfar}}, \bibinfo {author} {\bibfnamefont {B.}~\bibnamefont {Nord}}, \bibinfo {author} {\bibfnamefont {S.}~\bibnamefont {Sigurdsson}}, \bibinfo {author} {\bibfnamefont {J.}~\bibnamefont {Tricot}}, \bibinfo {author} {\bibfnamefont {N.}~\bibnamefont {Cappelluti}}, \bibinfo {author} {\bibfnamefont {D.}~\bibnamefont {George}},\ and\ \bibinfo {author} {\bibfnamefont {J.}~\bibnamefont {Hidary}},\ }\bibfield  {title} {\bibinfo {title} {Quotas: A new research platform for the data-driven discovery of black holes},\ }\href@noop {} {\bibfield  {journal} {\bibinfo  {journal} {The Astrophysical Journal}\ }\textbf {\bibinfo {volume} {952}},\ \bibinfo {pages} {146} (\bibinfo {year} {2023})}\BibitemShut {NoStop}%
\bibitem [{\citenamefont {McConnell}\ and\ \citenamefont {Ma}(2013)}]{mcconnell2013revisiting}%
  \BibitemOpen
  \bibfield  {author} {\bibinfo {author} {\bibfnamefont {N.~J.}\ \bibnamefont {McConnell}}\ and\ \bibinfo {author} {\bibfnamefont {C.-P.}\ \bibnamefont {Ma}},\ }\bibfield  {title} {\bibinfo {title} {Revisiting the scaling relations of black hole masses and host galaxy properties},\ }\href@noop {} {\bibfield  {journal} {\bibinfo  {journal} {The Astrophysical Journal}\ }\textbf {\bibinfo {volume} {764}},\ \bibinfo {pages} {184} (\bibinfo {year} {2013})}\BibitemShut {NoStop}%
\bibitem [{\citenamefont {Cook}\ \emph {et~al.}(2023)\citenamefont {Cook}, \citenamefont {Mazzarella}, \citenamefont {Helou}, \citenamefont {Alcala}, \citenamefont {Chen}, \citenamefont {Ebert}, \citenamefont {Frayer}, \citenamefont {Kim}, \citenamefont {Lo}, \citenamefont {Madore} \emph {et~al.}}]{cook2023completeness}%
  \BibitemOpen
  \bibfield  {author} {\bibinfo {author} {\bibfnamefont {D.}~\bibnamefont {Cook}}, \bibinfo {author} {\bibfnamefont {J.}~\bibnamefont {Mazzarella}}, \bibinfo {author} {\bibfnamefont {G.}~\bibnamefont {Helou}}, \bibinfo {author} {\bibfnamefont {A.}~\bibnamefont {Alcala}}, \bibinfo {author} {\bibfnamefont {T.}~\bibnamefont {Chen}}, \bibinfo {author} {\bibfnamefont {R.}~\bibnamefont {Ebert}}, \bibinfo {author} {\bibfnamefont {C.}~\bibnamefont {Frayer}}, \bibinfo {author} {\bibfnamefont {J.}~\bibnamefont {Kim}}, \bibinfo {author} {\bibfnamefont {T.}~\bibnamefont {Lo}}, \bibinfo {author} {\bibfnamefont {B.}~\bibnamefont {Madore}}, \emph {et~al.},\ }\bibfield  {title} {\bibinfo {title} {Completeness of the nasa/ipac extragalactic database (ned)--local volume sample},\ }\href@noop {} {\bibfield  {journal} {\bibinfo  {journal} {arXiv preprint arXiv:2306.06271}\ } (\bibinfo {year} {2023})}\BibitemShut {NoStop}%
\bibitem [{\citenamefont {Piotrovich}\ \emph {et~al.}(2023{\natexlab{b}})\citenamefont {Piotrovich}, \citenamefont {Shablovinskaya}, \citenamefont {Malygin}, \citenamefont {Buliga},\ and\ \citenamefont {Natsvlishvili}}]{piotrovich2023probing}%
  \BibitemOpen
  \bibfield  {author} {\bibinfo {author} {\bibfnamefont {M.~Y.}\ \bibnamefont {Piotrovich}}, \bibinfo {author} {\bibfnamefont {E.}~\bibnamefont {Shablovinskaya}}, \bibinfo {author} {\bibfnamefont {E.}~\bibnamefont {Malygin}}, \bibinfo {author} {\bibfnamefont {S.}~\bibnamefont {Buliga}},\ and\ \bibinfo {author} {\bibfnamefont {T.}~\bibnamefont {Natsvlishvili}},\ }\bibfield  {title} {\bibinfo {title} {Probing agn with spectropolarimetry: Accretion disk and smbh parameters},\ }\href@noop {} {\bibfield  {journal} {\bibinfo  {journal} {Monthly Notices of the Royal Astronomical Society}\ ,\ \bibinfo {pages} {stad2934}} (\bibinfo {year} {2023}{\natexlab{b}})}\BibitemShut {NoStop}%
\bibitem [{\citenamefont {Schmidt}(1963)}]{Schmidt1963wkp}%
  \BibitemOpen
  \bibfield  {author} {\bibinfo {author} {\bibfnamefont {M.}~\bibnamefont {Schmidt}},\ }\bibfield  {title} {\bibinfo {title} {{3C 273 : A Star-Like Object with Large Red-Shift}},\ }\href {https://doi.org/10.1038/1971040a0} {\bibfield  {journal} {\bibinfo  {journal} {Nature}\ }\textbf {\bibinfo {volume} {197}},\ \bibinfo {pages} {1040} (\bibinfo {year} {1963})}\BibitemShut {NoStop}%
\bibitem [{\citenamefont {Antonucci}\ and\ \citenamefont {Miller}(1985)}]{Antonucci1985aa}%
  \BibitemOpen
  \bibfield  {author} {\bibinfo {author} {\bibfnamefont {R.~R.~J.}\ \bibnamefont {Antonucci}}\ and\ \bibinfo {author} {\bibfnamefont {J.~S.}\ \bibnamefont {Miller}},\ }\bibfield  {title} {\bibinfo {title} {{Spectropolarimetry and the nature of NGC 1068}},\ }\href {https://doi.org/10.1086/163559} {\bibfield  {journal} {\bibinfo  {journal} {Astrophys. J.}\ }\textbf {\bibinfo {volume} {297}},\ \bibinfo {pages} {621} (\bibinfo {year} {1985})}\BibitemShut {NoStop}%
\bibitem [{\citenamefont {Ross}\ \emph {et~al.}(2013)\citenamefont {Ross}, \citenamefont {McGreer}, \citenamefont {White}, \citenamefont {Richards}, \citenamefont {Myers}, \citenamefont {Palanque-Delabrouille}, \citenamefont {Strauss}, \citenamefont {Anderson}, \citenamefont {Shen}, \citenamefont {Brandt} \emph {et~al.}}]{ross2013sdss}%
  \BibitemOpen
  \bibfield  {author} {\bibinfo {author} {\bibfnamefont {N.~P.}\ \bibnamefont {Ross}}, \bibinfo {author} {\bibfnamefont {I.~D.}\ \bibnamefont {McGreer}}, \bibinfo {author} {\bibfnamefont {M.}~\bibnamefont {White}}, \bibinfo {author} {\bibfnamefont {G.~T.}\ \bibnamefont {Richards}}, \bibinfo {author} {\bibfnamefont {A.~D.}\ \bibnamefont {Myers}}, \bibinfo {author} {\bibfnamefont {N.}~\bibnamefont {Palanque-Delabrouille}}, \bibinfo {author} {\bibfnamefont {M.~A.}\ \bibnamefont {Strauss}}, \bibinfo {author} {\bibfnamefont {S.~F.}\ \bibnamefont {Anderson}}, \bibinfo {author} {\bibfnamefont {Y.}~\bibnamefont {Shen}}, \bibinfo {author} {\bibfnamefont {W.}~\bibnamefont {Brandt}}, \emph {et~al.},\ }\bibfield  {title} {\bibinfo {title} {The sdss-iii baryon oscillation spectroscopic survey: the quasar luminosity function from data release nine},\ }\href@noop {} {\bibfield  {journal} {\bibinfo  {journal} {The Astrophysical Journal}\ }\textbf {\bibinfo {volume} {773}},\ \bibinfo {pages} {14} (\bibinfo {year}
  {2013})}\BibitemShut {NoStop}%
\bibitem [{\citenamefont {McGreer}\ \emph {et~al.}(2013)\citenamefont {McGreer}, \citenamefont {Jiang}, \citenamefont {Fan}, \citenamefont {Richards}, \citenamefont {Strauss}, \citenamefont {Ross}, \citenamefont {White}, \citenamefont {Shen}, \citenamefont {Schneider}, \citenamefont {Myers} \emph {et~al.}}]{mcgreer2013z}%
  \BibitemOpen
  \bibfield  {author} {\bibinfo {author} {\bibfnamefont {I.~D.}\ \bibnamefont {McGreer}}, \bibinfo {author} {\bibfnamefont {L.}~\bibnamefont {Jiang}}, \bibinfo {author} {\bibfnamefont {X.}~\bibnamefont {Fan}}, \bibinfo {author} {\bibfnamefont {G.~T.}\ \bibnamefont {Richards}}, \bibinfo {author} {\bibfnamefont {M.~A.}\ \bibnamefont {Strauss}}, \bibinfo {author} {\bibfnamefont {N.~P.}\ \bibnamefont {Ross}}, \bibinfo {author} {\bibfnamefont {M.}~\bibnamefont {White}}, \bibinfo {author} {\bibfnamefont {Y.}~\bibnamefont {Shen}}, \bibinfo {author} {\bibfnamefont {D.~P.}\ \bibnamefont {Schneider}}, \bibinfo {author} {\bibfnamefont {A.~D.}\ \bibnamefont {Myers}}, \emph {et~al.},\ }\bibfield  {title} {\bibinfo {title} {The z= 5 quasar luminosity function from sdss stripe 82},\ }\href@noop {} {\bibfield  {journal} {\bibinfo  {journal} {The Astrophysical Journal}\ }\textbf {\bibinfo {volume} {768}},\ \bibinfo {pages} {105} (\bibinfo {year} {2013})}\BibitemShut {NoStop}%
\bibitem [{\citenamefont {Glikman}\ \emph {et~al.}(2011)\citenamefont {Glikman}, \citenamefont {Djorgovski}, \citenamefont {Stern}, \citenamefont {Dey}, \citenamefont {Jannuzi},\ and\ \citenamefont {Lee}}]{glikman2011faint}%
  \BibitemOpen
  \bibfield  {author} {\bibinfo {author} {\bibfnamefont {E.}~\bibnamefont {Glikman}}, \bibinfo {author} {\bibfnamefont {S.}~\bibnamefont {Djorgovski}}, \bibinfo {author} {\bibfnamefont {D.}~\bibnamefont {Stern}}, \bibinfo {author} {\bibfnamefont {A.}~\bibnamefont {Dey}}, \bibinfo {author} {\bibfnamefont {B.~T.}\ \bibnamefont {Jannuzi}},\ and\ \bibinfo {author} {\bibfnamefont {K.-S.}\ \bibnamefont {Lee}},\ }\bibfield  {title} {\bibinfo {title} {The faint end of the quasar luminosity function at z~ 4: implications for ionization of the intergalactic medium and cosmic downsizing},\ }\href@noop {} {\bibfield  {journal} {\bibinfo  {journal} {The Astrophysical Journal Letters}\ }\textbf {\bibinfo {volume} {728}},\ \bibinfo {pages} {L26} (\bibinfo {year} {2011})}\BibitemShut {NoStop}%
\bibitem [{\citenamefont {Yang}\ \emph {et~al.}(2016)\citenamefont {Yang}, \citenamefont {Wang}, \citenamefont {Wu}, \citenamefont {Fan}, \citenamefont {McGreer}, \citenamefont {Bian}, \citenamefont {Yi}, \citenamefont {Yang}, \citenamefont {Ai}, \citenamefont {Dong} \emph {et~al.}}]{yang2016survey}%
  \BibitemOpen
  \bibfield  {author} {\bibinfo {author} {\bibfnamefont {J.}~\bibnamefont {Yang}}, \bibinfo {author} {\bibfnamefont {F.}~\bibnamefont {Wang}}, \bibinfo {author} {\bibfnamefont {X.-B.}\ \bibnamefont {Wu}}, \bibinfo {author} {\bibfnamefont {X.}~\bibnamefont {Fan}}, \bibinfo {author} {\bibfnamefont {I.~D.}\ \bibnamefont {McGreer}}, \bibinfo {author} {\bibfnamefont {F.}~\bibnamefont {Bian}}, \bibinfo {author} {\bibfnamefont {W.}~\bibnamefont {Yi}}, \bibinfo {author} {\bibfnamefont {Q.}~\bibnamefont {Yang}}, \bibinfo {author} {\bibfnamefont {Y.}~\bibnamefont {Ai}}, \bibinfo {author} {\bibfnamefont {X.}~\bibnamefont {Dong}}, \emph {et~al.},\ }\bibfield  {title} {\bibinfo {title} {A survey of luminous high-redshift quasars with sdss and wise. ii. the bright end of the quasar luminosity function at z~ 5},\ }\href@noop {} {\bibfield  {journal} {\bibinfo  {journal} {The Astrophysical Journal}\ }\textbf {\bibinfo {volume} {829}},\ \bibinfo {pages} {33} (\bibinfo {year} {2016})}\BibitemShut {NoStop}%
\bibitem [{\citenamefont {Jiang}\ \emph {et~al.}(2016)\citenamefont {Jiang}, \citenamefont {McGreer}, \citenamefont {Fan}, \citenamefont {Strauss}, \citenamefont {Ba{\~n}ados}, \citenamefont {Becker}, \citenamefont {Bian}, \citenamefont {Farnsworth}, \citenamefont {Shen}, \citenamefont {Wang} \emph {et~al.}}]{jiang2016final}%
  \BibitemOpen
  \bibfield  {author} {\bibinfo {author} {\bibfnamefont {L.}~\bibnamefont {Jiang}}, \bibinfo {author} {\bibfnamefont {I.~D.}\ \bibnamefont {McGreer}}, \bibinfo {author} {\bibfnamefont {X.}~\bibnamefont {Fan}}, \bibinfo {author} {\bibfnamefont {M.~A.}\ \bibnamefont {Strauss}}, \bibinfo {author} {\bibfnamefont {E.}~\bibnamefont {Ba{\~n}ados}}, \bibinfo {author} {\bibfnamefont {R.~H.}\ \bibnamefont {Becker}}, \bibinfo {author} {\bibfnamefont {F.}~\bibnamefont {Bian}}, \bibinfo {author} {\bibfnamefont {K.}~\bibnamefont {Farnsworth}}, \bibinfo {author} {\bibfnamefont {Y.}~\bibnamefont {Shen}}, \bibinfo {author} {\bibfnamefont {F.}~\bibnamefont {Wang}}, \emph {et~al.},\ }\bibfield  {title} {\bibinfo {title} {The final sdss high-redshift quasar sample of 52 quasars at z> 5.7},\ }\href@noop {} {\bibfield  {journal} {\bibinfo  {journal} {The Astrophysical Journal}\ }\textbf {\bibinfo {volume} {833}},\ \bibinfo {pages} {222} (\bibinfo {year} {2016})}\BibitemShut {NoStop}%
\bibitem [{\citenamefont {Willott}\ \emph {et~al.}(2010{\natexlab{a}})\citenamefont {Willott}, \citenamefont {Delorme}, \citenamefont {Reyl{\'e}}, \citenamefont {Albert}, \citenamefont {Bergeron}, \citenamefont {Crampton}, \citenamefont {Delfosse}, \citenamefont {Forveille}, \citenamefont {Hutchings}, \citenamefont {McLure} \emph {et~al.}}]{willott2010canada}%
  \BibitemOpen
  \bibfield  {author} {\bibinfo {author} {\bibfnamefont {C.~J.}\ \bibnamefont {Willott}}, \bibinfo {author} {\bibfnamefont {P.}~\bibnamefont {Delorme}}, \bibinfo {author} {\bibfnamefont {C.}~\bibnamefont {Reyl{\'e}}}, \bibinfo {author} {\bibfnamefont {L.}~\bibnamefont {Albert}}, \bibinfo {author} {\bibfnamefont {J.}~\bibnamefont {Bergeron}}, \bibinfo {author} {\bibfnamefont {D.}~\bibnamefont {Crampton}}, \bibinfo {author} {\bibfnamefont {X.}~\bibnamefont {Delfosse}}, \bibinfo {author} {\bibfnamefont {T.}~\bibnamefont {Forveille}}, \bibinfo {author} {\bibfnamefont {J.~B.}\ \bibnamefont {Hutchings}}, \bibinfo {author} {\bibfnamefont {R.~J.}\ \bibnamefont {McLure}}, \emph {et~al.},\ }\bibfield  {title} {\bibinfo {title} {The canada--france high-z quasar survey: nine new quasars and the luminosity function at redshift 6},\ }\href@noop {} {\bibfield  {journal} {\bibinfo  {journal} {The Astronomical Journal}\ }\textbf {\bibinfo {volume} {139}},\ \bibinfo {pages} {906} (\bibinfo {year}
  {2010}{\natexlab{a}})}\BibitemShut {NoStop}%
\bibitem [{\citenamefont {Kashikawa}\ \emph {et~al.}(2014)\citenamefont {Kashikawa}, \citenamefont {Ishizaki}, \citenamefont {Willott}, \citenamefont {Onoue}, \citenamefont {Im}, \citenamefont {Furusawa}, \citenamefont {Toshikawa}, \citenamefont {Ishikawa}, \citenamefont {Niino}, \citenamefont {Shimasaku} \emph {et~al.}}]{kashikawa2014subaru}%
  \BibitemOpen
  \bibfield  {author} {\bibinfo {author} {\bibfnamefont {N.}~\bibnamefont {Kashikawa}}, \bibinfo {author} {\bibfnamefont {Y.}~\bibnamefont {Ishizaki}}, \bibinfo {author} {\bibfnamefont {C.~J.}\ \bibnamefont {Willott}}, \bibinfo {author} {\bibfnamefont {M.}~\bibnamefont {Onoue}}, \bibinfo {author} {\bibfnamefont {M.}~\bibnamefont {Im}}, \bibinfo {author} {\bibfnamefont {H.}~\bibnamefont {Furusawa}}, \bibinfo {author} {\bibfnamefont {J.}~\bibnamefont {Toshikawa}}, \bibinfo {author} {\bibfnamefont {S.}~\bibnamefont {Ishikawa}}, \bibinfo {author} {\bibfnamefont {Y.}~\bibnamefont {Niino}}, \bibinfo {author} {\bibfnamefont {K.}~\bibnamefont {Shimasaku}}, \emph {et~al.},\ }\bibfield  {title} {\bibinfo {title} {The subaru high-z quasar survey: discovery of faint z~ 6 quasars},\ }\href@noop {} {\bibfield  {journal} {\bibinfo  {journal} {The Astrophysical Journal}\ }\textbf {\bibinfo {volume} {798}},\ \bibinfo {pages} {28} (\bibinfo {year} {2014})}\BibitemShut {NoStop}%
\bibitem [{\citenamefont {Giallongo}\ \emph {et~al.}(2015)\citenamefont {Giallongo}, \citenamefont {Grazian}, \citenamefont {Fiore}, \citenamefont {Fontana}, \citenamefont {Pentericci}, \citenamefont {Vanzella}, \citenamefont {Dickinson}, \citenamefont {Kocevski}, \citenamefont {Castellano}, \citenamefont {Cristiani} \emph {et~al.}}]{giallongo2015faint}%
  \BibitemOpen
  \bibfield  {author} {\bibinfo {author} {\bibfnamefont {E.}~\bibnamefont {Giallongo}}, \bibinfo {author} {\bibfnamefont {A.}~\bibnamefont {Grazian}}, \bibinfo {author} {\bibfnamefont {F.}~\bibnamefont {Fiore}}, \bibinfo {author} {\bibfnamefont {A.}~\bibnamefont {Fontana}}, \bibinfo {author} {\bibfnamefont {L.}~\bibnamefont {Pentericci}}, \bibinfo {author} {\bibfnamefont {E.}~\bibnamefont {Vanzella}}, \bibinfo {author} {\bibfnamefont {M.}~\bibnamefont {Dickinson}}, \bibinfo {author} {\bibfnamefont {D.}~\bibnamefont {Kocevski}}, \bibinfo {author} {\bibfnamefont {M.}~\bibnamefont {Castellano}}, \bibinfo {author} {\bibfnamefont {S.}~\bibnamefont {Cristiani}}, \emph {et~al.},\ }\bibfield  {title} {\bibinfo {title} {Faint agns at z> 4 in the candels goods-s field: looking for contributors to the reionization of the universe},\ }\href@noop {} {\bibfield  {journal} {\bibinfo  {journal} {Astronomy \& Astrophysics}\ }\textbf {\bibinfo {volume} {578}},\ \bibinfo {pages} {A83} (\bibinfo {year} {2015})}\BibitemShut
  {NoStop}%
\bibitem [{\citenamefont {Mortlock}\ \emph {et~al.}(2011)\citenamefont {Mortlock}, \citenamefont {Warren}, \citenamefont {Venemans}, \citenamefont {Patel}, \citenamefont {Hewett}, \citenamefont {McMahon}, \citenamefont {Simpson}, \citenamefont {Theuns}, \citenamefont {Gonz{\'a}les-Solares}, \citenamefont {Adamson} \emph {et~al.}}]{mortlock2011luminous}%
  \BibitemOpen
  \bibfield  {author} {\bibinfo {author} {\bibfnamefont {D.~J.}\ \bibnamefont {Mortlock}}, \bibinfo {author} {\bibfnamefont {S.~J.}\ \bibnamefont {Warren}}, \bibinfo {author} {\bibfnamefont {B.~P.}\ \bibnamefont {Venemans}}, \bibinfo {author} {\bibfnamefont {M.}~\bibnamefont {Patel}}, \bibinfo {author} {\bibfnamefont {P.~C.}\ \bibnamefont {Hewett}}, \bibinfo {author} {\bibfnamefont {R.~G.}\ \bibnamefont {McMahon}}, \bibinfo {author} {\bibfnamefont {C.}~\bibnamefont {Simpson}}, \bibinfo {author} {\bibfnamefont {T.}~\bibnamefont {Theuns}}, \bibinfo {author} {\bibfnamefont {E.~A.}\ \bibnamefont {Gonz{\'a}les-Solares}}, \bibinfo {author} {\bibfnamefont {A.}~\bibnamefont {Adamson}}, \emph {et~al.},\ }\bibfield  {title} {\bibinfo {title} {A luminous quasar at a redshift of z= 7.085},\ }\href@noop {} {\bibfield  {journal} {\bibinfo  {journal} {Nature}\ }\textbf {\bibinfo {volume} {474}},\ \bibinfo {pages} {616} (\bibinfo {year} {2011})}\BibitemShut {NoStop}%
\bibitem [{\citenamefont {Venemans}\ \emph {et~al.}(2015)\citenamefont {Venemans}, \citenamefont {Ba{\~n}ados}, \citenamefont {Decarli}, \citenamefont {Farina}, \citenamefont {Walter}, \citenamefont {Chambers}, \citenamefont {Fan}, \citenamefont {Rix}, \citenamefont {Schlafly}, \citenamefont {McMahon} \emph {et~al.}}]{venemans2015identification}%
  \BibitemOpen
  \bibfield  {author} {\bibinfo {author} {\bibfnamefont {B.}~\bibnamefont {Venemans}}, \bibinfo {author} {\bibfnamefont {E.}~\bibnamefont {Ba{\~n}ados}}, \bibinfo {author} {\bibfnamefont {R.}~\bibnamefont {Decarli}}, \bibinfo {author} {\bibfnamefont {E.}~\bibnamefont {Farina}}, \bibinfo {author} {\bibfnamefont {F.}~\bibnamefont {Walter}}, \bibinfo {author} {\bibfnamefont {K.}~\bibnamefont {Chambers}}, \bibinfo {author} {\bibfnamefont {X.}~\bibnamefont {Fan}}, \bibinfo {author} {\bibfnamefont {H.}~\bibnamefont {Rix}}, \bibinfo {author} {\bibfnamefont {E.}~\bibnamefont {Schlafly}}, \bibinfo {author} {\bibfnamefont {R.}~\bibnamefont {McMahon}}, \emph {et~al.},\ }\bibfield  {title} {\bibinfo {title} {The identification of z-dropouts in pan-starrs1: Three quasars at 6.5< z< 6.7},\ }\href@noop {} {\bibfield  {journal} {\bibinfo  {journal} {The Astrophysical Journal Letters}\ }\textbf {\bibinfo {volume} {801}},\ \bibinfo {pages} {L11} (\bibinfo {year} {2015})}\BibitemShut {NoStop}%
\bibitem [{\citenamefont {Banados}\ \emph {et~al.}(2018)\citenamefont {Banados}, \citenamefont {Venemans}, \citenamefont {Mazzucchelli}, \citenamefont {Farina}, \citenamefont {Walter}, \citenamefont {Wang}, \citenamefont {Decarli}, \citenamefont {Stern}, \citenamefont {Fan}, \citenamefont {Davies} \emph {et~al.}}]{banados2018800}%
  \BibitemOpen
  \bibfield  {author} {\bibinfo {author} {\bibfnamefont {E.}~\bibnamefont {Banados}}, \bibinfo {author} {\bibfnamefont {B.~P.}\ \bibnamefont {Venemans}}, \bibinfo {author} {\bibfnamefont {C.}~\bibnamefont {Mazzucchelli}}, \bibinfo {author} {\bibfnamefont {E.~P.}\ \bibnamefont {Farina}}, \bibinfo {author} {\bibfnamefont {F.}~\bibnamefont {Walter}}, \bibinfo {author} {\bibfnamefont {F.}~\bibnamefont {Wang}}, \bibinfo {author} {\bibfnamefont {R.}~\bibnamefont {Decarli}}, \bibinfo {author} {\bibfnamefont {D.}~\bibnamefont {Stern}}, \bibinfo {author} {\bibfnamefont {X.}~\bibnamefont {Fan}}, \bibinfo {author} {\bibfnamefont {F.~B.}\ \bibnamefont {Davies}}, \emph {et~al.},\ }\bibfield  {title} {\bibinfo {title} {An 800-million-solar-mass black hole in a significantly neutral universe at a redshift of 7.5},\ }\href@noop {} {\bibfield  {journal} {\bibinfo  {journal} {Nature}\ }\textbf {\bibinfo {volume} {553}},\ \bibinfo {pages} {473} (\bibinfo {year} {2018})}\BibitemShut {NoStop}%
\bibitem [{\citenamefont {Jiang}\ \emph {et~al.}(2007)\citenamefont {Jiang}, \citenamefont {Fan}, \citenamefont {Vestergaard}, \citenamefont {Kurk}, \citenamefont {Walter}, \citenamefont {Kelly},\ and\ \citenamefont {Strauss}}]{jiang2007gemini}%
  \BibitemOpen
  \bibfield  {author} {\bibinfo {author} {\bibfnamefont {L.}~\bibnamefont {Jiang}}, \bibinfo {author} {\bibfnamefont {X.}~\bibnamefont {Fan}}, \bibinfo {author} {\bibfnamefont {M.}~\bibnamefont {Vestergaard}}, \bibinfo {author} {\bibfnamefont {J.~D.}\ \bibnamefont {Kurk}}, \bibinfo {author} {\bibfnamefont {F.}~\bibnamefont {Walter}}, \bibinfo {author} {\bibfnamefont {B.~C.}\ \bibnamefont {Kelly}},\ and\ \bibinfo {author} {\bibfnamefont {M.~A.}\ \bibnamefont {Strauss}},\ }\bibfield  {title} {\bibinfo {title} {Gemini near-infrared spectroscopy of luminous z~ 6 quasars: Chemical abundances, black hole masses, and mg ii absorption},\ }\href@noop {} {\bibfield  {journal} {\bibinfo  {journal} {The Astronomical Journal}\ }\textbf {\bibinfo {volume} {134}},\ \bibinfo {pages} {1150} (\bibinfo {year} {2007})}\BibitemShut {NoStop}%
\bibitem [{\citenamefont {Kurk}\ \emph {et~al.}(2007)\citenamefont {Kurk}, \citenamefont {Walter}, \citenamefont {Fan}, \citenamefont {Jiang}, \citenamefont {Riechers}, \citenamefont {Rix}, \citenamefont {Pentericci}, \citenamefont {Strauss}, \citenamefont {Carilli},\ and\ \citenamefont {Wagner}}]{kurk2007black}%
  \BibitemOpen
  \bibfield  {author} {\bibinfo {author} {\bibfnamefont {J.~D.}\ \bibnamefont {Kurk}}, \bibinfo {author} {\bibfnamefont {F.}~\bibnamefont {Walter}}, \bibinfo {author} {\bibfnamefont {X.}~\bibnamefont {Fan}}, \bibinfo {author} {\bibfnamefont {L.}~\bibnamefont {Jiang}}, \bibinfo {author} {\bibfnamefont {D.~A.}\ \bibnamefont {Riechers}}, \bibinfo {author} {\bibfnamefont {H.-W.}\ \bibnamefont {Rix}}, \bibinfo {author} {\bibfnamefont {L.}~\bibnamefont {Pentericci}}, \bibinfo {author} {\bibfnamefont {M.~A.}\ \bibnamefont {Strauss}}, \bibinfo {author} {\bibfnamefont {C.}~\bibnamefont {Carilli}},\ and\ \bibinfo {author} {\bibfnamefont {S.}~\bibnamefont {Wagner}},\ }\bibfield  {title} {\bibinfo {title} {Black hole masses and enrichment of z\~{} 6 sdss quasars},\ }\href@noop {} {\bibfield  {journal} {\bibinfo  {journal} {The Astrophysical Journal}\ }\textbf {\bibinfo {volume} {669}},\ \bibinfo {pages} {32} (\bibinfo {year} {2007})}\BibitemShut {NoStop}%
\bibitem [{\citenamefont {Willott}\ \emph {et~al.}(2010{\natexlab{b}})\citenamefont {Willott}, \citenamefont {Albert}, \citenamefont {Arzoumanian}, \citenamefont {Bergeron}, \citenamefont {Crampton}, \citenamefont {Delorme}, \citenamefont {Hutchings}, \citenamefont {Omont}, \citenamefont {Reyl{\'e}},\ and\ \citenamefont {Schade}}]{willott2010eddington}%
  \BibitemOpen
  \bibfield  {author} {\bibinfo {author} {\bibfnamefont {C.~J.}\ \bibnamefont {Willott}}, \bibinfo {author} {\bibfnamefont {L.}~\bibnamefont {Albert}}, \bibinfo {author} {\bibfnamefont {D.}~\bibnamefont {Arzoumanian}}, \bibinfo {author} {\bibfnamefont {J.}~\bibnamefont {Bergeron}}, \bibinfo {author} {\bibfnamefont {D.}~\bibnamefont {Crampton}}, \bibinfo {author} {\bibfnamefont {P.}~\bibnamefont {Delorme}}, \bibinfo {author} {\bibfnamefont {J.~B.}\ \bibnamefont {Hutchings}}, \bibinfo {author} {\bibfnamefont {A.}~\bibnamefont {Omont}}, \bibinfo {author} {\bibfnamefont {C.}~\bibnamefont {Reyl{\'e}}},\ and\ \bibinfo {author} {\bibfnamefont {D.}~\bibnamefont {Schade}},\ }\bibfield  {title} {\bibinfo {title} {Eddington-limited accretion and the black hole mass function at redshift 6},\ }\href@noop {} {\bibfield  {journal} {\bibinfo  {journal} {The Astronomical Journal}\ }\textbf {\bibinfo {volume} {140}},\ \bibinfo {pages} {546} (\bibinfo {year} {2010}{\natexlab{b}})}\BibitemShut {NoStop}%
\bibitem [{\citenamefont {Shen}\ \emph {et~al.}(2011)\citenamefont {Shen}, \citenamefont {Richards}, \citenamefont {Strauss}, \citenamefont {Hall}, \citenamefont {Schneider}, \citenamefont {Snedden}, \citenamefont {Bizyaev}, \citenamefont {Brewington}, \citenamefont {Malanushenko}, \citenamefont {Malanushenko} \emph {et~al.}}]{shen2011catalog}%
  \BibitemOpen
  \bibfield  {author} {\bibinfo {author} {\bibfnamefont {Y.}~\bibnamefont {Shen}}, \bibinfo {author} {\bibfnamefont {G.~T.}\ \bibnamefont {Richards}}, \bibinfo {author} {\bibfnamefont {M.~A.}\ \bibnamefont {Strauss}}, \bibinfo {author} {\bibfnamefont {P.~B.}\ \bibnamefont {Hall}}, \bibinfo {author} {\bibfnamefont {D.~P.}\ \bibnamefont {Schneider}}, \bibinfo {author} {\bibfnamefont {S.}~\bibnamefont {Snedden}}, \bibinfo {author} {\bibfnamefont {D.}~\bibnamefont {Bizyaev}}, \bibinfo {author} {\bibfnamefont {H.}~\bibnamefont {Brewington}}, \bibinfo {author} {\bibfnamefont {V.}~\bibnamefont {Malanushenko}}, \bibinfo {author} {\bibfnamefont {E.}~\bibnamefont {Malanushenko}}, \emph {et~al.},\ }\bibfield  {title} {\bibinfo {title} {A catalog of quasar properties from sloan digital sky survey data release 7},\ }\href@noop {} {\bibfield  {journal} {\bibinfo  {journal} {The Astrophysical Journal Supplement Series}\ }\textbf {\bibinfo {volume} {194}},\ \bibinfo {pages} {45} (\bibinfo {year} {2011})}\BibitemShut
  {NoStop}%
\bibitem [{\citenamefont {De~Rosa}\ \emph {et~al.}(2011)\citenamefont {De~Rosa}, \citenamefont {Decarli}, \citenamefont {Walter}, \citenamefont {Fan}, \citenamefont {Jiang}, \citenamefont {Kurk}, \citenamefont {Pasquali},\ and\ \citenamefont {Rix}}]{de2011evidence}%
  \BibitemOpen
  \bibfield  {author} {\bibinfo {author} {\bibfnamefont {G.}~\bibnamefont {De~Rosa}}, \bibinfo {author} {\bibfnamefont {R.}~\bibnamefont {Decarli}}, \bibinfo {author} {\bibfnamefont {F.}~\bibnamefont {Walter}}, \bibinfo {author} {\bibfnamefont {X.}~\bibnamefont {Fan}}, \bibinfo {author} {\bibfnamefont {L.}~\bibnamefont {Jiang}}, \bibinfo {author} {\bibfnamefont {J.}~\bibnamefont {Kurk}}, \bibinfo {author} {\bibfnamefont {A.}~\bibnamefont {Pasquali}},\ and\ \bibinfo {author} {\bibfnamefont {H.-W.}\ \bibnamefont {Rix}},\ }\bibfield  {title} {\bibinfo {title} {Evidence for non-evolving fe ii/mg ii ratios in rapidly accreting z~ 6 qsos},\ }\href@noop {} {\bibfield  {journal} {\bibinfo  {journal} {The Astrophysical Journal}\ }\textbf {\bibinfo {volume} {739}},\ \bibinfo {pages} {56} (\bibinfo {year} {2011})}\BibitemShut {NoStop}%
\bibitem [{\citenamefont {Trakhtenbrot}\ \emph {et~al.}(2011)\citenamefont {Trakhtenbrot}, \citenamefont {Netzer}, \citenamefont {Lira},\ and\ \citenamefont {Shemmer}}]{trakhtenbrot2011black}%
  \BibitemOpen
  \bibfield  {author} {\bibinfo {author} {\bibfnamefont {B.}~\bibnamefont {Trakhtenbrot}}, \bibinfo {author} {\bibfnamefont {H.}~\bibnamefont {Netzer}}, \bibinfo {author} {\bibfnamefont {P.}~\bibnamefont {Lira}},\ and\ \bibinfo {author} {\bibfnamefont {O.}~\bibnamefont {Shemmer}},\ }\bibfield  {title} {\bibinfo {title} {Black hole mass and growth rate at $z \simeq 4.8$: a short episode of fast growth followed by short duty cycle activity},\ }\href@noop {} {\bibfield  {journal} {\bibinfo  {journal} {The Astrophysical Journal}\ }\textbf {\bibinfo {volume} {730}},\ \bibinfo {pages} {7} (\bibinfo {year} {2011})}\BibitemShut {NoStop}%
\bibitem [{\citenamefont {De~Rosa}\ \emph {et~al.}(2014)\citenamefont {De~Rosa}, \citenamefont {Venemans}, \citenamefont {Decarli}, \citenamefont {Gennaro}, \citenamefont {Simcoe}, \citenamefont {Dietrich}, \citenamefont {Peterson}, \citenamefont {Walter}, \citenamefont {Frank}, \citenamefont {McMahon} \emph {et~al.}}]{de2014black}%
  \BibitemOpen
  \bibfield  {author} {\bibinfo {author} {\bibfnamefont {G.}~\bibnamefont {De~Rosa}}, \bibinfo {author} {\bibfnamefont {B.~P.}\ \bibnamefont {Venemans}}, \bibinfo {author} {\bibfnamefont {R.}~\bibnamefont {Decarli}}, \bibinfo {author} {\bibfnamefont {M.}~\bibnamefont {Gennaro}}, \bibinfo {author} {\bibfnamefont {R.~A.}\ \bibnamefont {Simcoe}}, \bibinfo {author} {\bibfnamefont {M.}~\bibnamefont {Dietrich}}, \bibinfo {author} {\bibfnamefont {B.~M.}\ \bibnamefont {Peterson}}, \bibinfo {author} {\bibfnamefont {F.}~\bibnamefont {Walter}}, \bibinfo {author} {\bibfnamefont {S.}~\bibnamefont {Frank}}, \bibinfo {author} {\bibfnamefont {R.~G.}\ \bibnamefont {McMahon}}, \emph {et~al.},\ }\bibfield  {title} {\bibinfo {title} {Black hole mass estimates and emission-line properties of a sample of redshift z> 6.5 quasars},\ }\href@noop {} {\bibfield  {journal} {\bibinfo  {journal} {The Astrophysical Journal}\ }\textbf {\bibinfo {volume} {790}},\ \bibinfo {pages} {145} (\bibinfo {year} {2014})}\BibitemShut {NoStop}%
\bibitem [{\citenamefont {Zuo}\ \emph {et~al.}(2015)\citenamefont {Zuo}, \citenamefont {Wu}, \citenamefont {Fan}, \citenamefont {Green}, \citenamefont {Wang},\ and\ \citenamefont {Bian}}]{zuo2015black}%
  \BibitemOpen
  \bibfield  {author} {\bibinfo {author} {\bibfnamefont {W.}~\bibnamefont {Zuo}}, \bibinfo {author} {\bibfnamefont {X.-B.}\ \bibnamefont {Wu}}, \bibinfo {author} {\bibfnamefont {X.}~\bibnamefont {Fan}}, \bibinfo {author} {\bibfnamefont {R.}~\bibnamefont {Green}}, \bibinfo {author} {\bibfnamefont {R.}~\bibnamefont {Wang}},\ and\ \bibinfo {author} {\bibfnamefont {F.}~\bibnamefont {Bian}},\ }\bibfield  {title} {\bibinfo {title} {Black hole mass estimates and rapid growth of supermassive black holes in luminous z~ 3.5 quasars},\ }\href@noop {} {\bibfield  {journal} {\bibinfo  {journal} {The Astrophysical Journal}\ }\textbf {\bibinfo {volume} {799}},\ \bibinfo {pages} {189} (\bibinfo {year} {2015})}\BibitemShut {NoStop}%
\bibitem [{\citenamefont {Shao}\ \emph {et~al.}(2017)\citenamefont {Shao}, \citenamefont {Wang}, \citenamefont {Jones}, \citenamefont {Carilli}, \citenamefont {Walter}, \citenamefont {Fan}, \citenamefont {Riechers}, \citenamefont {Bertoldi}, \citenamefont {Wagg}, \citenamefont {Strauss} \emph {et~al.}}]{shao2017gas}%
  \BibitemOpen
  \bibfield  {author} {\bibinfo {author} {\bibfnamefont {Y.}~\bibnamefont {Shao}}, \bibinfo {author} {\bibfnamefont {R.}~\bibnamefont {Wang}}, \bibinfo {author} {\bibfnamefont {G.~C.}\ \bibnamefont {Jones}}, \bibinfo {author} {\bibfnamefont {C.~L.}\ \bibnamefont {Carilli}}, \bibinfo {author} {\bibfnamefont {F.}~\bibnamefont {Walter}}, \bibinfo {author} {\bibfnamefont {X.}~\bibnamefont {Fan}}, \bibinfo {author} {\bibfnamefont {D.~A.}\ \bibnamefont {Riechers}}, \bibinfo {author} {\bibfnamefont {F.}~\bibnamefont {Bertoldi}}, \bibinfo {author} {\bibfnamefont {J.}~\bibnamefont {Wagg}}, \bibinfo {author} {\bibfnamefont {M.~A.}\ \bibnamefont {Strauss}}, \emph {et~al.},\ }\bibfield  {title} {\bibinfo {title} {Gas dynamics of a luminous z= 6.13 quasar ulas j1319+ 0950 revealed by alma high-resolution observations},\ }\href@noop {} {\bibfield  {journal} {\bibinfo  {journal} {The Astrophysical Journal}\ }\textbf {\bibinfo {volume} {845}},\ \bibinfo {pages} {138} (\bibinfo {year} {2017})}\BibitemShut {NoStop}%
\bibitem [{\citenamefont {Koz{\l}owski}(2017)}]{kozlowski2017virial}%
  \BibitemOpen
  \bibfield  {author} {\bibinfo {author} {\bibfnamefont {S.}~\bibnamefont {Koz{\l}owski}},\ }\bibfield  {title} {\bibinfo {title} {Virial black hole mass estimates for 280,000 agns from the sdss broadband photometry and single-epoch spectra},\ }\href@noop {} {\bibfield  {journal} {\bibinfo  {journal} {The Astrophysical Journal Supplement Series}\ }\textbf {\bibinfo {volume} {228}},\ \bibinfo {pages} {9} (\bibinfo {year} {2017})}\BibitemShut {NoStop}%
\bibitem [{\citenamefont {Mazzucchelli}\ \emph {et~al.}(2017)\citenamefont {Mazzucchelli}, \citenamefont {Ba{\~n}ados}, \citenamefont {Venemans}, \citenamefont {Decarli}, \citenamefont {Farina}, \citenamefont {Walter}, \citenamefont {Eilers}, \citenamefont {Rix}, \citenamefont {Simcoe}, \citenamefont {Stern} \emph {et~al.}}]{mazzucchelli2017physical}%
  \BibitemOpen
  \bibfield  {author} {\bibinfo {author} {\bibfnamefont {C.}~\bibnamefont {Mazzucchelli}}, \bibinfo {author} {\bibfnamefont {E.}~\bibnamefont {Ba{\~n}ados}}, \bibinfo {author} {\bibfnamefont {B.}~\bibnamefont {Venemans}}, \bibinfo {author} {\bibfnamefont {R.}~\bibnamefont {Decarli}}, \bibinfo {author} {\bibfnamefont {E.}~\bibnamefont {Farina}}, \bibinfo {author} {\bibfnamefont {F.}~\bibnamefont {Walter}}, \bibinfo {author} {\bibfnamefont {A.-C.}\ \bibnamefont {Eilers}}, \bibinfo {author} {\bibfnamefont {H.-W.}\ \bibnamefont {Rix}}, \bibinfo {author} {\bibfnamefont {R.}~\bibnamefont {Simcoe}}, \bibinfo {author} {\bibfnamefont {D.}~\bibnamefont {Stern}}, \emph {et~al.},\ }\bibfield  {title} {\bibinfo {title} {Physical properties of 15 quasars at $z \gtrsim 6.5$},\ }\href@noop {} {\bibfield  {journal} {\bibinfo  {journal} {The Astrophysical Journal}\ }\textbf {\bibinfo {volume} {849}},\ \bibinfo {pages} {91} (\bibinfo {year} {2017})}\BibitemShut {NoStop}%
\bibitem [{\citenamefont {Eilers}\ \emph {et~al.}(2018)\citenamefont {Eilers}, \citenamefont {Hennawi},\ and\ \citenamefont {Davies}}]{eilers2018first}%
  \BibitemOpen
  \bibfield  {author} {\bibinfo {author} {\bibfnamefont {A.-C.}\ \bibnamefont {Eilers}}, \bibinfo {author} {\bibfnamefont {J.~F.}\ \bibnamefont {Hennawi}},\ and\ \bibinfo {author} {\bibfnamefont {F.~B.}\ \bibnamefont {Davies}},\ }\bibfield  {title} {\bibinfo {title} {First spectroscopic study of a young quasar},\ }\href@noop {} {\bibfield  {journal} {\bibinfo  {journal} {The Astrophysical Journal}\ }\textbf {\bibinfo {volume} {867}},\ \bibinfo {pages} {30} (\bibinfo {year} {2018})}\BibitemShut {NoStop}%
\bibitem [{\citenamefont {Shen}\ \emph {et~al.}(2019{\natexlab{b}})\citenamefont {Shen}, \citenamefont {Hall}, \citenamefont {Horne}, \citenamefont {Zhu}, \citenamefont {McGreer}, \citenamefont {Simm}, \citenamefont {Trump}, \citenamefont {Kinemuchi}, \citenamefont {Brandt}, \citenamefont {Green} \emph {et~al.}}]{shen2019sloan}%
  \BibitemOpen
  \bibfield  {author} {\bibinfo {author} {\bibfnamefont {Y.}~\bibnamefont {Shen}}, \bibinfo {author} {\bibfnamefont {P.~B.}\ \bibnamefont {Hall}}, \bibinfo {author} {\bibfnamefont {K.}~\bibnamefont {Horne}}, \bibinfo {author} {\bibfnamefont {G.}~\bibnamefont {Zhu}}, \bibinfo {author} {\bibfnamefont {I.}~\bibnamefont {McGreer}}, \bibinfo {author} {\bibfnamefont {T.}~\bibnamefont {Simm}}, \bibinfo {author} {\bibfnamefont {J.~R.}\ \bibnamefont {Trump}}, \bibinfo {author} {\bibfnamefont {K.}~\bibnamefont {Kinemuchi}}, \bibinfo {author} {\bibfnamefont {W.}~\bibnamefont {Brandt}}, \bibinfo {author} {\bibfnamefont {P.~J.}\ \bibnamefont {Green}}, \emph {et~al.},\ }\bibfield  {title} {\bibinfo {title} {The sloan digital sky survey reverberation mapping project: sample characterization},\ }\href@noop {} {\bibfield  {journal} {\bibinfo  {journal} {The Astrophysical Journal Supplement Series}\ }\textbf {\bibinfo {volume} {241}},\ \bibinfo {pages} {34} (\bibinfo {year} {2019}{\natexlab{b}})}\BibitemShut {NoStop}%
\bibitem [{\citenamefont {Matsuoka}\ \emph {et~al.}(2019)\citenamefont {Matsuoka}, \citenamefont {Onoue}, \citenamefont {Kashikawa}, \citenamefont {Strauss}, \citenamefont {Iwasawa}, \citenamefont {Lee}, \citenamefont {Imanishi}, \citenamefont {Nagao}, \citenamefont {Akiyama}, \citenamefont {Asami} \emph {et~al.}}]{matsuoka2019discovery}%
  \BibitemOpen
  \bibfield  {author} {\bibinfo {author} {\bibfnamefont {Y.}~\bibnamefont {Matsuoka}}, \bibinfo {author} {\bibfnamefont {M.}~\bibnamefont {Onoue}}, \bibinfo {author} {\bibfnamefont {N.}~\bibnamefont {Kashikawa}}, \bibinfo {author} {\bibfnamefont {M.~A.}\ \bibnamefont {Strauss}}, \bibinfo {author} {\bibfnamefont {K.}~\bibnamefont {Iwasawa}}, \bibinfo {author} {\bibfnamefont {C.-H.}\ \bibnamefont {Lee}}, \bibinfo {author} {\bibfnamefont {M.}~\bibnamefont {Imanishi}}, \bibinfo {author} {\bibfnamefont {T.}~\bibnamefont {Nagao}}, \bibinfo {author} {\bibfnamefont {M.}~\bibnamefont {Akiyama}}, \bibinfo {author} {\bibfnamefont {N.}~\bibnamefont {Asami}}, \emph {et~al.},\ }\bibfield  {title} {\bibinfo {title} {Discovery of the first low-luminosity quasar at z> 7},\ }\href@noop {} {\bibfield  {journal} {\bibinfo  {journal} {The Astrophysical Journal Letters}\ }\textbf {\bibinfo {volume} {872}},\ \bibinfo {pages} {L2} (\bibinfo {year} {2019})}\BibitemShut {NoStop}%
\bibitem [{\citenamefont {Farina}\ \emph {et~al.}(2022)\citenamefont {Farina}, \citenamefont {Schindler}, \citenamefont {Walter}, \citenamefont {Ba{\~n}ados}, \citenamefont {Davies}, \citenamefont {Decarli}, \citenamefont {Eilers}, \citenamefont {Fan}, \citenamefont {Hennawi}, \citenamefont {Mazzucchelli} \emph {et~al.}}]{farina2022x}%
  \BibitemOpen
  \bibfield  {author} {\bibinfo {author} {\bibfnamefont {E.~P.}\ \bibnamefont {Farina}}, \bibinfo {author} {\bibfnamefont {J.-T.}\ \bibnamefont {Schindler}}, \bibinfo {author} {\bibfnamefont {F.}~\bibnamefont {Walter}}, \bibinfo {author} {\bibfnamefont {E.}~\bibnamefont {Ba{\~n}ados}}, \bibinfo {author} {\bibfnamefont {F.~B.}\ \bibnamefont {Davies}}, \bibinfo {author} {\bibfnamefont {R.}~\bibnamefont {Decarli}}, \bibinfo {author} {\bibfnamefont {A.-C.}\ \bibnamefont {Eilers}}, \bibinfo {author} {\bibfnamefont {X.}~\bibnamefont {Fan}}, \bibinfo {author} {\bibfnamefont {J.~F.}\ \bibnamefont {Hennawi}}, \bibinfo {author} {\bibfnamefont {C.}~\bibnamefont {Mazzucchelli}}, \emph {et~al.},\ }\bibfield  {title} {\bibinfo {title} {The x--shooter/alma sample of quasars in the epoch of reionization. ii. black hole masses, eddington ratios, and the formation of the first quasars},\ }\href@noop {} {\bibfield  {journal} {\bibinfo  {journal} {The Astrophysical Journal}\ }\textbf {\bibinfo {volume} {941}},\ \bibinfo {pages}
  {106} (\bibinfo {year} {2022})}\BibitemShut {NoStop}%
\bibitem [{\citenamefont {Matsuoka}\ \emph {et~al.}(2022)\citenamefont {Matsuoka}, \citenamefont {Iwasawa}, \citenamefont {Onoue}, \citenamefont {Izumi}, \citenamefont {Kashikawa}, \citenamefont {Strauss}, \citenamefont {Imanishi}, \citenamefont {Nagao}, \citenamefont {Akiyama}, \citenamefont {Silverman} \emph {et~al.}}]{matsuoka2022subaru}%
  \BibitemOpen
  \bibfield  {author} {\bibinfo {author} {\bibfnamefont {Y.}~\bibnamefont {Matsuoka}}, \bibinfo {author} {\bibfnamefont {K.}~\bibnamefont {Iwasawa}}, \bibinfo {author} {\bibfnamefont {M.}~\bibnamefont {Onoue}}, \bibinfo {author} {\bibfnamefont {T.}~\bibnamefont {Izumi}}, \bibinfo {author} {\bibfnamefont {N.}~\bibnamefont {Kashikawa}}, \bibinfo {author} {\bibfnamefont {M.~A.}\ \bibnamefont {Strauss}}, \bibinfo {author} {\bibfnamefont {M.}~\bibnamefont {Imanishi}}, \bibinfo {author} {\bibfnamefont {T.}~\bibnamefont {Nagao}}, \bibinfo {author} {\bibfnamefont {M.}~\bibnamefont {Akiyama}}, \bibinfo {author} {\bibfnamefont {J.~D.}\ \bibnamefont {Silverman}}, \emph {et~al.},\ }\bibfield  {title} {\bibinfo {title} {Subaru high-z exploration of low-luminosity quasars (shellqs). xvi. 69 new quasars at 5.8< z< 7.0},\ }\href@noop {} {\bibfield  {journal} {\bibinfo  {journal} {The Astrophysical Journal Supplement Series}\ }\textbf {\bibinfo {volume} {259}},\ \bibinfo {pages} {18} (\bibinfo {year} {2022})}\BibitemShut
  {NoStop}%
\bibitem [{\citenamefont {Tachibana}\ \emph {et~al.}(2020)\citenamefont {Tachibana}, \citenamefont {Graham}, \citenamefont {Kawai}, \citenamefont {Djorgovski}, \citenamefont {Drake}, \citenamefont {Mahabal},\ and\ \citenamefont {Stern}}]{tachibana2020deep}%
  \BibitemOpen
  \bibfield  {author} {\bibinfo {author} {\bibfnamefont {Y.}~\bibnamefont {Tachibana}}, \bibinfo {author} {\bibfnamefont {M.~J.}\ \bibnamefont {Graham}}, \bibinfo {author} {\bibfnamefont {N.}~\bibnamefont {Kawai}}, \bibinfo {author} {\bibfnamefont {S.}~\bibnamefont {Djorgovski}}, \bibinfo {author} {\bibfnamefont {A.~J.}\ \bibnamefont {Drake}}, \bibinfo {author} {\bibfnamefont {A.~A.}\ \bibnamefont {Mahabal}},\ and\ \bibinfo {author} {\bibfnamefont {D.}~\bibnamefont {Stern}},\ }\bibfield  {title} {\bibinfo {title} {Deep modeling of quasar variability},\ }\href@noop {} {\bibfield  {journal} {\bibinfo  {journal} {The Astrophysical Journal}\ }\textbf {\bibinfo {volume} {903}},\ \bibinfo {pages} {54} (\bibinfo {year} {2020})}\BibitemShut {NoStop}%
\bibitem [{\citenamefont {Kaspi}\ \emph {et~al.}(2005)\citenamefont {Kaspi}, \citenamefont {Maoz}, \citenamefont {Netzer}, \citenamefont {Peterson}, \citenamefont {Vestergaard},\ and\ \citenamefont {Jannuzi}}]{kaspi2005relationship}%
  \BibitemOpen
  \bibfield  {author} {\bibinfo {author} {\bibfnamefont {S.}~\bibnamefont {Kaspi}}, \bibinfo {author} {\bibfnamefont {D.}~\bibnamefont {Maoz}}, \bibinfo {author} {\bibfnamefont {H.}~\bibnamefont {Netzer}}, \bibinfo {author} {\bibfnamefont {B.~M.}\ \bibnamefont {Peterson}}, \bibinfo {author} {\bibfnamefont {M.}~\bibnamefont {Vestergaard}},\ and\ \bibinfo {author} {\bibfnamefont {B.~T.}\ \bibnamefont {Jannuzi}},\ }\bibfield  {title} {\bibinfo {title} {The relationship between luminosity and broad-line region size in active galactic nuclei},\ }\href@noop {} {\bibfield  {journal} {\bibinfo  {journal} {The Astrophysical Journal}\ }\textbf {\bibinfo {volume} {629}},\ \bibinfo {pages} {61} (\bibinfo {year} {2005})}\BibitemShut {NoStop}%
\bibitem [{\citenamefont {Bentz}\ \emph {et~al.}(2009)\citenamefont {Bentz}, \citenamefont {Peterson}, \citenamefont {Netzer}, \citenamefont {Pogge},\ and\ \citenamefont {Vestergaard}}]{bentz2009radius}%
  \BibitemOpen
  \bibfield  {author} {\bibinfo {author} {\bibfnamefont {M.~C.}\ \bibnamefont {Bentz}}, \bibinfo {author} {\bibfnamefont {B.~M.}\ \bibnamefont {Peterson}}, \bibinfo {author} {\bibfnamefont {H.}~\bibnamefont {Netzer}}, \bibinfo {author} {\bibfnamefont {R.~W.}\ \bibnamefont {Pogge}},\ and\ \bibinfo {author} {\bibfnamefont {M.}~\bibnamefont {Vestergaard}},\ }\bibfield  {title} {\bibinfo {title} {The radius--luminosity relationship for active galactic nuclei: the effect of host-galaxy starlight on luminosity measurements. ii. the full sample of reverberation-mapped agns},\ }\href@noop {} {\bibfield  {journal} {\bibinfo  {journal} {The Astrophysical Journal}\ }\textbf {\bibinfo {volume} {697}},\ \bibinfo {pages} {160} (\bibinfo {year} {2009})}\BibitemShut {NoStop}%
\bibitem [{\citenamefont {Busca}\ and\ \citenamefont {Balland}(2018)}]{busca2018quasarnet}%
  \BibitemOpen
  \bibfield  {author} {\bibinfo {author} {\bibfnamefont {N.}~\bibnamefont {Busca}}\ and\ \bibinfo {author} {\bibfnamefont {C.}~\bibnamefont {Balland}},\ }\bibfield  {title} {\bibinfo {title} {Quasarnet: Human-level spectral classification and redshifting with deep neural networks},\ }\href@noop {} {\bibfield  {journal} {\bibinfo  {journal} {arXiv preprint arXiv:1808.09955}\ } (\bibinfo {year} {2018})}\BibitemShut {NoStop}%
\bibitem [{\citenamefont {Alam}\ \emph {et~al.}(2015)\citenamefont {Alam}, \citenamefont {Albareti}, \citenamefont {Prieto}, \citenamefont {Anders}, \citenamefont {Anderson}, \citenamefont {Anderton}, \citenamefont {Andrews}, \citenamefont {Armengaud}, \citenamefont {Aubourg}, \citenamefont {Bailey} \emph {et~al.}}]{alam2015eleventh}%
  \BibitemOpen
  \bibfield  {author} {\bibinfo {author} {\bibfnamefont {S.}~\bibnamefont {Alam}}, \bibinfo {author} {\bibfnamefont {F.~D.}\ \bibnamefont {Albareti}}, \bibinfo {author} {\bibfnamefont {C.~A.}\ \bibnamefont {Prieto}}, \bibinfo {author} {\bibfnamefont {F.}~\bibnamefont {Anders}}, \bibinfo {author} {\bibfnamefont {S.~F.}\ \bibnamefont {Anderson}}, \bibinfo {author} {\bibfnamefont {T.}~\bibnamefont {Anderton}}, \bibinfo {author} {\bibfnamefont {B.~H.}\ \bibnamefont {Andrews}}, \bibinfo {author} {\bibfnamefont {E.}~\bibnamefont {Armengaud}}, \bibinfo {author} {\bibfnamefont {{\'E}.}~\bibnamefont {Aubourg}}, \bibinfo {author} {\bibfnamefont {S.}~\bibnamefont {Bailey}}, \emph {et~al.},\ }\bibfield  {title} {\bibinfo {title} {The eleventh and twelfth data releases of the sloan digital sky survey: final data from sdss-iii},\ }\href@noop {} {\bibfield  {journal} {\bibinfo  {journal} {The Astrophysical Journal Supplement Series}\ }\textbf {\bibinfo {volume} {219}},\ \bibinfo {pages} {12} (\bibinfo {year}
  {2015})}\BibitemShut {NoStop}%
\bibitem [{\citenamefont {P{\^a}ris}\ \emph {et~al.}(2017)\citenamefont {P{\^a}ris}, \citenamefont {Petitjean}, \citenamefont {Ross}, \citenamefont {Myers}, \citenamefont {Aubourg}, \citenamefont {Streblyanska}, \citenamefont {Bailey}, \citenamefont {Armengaud}, \citenamefont {Palanque-Delabrouille}, \citenamefont {Y{\`e}che} \emph {et~al.}}]{paris2017sloan}%
  \BibitemOpen
  \bibfield  {author} {\bibinfo {author} {\bibfnamefont {I.}~\bibnamefont {P{\^a}ris}}, \bibinfo {author} {\bibfnamefont {P.}~\bibnamefont {Petitjean}}, \bibinfo {author} {\bibfnamefont {N.~P.}\ \bibnamefont {Ross}}, \bibinfo {author} {\bibfnamefont {A.~D.}\ \bibnamefont {Myers}}, \bibinfo {author} {\bibfnamefont {{\'E}.}~\bibnamefont {Aubourg}}, \bibinfo {author} {\bibfnamefont {A.}~\bibnamefont {Streblyanska}}, \bibinfo {author} {\bibfnamefont {S.}~\bibnamefont {Bailey}}, \bibinfo {author} {\bibfnamefont {{\'E}.}~\bibnamefont {Armengaud}}, \bibinfo {author} {\bibfnamefont {N.}~\bibnamefont {Palanque-Delabrouille}}, \bibinfo {author} {\bibfnamefont {C.}~\bibnamefont {Y{\`e}che}}, \emph {et~al.},\ }\bibfield  {title} {\bibinfo {title} {The sloan digital sky survey quasar catalog: twelfth data release},\ }\href@noop {} {\bibfield  {journal} {\bibinfo  {journal} {Astronomy \& Astrophysics}\ }\textbf {\bibinfo {volume} {597}},\ \bibinfo {pages} {A79} (\bibinfo {year} {2017})}\BibitemShut {NoStop}%
\bibitem [{\citenamefont {Weymann}\ \emph {et~al.}(1981)\citenamefont {Weymann}, \citenamefont {Carswell},\ and\ \citenamefont {Smith}}]{weymann1981absorption}%
  \BibitemOpen
  \bibfield  {author} {\bibinfo {author} {\bibfnamefont {R.~J.}\ \bibnamefont {Weymann}}, \bibinfo {author} {\bibfnamefont {R.~F.}\ \bibnamefont {Carswell}},\ and\ \bibinfo {author} {\bibfnamefont {M.~G.}\ \bibnamefont {Smith}},\ }\bibfield  {title} {\bibinfo {title} {Absorption lines in the spectra of quasistellar objects},\ }\href@noop {} {\bibfield  {journal} {\bibinfo  {journal} {Annual Review of Astronomy and Astrophysics}\ }\textbf {\bibinfo {volume} {19}},\ \bibinfo {pages} {41} (\bibinfo {year} {1981})}\BibitemShut {NoStop}%
\bibitem [{\citenamefont {Weymann}\ \emph {et~al.}(1991)\citenamefont {Weymann}, \citenamefont {Morris}, \citenamefont {Foltz},\ and\ \citenamefont {Hewett}}]{weymann1991comparisons}%
  \BibitemOpen
  \bibfield  {author} {\bibinfo {author} {\bibfnamefont {R.~J.}\ \bibnamefont {Weymann}}, \bibinfo {author} {\bibfnamefont {S.~L.}\ \bibnamefont {Morris}}, \bibinfo {author} {\bibfnamefont {C.~B.}\ \bibnamefont {Foltz}},\ and\ \bibinfo {author} {\bibfnamefont {P.~C.}\ \bibnamefont {Hewett}},\ }\bibfield  {title} {\bibinfo {title} {Comparisons of the emission-line and continuum properties of broad absorption line and normal quasi-stellar objects},\ }\href@noop {} {\bibfield  {journal} {\bibinfo  {journal} {Astrophysical Journal, Part 1 (ISSN 0004-637X), vol. 373, May 20, 1991, p. 23-53.}\ }\textbf {\bibinfo {volume} {373}},\ \bibinfo {pages} {23} (\bibinfo {year} {1991})}\BibitemShut {NoStop}%
\bibitem [{\citenamefont {Redmon}\ and\ \citenamefont {Farhadi}(2017)}]{redmon2017yolo9000}%
  \BibitemOpen
  \bibfield  {author} {\bibinfo {author} {\bibfnamefont {J.}~\bibnamefont {Redmon}}\ and\ \bibinfo {author} {\bibfnamefont {A.}~\bibnamefont {Farhadi}},\ }\bibfield  {title} {\bibinfo {title} {Yolo9000: better, faster, stronger},\ }in\ \href@noop {} {\emph {\bibinfo {booktitle} {Proceedings of the IEEE conference on computer vision and pattern recognition}}}\ (\bibinfo {year} {2017})\ pp.\ \bibinfo {pages} {7263--7271}\BibitemShut {NoStop}%
\bibitem [{\citenamefont {Cheng}\ \emph {et~al.}(2020)\citenamefont {Cheng}, \citenamefont {Chang},\ and\ \citenamefont {Bock}}]{cheng2020phase}%
  \BibitemOpen
  \bibfield  {author} {\bibinfo {author} {\bibfnamefont {Y.-T.}\ \bibnamefont {Cheng}}, \bibinfo {author} {\bibfnamefont {T.-C.}\ \bibnamefont {Chang}},\ and\ \bibinfo {author} {\bibfnamefont {J.~J.}\ \bibnamefont {Bock}},\ }\bibfield  {title} {\bibinfo {title} {Phase-space spectral line deconfusion in intensity mapping},\ }\href@noop {} {\bibfield  {journal} {\bibinfo  {journal} {The Astrophysical Journal}\ }\textbf {\bibinfo {volume} {901}},\ \bibinfo {pages} {142} (\bibinfo {year} {2020})}\BibitemShut {NoStop}%
\bibitem [{\citenamefont {Massara}\ \emph {et~al.}(2021)\citenamefont {Massara}, \citenamefont {Ho}, \citenamefont {Hirata}, \citenamefont {DeRose}, \citenamefont {Wechsler},\ and\ \citenamefont {Fang}}]{massara2021line}%
  \BibitemOpen
  \bibfield  {author} {\bibinfo {author} {\bibfnamefont {E.}~\bibnamefont {Massara}}, \bibinfo {author} {\bibfnamefont {S.}~\bibnamefont {Ho}}, \bibinfo {author} {\bibfnamefont {C.~M.}\ \bibnamefont {Hirata}}, \bibinfo {author} {\bibfnamefont {J.}~\bibnamefont {DeRose}}, \bibinfo {author} {\bibfnamefont {R.~H.}\ \bibnamefont {Wechsler}},\ and\ \bibinfo {author} {\bibfnamefont {X.}~\bibnamefont {Fang}},\ }\bibfield  {title} {\bibinfo {title} {Line confusion in spectroscopic surveys and its possible effects: shifts in baryon acoustic oscillations position},\ }\href@noop {} {\bibfield  {journal} {\bibinfo  {journal} {Monthly Notices of the Royal Astronomical Society}\ }\textbf {\bibinfo {volume} {508}},\ \bibinfo {pages} {4193} (\bibinfo {year} {2021})}\BibitemShut {NoStop}%
\bibitem [{\citenamefont {Dawson}\ \emph {et~al.}(2016)\citenamefont {Dawson}, \citenamefont {Kneib}, \citenamefont {Percival}, \citenamefont {Alam}, \citenamefont {Albareti}, \citenamefont {Anderson}, \citenamefont {Armengaud}, \citenamefont {Aubourg}, \citenamefont {Bailey}, \citenamefont {Bautista} \emph {et~al.}}]{dawson2016sdss}%
  \BibitemOpen
  \bibfield  {author} {\bibinfo {author} {\bibfnamefont {K.~S.}\ \bibnamefont {Dawson}}, \bibinfo {author} {\bibfnamefont {J.-P.}\ \bibnamefont {Kneib}}, \bibinfo {author} {\bibfnamefont {W.~J.}\ \bibnamefont {Percival}}, \bibinfo {author} {\bibfnamefont {S.}~\bibnamefont {Alam}}, \bibinfo {author} {\bibfnamefont {F.~D.}\ \bibnamefont {Albareti}}, \bibinfo {author} {\bibfnamefont {S.~F.}\ \bibnamefont {Anderson}}, \bibinfo {author} {\bibfnamefont {E.}~\bibnamefont {Armengaud}}, \bibinfo {author} {\bibfnamefont {{\'E}.}~\bibnamefont {Aubourg}}, \bibinfo {author} {\bibfnamefont {S.}~\bibnamefont {Bailey}}, \bibinfo {author} {\bibfnamefont {J.~E.}\ \bibnamefont {Bautista}}, \emph {et~al.},\ }\bibfield  {title} {\bibinfo {title} {The sdss-iv extended baryon oscillation spectroscopic survey: overview and early data},\ }\href@noop {} {\bibfield  {journal} {\bibinfo  {journal} {The Astronomical Journal}\ }\textbf {\bibinfo {volume} {151}},\ \bibinfo {pages} {44} (\bibinfo {year} {2016})}\BibitemShut {NoStop}%
\bibitem [{\citenamefont {Aghamousa}\ \emph {et~al.}(2016{\natexlab{a}})\citenamefont {Aghamousa}, \citenamefont {Aguilar}, \citenamefont {Ahlen}, \citenamefont {Alam}, \citenamefont {Allen}, \citenamefont {Prieto}, \citenamefont {Annis}, \citenamefont {Bailey}, \citenamefont {Balland}, \citenamefont {Ballester} \emph {et~al.}}]{aghamousa2016desia}%
  \BibitemOpen
  \bibfield  {author} {\bibinfo {author} {\bibfnamefont {A.}~\bibnamefont {Aghamousa}}, \bibinfo {author} {\bibfnamefont {J.}~\bibnamefont {Aguilar}}, \bibinfo {author} {\bibfnamefont {S.}~\bibnamefont {Ahlen}}, \bibinfo {author} {\bibfnamefont {S.}~\bibnamefont {Alam}}, \bibinfo {author} {\bibfnamefont {L.~E.}\ \bibnamefont {Allen}}, \bibinfo {author} {\bibfnamefont {C.~A.}\ \bibnamefont {Prieto}}, \bibinfo {author} {\bibfnamefont {J.}~\bibnamefont {Annis}}, \bibinfo {author} {\bibfnamefont {S.}~\bibnamefont {Bailey}}, \bibinfo {author} {\bibfnamefont {C.}~\bibnamefont {Balland}}, \bibinfo {author} {\bibfnamefont {O.}~\bibnamefont {Ballester}}, \emph {et~al.},\ }\bibfield  {title} {\bibinfo {title} {The desi experiment part i: science, targeting, and survey design},\ }\href@noop {} {\bibfield  {journal} {\bibinfo  {journal} {arXiv preprint arXiv:1611.00036}\ } (\bibinfo {year} {2016}{\natexlab{a}})}\BibitemShut {NoStop}%
\bibitem [{\citenamefont {Aghamousa}\ \emph {et~al.}(2016{\natexlab{b}})\citenamefont {Aghamousa}, \citenamefont {Aguilar}, \citenamefont {Ahlen}, \citenamefont {Alam}, \citenamefont {Allen}, \citenamefont {Prieto}, \citenamefont {Annis}, \citenamefont {Bailey}, \citenamefont {Balland}, \citenamefont {Ballester} \emph {et~al.}}]{aghamousa2016desib}%
  \BibitemOpen
  \bibfield  {author} {\bibinfo {author} {\bibfnamefont {A.}~\bibnamefont {Aghamousa}}, \bibinfo {author} {\bibfnamefont {J.}~\bibnamefont {Aguilar}}, \bibinfo {author} {\bibfnamefont {S.}~\bibnamefont {Ahlen}}, \bibinfo {author} {\bibfnamefont {S.}~\bibnamefont {Alam}}, \bibinfo {author} {\bibfnamefont {L.~E.}\ \bibnamefont {Allen}}, \bibinfo {author} {\bibfnamefont {C.~A.}\ \bibnamefont {Prieto}}, \bibinfo {author} {\bibfnamefont {J.}~\bibnamefont {Annis}}, \bibinfo {author} {\bibfnamefont {S.}~\bibnamefont {Bailey}}, \bibinfo {author} {\bibfnamefont {C.}~\bibnamefont {Balland}}, \bibinfo {author} {\bibfnamefont {O.}~\bibnamefont {Ballester}}, \emph {et~al.},\ }\bibfield  {title} {\bibinfo {title} {The desi experiment part ii: instrument design},\ }\href@noop {} {\bibfield  {journal} {\bibinfo  {journal} {arXiv preprint arXiv:1611.00037}\ } (\bibinfo {year} {2016}{\natexlab{b}})}\BibitemShut {NoStop}%
\bibitem [{\citenamefont {de~Jong}\ \emph {et~al.}(2012)\citenamefont {de~Jong}, \citenamefont {Chiappini},\ and\ \citenamefont {Schnurr}}]{de20124most}%
  \BibitemOpen
  \bibfield  {author} {\bibinfo {author} {\bibfnamefont {R.~S.}\ \bibnamefont {de~Jong}}, \bibinfo {author} {\bibfnamefont {C.}~\bibnamefont {Chiappini}},\ and\ \bibinfo {author} {\bibfnamefont {O.}~\bibnamefont {Schnurr}},\ }\bibfield  {title} {\bibinfo {title} {4most--4-meter multi-object spectroscopic telescope},\ }in\ \href@noop {} {\emph {\bibinfo {booktitle} {EPJ Web of Conferences}}},\ Vol.~\bibinfo {volume} {19}\ (\bibinfo {organization} {EDP Sciences},\ \bibinfo {year} {2012})\ p.\ \bibinfo {pages} {09004}\BibitemShut {NoStop}%
\bibitem [{\citenamefont {Alonso}\ \emph {et~al.}(2016)\citenamefont {Alonso}, \citenamefont {Louis}, \citenamefont {Bull},\ and\ \citenamefont {Ferreira}}]{alonso2016reconstructing}%
  \BibitemOpen
  \bibfield  {author} {\bibinfo {author} {\bibfnamefont {D.}~\bibnamefont {Alonso}}, \bibinfo {author} {\bibfnamefont {T.}~\bibnamefont {Louis}}, \bibinfo {author} {\bibfnamefont {P.}~\bibnamefont {Bull}},\ and\ \bibinfo {author} {\bibfnamefont {P.~G.}\ \bibnamefont {Ferreira}},\ }\bibfield  {title} {\bibinfo {title} {Reconstructing cosmic growth with kinetic sunyaev-zel’dovich observations in the era of stage iv experiments},\ }\href@noop {} {\bibfield  {journal} {\bibinfo  {journal} {Physical Review D}\ }\textbf {\bibinfo {volume} {94}},\ \bibinfo {pages} {043522} (\bibinfo {year} {2016})}\BibitemShut {NoStop}%
\bibitem [{\citenamefont {Breiman}(2001)}]{breiman2001random}%
  \BibitemOpen
  \bibfield  {author} {\bibinfo {author} {\bibfnamefont {L.}~\bibnamefont {Breiman}},\ }\bibfield  {title} {\bibinfo {title} {Random forests},\ }\href@noop {} {\bibfield  {journal} {\bibinfo  {journal} {Machine learning}\ }\textbf {\bibinfo {volume} {45}},\ \bibinfo {pages} {5} (\bibinfo {year} {2001})}\BibitemShut {NoStop}%
\bibitem [{\citenamefont {Ioffe}\ and\ \citenamefont {Szegedy}(2015)}]{ioffe2015batch}%
  \BibitemOpen
  \bibfield  {author} {\bibinfo {author} {\bibfnamefont {S.}~\bibnamefont {Ioffe}}\ and\ \bibinfo {author} {\bibfnamefont {C.}~\bibnamefont {Szegedy}},\ }\bibfield  {title} {\bibinfo {title} {Batch normalization: Accelerating deep network training by reducing internal covariate shift},\ }in\ \href@noop {} {\emph {\bibinfo {booktitle} {International conference on machine learning}}}\ (\bibinfo {organization} {pmlr},\ \bibinfo {year} {2015})\ pp.\ \bibinfo {pages} {448--456}\BibitemShut {NoStop}%
\bibitem [{\citenamefont {Neugebauer}\ \emph {et~al.}(1987)\citenamefont {Neugebauer}, \citenamefont {Green}, \citenamefont {Matthews}, \citenamefont {Schmidt}, \citenamefont {Soifer},\ and\ \citenamefont {Bennett}}]{neugebauer1987continuum}%
  \BibitemOpen
  \bibfield  {author} {\bibinfo {author} {\bibfnamefont {G.}~\bibnamefont {Neugebauer}}, \bibinfo {author} {\bibfnamefont {R.}~\bibnamefont {Green}}, \bibinfo {author} {\bibfnamefont {K.}~\bibnamefont {Matthews}}, \bibinfo {author} {\bibfnamefont {M.}~\bibnamefont {Schmidt}}, \bibinfo {author} {\bibfnamefont {B.}~\bibnamefont {Soifer}},\ and\ \bibinfo {author} {\bibfnamefont {J.}~\bibnamefont {Bennett}},\ }\bibfield  {title} {\bibinfo {title} {Continuum energy distributions of quasars in the palomar-green survey},\ }\href@noop {} {\bibfield  {journal} {\bibinfo  {journal} {Astrophysical Journal Supplement Series (ISSN 0067-0049), vol. 63, March 1987, p. 615-644. NSF-supported research.}\ }\textbf {\bibinfo {volume} {63}},\ \bibinfo {pages} {615} (\bibinfo {year} {1987})}\BibitemShut {NoStop}%
\bibitem [{\citenamefont {Scott}\ \emph {et~al.}(2004)\citenamefont {Scott}, \citenamefont {Kriss}, \citenamefont {Brotherton}, \citenamefont {Green}, \citenamefont {Hutchings}, \citenamefont {Shull},\ and\ \citenamefont {Zheng}}]{scott2004composite}%
  \BibitemOpen
  \bibfield  {author} {\bibinfo {author} {\bibfnamefont {J.~E.}\ \bibnamefont {Scott}}, \bibinfo {author} {\bibfnamefont {G.~A.}\ \bibnamefont {Kriss}}, \bibinfo {author} {\bibfnamefont {M.}~\bibnamefont {Brotherton}}, \bibinfo {author} {\bibfnamefont {R.~F.}\ \bibnamefont {Green}}, \bibinfo {author} {\bibfnamefont {J.}~\bibnamefont {Hutchings}}, \bibinfo {author} {\bibfnamefont {J.~M.}\ \bibnamefont {Shull}},\ and\ \bibinfo {author} {\bibfnamefont {W.}~\bibnamefont {Zheng}},\ }\bibfield  {title} {\bibinfo {title} {A composite extreme-ultraviolet qso spectrum from fuse},\ }\href@noop {} {\bibfield  {journal} {\bibinfo  {journal} {The Astrophysical Journal}\ }\textbf {\bibinfo {volume} {615}},\ \bibinfo {pages} {135} (\bibinfo {year} {2004})}\BibitemShut {NoStop}%
\bibitem [{\citenamefont {Brandt}\ \emph {et~al.}(2000)\citenamefont {Brandt}, \citenamefont {Laor},\ and\ \citenamefont {Wills}}]{brandt2000nature}%
  \BibitemOpen
  \bibfield  {author} {\bibinfo {author} {\bibfnamefont {W.}~\bibnamefont {Brandt}}, \bibinfo {author} {\bibfnamefont {A.}~\bibnamefont {Laor}},\ and\ \bibinfo {author} {\bibfnamefont {B.~J.}\ \bibnamefont {Wills}},\ }\bibfield  {title} {\bibinfo {title} {On the nature of soft x-ray weak quasi-stellar objects},\ }\href@noop {} {\bibfield  {journal} {\bibinfo  {journal} {The Astrophysical Journal}\ }\textbf {\bibinfo {volume} {528}},\ \bibinfo {pages} {637} (\bibinfo {year} {2000})}\BibitemShut {NoStop}%
\bibitem [{\citenamefont {Urry}\ and\ \citenamefont {Padovani}(1995)}]{urry1995unified}%
  \BibitemOpen
  \bibfield  {author} {\bibinfo {author} {\bibfnamefont {C.~M.}\ \bibnamefont {Urry}}\ and\ \bibinfo {author} {\bibfnamefont {P.}~\bibnamefont {Padovani}},\ }\bibfield  {title} {\bibinfo {title} {Unified schemes for radio-loud active galactic nuclei},\ }\href@noop {} {\bibfield  {journal} {\bibinfo  {journal} {Publications of the Astronomical Society of the Pacific}\ }\textbf {\bibinfo {volume} {107}},\ \bibinfo {pages} {803} (\bibinfo {year} {1995})}\BibitemShut {NoStop}%
\bibitem [{\citenamefont {Paliya}\ \emph {et~al.}(2020)\citenamefont {Paliya}, \citenamefont {Ajello}, \citenamefont {Cao}, \citenamefont {Giroletti}, \citenamefont {Kaur}, \citenamefont {Madejski}, \citenamefont {Lott},\ and\ \citenamefont {Hartmann}}]{paliya2020blazars}%
  \BibitemOpen
  \bibfield  {author} {\bibinfo {author} {\bibfnamefont {V.~S.}\ \bibnamefont {Paliya}}, \bibinfo {author} {\bibfnamefont {M.}~\bibnamefont {Ajello}}, \bibinfo {author} {\bibfnamefont {H.-M.}\ \bibnamefont {Cao}}, \bibinfo {author} {\bibfnamefont {M.}~\bibnamefont {Giroletti}}, \bibinfo {author} {\bibfnamefont {A.}~\bibnamefont {Kaur}}, \bibinfo {author} {\bibfnamefont {G.}~\bibnamefont {Madejski}}, \bibinfo {author} {\bibfnamefont {B.}~\bibnamefont {Lott}},\ and\ \bibinfo {author} {\bibfnamefont {D.}~\bibnamefont {Hartmann}},\ }\bibfield  {title} {\bibinfo {title} {Blazars at the cosmic dawn},\ }\href@noop {} {\bibfield  {journal} {\bibinfo  {journal} {The Astrophysical Journal}\ }\textbf {\bibinfo {volume} {897}},\ \bibinfo {pages} {177} (\bibinfo {year} {2020})}\BibitemShut {NoStop}%
\bibitem [{\citenamefont {Burke}\ \emph {et~al.}(2024)\citenamefont {Burke}, \citenamefont {Liu},\ and\ \citenamefont {Shen}}]{burke2024gemini}%
  \BibitemOpen
  \bibfield  {author} {\bibinfo {author} {\bibfnamefont {C.~J.}\ \bibnamefont {Burke}}, \bibinfo {author} {\bibfnamefont {X.}~\bibnamefont {Liu}},\ and\ \bibinfo {author} {\bibfnamefont {Y.}~\bibnamefont {Shen}},\ }\bibfield  {title} {\bibinfo {title} {Gemini near-infrared spectroscopy of high-redshift fermi blazars: jetted black holes in the early universe were overly massive},\ }\href@noop {} {\bibfield  {journal} {\bibinfo  {journal} {Monthly Notices of the Royal Astronomical Society}\ }\textbf {\bibinfo {volume} {527}},\ \bibinfo {pages} {5356} (\bibinfo {year} {2024})}\BibitemShut {NoStop}%
\bibitem [{\citenamefont {Dietrich}\ and\ \citenamefont {Hamann}(2004)}]{dietrich2004implications}%
  \BibitemOpen
  \bibfield  {author} {\bibinfo {author} {\bibfnamefont {M.}~\bibnamefont {Dietrich}}\ and\ \bibinfo {author} {\bibfnamefont {F.}~\bibnamefont {Hamann}},\ }\bibfield  {title} {\bibinfo {title} {Implications of quasar black hole masses at high redshifts},\ }\href@noop {} {\bibfield  {journal} {\bibinfo  {journal} {The Astrophysical Journal}\ }\textbf {\bibinfo {volume} {611}},\ \bibinfo {pages} {761} (\bibinfo {year} {2004})}\BibitemShut {NoStop}%
\bibitem [{\citenamefont {Dietrich}\ \emph {et~al.}(1999)\citenamefont {Dietrich}, \citenamefont {Appenzeller}, \citenamefont {Wagner}, \citenamefont {G{\"a}ssler}, \citenamefont {H{\"a}fner}, \citenamefont {Hess}, \citenamefont {Hummel}, \citenamefont {Muschielok}, \citenamefont {Nicklas}, \citenamefont {Rupprecht} \emph {et~al.}}]{dietrich1999spectroscopic}%
  \BibitemOpen
  \bibfield  {author} {\bibinfo {author} {\bibfnamefont {M.}~\bibnamefont {Dietrich}}, \bibinfo {author} {\bibfnamefont {I.}~\bibnamefont {Appenzeller}}, \bibinfo {author} {\bibfnamefont {S.}~\bibnamefont {Wagner}}, \bibinfo {author} {\bibfnamefont {W.}~\bibnamefont {G{\"a}ssler}}, \bibinfo {author} {\bibfnamefont {R.}~\bibnamefont {H{\"a}fner}}, \bibinfo {author} {\bibfnamefont {H.-J.}\ \bibnamefont {Hess}}, \bibinfo {author} {\bibfnamefont {W.}~\bibnamefont {Hummel}}, \bibinfo {author} {\bibfnamefont {B.}~\bibnamefont {Muschielok}}, \bibinfo {author} {\bibfnamefont {H.}~\bibnamefont {Nicklas}}, \bibinfo {author} {\bibfnamefont {G.}~\bibnamefont {Rupprecht}}, \emph {et~al.},\ }\bibfield  {title} {\bibinfo {title} {Spectroscopic study of high redshift quasars},\ }\href@noop {} {\bibfield  {journal} {\bibinfo  {journal} {Astronomy and Astrophysics, v. 352, p. L1-L4 (1999)}\ }\textbf {\bibinfo {volume} {352}},\ \bibinfo {pages} {L1} (\bibinfo {year} {1999})}\BibitemShut {NoStop}%
\bibitem [{\citenamefont {Dietrich}\ and\ \citenamefont {Wilhelm-Erkens}(2000)}]{dietrich2000elemental}%
  \BibitemOpen
  \bibfield  {author} {\bibinfo {author} {\bibfnamefont {M.}~\bibnamefont {Dietrich}}\ and\ \bibinfo {author} {\bibfnamefont {U.}~\bibnamefont {Wilhelm-Erkens}},\ }\bibfield  {title} {\bibinfo {title} {Elemental abundances of high redshift quasars},\ }\href@noop {} {\bibfield  {journal} {\bibinfo  {journal} {Astronomy and Astrophysics, v. 354, p. 17-27 (2000)}\ }\textbf {\bibinfo {volume} {354}},\ \bibinfo {pages} {17} (\bibinfo {year} {2000})}\BibitemShut {NoStop}%
\bibitem [{\citenamefont {Dietrich}\ \emph {et~al.}(2002)\citenamefont {Dietrich}, \citenamefont {Appenzeller}, \citenamefont {Vestergaard},\ and\ \citenamefont {Wagner}}]{dietrich2002high}%
  \BibitemOpen
  \bibfield  {author} {\bibinfo {author} {\bibfnamefont {M.}~\bibnamefont {Dietrich}}, \bibinfo {author} {\bibfnamefont {I.}~\bibnamefont {Appenzeller}}, \bibinfo {author} {\bibfnamefont {M.}~\bibnamefont {Vestergaard}},\ and\ \bibinfo {author} {\bibfnamefont {S.}~\bibnamefont {Wagner}},\ }\bibfield  {title} {\bibinfo {title} {High-redshift quasars and star formation in the early universe},\ }\href@noop {} {\bibfield  {journal} {\bibinfo  {journal} {The Astrophysical Journal}\ }\textbf {\bibinfo {volume} {564}},\ \bibinfo {pages} {581} (\bibinfo {year} {2002})}\BibitemShut {NoStop}%
\bibitem [{\citenamefont {Dietrich}\ \emph {et~al.}(2003)\citenamefont {Dietrich}, \citenamefont {Appenzeller}, \citenamefont {Hamann}, \citenamefont {Heidt}, \citenamefont {J{\"a}ger}, \citenamefont {Vestergaard},\ and\ \citenamefont {Wagner}}]{dietrich2003elemental}%
  \BibitemOpen
  \bibfield  {author} {\bibinfo {author} {\bibfnamefont {M.}~\bibnamefont {Dietrich}}, \bibinfo {author} {\bibfnamefont {I.}~\bibnamefont {Appenzeller}}, \bibinfo {author} {\bibfnamefont {F.}~\bibnamefont {Hamann}}, \bibinfo {author} {\bibfnamefont {J.}~\bibnamefont {Heidt}}, \bibinfo {author} {\bibfnamefont {K.}~\bibnamefont {J{\"a}ger}}, \bibinfo {author} {\bibfnamefont {M.}~\bibnamefont {Vestergaard}},\ and\ \bibinfo {author} {\bibfnamefont {S.}~\bibnamefont {Wagner}},\ }\bibfield  {title} {\bibinfo {title} {Elemental abundances in the broad emission line region of quasars at redshifts larger than 4},\ }\href@noop {} {\bibfield  {journal} {\bibinfo  {journal} {Astronomy \& Astrophysics}\ }\textbf {\bibinfo {volume} {398}},\ \bibinfo {pages} {891} (\bibinfo {year} {2003})}\BibitemShut {NoStop}%
\bibitem [{\citenamefont {Dietrich}\ \emph {et~al.}(2009)\citenamefont {Dietrich}, \citenamefont {Mathur}, \citenamefont {Grupe},\ and\ \citenamefont {Komossa}}]{dietrich2009black}%
  \BibitemOpen
  \bibfield  {author} {\bibinfo {author} {\bibfnamefont {M.}~\bibnamefont {Dietrich}}, \bibinfo {author} {\bibfnamefont {S.}~\bibnamefont {Mathur}}, \bibinfo {author} {\bibfnamefont {D.}~\bibnamefont {Grupe}},\ and\ \bibinfo {author} {\bibfnamefont {S.}~\bibnamefont {Komossa}},\ }\bibfield  {title} {\bibinfo {title} {Black hole masses of intermediate-redshift quasars: Near-infrared spectroscopy},\ }\href@noop {} {\bibfield  {journal} {\bibinfo  {journal} {The Astrophysical Journal}\ }\textbf {\bibinfo {volume} {696}},\ \bibinfo {pages} {1998} (\bibinfo {year} {2009})}\BibitemShut {NoStop}%
\bibitem [{\citenamefont {Duras}\ \emph {et~al.}(2020)\citenamefont {Duras}, \citenamefont {Bongiorno}, \citenamefont {Ricci}, \citenamefont {Piconcelli}, \citenamefont {Shankar}, \citenamefont {Lusso}, \citenamefont {Bianchi}, \citenamefont {Fiore}, \citenamefont {Maiolino}, \citenamefont {Marconi} \emph {et~al.}}]{duras2020universal}%
  \BibitemOpen
  \bibfield  {author} {\bibinfo {author} {\bibfnamefont {F.}~\bibnamefont {Duras}}, \bibinfo {author} {\bibfnamefont {A.}~\bibnamefont {Bongiorno}}, \bibinfo {author} {\bibfnamefont {F.}~\bibnamefont {Ricci}}, \bibinfo {author} {\bibfnamefont {E.}~\bibnamefont {Piconcelli}}, \bibinfo {author} {\bibfnamefont {F.}~\bibnamefont {Shankar}}, \bibinfo {author} {\bibfnamefont {E.}~\bibnamefont {Lusso}}, \bibinfo {author} {\bibfnamefont {S.}~\bibnamefont {Bianchi}}, \bibinfo {author} {\bibfnamefont {F.}~\bibnamefont {Fiore}}, \bibinfo {author} {\bibfnamefont {R.}~\bibnamefont {Maiolino}}, \bibinfo {author} {\bibfnamefont {A.}~\bibnamefont {Marconi}}, \emph {et~al.},\ }\bibfield  {title} {\bibinfo {title} {Universal bolometric corrections for active galactic nuclei over seven luminosity decades},\ }\href@noop {} {\bibfield  {journal} {\bibinfo  {journal} {Astronomy \& Astrophysics}\ }\textbf {\bibinfo {volume} {636}},\ \bibinfo {pages} {A73} (\bibinfo {year} {2020})}\BibitemShut {NoStop}%
\bibitem [{\citenamefont {Pierre}\ \emph {et~al.}(2016)\citenamefont {Pierre}, \citenamefont {Pacaud}, \citenamefont {Adami}, \citenamefont {Alis}, \citenamefont {Altieri}, \citenamefont {Baran}, \citenamefont {Benoist}, \citenamefont {Birkinshaw}, \citenamefont {Bongiorno}, \citenamefont {Bremer} \emph {et~al.}}]{pierre2016xxl}%
  \BibitemOpen
  \bibfield  {author} {\bibinfo {author} {\bibfnamefont {M.}~\bibnamefont {Pierre}}, \bibinfo {author} {\bibfnamefont {F.}~\bibnamefont {Pacaud}}, \bibinfo {author} {\bibfnamefont {C.}~\bibnamefont {Adami}}, \bibinfo {author} {\bibfnamefont {S.}~\bibnamefont {Alis}}, \bibinfo {author} {\bibfnamefont {B.}~\bibnamefont {Altieri}}, \bibinfo {author} {\bibfnamefont {N.}~\bibnamefont {Baran}}, \bibinfo {author} {\bibfnamefont {C.}~\bibnamefont {Benoist}}, \bibinfo {author} {\bibfnamefont {M.}~\bibnamefont {Birkinshaw}}, \bibinfo {author} {\bibfnamefont {A.}~\bibnamefont {Bongiorno}}, \bibinfo {author} {\bibfnamefont {M.}~\bibnamefont {Bremer}}, \emph {et~al.},\ }\bibfield  {title} {\bibinfo {title} {The xxl survey-i. scientific motivations- xmm-newton observing plan- follow-up observations and simulation programme},\ }\href@noop {} {\bibfield  {journal} {\bibinfo  {journal} {Astronomy \& Astrophysics}\ }\textbf {\bibinfo {volume} {592}},\ \bibinfo {pages} {A1} (\bibinfo {year} {2016})}\BibitemShut {NoStop}%
\bibitem [{\citenamefont {Liu}\ \emph {et~al.}(2016)\citenamefont {Liu}, \citenamefont {Merloni}, \citenamefont {Georgakakis}, \citenamefont {Menzel}, \citenamefont {Buchner}, \citenamefont {Nandra}, \citenamefont {Salvato}, \citenamefont {Shen}, \citenamefont {Brusa},\ and\ \citenamefont {Streblyanska}}]{liu2016x}%
  \BibitemOpen
  \bibfield  {author} {\bibinfo {author} {\bibfnamefont {Z.}~\bibnamefont {Liu}}, \bibinfo {author} {\bibfnamefont {A.}~\bibnamefont {Merloni}}, \bibinfo {author} {\bibfnamefont {A.}~\bibnamefont {Georgakakis}}, \bibinfo {author} {\bibfnamefont {M.-L.}\ \bibnamefont {Menzel}}, \bibinfo {author} {\bibfnamefont {J.}~\bibnamefont {Buchner}}, \bibinfo {author} {\bibfnamefont {K.}~\bibnamefont {Nandra}}, \bibinfo {author} {\bibfnamefont {M.}~\bibnamefont {Salvato}}, \bibinfo {author} {\bibfnamefont {Y.}~\bibnamefont {Shen}}, \bibinfo {author} {\bibfnamefont {M.}~\bibnamefont {Brusa}},\ and\ \bibinfo {author} {\bibfnamefont {A.}~\bibnamefont {Streblyanska}},\ }\bibfield  {title} {\bibinfo {title} {X-ray spectral properties of the agn sample in the northern xmm-xxl field},\ }\href@noop {} {\bibfield  {journal} {\bibinfo  {journal} {Monthly Notices of the Royal Astronomical Society}\ }\textbf {\bibinfo {volume} {459}},\ \bibinfo {pages} {1602} (\bibinfo {year} {2016})}\BibitemShut {NoStop}%
\bibitem [{\citenamefont {Diana}\ \emph {et~al.}(2022)\citenamefont {Diana}, \citenamefont {Caccianiga}, \citenamefont {Ighina}, \citenamefont {Belladitta}, \citenamefont {Moretti},\ and\ \citenamefont {Della~Ceca}}]{diana2022evolution}%
  \BibitemOpen
  \bibfield  {author} {\bibinfo {author} {\bibfnamefont {A.}~\bibnamefont {Diana}}, \bibinfo {author} {\bibfnamefont {A.}~\bibnamefont {Caccianiga}}, \bibinfo {author} {\bibfnamefont {L.}~\bibnamefont {Ighina}}, \bibinfo {author} {\bibfnamefont {S.}~\bibnamefont {Belladitta}}, \bibinfo {author} {\bibfnamefont {A.}~\bibnamefont {Moretti}},\ and\ \bibinfo {author} {\bibfnamefont {R.}~\bibnamefont {Della~Ceca}},\ }\bibfield  {title} {\bibinfo {title} {The evolution of the heaviest supermassive black holes in jetted agns},\ }\href@noop {} {\bibfield  {journal} {\bibinfo  {journal} {Monthly Notices of the Royal Astronomical Society}\ }\textbf {\bibinfo {volume} {511}},\ \bibinfo {pages} {5436} (\bibinfo {year} {2022})}\BibitemShut {NoStop}%
\bibitem [{\citenamefont {Blanton}\ \emph {et~al.}(2017)\citenamefont {Blanton}, \citenamefont {Bershady}, \citenamefont {Abolfathi}, \citenamefont {Albareti}, \citenamefont {Prieto}, \citenamefont {Almeida}, \citenamefont {Alonso-Garc{\'\i}a}, \citenamefont {Anders}, \citenamefont {Anderson}, \citenamefont {Andrews} \emph {et~al.}}]{blanton2017sloan}%
  \BibitemOpen
  \bibfield  {author} {\bibinfo {author} {\bibfnamefont {M.~R.}\ \bibnamefont {Blanton}}, \bibinfo {author} {\bibfnamefont {M.~A.}\ \bibnamefont {Bershady}}, \bibinfo {author} {\bibfnamefont {B.}~\bibnamefont {Abolfathi}}, \bibinfo {author} {\bibfnamefont {F.~D.}\ \bibnamefont {Albareti}}, \bibinfo {author} {\bibfnamefont {C.~A.}\ \bibnamefont {Prieto}}, \bibinfo {author} {\bibfnamefont {A.}~\bibnamefont {Almeida}}, \bibinfo {author} {\bibfnamefont {J.}~\bibnamefont {Alonso-Garc{\'\i}a}}, \bibinfo {author} {\bibfnamefont {F.}~\bibnamefont {Anders}}, \bibinfo {author} {\bibfnamefont {S.~F.}\ \bibnamefont {Anderson}}, \bibinfo {author} {\bibfnamefont {B.}~\bibnamefont {Andrews}}, \emph {et~al.},\ }\bibfield  {title} {\bibinfo {title} {Sloan digital sky survey iv: Mapping the milky way, nearby galaxies, and the distant universe},\ }\href@noop {} {\bibfield  {journal} {\bibinfo  {journal} {The Astronomical Journal}\ }\textbf {\bibinfo {volume} {154}},\ \bibinfo {pages} {28} (\bibinfo {year} {2017})}\BibitemShut
  {NoStop}%
\bibitem [{\citenamefont {Browne}\ \emph {et~al.}(2003)\citenamefont {Browne}, \citenamefont {Wilkinson}, \citenamefont {Jackson}, \citenamefont {Myers}, \citenamefont {Fassnacht}, \citenamefont {Koopmans}, \citenamefont {Marlow}, \citenamefont {Norbury}, \citenamefont {Rusin}, \citenamefont {Sykes} \emph {et~al.}}]{browne2003cosmic}%
  \BibitemOpen
  \bibfield  {author} {\bibinfo {author} {\bibfnamefont {I.}~\bibnamefont {Browne}}, \bibinfo {author} {\bibfnamefont {P.}~\bibnamefont {Wilkinson}}, \bibinfo {author} {\bibfnamefont {N.}~\bibnamefont {Jackson}}, \bibinfo {author} {\bibfnamefont {S.}~\bibnamefont {Myers}}, \bibinfo {author} {\bibfnamefont {C.}~\bibnamefont {Fassnacht}}, \bibinfo {author} {\bibfnamefont {L.}~\bibnamefont {Koopmans}}, \bibinfo {author} {\bibfnamefont {D.~R.}\ \bibnamefont {Marlow}}, \bibinfo {author} {\bibfnamefont {M.}~\bibnamefont {Norbury}}, \bibinfo {author} {\bibfnamefont {D.}~\bibnamefont {Rusin}}, \bibinfo {author} {\bibfnamefont {C.}~\bibnamefont {Sykes}}, \emph {et~al.},\ }\bibfield  {title} {\bibinfo {title} {The cosmic lens all-sky survey-ii. gravitational lens candidate selection and follow-up},\ }\href@noop {} {\bibfield  {journal} {\bibinfo  {journal} {Monthly Notices of the Royal Astronomical Society}\ }\textbf {\bibinfo {volume} {341}},\ \bibinfo {pages} {13} (\bibinfo {year} {2003})}\BibitemShut {NoStop}%
\bibitem [{\citenamefont {Myers}\ \emph {et~al.}(2003)\citenamefont {Myers}, \citenamefont {Jackson}, \citenamefont {Browne}, \citenamefont {De~Bruyn}, \citenamefont {Pearson}, \citenamefont {Readhead}, \citenamefont {Wilkinson}, \citenamefont {Biggs}, \citenamefont {Blandford}, \citenamefont {Fassnacht} \emph {et~al.}}]{myers2003cosmic}%
  \BibitemOpen
  \bibfield  {author} {\bibinfo {author} {\bibfnamefont {S.}~\bibnamefont {Myers}}, \bibinfo {author} {\bibfnamefont {N.}~\bibnamefont {Jackson}}, \bibinfo {author} {\bibfnamefont {I.}~\bibnamefont {Browne}}, \bibinfo {author} {\bibfnamefont {A.}~\bibnamefont {De~Bruyn}}, \bibinfo {author} {\bibfnamefont {T.}~\bibnamefont {Pearson}}, \bibinfo {author} {\bibfnamefont {A.}~\bibnamefont {Readhead}}, \bibinfo {author} {\bibfnamefont {P.}~\bibnamefont {Wilkinson}}, \bibinfo {author} {\bibfnamefont {A.}~\bibnamefont {Biggs}}, \bibinfo {author} {\bibfnamefont {R.}~\bibnamefont {Blandford}}, \bibinfo {author} {\bibfnamefont {C.}~\bibnamefont {Fassnacht}}, \emph {et~al.},\ }\bibfield  {title} {\bibinfo {title} {the cosmic lens all-sky survey-i. source selection and observations},\ }\href@noop {} {\bibfield  {journal} {\bibinfo  {journal} {Monthly Notices of the Royal Astronomical Society}\ }\textbf {\bibinfo {volume} {341}},\ \bibinfo {pages} {1} (\bibinfo {year} {2003})}\BibitemShut {NoStop}%
\bibitem [{\citenamefont {Green}\ \emph {et~al.}(1986)\citenamefont {Green}, \citenamefont {Schmidt},\ and\ \citenamefont {Liebert}}]{green1986palomar}%
  \BibitemOpen
  \bibfield  {author} {\bibinfo {author} {\bibfnamefont {R.~F.}\ \bibnamefont {Green}}, \bibinfo {author} {\bibfnamefont {M.}~\bibnamefont {Schmidt}},\ and\ \bibinfo {author} {\bibfnamefont {J.}~\bibnamefont {Liebert}},\ }\bibfield  {title} {\bibinfo {title} {The palomar-green catalog of ultraviolet-excess stellar objects},\ }\href@noop {} {\bibfield  {journal} {\bibinfo  {journal} {Astrophysical Journal Supplement Series (ISSN 0067-0049), vol. 61, June 1986, p. 305-352. Research supported by the California Institute of Technology.}\ }\textbf {\bibinfo {volume} {61}},\ \bibinfo {pages} {305} (\bibinfo {year} {1986})}\BibitemShut {NoStop}%
\bibitem [{\citenamefont {Hewett}\ \emph {et~al.}(1995)\citenamefont {Hewett}, \citenamefont {Foltz},\ and\ \citenamefont {Chaffee}}]{hewett1995large}%
  \BibitemOpen
  \bibfield  {author} {\bibinfo {author} {\bibfnamefont {P.~C.}\ \bibnamefont {Hewett}}, \bibinfo {author} {\bibfnamefont {C.~B.}\ \bibnamefont {Foltz}},\ and\ \bibinfo {author} {\bibfnamefont {F.~H.}\ \bibnamefont {Chaffee}},\ }\bibfield  {title} {\bibinfo {title} {The large bright quasar survey. 6: Quasar catalog and survey parameters},\ }\href@noop {} {\bibfield  {journal} {\bibinfo  {journal} {Astronomical Journal (ISSN 0004-6256), vol. 109, no. 4, p. 1498-1521}\ }\textbf {\bibinfo {volume} {109}},\ \bibinfo {pages} {1498} (\bibinfo {year} {1995})}\BibitemShut {NoStop}%
\bibitem [{\citenamefont {York}\ \emph {et~al.}(2000)\citenamefont {York}, \citenamefont {Adelman}, \citenamefont {Anderson~Jr}, \citenamefont {Anderson}, \citenamefont {Annis}, \citenamefont {Bahcall}, \citenamefont {Bakken}, \citenamefont {Barkhouser}, \citenamefont {Bastian}, \citenamefont {Berman} \emph {et~al.}}]{york2000sloan}%
  \BibitemOpen
  \bibfield  {author} {\bibinfo {author} {\bibfnamefont {D.~G.}\ \bibnamefont {York}}, \bibinfo {author} {\bibfnamefont {J.}~\bibnamefont {Adelman}}, \bibinfo {author} {\bibfnamefont {J.~E.}\ \bibnamefont {Anderson~Jr}}, \bibinfo {author} {\bibfnamefont {S.~F.}\ \bibnamefont {Anderson}}, \bibinfo {author} {\bibfnamefont {J.}~\bibnamefont {Annis}}, \bibinfo {author} {\bibfnamefont {N.~A.}\ \bibnamefont {Bahcall}}, \bibinfo {author} {\bibfnamefont {J.}~\bibnamefont {Bakken}}, \bibinfo {author} {\bibfnamefont {R.}~\bibnamefont {Barkhouser}}, \bibinfo {author} {\bibfnamefont {S.}~\bibnamefont {Bastian}}, \bibinfo {author} {\bibfnamefont {E.}~\bibnamefont {Berman}}, \emph {et~al.},\ }\bibfield  {title} {\bibinfo {title} {The sloan digital sky survey: Technical summary},\ }\href@noop {} {\bibfield  {journal} {\bibinfo  {journal} {The Astronomical Journal}\ }\textbf {\bibinfo {volume} {120}},\ \bibinfo {pages} {1579} (\bibinfo {year} {2000})}\BibitemShut {NoStop}%
\bibitem [{\citenamefont {Boyle}\ \emph {et~al.}(2000)\citenamefont {Boyle}, \citenamefont {Shanks}, \citenamefont {Croom}, \citenamefont {Smith}, \citenamefont {Miller}, \citenamefont {Loaring},\ and\ \citenamefont {Heymans}}]{boyle20002df}%
  \BibitemOpen
  \bibfield  {author} {\bibinfo {author} {\bibfnamefont {B.~J.}\ \bibnamefont {Boyle}}, \bibinfo {author} {\bibfnamefont {T.}~\bibnamefont {Shanks}}, \bibinfo {author} {\bibfnamefont {S.}~\bibnamefont {Croom}}, \bibinfo {author} {\bibfnamefont {R.}~\bibnamefont {Smith}}, \bibinfo {author} {\bibfnamefont {L.}~\bibnamefont {Miller}}, \bibinfo {author} {\bibfnamefont {N.}~\bibnamefont {Loaring}},\ and\ \bibinfo {author} {\bibfnamefont {C.}~\bibnamefont {Heymans}},\ }\bibfield  {title} {\bibinfo {title} {The 2df qso redshift survey—i. the optical luminosity function of quasi-stellar objects},\ }\href@noop {} {\bibfield  {journal} {\bibinfo  {journal} {Monthly Notices of the Royal Astronomical Society}\ }\textbf {\bibinfo {volume} {317}},\ \bibinfo {pages} {1014} (\bibinfo {year} {2000})}\BibitemShut {NoStop}%
\bibitem [{\citenamefont {Blakeslee}\ \emph {et~al.}(2002)\citenamefont {Blakeslee}, \citenamefont {Lucey}, \citenamefont {Tonry}, \citenamefont {Hudson}, \citenamefont {Narayanan},\ and\ \citenamefont {Barris}}]{blakeslee2002early}%
  \BibitemOpen
  \bibfield  {author} {\bibinfo {author} {\bibfnamefont {J.~P.}\ \bibnamefont {Blakeslee}}, \bibinfo {author} {\bibfnamefont {J.~R.}\ \bibnamefont {Lucey}}, \bibinfo {author} {\bibfnamefont {J.~L.}\ \bibnamefont {Tonry}}, \bibinfo {author} {\bibfnamefont {M.~J.}\ \bibnamefont {Hudson}}, \bibinfo {author} {\bibfnamefont {V.~K.}\ \bibnamefont {Narayanan}},\ and\ \bibinfo {author} {\bibfnamefont {B.~J.}\ \bibnamefont {Barris}},\ }\bibfield  {title} {\bibinfo {title} {Early-type galaxy distances from the fundamental plane and surface brightness fluctuations},\ }\href@noop {} {\bibfield  {journal} {\bibinfo  {journal} {Monthly Notices of the Royal Astronomical Society}\ }\textbf {\bibinfo {volume} {330}},\ \bibinfo {pages} {443} (\bibinfo {year} {2002})}\BibitemShut {NoStop}%
\bibitem [{\citenamefont {Fan}(1999)}]{fan1999simulation}%
  \BibitemOpen
  \bibfield  {author} {\bibinfo {author} {\bibfnamefont {X.}~\bibnamefont {Fan}},\ }\bibfield  {title} {\bibinfo {title} {Simulation of stellar objects in sdss color space},\ }\href@noop {} {\bibfield  {journal} {\bibinfo  {journal} {The Astronomical Journal}\ }\textbf {\bibinfo {volume} {117}},\ \bibinfo {pages} {2528} (\bibinfo {year} {1999})}\BibitemShut {NoStop}%
\bibitem [{\citenamefont {Skrutskie}\ \emph {et~al.}(2006)\citenamefont {Skrutskie}, \citenamefont {Cutri}, \citenamefont {Stiening}, \citenamefont {Weinberg}, \citenamefont {Schneider}, \citenamefont {Carpenter}, \citenamefont {Beichman}, \citenamefont {Capps}, \citenamefont {Chester}, \citenamefont {Elias} \emph {et~al.}}]{skrutskie2006two}%
  \BibitemOpen
  \bibfield  {author} {\bibinfo {author} {\bibfnamefont {M.}~\bibnamefont {Skrutskie}}, \bibinfo {author} {\bibfnamefont {R.}~\bibnamefont {Cutri}}, \bibinfo {author} {\bibfnamefont {R.}~\bibnamefont {Stiening}}, \bibinfo {author} {\bibfnamefont {M.}~\bibnamefont {Weinberg}}, \bibinfo {author} {\bibfnamefont {S.}~\bibnamefont {Schneider}}, \bibinfo {author} {\bibfnamefont {J.}~\bibnamefont {Carpenter}}, \bibinfo {author} {\bibfnamefont {C.}~\bibnamefont {Beichman}}, \bibinfo {author} {\bibfnamefont {R.}~\bibnamefont {Capps}}, \bibinfo {author} {\bibfnamefont {T.}~\bibnamefont {Chester}}, \bibinfo {author} {\bibfnamefont {J.}~\bibnamefont {Elias}}, \emph {et~al.},\ }\bibfield  {title} {\bibinfo {title} {The two micron all sky survey (2mass)},\ }\href@noop {} {\bibfield  {journal} {\bibinfo  {journal} {The Astronomical Journal}\ }\textbf {\bibinfo {volume} {131}},\ \bibinfo {pages} {1163} (\bibinfo {year} {2006})}\BibitemShut {NoStop}%
\bibitem [{\citenamefont {Tago}\ \emph {et~al.}(2010)\citenamefont {Tago}, \citenamefont {Saar}, \citenamefont {Tempel}, \citenamefont {Einasto}, \citenamefont {Einasto}, \citenamefont {Nurmi},\ and\ \citenamefont {Hein{\"a}m{\"a}ki}}]{tago2010groups}%
  \BibitemOpen
  \bibfield  {author} {\bibinfo {author} {\bibfnamefont {E.}~\bibnamefont {Tago}}, \bibinfo {author} {\bibfnamefont {E.}~\bibnamefont {Saar}}, \bibinfo {author} {\bibfnamefont {E.}~\bibnamefont {Tempel}}, \bibinfo {author} {\bibfnamefont {J.}~\bibnamefont {Einasto}}, \bibinfo {author} {\bibfnamefont {M.}~\bibnamefont {Einasto}}, \bibinfo {author} {\bibfnamefont {P.}~\bibnamefont {Nurmi}},\ and\ \bibinfo {author} {\bibfnamefont {P.}~\bibnamefont {Hein{\"a}m{\"a}ki}},\ }\bibfield  {title} {\bibinfo {title} {Groups of galaxies in the sdss data release 7-flux-and volume-limited samples},\ }\href@noop {} {\bibfield  {journal} {\bibinfo  {journal} {Astronomy \& Astrophysics}\ }\textbf {\bibinfo {volume} {514}},\ \bibinfo {pages} {A102} (\bibinfo {year} {2010})}\BibitemShut {NoStop}%
\bibitem [{\citenamefont {Croom}\ \emph {et~al.}(2004)\citenamefont {Croom}, \citenamefont {Smith}, \citenamefont {Boyle}, \citenamefont {Shanks}, \citenamefont {Miller}, \citenamefont {Outram},\ and\ \citenamefont {Loaring}}]{croom20042df}%
  \BibitemOpen
  \bibfield  {author} {\bibinfo {author} {\bibfnamefont {S.~M.}\ \bibnamefont {Croom}}, \bibinfo {author} {\bibfnamefont {R.}~\bibnamefont {Smith}}, \bibinfo {author} {\bibfnamefont {B.}~\bibnamefont {Boyle}}, \bibinfo {author} {\bibfnamefont {T.}~\bibnamefont {Shanks}}, \bibinfo {author} {\bibfnamefont {L.}~\bibnamefont {Miller}}, \bibinfo {author} {\bibfnamefont {P.}~\bibnamefont {Outram}},\ and\ \bibinfo {author} {\bibfnamefont {N.}~\bibnamefont {Loaring}},\ }\bibfield  {title} {\bibinfo {title} {The 2df qso redshift survey--xii. the spectroscopic catalogue and luminosity function},\ }\href@noop {} {\bibfield  {journal} {\bibinfo  {journal} {Monthly Notices of the Royal Astronomical Society}\ }\textbf {\bibinfo {volume} {349}},\ \bibinfo {pages} {1397} (\bibinfo {year} {2004})}\BibitemShut {NoStop}%
\bibitem [{\citenamefont {Ilbert}\ \emph {et~al.}(2004)\citenamefont {Ilbert}, \citenamefont {Tresse}, \citenamefont {Arnouts}, \citenamefont {Zucca}, \citenamefont {Bardelli}, \citenamefont {Zamorani}, \citenamefont {Adami}, \citenamefont {Cappi}, \citenamefont {Garilli}, \citenamefont {Le~Fevre} \emph {et~al.}}]{ilbert2004bias}%
  \BibitemOpen
  \bibfield  {author} {\bibinfo {author} {\bibfnamefont {O.}~\bibnamefont {Ilbert}}, \bibinfo {author} {\bibfnamefont {L.}~\bibnamefont {Tresse}}, \bibinfo {author} {\bibfnamefont {S.}~\bibnamefont {Arnouts}}, \bibinfo {author} {\bibfnamefont {E.}~\bibnamefont {Zucca}}, \bibinfo {author} {\bibfnamefont {S.}~\bibnamefont {Bardelli}}, \bibinfo {author} {\bibfnamefont {G.}~\bibnamefont {Zamorani}}, \bibinfo {author} {\bibfnamefont {C.}~\bibnamefont {Adami}}, \bibinfo {author} {\bibfnamefont {A.}~\bibnamefont {Cappi}}, \bibinfo {author} {\bibfnamefont {B.}~\bibnamefont {Garilli}}, \bibinfo {author} {\bibfnamefont {O.}~\bibnamefont {Le~Fevre}}, \emph {et~al.},\ }\bibfield  {title} {\bibinfo {title} {Bias in the estimation of global luminosity functions},\ }\href@noop {} {\bibfield  {journal} {\bibinfo  {journal} {Monthly Notices of the Royal Astronomical Society}\ }\textbf {\bibinfo {volume} {351}},\ \bibinfo {pages} {541} (\bibinfo {year} {2004})}\BibitemShut {NoStop}%
\bibitem [{\citenamefont {Cole}(2011)}]{cole2011maximum}%
  \BibitemOpen
  \bibfield  {author} {\bibinfo {author} {\bibfnamefont {S.}~\bibnamefont {Cole}},\ }\bibfield  {title} {\bibinfo {title} {Maximum likelihood random galaxy catalogues and luminosity function estimation},\ }\href@noop {} {\bibfield  {journal} {\bibinfo  {journal} {Monthly Notices of the Royal Astronomical Society}\ }\textbf {\bibinfo {volume} {416}},\ \bibinfo {pages} {739} (\bibinfo {year} {2011})}\BibitemShut {NoStop}%
\bibitem [{\citenamefont {Turner}\ and\ \citenamefont {Gott~III}(1976)}]{turner1976groups}%
  \BibitemOpen
  \bibfield  {author} {\bibinfo {author} {\bibfnamefont {E.~L.}\ \bibnamefont {Turner}}\ and\ \bibinfo {author} {\bibfnamefont {J.~R.}\ \bibnamefont {Gott~III}},\ }\bibfield  {title} {\bibinfo {title} {Groups of galaxies. i. a catalog.},\ }\href@noop {} {\bibfield  {journal} {\bibinfo  {journal} {Astrophysical Journal, Suppl. Ser., Vol. 32, p. 409-427}\ }\textbf {\bibinfo {volume} {32}},\ \bibinfo {pages} {409} (\bibinfo {year} {1976})}\BibitemShut {NoStop}%
\bibitem [{\citenamefont {Cunningham}\ \emph {et~al.}(2020)\citenamefont {Cunningham}, \citenamefont {Tremblay}, \citenamefont {Gentile~Fusillo}, \citenamefont {Hollands},\ and\ \citenamefont {Cukanovaite}}]{cunningham2020hydrogen}%
  \BibitemOpen
  \bibfield  {author} {\bibinfo {author} {\bibfnamefont {T.}~\bibnamefont {Cunningham}}, \bibinfo {author} {\bibfnamefont {P.-E.}\ \bibnamefont {Tremblay}}, \bibinfo {author} {\bibfnamefont {N.~P.}\ \bibnamefont {Gentile~Fusillo}}, \bibinfo {author} {\bibfnamefont {M.}~\bibnamefont {Hollands}},\ and\ \bibinfo {author} {\bibfnamefont {E.}~\bibnamefont {Cukanovaite}},\ }\bibfield  {title} {\bibinfo {title} {From hydrogen to helium: the spectral evolution of white dwarfs as evidence for convective mixing},\ }\href@noop {} {\bibfield  {journal} {\bibinfo  {journal} {Monthly Notices of the Royal Astronomical Society}\ }\textbf {\bibinfo {volume} {492}},\ \bibinfo {pages} {3540} (\bibinfo {year} {2020})}\BibitemShut {NoStop}%
\bibitem [{\citenamefont {Tempel}\ \emph {et~al.}(2014)\citenamefont {Tempel}, \citenamefont {Tamm}, \citenamefont {Gramann}, \citenamefont {Tuvikene}, \citenamefont {Liivam{\"a}gi}, \citenamefont {Suhhonenko}, \citenamefont {Kipper}, \citenamefont {Einasto},\ and\ \citenamefont {Saar}}]{tempel2014flux}%
  \BibitemOpen
  \bibfield  {author} {\bibinfo {author} {\bibfnamefont {E.}~\bibnamefont {Tempel}}, \bibinfo {author} {\bibfnamefont {A.}~\bibnamefont {Tamm}}, \bibinfo {author} {\bibfnamefont {M.}~\bibnamefont {Gramann}}, \bibinfo {author} {\bibfnamefont {T.}~\bibnamefont {Tuvikene}}, \bibinfo {author} {\bibfnamefont {L.}~\bibnamefont {Liivam{\"a}gi}}, \bibinfo {author} {\bibfnamefont {I.}~\bibnamefont {Suhhonenko}}, \bibinfo {author} {\bibfnamefont {R.}~\bibnamefont {Kipper}}, \bibinfo {author} {\bibfnamefont {M.}~\bibnamefont {Einasto}},\ and\ \bibinfo {author} {\bibfnamefont {E.}~\bibnamefont {Saar}},\ }\bibfield  {title} {\bibinfo {title} {Flux-and volume-limited groups/clusters for the sdss galaxies: catalogues and mass estimation},\ }\href@noop {} {\bibfield  {journal} {\bibinfo  {journal} {Astronomy \& Astrophysics}\ }\textbf {\bibinfo {volume} {566}},\ \bibinfo {pages} {A1} (\bibinfo {year} {2014})}\BibitemShut {NoStop}%
\bibitem [{\citenamefont {Duarte}\ and\ \citenamefont {Mamon}(2014)}]{duarte2014well}%
  \BibitemOpen
  \bibfield  {author} {\bibinfo {author} {\bibfnamefont {M.}~\bibnamefont {Duarte}}\ and\ \bibinfo {author} {\bibfnamefont {G.~A.}\ \bibnamefont {Mamon}},\ }\bibfield  {title} {\bibinfo {title} {How well does the friends-of-friends algorithm recover group properties from galaxy catalogues limited in both distance and luminosity?},\ }\href@noop {} {\bibfield  {journal} {\bibinfo  {journal} {Monthly Notices of the Royal Astronomical Society}\ }\textbf {\bibinfo {volume} {440}},\ \bibinfo {pages} {1763} (\bibinfo {year} {2014})}\BibitemShut {NoStop}%
\bibitem [{\citenamefont {Etherington}\ and\ \citenamefont {Thomas}(2015)}]{etherington2015measuring}%
  \BibitemOpen
  \bibfield  {author} {\bibinfo {author} {\bibfnamefont {J.}~\bibnamefont {Etherington}}\ and\ \bibinfo {author} {\bibfnamefont {D.}~\bibnamefont {Thomas}},\ }\bibfield  {title} {\bibinfo {title} {Measuring galaxy environments in large-scale photometric surveys},\ }\href@noop {} {\bibfield  {journal} {\bibinfo  {journal} {Monthly Notices of the Royal Astronomical Society}\ }\textbf {\bibinfo {volume} {451}},\ \bibinfo {pages} {660} (\bibinfo {year} {2015})}\BibitemShut {NoStop}%
\bibitem [{\citenamefont {Ghirlanda}\ \emph {et~al.}(2012)\citenamefont {Ghirlanda}, \citenamefont {Ghisellini}, \citenamefont {Nava}, \citenamefont {Salvaterra}, \citenamefont {Tagliaferri}, \citenamefont {Campana}, \citenamefont {Covino}, \citenamefont {D’Avanzo}, \citenamefont {Fugazza}, \citenamefont {Melandri} \emph {et~al.}}]{ghirlanda2012impact}%
  \BibitemOpen
  \bibfield  {author} {\bibinfo {author} {\bibfnamefont {G.}~\bibnamefont {Ghirlanda}}, \bibinfo {author} {\bibfnamefont {G.}~\bibnamefont {Ghisellini}}, \bibinfo {author} {\bibfnamefont {L.}~\bibnamefont {Nava}}, \bibinfo {author} {\bibfnamefont {R.}~\bibnamefont {Salvaterra}}, \bibinfo {author} {\bibfnamefont {G.}~\bibnamefont {Tagliaferri}}, \bibinfo {author} {\bibfnamefont {S.}~\bibnamefont {Campana}}, \bibinfo {author} {\bibfnamefont {S.}~\bibnamefont {Covino}}, \bibinfo {author} {\bibfnamefont {P.}~\bibnamefont {D’Avanzo}}, \bibinfo {author} {\bibfnamefont {D.}~\bibnamefont {Fugazza}}, \bibinfo {author} {\bibfnamefont {A.}~\bibnamefont {Melandri}}, \emph {et~al.},\ }\bibfield  {title} {\bibinfo {title} {The impact of selection biases on the e peak--l iso correlation of gamma-ray bursts},\ }\href@noop {} {\bibfield  {journal} {\bibinfo  {journal} {Monthly Notices of the Royal Astronomical Society}\ }\textbf {\bibinfo {volume} {422}},\ \bibinfo {pages} {2553} (\bibinfo {year} {2012})}\BibitemShut
  {NoStop}%
\bibitem [{\citenamefont {Best}\ \emph {et~al.}(2024)\citenamefont {Best}, \citenamefont {Sanghi}, \citenamefont {Liu}, \citenamefont {Magnier},\ and\ \citenamefont {Dupuy}}]{best2024volume}%
  \BibitemOpen
  \bibfield  {author} {\bibinfo {author} {\bibfnamefont {W.~M.}\ \bibnamefont {Best}}, \bibinfo {author} {\bibfnamefont {A.}~\bibnamefont {Sanghi}}, \bibinfo {author} {\bibfnamefont {M.~C.}\ \bibnamefont {Liu}}, \bibinfo {author} {\bibfnamefont {E.~A.}\ \bibnamefont {Magnier}},\ and\ \bibinfo {author} {\bibfnamefont {T.~J.}\ \bibnamefont {Dupuy}},\ }\bibfield  {title} {\bibinfo {title} {A volume-limited sample of ultracool dwarfs. ii. the substellar age and mass functions in the solar neighborhood},\ }\href@noop {} {\bibfield  {journal} {\bibinfo  {journal} {arXiv preprint arXiv:2401.09535}\ } (\bibinfo {year} {2024})}\BibitemShut {NoStop}%
\bibitem [{\citenamefont {Ross}\ \emph {et~al.}(2012)\citenamefont {Ross}, \citenamefont {Myers}, \citenamefont {Sheldon}, \citenamefont {Y{\`e}che}, \citenamefont {Strauss}, \citenamefont {Bovy}, \citenamefont {Kirkpatrick}, \citenamefont {Richards}, \citenamefont {Aubourg}, \citenamefont {Blanton} \emph {et~al.}}]{ross2012sdss}%
  \BibitemOpen
  \bibfield  {author} {\bibinfo {author} {\bibfnamefont {N.~P.}\ \bibnamefont {Ross}}, \bibinfo {author} {\bibfnamefont {A.~D.}\ \bibnamefont {Myers}}, \bibinfo {author} {\bibfnamefont {E.~S.}\ \bibnamefont {Sheldon}}, \bibinfo {author} {\bibfnamefont {C.}~\bibnamefont {Y{\`e}che}}, \bibinfo {author} {\bibfnamefont {M.~A.}\ \bibnamefont {Strauss}}, \bibinfo {author} {\bibfnamefont {J.}~\bibnamefont {Bovy}}, \bibinfo {author} {\bibfnamefont {J.~A.}\ \bibnamefont {Kirkpatrick}}, \bibinfo {author} {\bibfnamefont {G.~T.}\ \bibnamefont {Richards}}, \bibinfo {author} {\bibfnamefont {{\'E}.}~\bibnamefont {Aubourg}}, \bibinfo {author} {\bibfnamefont {M.~R.}\ \bibnamefont {Blanton}}, \emph {et~al.},\ }\bibfield  {title} {\bibinfo {title} {The sdss-iii baryon oscillation spectroscopic survey: quasar target selection for data release nine},\ }\href@noop {} {\bibfield  {journal} {\bibinfo  {journal} {The Astrophysical Journal Supplement Series}\ }\textbf {\bibinfo {volume} {199}},\ \bibinfo {pages} {3} (\bibinfo {year}
  {2012})}\BibitemShut {NoStop}%
\bibitem [{\citenamefont {Zhang}\ and\ \citenamefont {Lu}(2020)}]{zhang2020extracting}%
  \BibitemOpen
  \bibfield  {author} {\bibinfo {author} {\bibfnamefont {F.}~\bibnamefont {Zhang}}\ and\ \bibinfo {author} {\bibfnamefont {Y.}~\bibnamefont {Lu}},\ }\bibfield  {title} {\bibinfo {title} {Extracting the possible intrinsic relation between the radiative efficiency and mass of qsos: A maximum likelihood method and its application to the sdss dr7 qsos},\ }\href@noop {} {\bibfield  {journal} {\bibinfo  {journal} {The Astrophysical Journal}\ }\textbf {\bibinfo {volume} {902}},\ \bibinfo {pages} {52} (\bibinfo {year} {2020})}\BibitemShut {NoStop}%
\bibitem [{\citenamefont {White}\ and\ \citenamefont {Ghosh}(1998)}]{white1998low}%
  \BibitemOpen
  \bibfield  {author} {\bibinfo {author} {\bibfnamefont {N.~E.}\ \bibnamefont {White}}\ and\ \bibinfo {author} {\bibfnamefont {P.}~\bibnamefont {Ghosh}},\ }\bibfield  {title} {\bibinfo {title} {Low-mass x-ray binaries, millisecond radio pulsars, and the cosmic star formation rate},\ }\href@noop {} {\bibfield  {journal} {\bibinfo  {journal} {The Astrophysical Journal}\ }\textbf {\bibinfo {volume} {504}},\ \bibinfo {pages} {L31} (\bibinfo {year} {1998})}\BibitemShut {NoStop}%
\bibitem [{\citenamefont {Atek}\ \emph {et~al.}(2014)\citenamefont {Atek}, \citenamefont {Kneib}, \citenamefont {Pacifici}, \citenamefont {Malkan}, \citenamefont {Charlot}, \citenamefont {Lee}, \citenamefont {Bedregal}, \citenamefont {Bunker}, \citenamefont {Colbert}, \citenamefont {Dressler} \emph {et~al.}}]{atek2014hubble}%
  \BibitemOpen
  \bibfield  {author} {\bibinfo {author} {\bibfnamefont {H.}~\bibnamefont {Atek}}, \bibinfo {author} {\bibfnamefont {J.-P.}\ \bibnamefont {Kneib}}, \bibinfo {author} {\bibfnamefont {C.}~\bibnamefont {Pacifici}}, \bibinfo {author} {\bibfnamefont {M.}~\bibnamefont {Malkan}}, \bibinfo {author} {\bibfnamefont {S.}~\bibnamefont {Charlot}}, \bibinfo {author} {\bibfnamefont {J.}~\bibnamefont {Lee}}, \bibinfo {author} {\bibfnamefont {A.}~\bibnamefont {Bedregal}}, \bibinfo {author} {\bibfnamefont {A.~J.}\ \bibnamefont {Bunker}}, \bibinfo {author} {\bibfnamefont {J.~W.}\ \bibnamefont {Colbert}}, \bibinfo {author} {\bibfnamefont {A.}~\bibnamefont {Dressler}}, \emph {et~al.},\ }\bibfield  {title} {\bibinfo {title} {Hubble space telescope grism spectroscopy of extreme starbursts across cosmic time: The role of dwarf galaxies in the star formation history of the universe},\ }\href@noop {} {\bibfield  {journal} {\bibinfo  {journal} {The Astrophysical Journal}\ }\textbf {\bibinfo {volume} {789}},\ \bibinfo {pages} {96}
  (\bibinfo {year} {2014})}\BibitemShut {NoStop}%
\bibitem [{\citenamefont {Zolotov}\ \emph {et~al.}(2015)\citenamefont {Zolotov}, \citenamefont {Dekel}, \citenamefont {Mandelker}, \citenamefont {Tweed}, \citenamefont {Inoue}, \citenamefont {DeGraf}, \citenamefont {Ceverino}, \citenamefont {Primack}, \citenamefont {Barro},\ and\ \citenamefont {Faber}}]{zolotov2015compaction}%
  \BibitemOpen
  \bibfield  {author} {\bibinfo {author} {\bibfnamefont {A.}~\bibnamefont {Zolotov}}, \bibinfo {author} {\bibfnamefont {A.}~\bibnamefont {Dekel}}, \bibinfo {author} {\bibfnamefont {N.}~\bibnamefont {Mandelker}}, \bibinfo {author} {\bibfnamefont {D.}~\bibnamefont {Tweed}}, \bibinfo {author} {\bibfnamefont {S.}~\bibnamefont {Inoue}}, \bibinfo {author} {\bibfnamefont {C.}~\bibnamefont {DeGraf}}, \bibinfo {author} {\bibfnamefont {D.}~\bibnamefont {Ceverino}}, \bibinfo {author} {\bibfnamefont {J.~R.}\ \bibnamefont {Primack}}, \bibinfo {author} {\bibfnamefont {G.}~\bibnamefont {Barro}},\ and\ \bibinfo {author} {\bibfnamefont {S.~M.}\ \bibnamefont {Faber}},\ }\bibfield  {title} {\bibinfo {title} {Compaction and quenching of high-z galaxies in cosmological simulations: blue and red nuggets},\ }\href@noop {} {\bibfield  {journal} {\bibinfo  {journal} {Monthly Notices of the Royal Astronomical Society}\ }\textbf {\bibinfo {volume} {450}},\ \bibinfo {pages} {2327} (\bibinfo {year} {2015})}\BibitemShut {NoStop}%
\bibitem [{\citenamefont {Driver}\ \emph {et~al.}(2018)\citenamefont {Driver}, \citenamefont {Andrews}, \citenamefont {Da~Cunha}, \citenamefont {Davies}, \citenamefont {Lagos}, \citenamefont {Robotham}, \citenamefont {Vinsen}, \citenamefont {Wright}, \citenamefont {Alpaslan}, \citenamefont {Bland-Hawthorn} \emph {et~al.}}]{driver2018gama}%
  \BibitemOpen
  \bibfield  {author} {\bibinfo {author} {\bibfnamefont {S.~P.}\ \bibnamefont {Driver}}, \bibinfo {author} {\bibfnamefont {S.~K.}\ \bibnamefont {Andrews}}, \bibinfo {author} {\bibfnamefont {E.}~\bibnamefont {Da~Cunha}}, \bibinfo {author} {\bibfnamefont {L.~J.}\ \bibnamefont {Davies}}, \bibinfo {author} {\bibfnamefont {C.}~\bibnamefont {Lagos}}, \bibinfo {author} {\bibfnamefont {A.~S.}\ \bibnamefont {Robotham}}, \bibinfo {author} {\bibfnamefont {K.}~\bibnamefont {Vinsen}}, \bibinfo {author} {\bibfnamefont {A.~H.}\ \bibnamefont {Wright}}, \bibinfo {author} {\bibfnamefont {M.}~\bibnamefont {Alpaslan}}, \bibinfo {author} {\bibfnamefont {J.}~\bibnamefont {Bland-Hawthorn}}, \emph {et~al.},\ }\bibfield  {title} {\bibinfo {title} {Gama/g10-cosmos/3d-hst: the 0< z< 5 cosmic star formation history, stellar-mass, and dust-mass densities},\ }\href@noop {} {\bibfield  {journal} {\bibinfo  {journal} {Monthly Notices of the Royal Astronomical Society}\ }\textbf {\bibinfo {volume} {475}},\ \bibinfo {pages} {2891} (\bibinfo
  {year} {2018})}\BibitemShut {NoStop}%
\bibitem [{\citenamefont {Hopkins}\ and\ \citenamefont {Beacom}(2006)}]{hopkins2006normalization}%
  \BibitemOpen
  \bibfield  {author} {\bibinfo {author} {\bibfnamefont {A.~M.}\ \bibnamefont {Hopkins}}\ and\ \bibinfo {author} {\bibfnamefont {J.~F.}\ \bibnamefont {Beacom}},\ }\bibfield  {title} {\bibinfo {title} {On the normalization of the cosmic star formation history},\ }\href@noop {} {\bibfield  {journal} {\bibinfo  {journal} {The Astrophysical Journal}\ }\textbf {\bibinfo {volume} {651}},\ \bibinfo {pages} {142} (\bibinfo {year} {2006})}\BibitemShut {NoStop}%
\bibitem [{\citenamefont {Vangioni}\ \emph {et~al.}(2015)\citenamefont {Vangioni}, \citenamefont {Olive}, \citenamefont {Prestegard}, \citenamefont {Silk}, \citenamefont {Petitjean},\ and\ \citenamefont {Mandic}}]{vangioni2015impact}%
  \BibitemOpen
  \bibfield  {author} {\bibinfo {author} {\bibfnamefont {E.}~\bibnamefont {Vangioni}}, \bibinfo {author} {\bibfnamefont {K.~A.}\ \bibnamefont {Olive}}, \bibinfo {author} {\bibfnamefont {T.}~\bibnamefont {Prestegard}}, \bibinfo {author} {\bibfnamefont {J.}~\bibnamefont {Silk}}, \bibinfo {author} {\bibfnamefont {P.}~\bibnamefont {Petitjean}},\ and\ \bibinfo {author} {\bibfnamefont {V.}~\bibnamefont {Mandic}},\ }\bibfield  {title} {\bibinfo {title} {The impact of star formation and gamma-ray burst rates at high redshift on cosmic chemical evolution and reionization},\ }\href@noop {} {\bibfield  {journal} {\bibinfo  {journal} {Monthly Notices of the Royal Astronomical Society}\ }\textbf {\bibinfo {volume} {447}},\ \bibinfo {pages} {2575} (\bibinfo {year} {2015})}\BibitemShut {NoStop}%
\bibitem [{\citenamefont {Cucciati}\ \emph {et~al.}(2012)\citenamefont {Cucciati}, \citenamefont {Tresse}, \citenamefont {Ilbert}, \citenamefont {Le~Fevre}, \citenamefont {Garilli}, \citenamefont {Le~Brun}, \citenamefont {Cassata}, \citenamefont {Franzetti}, \citenamefont {Maccagni}, \citenamefont {Scodeggio} \emph {et~al.}}]{cucciati2012star}%
  \BibitemOpen
  \bibfield  {author} {\bibinfo {author} {\bibfnamefont {O.}~\bibnamefont {Cucciati}}, \bibinfo {author} {\bibfnamefont {L.}~\bibnamefont {Tresse}}, \bibinfo {author} {\bibfnamefont {O.}~\bibnamefont {Ilbert}}, \bibinfo {author} {\bibfnamefont {O.}~\bibnamefont {Le~Fevre}}, \bibinfo {author} {\bibfnamefont {B.}~\bibnamefont {Garilli}}, \bibinfo {author} {\bibfnamefont {V.}~\bibnamefont {Le~Brun}}, \bibinfo {author} {\bibfnamefont {P.}~\bibnamefont {Cassata}}, \bibinfo {author} {\bibfnamefont {P.}~\bibnamefont {Franzetti}}, \bibinfo {author} {\bibfnamefont {D.}~\bibnamefont {Maccagni}}, \bibinfo {author} {\bibfnamefont {M.}~\bibnamefont {Scodeggio}}, \emph {et~al.},\ }\bibfield  {title} {\bibinfo {title} {The star formation rate density and dust attenuation evolution over 12 gyr with the vvds surveys},\ }\href@noop {} {\bibfield  {journal} {\bibinfo  {journal} {Astronomy \& Astrophysics}\ }\textbf {\bibinfo {volume} {539}},\ \bibinfo {pages} {A31} (\bibinfo {year} {2012})}\BibitemShut {NoStop}%
\bibitem [{\citenamefont {Jo}\ \emph {et~al.}(2021)\citenamefont {Jo}, \citenamefont {Youn}, \citenamefont {Kim}, \citenamefont {Park}, \citenamefont {Hwang}, \citenamefont {Lee},\ and\ \citenamefont {Kim}}]{jo2021star}%
  \BibitemOpen
  \bibfield  {author} {\bibinfo {author} {\bibfnamefont {J.~U.}\ \bibnamefont {Jo}}, \bibinfo {author} {\bibfnamefont {S.}~\bibnamefont {Youn}}, \bibinfo {author} {\bibfnamefont {S.}~\bibnamefont {Kim}}, \bibinfo {author} {\bibfnamefont {Y.}~\bibnamefont {Park}}, \bibinfo {author} {\bibfnamefont {J.}~\bibnamefont {Hwang}}, \bibinfo {author} {\bibfnamefont {J.~H.}\ \bibnamefont {Lee}},\ and\ \bibinfo {author} {\bibfnamefont {G.}~\bibnamefont {Kim}},\ }\bibfield  {title} {\bibinfo {title} {Star formation rate density across the cosmic time},\ }\href@noop {} {\bibfield  {journal} {\bibinfo  {journal} {Astrophysics and Space Science}\ }\textbf {\bibinfo {volume} {366}},\ \bibinfo {pages} {18} (\bibinfo {year} {2021})}\BibitemShut {NoStop}%
\bibitem [{\citenamefont {Schinnerer}\ \emph {et~al.}(2016)\citenamefont {Schinnerer}, \citenamefont {Groves}, \citenamefont {Sargent}, \citenamefont {Karim}, \citenamefont {Oesch}, \citenamefont {Magnelli}, \citenamefont {LeFevre}, \citenamefont {Tasca}, \citenamefont {Civano}, \citenamefont {Cassata} \emph {et~al.}}]{schinnerer2016gas}%
  \BibitemOpen
  \bibfield  {author} {\bibinfo {author} {\bibfnamefont {E.}~\bibnamefont {Schinnerer}}, \bibinfo {author} {\bibfnamefont {B.}~\bibnamefont {Groves}}, \bibinfo {author} {\bibfnamefont {M.}~\bibnamefont {Sargent}}, \bibinfo {author} {\bibfnamefont {A.}~\bibnamefont {Karim}}, \bibinfo {author} {\bibfnamefont {P.}~\bibnamefont {Oesch}}, \bibinfo {author} {\bibfnamefont {B.}~\bibnamefont {Magnelli}}, \bibinfo {author} {\bibfnamefont {O.}~\bibnamefont {LeFevre}}, \bibinfo {author} {\bibfnamefont {L.}~\bibnamefont {Tasca}}, \bibinfo {author} {\bibfnamefont {F.}~\bibnamefont {Civano}}, \bibinfo {author} {\bibfnamefont {P.}~\bibnamefont {Cassata}}, \emph {et~al.},\ }\bibfield  {title} {\bibinfo {title} {Gas fraction and depletion time of massive star-forming galaxies at z~ 3.2: no change in global star formation process out to z> 3},\ }\href@noop {} {\bibfield  {journal} {\bibinfo  {journal} {The Astrophysical Journal}\ }\textbf {\bibinfo {volume} {833}},\ \bibinfo {pages} {112} (\bibinfo {year} {2016})}\BibitemShut
  {NoStop}%
\bibitem [{\citenamefont {Scoville}\ \emph {et~al.}(2017)\citenamefont {Scoville}, \citenamefont {Lee}, \citenamefont {Bout}, \citenamefont {Diaz-Santos}, \citenamefont {Sanders}, \citenamefont {Darvish}, \citenamefont {Bongiorno}, \citenamefont {Casey}, \citenamefont {Murchikova}, \citenamefont {Koda} \emph {et~al.}}]{scoville2017evolution}%
  \BibitemOpen
  \bibfield  {author} {\bibinfo {author} {\bibfnamefont {N.}~\bibnamefont {Scoville}}, \bibinfo {author} {\bibfnamefont {N.}~\bibnamefont {Lee}}, \bibinfo {author} {\bibfnamefont {P.~V.}\ \bibnamefont {Bout}}, \bibinfo {author} {\bibfnamefont {T.}~\bibnamefont {Diaz-Santos}}, \bibinfo {author} {\bibfnamefont {D.}~\bibnamefont {Sanders}}, \bibinfo {author} {\bibfnamefont {B.}~\bibnamefont {Darvish}}, \bibinfo {author} {\bibfnamefont {A.}~\bibnamefont {Bongiorno}}, \bibinfo {author} {\bibfnamefont {C.}~\bibnamefont {Casey}}, \bibinfo {author} {\bibfnamefont {L.}~\bibnamefont {Murchikova}}, \bibinfo {author} {\bibfnamefont {J.}~\bibnamefont {Koda}}, \emph {et~al.},\ }\bibfield  {title} {\bibinfo {title} {Evolution of interstellar medium, star formation, and accretion at high redshift},\ }\href@noop {} {\bibfield  {journal} {\bibinfo  {journal} {The Astrophysical Journal}\ }\textbf {\bibinfo {volume} {837}},\ \bibinfo {pages} {150} (\bibinfo {year} {2017})}\BibitemShut {NoStop}%
\bibitem [{\citenamefont {Tacconi}\ \emph {et~al.}(2018)\citenamefont {Tacconi}, \citenamefont {Genzel}, \citenamefont {Saintonge}, \citenamefont {Combes}, \citenamefont {Garc{\'\i}a-Burillo}, \citenamefont {Neri}, \citenamefont {Bolatto}, \citenamefont {Contini}, \citenamefont {Schreiber}, \citenamefont {Lilly} \emph {et~al.}}]{tacconi2018phibss}%
  \BibitemOpen
  \bibfield  {author} {\bibinfo {author} {\bibfnamefont {L.~J.}\ \bibnamefont {Tacconi}}, \bibinfo {author} {\bibfnamefont {R.}~\bibnamefont {Genzel}}, \bibinfo {author} {\bibfnamefont {A.}~\bibnamefont {Saintonge}}, \bibinfo {author} {\bibfnamefont {F.}~\bibnamefont {Combes}}, \bibinfo {author} {\bibfnamefont {S.}~\bibnamefont {Garc{\'\i}a-Burillo}}, \bibinfo {author} {\bibfnamefont {R.}~\bibnamefont {Neri}}, \bibinfo {author} {\bibfnamefont {A.}~\bibnamefont {Bolatto}}, \bibinfo {author} {\bibfnamefont {T.}~\bibnamefont {Contini}}, \bibinfo {author} {\bibfnamefont {N.~F.}\ \bibnamefont {Schreiber}}, \bibinfo {author} {\bibfnamefont {S.}~\bibnamefont {Lilly}}, \emph {et~al.},\ }\bibfield  {title} {\bibinfo {title} {Phibss: unified scaling relations of gas depletion time and molecular gas fractions},\ }\href@noop {} {\bibfield  {journal} {\bibinfo  {journal} {The Astrophysical Journal}\ }\textbf {\bibinfo {volume} {853}},\ \bibinfo {pages} {179} (\bibinfo {year} {2018})}\BibitemShut {NoStop}%
\bibitem [{\citenamefont {Reddy}\ \emph {et~al.}(2007)\citenamefont {Reddy}, \citenamefont {Steidel}, \citenamefont {Pettini}, \citenamefont {Adelberger}, \citenamefont {Shapley}, \citenamefont {Erb},\ and\ \citenamefont {Dickinson}}]{reddy2007multi}%
  \BibitemOpen
  \bibfield  {author} {\bibinfo {author} {\bibfnamefont {N.~A.}\ \bibnamefont {Reddy}}, \bibinfo {author} {\bibfnamefont {C.~C.}\ \bibnamefont {Steidel}}, \bibinfo {author} {\bibfnamefont {M.}~\bibnamefont {Pettini}}, \bibinfo {author} {\bibfnamefont {K.~L.}\ \bibnamefont {Adelberger}}, \bibinfo {author} {\bibfnamefont {A.~E.}\ \bibnamefont {Shapley}}, \bibinfo {author} {\bibfnamefont {D.~K.}\ \bibnamefont {Erb}},\ and\ \bibinfo {author} {\bibfnamefont {M.}~\bibnamefont {Dickinson}},\ }\bibfield  {title} {\bibinfo {title} {Multi-wavelength constraints on the cosmic star formation history from spectroscopy: the rest-frame uv, h-alpha, and infrared luminosity functions at redshifts 1.9< z< 3.4},\ }\href@noop {} {\bibfield  {journal} {\bibinfo  {journal} {arXiv preprint arXiv:0706.4091}\ } (\bibinfo {year} {2007})}\BibitemShut {NoStop}%
\bibitem [{\citenamefont {Madau}\ and\ \citenamefont {Dickinson}(2014)}]{madau2014cosmic}%
  \BibitemOpen
  \bibfield  {author} {\bibinfo {author} {\bibfnamefont {P.}~\bibnamefont {Madau}}\ and\ \bibinfo {author} {\bibfnamefont {M.}~\bibnamefont {Dickinson}},\ }\bibfield  {title} {\bibinfo {title} {Cosmic star-formation history},\ }\href@noop {} {\bibfield  {journal} {\bibinfo  {journal} {Annual Review of Astronomy and Astrophysics}\ }\textbf {\bibinfo {volume} {52}},\ \bibinfo {pages} {415} (\bibinfo {year} {2014})}\BibitemShut {NoStop}%
\bibitem [{\citenamefont {Hodge}\ and\ \citenamefont {da~Cunha}(2020)}]{hodge2020high}%
  \BibitemOpen
  \bibfield  {author} {\bibinfo {author} {\bibfnamefont {J.~A.}\ \bibnamefont {Hodge}}\ and\ \bibinfo {author} {\bibfnamefont {E.}~\bibnamefont {da~Cunha}},\ }\bibfield  {title} {\bibinfo {title} {High-redshift star formation in the atacama large millimetre/submillimetre array era},\ }\href@noop {} {\bibfield  {journal} {\bibinfo  {journal} {Royal Society Open Science}\ }\textbf {\bibinfo {volume} {7}},\ \bibinfo {pages} {200556} (\bibinfo {year} {2020})}\BibitemShut {NoStop}%
\bibitem [{\citenamefont {Cowie}\ \emph {et~al.}(1996)\citenamefont {Cowie}, \citenamefont {Songaila}, \citenamefont {Hu},\ and\ \citenamefont {Cohen}}]{cowie1996new}%
  \BibitemOpen
  \bibfield  {author} {\bibinfo {author} {\bibfnamefont {L.~L.}\ \bibnamefont {Cowie}}, \bibinfo {author} {\bibfnamefont {A.}~\bibnamefont {Songaila}}, \bibinfo {author} {\bibfnamefont {E.~M.}\ \bibnamefont {Hu}},\ and\ \bibinfo {author} {\bibfnamefont {J.}~\bibnamefont {Cohen}},\ }\bibfield  {title} {\bibinfo {title} {New insight on galaxy formation and evolution from keck spectroscopy of the hawaii deep fields},\ }\href@noop {} {\bibfield  {journal} {\bibinfo  {journal} {arXiv preprint astro-ph/9606079}\ } (\bibinfo {year} {1996})}\BibitemShut {NoStop}%
\bibitem [{\citenamefont {Firmani}\ \emph {et~al.}(2010)\citenamefont {Firmani}, \citenamefont {Avila-Reese},\ and\ \citenamefont {Rodr{\'\i}guez-Puebla}}]{firmani2010can}%
  \BibitemOpen
  \bibfield  {author} {\bibinfo {author} {\bibfnamefont {C.}~\bibnamefont {Firmani}}, \bibinfo {author} {\bibfnamefont {V.}~\bibnamefont {Avila-Reese}},\ and\ \bibinfo {author} {\bibfnamefont {A.}~\bibnamefont {Rodr{\'\i}guez-Puebla}},\ }\bibfield  {title} {\bibinfo {title} {Can galaxy outflows and re-accretion produce a downsizing in the specific star-formation rate of late-type galaxies?},\ }\href@noop {} {\bibfield  {journal} {\bibinfo  {journal} {Monthly Notices of the Royal Astronomical Society}\ }\textbf {\bibinfo {volume} {404}},\ \bibinfo {pages} {1100} (\bibinfo {year} {2010})}\BibitemShut {NoStop}%
\bibitem [{\citenamefont {Villar}\ \emph {et~al.}(2011)\citenamefont {Villar}, \citenamefont {Gallego}, \citenamefont {P{\'e}rez-Gonz{\'a}lez}, \citenamefont {Barro}, \citenamefont {Zamorano}, \citenamefont {Noeske},\ and\ \citenamefont {Koo}}]{villar2011star}%
  \BibitemOpen
  \bibfield  {author} {\bibinfo {author} {\bibfnamefont {V.}~\bibnamefont {Villar}}, \bibinfo {author} {\bibfnamefont {J.}~\bibnamefont {Gallego}}, \bibinfo {author} {\bibfnamefont {P.~G.}\ \bibnamefont {P{\'e}rez-Gonz{\'a}lez}}, \bibinfo {author} {\bibfnamefont {G.}~\bibnamefont {Barro}}, \bibinfo {author} {\bibfnamefont {J.}~\bibnamefont {Zamorano}}, \bibinfo {author} {\bibfnamefont {K.}~\bibnamefont {Noeske}},\ and\ \bibinfo {author} {\bibfnamefont {D.~C.}\ \bibnamefont {Koo}},\ }\bibfield  {title} {\bibinfo {title} {Star formation rates and stellar masses of h$\alpha$ selected star-forming galaxies at z= 0.84: a quantification of the downsizing},\ }\href@noop {} {\bibfield  {journal} {\bibinfo  {journal} {The Astrophysical Journal}\ }\textbf {\bibinfo {volume} {740}},\ \bibinfo {pages} {47} (\bibinfo {year} {2011})}\BibitemShut {NoStop}%
\bibitem [{\citenamefont {De~Lucia}\ and\ \citenamefont {Blaizot}(2007)}]{de2007hierarchical}%
  \BibitemOpen
  \bibfield  {author} {\bibinfo {author} {\bibfnamefont {G.}~\bibnamefont {De~Lucia}}\ and\ \bibinfo {author} {\bibfnamefont {J.}~\bibnamefont {Blaizot}},\ }\bibfield  {title} {\bibinfo {title} {The hierarchical formation of the brightest cluster galaxies},\ }\href@noop {} {\bibfield  {journal} {\bibinfo  {journal} {Monthly Notices of the Royal Astronomical Society}\ }\textbf {\bibinfo {volume} {375}},\ \bibinfo {pages} {2} (\bibinfo {year} {2007})}\BibitemShut {NoStop}%
\bibitem [{\citenamefont {Firmani}\ and\ \citenamefont {Avila-Reese}(2010)}]{firmani2010galaxy}%
  \BibitemOpen
  \bibfield  {author} {\bibinfo {author} {\bibfnamefont {C.}~\bibnamefont {Firmani}}\ and\ \bibinfo {author} {\bibfnamefont {V.}~\bibnamefont {Avila-Reese}},\ }\bibfield  {title} {\bibinfo {title} {Galaxy downsizing evidenced by hybrid evolutionary tracks},\ }\href@noop {} {\bibfield  {journal} {\bibinfo  {journal} {The Astrophysical Journal}\ }\textbf {\bibinfo {volume} {723}},\ \bibinfo {pages} {755} (\bibinfo {year} {2010})}\BibitemShut {NoStop}%
\bibitem [{\citenamefont {Conselice}\ \emph {et~al.}(2018)\citenamefont {Conselice}, \citenamefont {Twite}, \citenamefont {Palamara},\ and\ \citenamefont {Hartley}}]{conselice2018halo}%
  \BibitemOpen
  \bibfield  {author} {\bibinfo {author} {\bibfnamefont {C.~J.}\ \bibnamefont {Conselice}}, \bibinfo {author} {\bibfnamefont {J.~W.}\ \bibnamefont {Twite}}, \bibinfo {author} {\bibfnamefont {D.~P.}\ \bibnamefont {Palamara}},\ and\ \bibinfo {author} {\bibfnamefont {W.}~\bibnamefont {Hartley}},\ }\bibfield  {title} {\bibinfo {title} {The halo masses of galaxies to z~ 3: A hybrid observational and theoretical approach},\ }\href@noop {} {\bibfield  {journal} {\bibinfo  {journal} {The Astrophysical Journal}\ }\textbf {\bibinfo {volume} {863}},\ \bibinfo {pages} {42} (\bibinfo {year} {2018})}\BibitemShut {NoStop}%
\bibitem [{\citenamefont {Somerville}\ \emph {et~al.}(2018)\citenamefont {Somerville}, \citenamefont {Behroozi}, \citenamefont {Pandya}, \citenamefont {Dekel}, \citenamefont {Faber}, \citenamefont {Fontana}, \citenamefont {Koekemoer}, \citenamefont {Koo}, \citenamefont {P{\'e}rez-Gonz{\'a}lez}, \citenamefont {Primack} \emph {et~al.}}]{somerville2018relationship}%
  \BibitemOpen
  \bibfield  {author} {\bibinfo {author} {\bibfnamefont {R.~S.}\ \bibnamefont {Somerville}}, \bibinfo {author} {\bibfnamefont {P.}~\bibnamefont {Behroozi}}, \bibinfo {author} {\bibfnamefont {V.}~\bibnamefont {Pandya}}, \bibinfo {author} {\bibfnamefont {A.}~\bibnamefont {Dekel}}, \bibinfo {author} {\bibfnamefont {S.}~\bibnamefont {Faber}}, \bibinfo {author} {\bibfnamefont {A.}~\bibnamefont {Fontana}}, \bibinfo {author} {\bibfnamefont {A.~M.}\ \bibnamefont {Koekemoer}}, \bibinfo {author} {\bibfnamefont {D.~C.}\ \bibnamefont {Koo}}, \bibinfo {author} {\bibfnamefont {P.}~\bibnamefont {P{\'e}rez-Gonz{\'a}lez}}, \bibinfo {author} {\bibfnamefont {J.~R.}\ \bibnamefont {Primack}}, \emph {et~al.},\ }\bibfield  {title} {\bibinfo {title} {The relationship between galaxy and dark matter halo size from z~ 3 to the present},\ }\href@noop {} {\bibfield  {journal} {\bibinfo  {journal} {Monthly Notices of the Royal Astronomical Society}\ }\textbf {\bibinfo {volume} {473}},\ \bibinfo {pages} {2714} (\bibinfo {year}
  {2018})}\BibitemShut {NoStop}%
\bibitem [{\citenamefont {Mutch}\ \emph {et~al.}(2013)\citenamefont {Mutch}, \citenamefont {Croton},\ and\ \citenamefont {Poole}}]{mutch2013simplest}%
  \BibitemOpen
  \bibfield  {author} {\bibinfo {author} {\bibfnamefont {S.~J.}\ \bibnamefont {Mutch}}, \bibinfo {author} {\bibfnamefont {D.~J.}\ \bibnamefont {Croton}},\ and\ \bibinfo {author} {\bibfnamefont {G.~B.}\ \bibnamefont {Poole}},\ }\bibfield  {title} {\bibinfo {title} {The simplest model of galaxy formation--i. a formation history model of galaxy stellar mass growth},\ }\href@noop {} {\bibfield  {journal} {\bibinfo  {journal} {Monthly Notices of the Royal Astronomical Society}\ }\textbf {\bibinfo {volume} {435}},\ \bibinfo {pages} {2445} (\bibinfo {year} {2013})}\BibitemShut {NoStop}%
\bibitem [{\citenamefont {Kroupa}\ \emph {et~al.}(2020)\citenamefont {Kroupa}, \citenamefont {Subr}, \citenamefont {Jerabkova},\ and\ \citenamefont {Wang}}]{kroupa2020very}%
  \BibitemOpen
  \bibfield  {author} {\bibinfo {author} {\bibfnamefont {P.}~\bibnamefont {Kroupa}}, \bibinfo {author} {\bibfnamefont {L.}~\bibnamefont {Subr}}, \bibinfo {author} {\bibfnamefont {T.}~\bibnamefont {Jerabkova}},\ and\ \bibinfo {author} {\bibfnamefont {L.}~\bibnamefont {Wang}},\ }\bibfield  {title} {\bibinfo {title} {Very high redshift quasars and the rapid emergence of supermassive black holes},\ }\href@noop {} {\bibfield  {journal} {\bibinfo  {journal} {Monthly Notices of the Royal Astronomical Society}\ }\textbf {\bibinfo {volume} {498}},\ \bibinfo {pages} {5652} (\bibinfo {year} {2020})}\BibitemShut {NoStop}%
\bibitem [{\citenamefont {Fontanot}\ \emph {et~al.}(2009)\citenamefont {Fontanot}, \citenamefont {De~Lucia}, \citenamefont {Monaco}, \citenamefont {Somerville},\ and\ \citenamefont {Santini}}]{fontanot2009many}%
  \BibitemOpen
  \bibfield  {author} {\bibinfo {author} {\bibfnamefont {F.}~\bibnamefont {Fontanot}}, \bibinfo {author} {\bibfnamefont {G.}~\bibnamefont {De~Lucia}}, \bibinfo {author} {\bibfnamefont {P.}~\bibnamefont {Monaco}}, \bibinfo {author} {\bibfnamefont {R.~S.}\ \bibnamefont {Somerville}},\ and\ \bibinfo {author} {\bibfnamefont {P.}~\bibnamefont {Santini}},\ }\bibfield  {title} {\bibinfo {title} {The many manifestations of downsizing: hierarchical galaxy formation models confront observations},\ }\href@noop {} {\bibfield  {journal} {\bibinfo  {journal} {Monthly Notices of the Royal Astronomical Society}\ }\textbf {\bibinfo {volume} {397}},\ \bibinfo {pages} {1776} (\bibinfo {year} {2009})}\BibitemShut {NoStop}%
\bibitem [{\citenamefont {Li}\ \emph {et~al.}(2012)\citenamefont {Li}, \citenamefont {Wang},\ and\ \citenamefont {Ho}}]{li2012cosmological}%
  \BibitemOpen
  \bibfield  {author} {\bibinfo {author} {\bibfnamefont {Y.-R.}\ \bibnamefont {Li}}, \bibinfo {author} {\bibfnamefont {J.-M.}\ \bibnamefont {Wang}},\ and\ \bibinfo {author} {\bibfnamefont {L.~C.}\ \bibnamefont {Ho}},\ }\bibfield  {title} {\bibinfo {title} {Cosmological evolution of supermassive black holes. ii. evidence for downsizing of spin evolution},\ }\href@noop {} {\bibfield  {journal} {\bibinfo  {journal} {The Astrophysical Journal}\ }\textbf {\bibinfo {volume} {749}},\ \bibinfo {pages} {187} (\bibinfo {year} {2012})}\BibitemShut {NoStop}%
\bibitem [{\citenamefont {Izumi}(2018)}]{izumi2018supermassive}%
  \BibitemOpen
  \bibfield  {author} {\bibinfo {author} {\bibfnamefont {T.}~\bibnamefont {Izumi}},\ }\bibfield  {title} {\bibinfo {title} {Supermassive black holes with higher eddington ratios preferentially form in gas-rich galaxies},\ }\href@noop {} {\bibfield  {journal} {\bibinfo  {journal} {Publications of the Astronomical Society of Japan}\ }\textbf {\bibinfo {volume} {70}},\ \bibinfo {pages} {L2} (\bibinfo {year} {2018})}\BibitemShut {NoStop}%
\bibitem [{\citenamefont {Neistein}\ \emph {et~al.}(2006)\citenamefont {Neistein}, \citenamefont {Van Den~Bosch},\ and\ \citenamefont {Dekel}}]{neistein2006natural}%
  \BibitemOpen
  \bibfield  {author} {\bibinfo {author} {\bibfnamefont {E.}~\bibnamefont {Neistein}}, \bibinfo {author} {\bibfnamefont {F.~C.}\ \bibnamefont {Van Den~Bosch}},\ and\ \bibinfo {author} {\bibfnamefont {A.}~\bibnamefont {Dekel}},\ }\bibfield  {title} {\bibinfo {title} {Natural downsizing in hierarchical galaxy formation},\ }\href@noop {} {\bibfield  {journal} {\bibinfo  {journal} {Monthly Notices of the Royal Astronomical Society}\ }\textbf {\bibinfo {volume} {372}},\ \bibinfo {pages} {933} (\bibinfo {year} {2006})}\BibitemShut {NoStop}%
\bibitem [{\citenamefont {Tabasi}\ \emph {et~al.}(2023)\citenamefont {Tabasi}, \citenamefont {Salmani}, \citenamefont {Khaliliyan},\ and\ \citenamefont {Firouzjaee}}]{tabasi2023modeling}%
  \BibitemOpen
  \bibfield  {author} {\bibinfo {author} {\bibfnamefont {S.~S.}\ \bibnamefont {Tabasi}}, \bibinfo {author} {\bibfnamefont {R.~V.}\ \bibnamefont {Salmani}}, \bibinfo {author} {\bibfnamefont {P.}~\bibnamefont {Khaliliyan}},\ and\ \bibinfo {author} {\bibfnamefont {J.~T.}\ \bibnamefont {Firouzjaee}},\ }\bibfield  {title} {\bibinfo {title} {Modeling the central supermassive black holes mass of quasars via lstm approach},\ }\href@noop {} {\bibfield  {journal} {\bibinfo  {journal} {arXiv preprint arXiv:2301.01459}\ } (\bibinfo {year} {2023})}\BibitemShut {NoStop}%
\end{thebibliography}%

\LTcapwidth=\textwidth
\clearpage
\onecolumngrid
\setlength{\tabcolsep}{0.5cm}
\begin{longtable}{|c|c|c|c|c|c|c|}
\caption{
This table represents the flux- and volume-limited QUOTAS+QuasarNET dataset, which has been used in this work.}
\label{tab:qn-tab-fin}\\
\hline
\textbf{Object} & \textbf{$z^a$} & \textbf{Log $M_{BH}^b$} & \textbf{Log $L_{bol}^c$} & \textbf{Log $\lambda_{edd}^d$} & \textbf{Log $L_{opt}^e$} & \textbf{Log $\epsilon_{opt}^f$} \\ \hline
\endfirsthead

\multicolumn{7}{c}
{{\bfseries Table \thetable\ (Continued)}} \\
\hline

\endhead

SDSS J0005-0006 & 5.844 & 8.0 & 46.67 & 0.556 & 45.958 & -1.975 \\ \hline
SDSS J1411+1217 & 5.854 & 8.954 & 47.182 & 0.114 & 46.47 & -1.387 \\ \hline
SDSS J1411+1217 & 5.903 & 9.204 & 47.272 & -0.046 & 46.56 & -1.21 \\ \hline
SDSS J1306+0356 & 6.017 & 9.23 & 47.123 & -0.222 & 46.411 & -1.11 \\ \hline
SDSS J1306+0356 & 6.018 & 9.462 & 47.053 & -0.523 & 46.342 & -0.869 \\ \hline
SDSS J1630+4012 & 6.058 & 9.23 & 47.043 & -0.301 & 46.332 & -1.07 \\ \hline
SDSS J0303-0019 & 6.079 & 8.699 & 46.591 & -0.222 & 45.879 & -1.312 \\ \hline
SDSS J1623+3112 & 6.211 & 9.342 & 47.155 & -0.301 & 46.444 & -1.027 \\ \hline
SDSS J1030+0524 & 6.302 & 9.301 & 47.114 & -0.301 & 46.402 & -1.043 \\ \hline
SDSS J1030+0524 & 6.299 & 9.38 & 47.193 & -0.301 & 46.481 & -1.013 \\ \hline
SDSS J0842+1218 & 6.069 & 9.462 & 47.178 & -0.398 & 46.467 & -0.932 \\ \hline
SDSS J0005-0006 & 5.844 & 7.778 & 46.67 & 0.851 & 45.958 & -2.172 \\ \hline
SDSS J1411+1217 & 5.854 & 8.699 & 47.182 & 0.342 & 46.47 & -1.614 \\ \hline
SDSS J1411+1217 & 5.903 & 9.0 & 47.272 & 0.176 & 46.56 & -1.392 \\ \hline
SDSS J1306+0356 & 6.018 & 9.23 & 47.053 & -0.222 & 46.342 & -1.075 \\ \hline
SDSS J1630+4012 & 6.058 & 8.954 & 47.043 & -0.097 & 46.332 & -1.316 \\ \hline
SDSS J0303-0019 & 6.079 & 8.477 & 46.591 & 0.114 & 45.879 & -1.51 \\ \hline
SDSS J1623+3112 & 6.211 & 9.079 & 47.155 & -0.097 & 46.444 & -1.262 \\ \hline
SDSS J1030+0524 & 6.302 & 9.041 & 47.114 & -0.046 & 46.402 & -1.274 \\ \hline
SDSS J1030+0524 & 6.299 & 9.146 & 47.193 & -0.097 & 46.481 & -1.221 \\ \hline
SDSS J0842+1218 & 6.069 & 9.23 & 47.178 & -0.155 & 46.467 & -1.139 \\ \hline
SDSS J0002+2550 & 5.818 & 9.684 & 47.296 & -0.502 & 46.584 & -1.105 \\ \hline
SDSS J000825.77-062604.6 & 5.929 & 9.157 & 47.028 & 0.04 & 46.272 & -1.128 \\ \hline
SDSS J002806.57+045725.3 & 5.982 & 9.59 & 47.004 & -0.541 & 46.292 & -1.067 \\ \hline
SDSS J003311.40-012524.9 & 5.978 & 8.452 & 46.5 & -0.065 & 45.788 & -1.574 \\ \hline
CFHQS J005006+344522 & 6.251 & 9.639 & 47.257 & -0.639 & 46.494 & -0.815 \\ \hline
CFHQS J022122-080251 & 6.161 & 9.207 & 46.663 & -0.658 & 44.976 & -0.897 \\ \hline
SDSS J0353+0104 & 6.057 & 9.066 & 47.107 & -0.25 & 46.217 & -1.249 \\ \hline
SDSS J0810+5105 & 5.805 & 9.537 & 47.226 & -0.426 & 46.514 & -1.013 \\ \hline
SDSS J083525.76+321752.6 & 5.902 & 8.948 & 46.481 & -0.58 & 45.77 & -1.301 \\ \hline
SDSS J0840+5624 & 5.816 & 9.223 & 46.707 & -0.63 & 45.995 & -1.179 \\ \hline
SDSS J0841+2905 & 5.954 & 9.468 & 47.036 & -0.234 & 46.324 & -0.937 \\ \hline
SDSS J0842+1218 & 6.069 & 9.595 & 47.238 & -0.418 & 46.526 & -0.983 \\ \hline
SDSS J1143+3808 & 5.8 & 9.654 & 47.103 & -0.665 & 46.391 & -0.969 \\ \hline
SDSS J114803.28+070208.3 & 6.344 & 9.499 & 47.136 & -0.323 & 46.408 & -0.895 \\ \hline
SDSS J120737.43+063010.1 & 6.028 & 10.038 & 46.964 & -0.922 & 46.253 & -0.585 \\ \hline
SDSS J124340.81+252923.9 & 5.842 & 9.757 & 47.23 & -0.641 & 46.326 & -0.696 \\ \hline
SDSS J1250+3130 & 6.138 & 9.159 & 47.117 & -0.059 & 46.405 & -1.242 \\ \hline
SDSS J1257+6349 & 5.992 & 9.521 & 46.808 & -0.642 & 46.096 & -0.921 \\ \hline
SDSS J1335+3533 & 5.87 & 9.777 & 47.194 & -0.697 & 46.483 & -0.843 \\ \hline
SDSS J142516.30+325409.0 & 5.862 & 9.433 & 46.981 & -0.566 & 46.269 & -1.176 \\ \hline
SDSS J142738.59+331242.0 & 6.118 & 8.838 & 47.056 & -0.038 & 46.151 & -1.426 \\ \hline
CFHQS J142952.17+554717.6W & 6.119 & 9.282 & 46.926 & -0.47 & 46.214 & -1.091 \\ \hline
SDSS J1436+5007 & 5.809 & 9.665 & 47.133 & -0.646 & 46.421 & -0.865 \\ \hline
SDSS J1545+6028 & 5.794 & 9.213 & 46.677 & -0.65 & 45.965 & -1.087 \\ \hline
SDSS J1602+4228 & 6.083 & 9.553 & 47.269 & -0.282 & 46.557 & -0.965 \\ \hline
SDSS J160937.27+304147.7 & 6.146 & 9.623 & 46.784 & -0.89 & 46.073 & -0.959 \\ \hline
SDSS J1623+3112 & 6.254 & 9.39 & 47.12 & -0.36 & 46.409 & -1.082 \\ \hline
SDSS J1630+4012 & 6.066 & 9.349 & 46.85 & -0.665 & 46.138 & -1.03 \\ \hline
SDSS J2307+0031 & 5.9 & 8.479 & 46.808 & -0.092 & 45.792 & -1.619 \\ \hline
CFHQS J232914-040324 & 5.883 & 8.808 & 46.363 & -0.559 & 45.637 & -1.099 \\ \hline
PSO J060.5529+24.8567 & 6.17 & 9.324 & 47.149 & -0.24 & 46.437 & -1.159 \\ \hline
PSO J210.4472+27.8263 & 6.166 & 9.541 & 46.962 & -0.638 & 46.25 & -0.839 \\ \hline
PSO J228.6871+21.2388 & 5.893 & 9.081 & 46.732 & -0.464 & 46.02 & -1.301 \\ \hline
PSO J333.9859+26.1081 & 6.027 & 9.464 & 46.899 & -0.288 & 46.187 & -0.868 \\ \hline
SDSS J232908.28-030158.8 & 6.417 & 8.398 & 46.563 & 0.114 & 45.851 & -1.566 \\ \hline
CFHQS J005006+344522 & 6.253 & 9.415 & 47.246 & -0.208 & 46.534 & -1.009 \\ \hline
CFHQS J022122-080251 & 6.161 & 8.845 & 46.411 & -0.481 & 45.699 & -1.09 \\ \hline
CFHQS J222901+145709 & 6.152 & 8.079 & 46.49 & 0.38 & 45.778 & -1.812 \\ \hline
CFHQS J210054-171522 & 6.087 & 8.973 & 46.703 & -0.31 & 45.991 & -1.125 \\ \hline
SDSS J164121.64+375520.5 & 6.047 & 8.38 & 46.791 & 0.362 & 46.079 & -1.698 \\ \hline
CFHQS J005502+014618 & 5.983 & 8.38 & 46.511 & 0.079 & 45.799 & -1.555 \\ \hline
VIKING J010953.13-304726.3 & 6.791 & 9.124 & 46.708 & -0.538 & 46.0 & -1.0 \\ \hline
VIKING J030516.92-315056.0 & 6.614 & 8.954 & 46.875 & -0.194 & 46.176 & -1.249 \\ \hline
PSO J167.6415-13.4960 & 6.515 & 8.477 & 46.672 & 0.086 & 45.954 & -1.542 \\ \hline
ULAS J1120+0641 & 7.084 & 9.393 & 47.262 & -0.244 & 46.556 & -1.046 \\ \hline
HSC J1205-0000 & 6.73 & 9.672 & 46.556 & -1.222 & 45.845 & -0.429 \\ \hline
PSO J231.6576-20.8335 & 6.586 & 9.484 & 47.276 & -0.319 & 46.568 & -0.968 \\ \hline
PSO J247.2970+24.1277 & 6.476 & 8.716 & 47.248 & 0.415 & 46.531 & -1.625 \\ \hline
PSO J323.1382+12.2986 & 6.588 & 9.143 & 46.908 & -0.357 & 46.204 & -1.09 \\ \hline
VIKING J234833.34-305410.0 & 6.902 & 9.297 & 46.633 & -0.77 & 45.954 & -0.851 \\ \hline
ULAS J1120+0641 & 7.1 & 9.209 & 47.174 & -0.319 & 46.462 & -1.397 \\ \hline
VIKING J234833.34-305410.0 & 6.9 & 9.322 & 46.685 & -0.745 & 45.973 & -0.805 \\ \hline
VIKING J010953.13-304726.3 & 6.7 & 9.031 & 46.741 & -0.62 & 46.029 & -1.264 \\ \hline
VIKING J030516.92-315056.0 & 6.6 & 9.028 & 46.932 & -0.167 & 46.22 & -1.382 \\ \hline

\multicolumn{7}{c}
{\parbox{\dimexpr\textwidth-2\tabcolsep}
{

\vspace{+0.2cm}
\hspace*{-0.5cm}
\textbf{Notes:} We report accretion efficiency using optical luminosity, bolometric luminosity, and SMBH mass utilizing Eq. (\ref{epsilon_opt}).

\hspace{-0.5cm}
$^a$ Redshift.

\hspace{-0.5cm}
$^b$ Logarithmic mass of the SMBH in units of $M_\odot$.

\hspace{-0.5cm}
$^c$ Logarithmic bolometric luminosity in units of erg $s^{-1}$.

\hspace{-0.5cm}
$^d$ Logarithmic eddington ratio in units of erg $s^{-1}$.

\hspace{-0.5cm}
$^e$ Logarithmic optical luminosity in units of erg $s^{-1}$.

\hspace{-0.5cm}
$^f$ Logarithmic accretion efficiency.}}
\end{longtable}
\clearpage
\twocolumngrid

\clearpage
\onecolumngrid
\LTcapwidth=\textwidth
\setlength{\tabcolsep}{0.5cm}
\begin{longtable}{|c|c|c|c|c|c|c|c|}
\caption{This table represents the flux- and volume-limited DL11 dataset for low redshift PG quasars.}
\label{D11_tab}
\\
\hline
\textbf{Object} & \textbf{$z^a$} & \textbf{Log $M_{BH}^b$} & \textbf{Log $L_{bol}^c$} & \textbf{Log $\lambda_{edd}^d$} & \textbf{Log $L_{opt}^e$} & \textbf{Log $\dot M^f$} & \textbf{ Log $\epsilon^g$} \\ \hline
\endfirsthead

\multicolumn{8}{c}
{{\bfseries Table \thetable\ (Continued)}} \\
\hline

\endhead
PG0003 + 158 & 0.45 & 9.055 & 46.92 & -0.358 & 45.87 & 0.79 & -0.52 \\ \hline
PG0003 + 199 & 0.026 & 7.22 & 45.13 & -0.342 & 43.91 & -0.06 & -1.47 \\ \hline
PG0007 + 106 & 0.089 & 8.561 & 45.52 & -0.972 & 44.55 & -0.42 & -0.72 \\ \hline
PG0026 + 129 & 0.145 & 7.833 & 46.15 & 0.053 & 44.99 & 0.8 & -1.3 \\ \hline
PG0043 + 039 & 0.386 & 8.952 & 45.98 & -0.648 & 45.47 & 0.36 & -1.04 \\ \hline
PG0050 + 124 & 0.059 & 7.238 & 45.12 & 0.162 & 44.41 & 0.58 & -2.12 \\ \hline
PG0052 + 251 & 0.154 & 8.745 & 46.06 & -0.822 & 45.0 & -0.04 & -0.55 \\ \hline
PG0157 + 001 & 0.163 & 8.006 & 45.93 & -0.261 & 45.02 & 0.59 & -1.31 \\ \hline
PG0804 + 761 & 0.1 & 8.352 & 45.82 & -0.3 & 44.79 & 0.2 & -1.03 \\ \hline
PG0838 + 770 & 0.132 & 7.992 & 45.22 & -0.493 & 44.56 & 0.08 & -1.51 \\ \hline
PG0844 + 349 & 0.064 & 7.759 & 45.4 & -0.479 & 44.31 & -0.01 & -1.24 \\ \hline
PG0921 + 525 & 0.035 & 7.206 & 44.47 & -0.802 & 43.56 & -0.55 & -1.64 \\ \hline
PG0923 + 129 & 0.029 & 7.233 & 44.53 & -0.665 & 43.58 & -0.49 & -1.63 \\ \hline
PG0923 + 201 & 0.193 & 9.094 & 45.68 & -1.134 & 44.81 & -0.47 & -0.51 \\ \hline
PG0947 + 396 & 0.206 & 8.53 & 46.2 & -0.909 & 45.2 & 0.19 & -0.65 \\ \hline
PG1001 + 054 & 0.161 & 7.645 & 45.36 & -0.02 & 44.69 & 0.59 & -1.88 \\ \hline
PG1011 - 040 & 0.058 & 7.19 & 45.02 & -0.146 & 44.08 & 0.17 & -1.81 \\ \hline
PG1012 + 008 & 0.186 & 8.0 & 45.53 & 0.0 & 44.95 & 0.46 & -1.59 \\ \hline
PG1022 + 519 & 0.045 & 6.94 & 45.1 & -0.6 & 43.56 & -0.36 & -1.19 \\ \hline
PG1048 - 090 & 0.346 & 9.022 & 46.57 & -0.679 & 45.45 & 0.3 & -0.38 \\ \hline
PG1048 + 342 & 0.167 & 8.241 & 45.7 & -0.687 & 44.74 & 0.02 & -0.98 \\ \hline
PG1049 - 006 & 0.36 & 8.989 & 46.29 & -0.63 & 45.46 & 0.34 & -0.71 \\ \hline
PG1100 + 772 & 0.312 & 9.112 & 46.61 & -0.749 & 45.51 & 0.29 & -0.34 \\ \hline
PG1103 - 006 & 0.423 & 9.132 & 46.19 & -0.737 & 45.43 & 0.21 & -0.68 \\ \hline
PG1114 + 445 & 0.144 & 8.415 & 45.92 & -0.927 & 44.75 & -0.16 & -0.58 \\ \hline
PG1115 + 407 & 0.154 & 7.505 & 45.59 & -0.139 & 44.58 & 0.49 & -1.56 \\ \hline
PG1116 + 215 & 0.176 & 8.425 & 46.27 & -0.139 & 45.31 & 0.69 & -1.08 \\ \hline
PG1119 + 120 & 0.05 & 7.28 & 45.18 & -0.462 & 44.01 & -0.06 & -1.42 \\ \hline
PG1121 + 422 & 0.225 & 7.856 & 45.87 & -0.232 & 44.8 & 0.48 & -1.27 \\ \hline
PG1126 - 041 & 0.06 & 7.598 & 45.16 & -0.434 & 44.19 & -0.02 & -1.47 \\ \hline
PG1149 - 110 & 0.049 & 7.729 & 44.75 & -0.916 & 43.79 & -0.66 & -1.24 \\ \hline
PG1151 + 117 & 0.176 & 8.435 & 45.43 & -0.801 & 44.65 & -0.2 & -1.02 \\ \hline
PG1202 + 281 & 0.165 & 8.462 & 45.39 & -1.053 & 44.58 & -0.38 & -0.89 \\ \hline
PG1211 + 143 & 0.081 & 7.831 & 46.41 & 0.051 & 44.85 & 0.68 & -0.93 \\ \hline
PG1216 + 069 & 0.332 & 9.0 & 46.61 & -1.0 & 45.62 & 0.51 & -0.55 \\ \hline
PG1226 + 023 & 0.158 & 8.876 & 47.09 & -0.012 & 46.03 & 1.18 & -0.74 \\ \hline
PG1229 + 204 & 0.064 & 8.004 & 45.06 & -0.804 & 44.24 & -0.35 & -1.25 \\ \hline
PG1244 + 026 & 0.048 & 6.614 & 44.74 & 0.235 & 43.7 & 0.15 & -2.07 \\ \hline
PG1259 + 593 & 0.477 & 8.738 & 47.04 & -0.085 & 45.79 & 0.99 & -0.61 \\ \hline
PG1302 - 102 & 0.278 & 8.749 & 46.51 & -0.08 & 45.71 & 0.92 & -1.06 \\ \hline
PG1307 + 085 & 0.154 & 8.541 & 45.93 & -0.651 & 44.92 & 0.05 & -0.78 \\ \hline
PG1309 + 355 & 0.182 & 8.155 & 45.63 & -0.421 & 44.95 & 0.37 & -1.4 \\ \hline
PG1310 - 108 & 0.034 & 7.759 & 44.37 & -1.183 & 43.56 & -1.0 & -1.28 \\ \hline
PG1322 + 659 & 0.168 & 8.076 & 45.92 & -0.409 & 44.78 & 0.27 & -1.01 \\ \hline
PG1341 + 258 & 0.086 & 7.878 & 44.94 & -0.756 & 44.13 & -0.37 & -1.35 \\ \hline
PG1351 + 236 & 0.055 & 8.216 & 44.57 & -1.748 & 43.93 & -1.14 & -0.95 \\ \hline
PG1351 + 640 & 0.088 & 8.656 & 45.31 & -1.058 & 44.69 & -0.38 & -0.97 \\ \hline
PG1352 + 183 & 0.151 & 8.299 & 45.72 & -0.629 & 44.65 & -0.06 & -0.88 \\ \hline
PG1402 + 261 & 0.164 & 7.845 & 46.07 & 0.018 & 44.82 & 0.63 & -1.22 \\ \hline
PG1404 + 226 & 0.098 & 6.713 & 45.21 & 0.232 & 44.16 & 0.55 & -2.0 \\ \hline
PG1411 + 442 & 0.09 & 7.874 & 45.06 & -0.535 & 44.45 & 0.02 & -1.61 \\ \hline
PG1415 + 451 & 0.113 & 7.797 & 45.6 & -0.579 & 44.34 & -0.06 & -1.0 \\ \hline
PG1416 - 129 & 0.129 & 9.002 & 45.82 & -0.845 & 44.94 & -0.21 & -0.63 \\ \hline
PG1425 + 267 & 0.364 & 9.317 & 46.35 & -1.28 & 45.55 & 0.07 & -0.38 \\ \hline
PG1426 + 015 & 0.086 & 8.921 & 45.84 & -1.117 & 44.71 & -0.49 & -0.32 \\ \hline
PG1427 + 480 & 0.22 & 7.978 & 45.64 & -0.344 & 44.69 & 0.27 & -1.29 \\ \hline
PG1435 - 067 & 0.129 & 8.3 & 45.6 & -0.412 & 44.9 & 0.26 & -1.32 \\ \hline
PG1440 + 356 & 0.078 & 7.335 & 45.62 & -0.013 & 44.37 & 0.43 & -1.47 \\ \hline
PG1444 + 407 & 0.268 & 8.158 & 46.28 & -0.122 & 45.11 & 0.66 & -1.04 \\ \hline
PG1501 + 106 & 0.036 & 8.482 & 44.9 & -1.172 & 44.18 & -0.79 & -0.97 \\ \hline
PG1512 + 370 & 0.371 & 9.168 & 47.11 & -0.867 & 45.48 & 0.2 & 0.26 \\ \hline
PG1519 + 226 & 0.136 & 7.777 & 45.98 & -0.311 & 44.45 & 0.18 & -0.86 \\ \hline
PG1534 + 580 & 0.03 & 8.047 & 44.49 & -1.565 & 43.63 & -1.24 & -0.92 \\ \hline
PG1535 + 547 & 0.039 & 7.01 & 44.34 & -0.373 & 43.9 & -0.01 & -2.3 \\ \hline
PG1543 + 489 & 0.401 & 7.844 & 46.43 & 0.369 & 45.27 & 1.18 & -1.41 \\ \hline
PG1545 + 210 & 0.264 & 9.0 & 46.14 & -1.0 & 45.29 & 0.01 & -0.53 \\ \hline
PG1552 + 085 & 0.119 & 7.364 & 45.04 & 0.04 & 44.5 & 0.56 & -2.18 \\ \hline
PG1612 + 261 & 0.131 & 7.913 & 45.38 & -0.395 & 44.54 & 0.15 & -1.42 \\ \hline
PG1613 + 658 & 0.129 & 8.953 & 45.89 & -1.457 & 44.75 & -0.59 & -0.17 \\ \hline
PG1617 + 175 & 0.114 & 8.729 & 45.44 & -0.88 & 44.63 & -0.38 & -0.84 \\ \hline
PG1626 + 554 & 0.132 & 8.371 & 45.53 & -0.94 & 44.46 & -0.4 & -0.73 \\ \hline
PG1704 + 608 & 0.372 & 9.198 & 46.67 & -0.772 & 45.65 & 0.38 & -0.36 \\ \hline
PG2112 + 059 & 0.46 & 8.834 & 46.47 & 0.116 & 45.92 & 1.16 & -1.34 \\ \hline
PG2130 + 099 & 0.063 & 7.805 & 45.52 & -0.367 & 44.35 & 0.05 & -1.19 \\ \hline
PG2209 + 184 & 0.07 & 8.601 & 46.02 & -1.353 & 44.11 & -0.98 & 0.34 \\ \hline
PG2214 + 139 & 0.066 & 8.308 & 45.15 & -1.027 & 44.36 & -0.5 & -1.01 \\ \hline
PG2251 + 113 & 0.326 & 8.816 & 46.13 & -0.363 & 45.6 & 0.66 & -1.18 \\ \hline
PG2304 + 042 & 0.043 & 8.32 & 44.49 & -1.633 & 43.67 & -1.35 & -0.81 \\ \hline
PG2308 + 098 & 0.434 & 9.372 & 46.61 & -0.936 & 45.62 & 0.22 & -0.27 \\ \hline
\multicolumn{8}{c}{\parbox{\dimexpr\textwidth-2\tabcolsep}{

\vspace{+0.2cm}
\hspace{-0.5cm}
\textbf{Notes:} We report accretion efficiency using optical luminosity, bolometric luminosity, and SMBH mass.

\hspace{-0.5cm}
$^a$ Redshift.

\hspace{-0.5cm}
$^b$ Logarithmic mass of the SMBH in units of $M_\odot$.

\hspace{-0.5cm}
$^c$ Logarithmic bolometric luminosity in units of erg $s^{-1}$.

\hspace{-0.5cm}
$^d$ Logarithmic eddington ratio in units of erg $s^{-1}$.

\hspace{-0.5cm}
$^e$ Logarithmic optical luminosity in units of erg $s^{-1}$.

\hspace{-0.5cm}
$^f$ Logarithmic accretion rate, in units of $M_\odot yr^{-1}$.

\hspace{-0.5cm}
$^g$ Logarithmic accretion efficiency.}}
\end{longtable}

\clearpage
\twocolumngrid

\clearpage
\onecolumngrid
\LTcapwidth=\textwidth
\setlength{\tabcolsep}{0.6cm}
\begin{longtable}{|c|c|c|c|c|c|}\caption{This table represents the astronomical characteristics of validation objects.}
\label{tab:val_dat}\\
\hline
\textbf{Object} & \textbf{$z^a$} & \textbf{Log $M_{BH}^b$} & \textbf{Log $L_{bol}^c$} & \textbf{Log $L_{opt}^d$} & \textbf{ Log $\epsilon^e$} \\ \hline
\endfirsthead

\multicolumn{6}{c} {{\bfseries Table \thetable\ (Continued)}} \\[12pt]     
\hline

\endhead

J0337-1204 & 3.44 & 9.0 & 46.36 & 46.055 & 0.029 \\ \hline
J0539-2839 & 3.14 & 9.3 & 46.7 & 46.843 & 0.008 \\ \hline
J0733+0456 & 3.01 & 8.7 & 46.6 & 46.034 & 0.029 \\ \hline
J0805+6144 & 3.03 & 9.0 & 46.34 & 46.193 & 0.017 \\ \hline
J0833-0454 & 3.45 & 9.5 & 47.15 & 46.497 & 0.106 \\ \hline
J1354-0206 & 3.72 & 9.0 & 46.78 & 46.151 & 0.054 \\ \hline
J1429+5406 & 3.01 & 9.0 & 46.26 & 45.956 & 0.033 \\ \hline
J1510+5702 & 4.31 & 9.6 & 46.63 & 46.615 & 0.027 \\ \hline
J1635+3629 & 3.6 & 9.5 & 46.3 & 46.145 & 0.052 \\ \hline
BRI 0019-1522 & 4.53 & 9.23 & 47.56 & 46.92 & 0.036 \\ \hline
BR 0103+0032 & 4.44 & 9.0 & 47.63 & 46.99 & 0.021 \\ \hline
PSS J0248+1802 & 4.44 & 9.82 & 47.88 & 47.24 & 0.083 \\ \hline
Q2050-259 & 3.51 & 9.839 & 48.08 & 47.44 & 0.068 \\ \hline
BRI 2237-0607 & 4.56 & 9.462 & 47.91 & 47.27 & 0.039 \\ \hline
PC 1158+4635 & 4.73 & 9.799 & 48.03 & 47.39 & 0.067 \\ \hline
PKS 2126-158 & 3.27 & 10.127 & 48.25 & 47.61 & 0.101 \\ \hline
Q0105-2634 & 3.49 & 10.35 & 48.1 & 47.46 & 0.191 \\ \hline
Q0256-0000 & 3.38 & 9.681 & 47.78 & 47.14 & 0.07 \\ \hline
Q0302-0019 & 3.29 & 9.301 & 47.88 & 47.24 & 0.029 \\ \hline
Q2227-3928 & 3.44 & 9.863 & 47.71 & 47.07 & 0.111 \\ \hline
Q2348-4025 & 3.31 & 9.0 & 47.91 & 47.27 & 0.015 \\ \hline
Q0103-260 & 3.38 & 9.672 & 47.47 & 46.83 & 0.099 \\ \hline
SDSS 0338+0021 & 5.0 & 9.398 & 47.25 & 46.61 & 0.073 \\ \hline
SDSS 1204-0021 & 5.09 & 9.322 & 47.42 & 46.79 & 0.05 \\ \hline
J011521.20+152453.3g & 3.443 & 9.43 & 47.57 & 46.61 & 0.119 \\ \hline
J014214.75+002324.2g & 3.379 & 9.8 & 47.58 & 46.61 & 0.15 \\ \hline
J015741.57-010629.6g & 3.572 & 9.99 & 47.68 & 46.72 & 0.234 \\ \hline
J025021.76-075749.9g & 3.337 & 9.49 & 47.6 & 46.63 & 0.086 \\ \hline
J025905.63+001121.9g & 3.373 & 9.5 & 47.97 & 47.01 & 0.086 \\ \hline
J030341.04-002321.9g & 3.233 & 9.47 & 47.79 & 46.82 & 0.076 \\ \hline
J030449.85-000813.4g & 3.287 & 9.12 & 47.79 & 46.82 & 0.033 \\ \hline
J075303.34+423130.8g & 3.59 & 10.09 & 47.71 & 46.74 & 0.391 \\ \hline
J075819.70+202300.9m & 3.761 & 9.74 & 47.64 & 46.67 & 0.168 \\ \hline
J080430.56+542041.1g & 3.759 & 10.13 & 47.87 & 46.91 & 0.267 \\ \hline
J080819.69+373047.3g & 3.48 & 10.18 & 47.64 & 46.67 & 0.294 \\ \hline
J080956.02+502000.9g & 3.281 & 9.52 & 47.7 & 46.74 & 0.111 \\ \hline
J081855.77+095848.0g & 3.7 & 10.08 & 47.71 & 46.74 & 0.311 \\ \hline
J090033.50+421547.0g & 3.29 & 9.67 & 48.22 & 47.25 & 0.076 \\ \hline
J094202.04+042244.5g & 3.276 & 9.31 & 48.0 & 47.03 & 0.054 \\ \hline
J102325.31+514251.0m & 3.477 & 10.52 & 47.77 & 46.8 & 1.083 \\ \hline
J115954.33+201921.1g & 3.426 & 10.15 & 47.96 & 46.99 & 0.226 \\ \hline
J173352.23+540030.4g & 3.432 & 9.53 & 47.87 & 46.9 & 0.061 \\ \hline
J213023.61+122252.0g & 3.272 & 9.0 & 47.71 & 46.74 & 0.04 \\ \hline
J224956.08+000218.0g & 3.311 & 9.24 & 47.92 & 46.95 & 0.068 \\ \hline
J230301.45-093930.7g & 3.492 & 9.96 & 47.64 & 46.68 & 0.223 \\ \hline
J234625.66-001600.4m & 3.507 & 8.59 & 47.72 & 46.75 & 0.021 \\ \hline
J074521.78+734336.1g & 3.22 & 10.29 & 48.29 & 47.33 & 0.251 \\ \hline
GB6J001115+144608 & 4.96 & 9.53 & 47.979 & 47.326 & 0.043 \\ \hline
GB6J083548+182519 & 4.41 & 8.53 & 46.757 & 46.104 & 0.023 \\ \hline
GB6J083945+511206 & 4.4 & 9.69 & 47.448 & 46.795 & 0.111 \\ \hline
GB6J091825+063722 & 4.22 & 9.13 & 47.355 & 46.702 & 0.039 \\ \hline
GB6J102107+220904 & 4.26 & 8.94 & 46.66 & 46.007 & 0.06 \\ \hline
GB6J102623+254255 & 5.28 & 9.4 & 47.345 & 46.692 & 0.069 \\ \hline
GB6J132512+112338 & 4.42 & 9.17 & 47.221 & 46.568 & 0.05 \\ \hline
GB6J134811+193520 & 4.4 & 9.43 & 46.99 & 46.337 & 0.111 \\ \hline
GB6J141212+062408 & 4.47 & 9.02 & 47.147 & 46.494 & 0.04 \\ \hline
GB6J143023+420450 & 4.71 & 9.22 & 47.28 & 46.627 & 0.052 \\ \hline
GB6J151002+570256 & 4.31 & 8.88 & 47.185 & 46.532 & 0.029 \\ \hline
GB6J153533+025419 & 4.39 & 9.06 & 47.016 & 46.363 & 0.051 \\ \hline
GB6J162956+095959 & 5.0 & 8.73 & 46.967 & 46.314 & 0.027 \\ \hline
GB6J164856+460341 & 5.36 & 8.76 & 46.913 & 46.26 & 0.031 \\ \hline
GB6J171103+383016 & 4.0 & 8.62 & 46.933 & 46.28 & 0.023 \\ \hline
GB6J231449+020146 & 4.11 & 9.15 & 46.96 & 46.307 & 0.065 \\ \hline
GB6J235758+140205 & 4.35 & 8.94 & 46.92 & 46.267 & 0.044 \\ \hline
Q 0019+0107 & 2.131 & 9.978 & 46.422 & 46.35 & 0.089 \\ \hline
Q 0020-0154 & 1.465 & 9.255 & 46.43 & 46.238 & 0.031 \\ \hline
Q 0150-202 & 2.147 & 9.806 & 46.637 & 46.64 & 0.038 \\ \hline
Q 0302-2223 & 1.406 & 9.959 & 46.436 & 46.436 & 0.066 \\ \hline
Q 2116-4439 & 1.504 & 9.301 & 46.265 & 46.117 & 0.035 \\ \hline
Q 2154-2005 & 2.042 & 9.643 & 46.301 & 46.248 & 0.049 \\ \hline
Q 2209-1842 & 2.098 & 9.505 & 46.431 & 46.387 & 0.031 \\ \hline
Q 2221-1759 & 2.23 & 9.716 & 46.394 & 46.438 & 0.036 \\ \hline
Q 2230+0232 & 2.215 & 9.531 & 46.303 & 46.199 & 0.046 \\ \hline
Q 2302+0255 & 1.062 & 9.505 & 46.512 & 46.431 & 0.032 \\ \hline
J022112.62-034252.2 & 5.011 & 9.42 & 47.1 & 46.38 & 0.121 \\ \hline
J021438.15-052024.5 & 3.198 & 8.6 & 46.71 & 45.82 & 0.064 \\ \hline
J022906.04-051428.9 & 3.174 & 8.24 & 46.54 & 45.75 & 0.026 \\ \hline
J021830.60-050125.6 & 3.001 & 8.19 & 46.58 & 45.51 & 0.06 \\ \hline
J023036.81-040936.5 & 2.765 & 9.41 & 46.7 & 46.08 & 0.134 \\ \hline
J021645.89-033140.9 & 2.71 & 9.11 & 46.94 & 46.27 & 0.065 \\ \hline
J021719.46-052305.3 & 2.707 & 9.43 & 46.64 & 45.97 & 0.178 \\ \hline
J021806.16-041640.6 & 2.647 & 8.76 & 46.44 & 45.95 & 0.031 \\ \hline
J021457.22-043011.5 & 2.635 & 9.81 & 47.07 & 46.44 & 0.204 \\ \hline
J022740.55-040251.2 & 2.602 & 9.7 & 47.06 & 46.4 & 0.183 \\ \hline
J022456.47-045517.1 & 2.327 & 8.76 & 46.31 & 45.59 & 0.079 \\ \hline
J022223.71-044125.7 & 2.221 & 7.95 & 46.07 & 45.23 & 0.03 \\ \hline
J021850.08-050954.2 & 2.09 & 8.77 & 46.5 & 45.94 & 0.037 \\ \hline
J022622.16-042221.8 & 2.009 & 9.09 & 46.41 & 45.85 & 0.079 \\ \hline
J022119.24-031442.2 & 1.957 & 7.91 & 46.27 & 45.36 & 0.028 \\ \hline
J022328.89-040134.8 & 1.915 & 8.85 & 46.25 & 45.64 & 0.07 \\ \hline
J021529.28-034713.1 & 1.876 & 9.24 & 46.89 & 46.01 & 0.186 \\ \hline
J022845.57-043350.4 & 1.871 & 9.91 & 47.22 & 46.67 & 0.159 \\ \hline
J021643.70-052236.3 & 1.851 & 8.36 & 46.17 & 45.38 & 0.052 \\ \hline
J022900.21-043752.8 & 1.704 & 9.59 & 46.97 & 46.41 & 0.115 \\ \hline
J022438.84-033306.2 & 1.676 & 8.95 & 46.53 & 45.92 & 0.062 \\ \hline
J021230.34-042802.9 & 1.642 & 8.39 & 46.28 & 45.65 & 0.028 \\ \hline
J022059.49-044917.1 & 1.545 & 9.34 & 46.59 & 45.9 & 0.168 \\ \hline
J021830.59-045622.9 & 1.397 & 9.33 & 46.73 & 46.04 & 0.14 \\ \hline
J021238.53-043404.6 & 1.371 & 9.39 & 46.62 & 46.01 & 0.137 \\ \hline
J022422.27-031054.6 & 1.225 & 9.19 & 46.69 & 46.02 & 0.103 \\ \hline
J022823.19-041223.8 & 1.182 & 9.27 & 46.68 & 46.12 & 0.084 \\ \hline
J023004.07-044232.8 & 1.164 & 8.87 & 46.63 & 45.84 & 0.087 \\ \hline
J020326.22-051020.5 & 0.891 & 8.3 & 46.09 & 45.53 & 0.023 \\ \hline
\multicolumn{6}{c}
{\parbox{\dimexpr\textwidth-2\tabcolsep}{

\vspace{+0.2cm}
\hspace{-0.6cm}
\textbf{Notes:} We report accretion efficiency using optical luminosity, bolometric luminosity, and SMBH mass utilizing Eq. (\ref{epsilon_opt}).

\hspace{-0.6cm}
$^a$ Redshift.

\hspace{-0.6cm}
$^b$ Logarithmic mass of the SMBH in units of $M_\odot$.

\hspace{-0.6cm}
$^c$ Logarithmic bolometric luminosity in units of erg $s^{-1}$.

\hspace{-0.6cm}
$^d$ Logarithmic optical luminosity in units of erg $s^{-1}$.

\hspace{-0.6cm}
$^e$ Logarithmic accretion efficiency.}}
\end{longtable}

\end{document}